\documentclass[11pt,a4paper]{article}

\pdfoutput=1
\usepackage{amsmath,amssymb,mathrsfs}
\usepackage{epsfig,graphicx}
\usepackage{subfigure}
\usepackage{rotating}
\usepackage{cancel}
\usepackage{bm}
\usepackage{color}
\usepackage[dvipsnames]{xcolor}
\usepackage{comment}
\usepackage{cite}
\usepackage{psfrag}
\usepackage[colorlinks=true,linkcolor=blue,urlcolor=magenta,citecolor=blue]{hyperref}
\hypersetup{%
  colorlinks = true,
  linkcolor  = black
}
\usepackage{morefloats}
\usepackage{bm}

\renewcommand\({\left(}
\renewcommand\){\right)}
\renewcommand\[{\left[}
\renewcommand\]{\right]}

\newcommand{\be}{\begin{equation}}
\newcommand{\ee}{\end{equation}}
\newcommand{\bea}{\begin{eqnarray}}
\newcommand{\eea}{\end{eqnarray}}

\definecolor{gbcolor}{rgb}{.8,.3,.1}

\newcommand{\exclude}[1]{}

\def\beq{\begin{equation}}
\def\eeq{\end{equation}}

\newcommand{\eq}[1]{(\ref{#1})}

\providecommand{\aver}[1]{\langle #1 \rangle}

\definecolor{gbcolor}{rgb}{.8,.3,.1}
\definecolor{gbcolor2}{rgb}{.8,.1,.7}

\begin{document}

\thispagestyle{empty}
\enlargethispage{3cm}
\vspace*{-2.5cm}
\begin{minipage}{.45\linewidth}
\begin{flushleft}                          
{DESY 16-184}
\end{flushleft} 
\end{minipage}
\hfill
\begin{minipage}{.45\linewidth}
\begin{flushright}  
IPPP/16/79
\end{flushright} 
\end{minipage}

{\flushright 
\vspace{1.8cm}}

\begin{center}
{\Large \textbf{Standard Model -- Axion -- Seesaw -- Higgs Portal Inflation.\\\vspace{0.13cm} Five problems of particle physics and cosmology\\\vspace{0.1cm} solved in one stroke}}\\
\bigskip
\vspace{1cm}
{ \large \bf Guillermo Ballesteros$^1$, Javier Redondo$^{2,3}$, Andreas Ringwald$^4$, Carlos Tamarit$^5$}\\[7mm]
{$^1$ \it Institut de Physique Th\'eorique, Universit\'e Paris Saclay, CEA, CNRS}\\[-0.7mm]
{\it 91191 Gif-sur-Yvette, France}\\[1.4mm]
{ $^2$  \it Departamento de F\'isica Te\'orica, Universidad de Zaragoza}\\[-0.7mm]
{\it Pedro Cerbuna 12, E-50009, Zaragoza, Spain}\\[1.4mm]
{$^3$ \it Max-Planck-Institut f\"ur Physik, F\"ohringer Ring 6, 80805 M\"unchen, Germany}\\[1.4mm]
{$^4$ \it DESY, Notkestr. 85, 22607 Hamburg, Germany}\\[1.4mm]
{$^5$\it Institute for Particle Physics Phenomenology, Durham University\\ South Road, DH1 3LE, United Kingdom}\\
[17mm]
{\bf Abstract}
\end{center}\vspace{0.3cm}
\noindent \noindent
We present a minimal extension of the Standard Model (SM) providing a consistent picture of particle physics 
from the electroweak scale to the Planck scale and of cosmology from inflation until today.
Three right-handed neutrinos $N_i$,  a new color triplet $Q$ and a complex SM-singlet 
scalar $\sigma$, whose vacuum expectation value $v_\sigma \sim 10^{11}$ GeV breaks lepton number and a 
Peccei-Quinn symmetry simultaneously, are added to the SM.
At low energies, the model reduces to the SM, augmented by seesaw generated neutrino masses and mixing, plus the axion. The latter solves the strong CP problem and accounts for the cold dark matter in the Universe. The inflaton is comprised by a mixture of $\sigma$ and the SM Higgs, and reheating of the Universe after inflation proceeds via the Higgs portal. 
Baryogenesis occurs via thermal leptogenesis. Thus, five fundamental problems of particle physics and cosmology are solved at one stroke in this unified Standard Model - Axion - seesaw - Higgs portal inflation (SMASH) model. It can be probed decisively by upcoming cosmic microwave
background and axion dark matter experiments.

\begin{center}

\vfill\flushleft
\noindent\rule{6cm}{0.4pt}\\
{\small  E-mail addresses: \tt guillermo.ballesteros@cea.fr, jredondo@unizar.es, andreas.ringwald@desy.de, carlos.tamarit@durham.ac.uk}

\end{center}
\bigskip

\newpage

\renewcommand{\baselinestretch}{0.94}\normalsize
\tableofcontents
\renewcommand{\baselinestretch}{1.0}\normalsize

\newpage

\section{Introduction}

The discovery of the Higgs boson at the LHC \cite{Aad:2012tfa,Chatrchyan:2012xdj} marks the completion of the particle content of the Standard Model (SM). 
However, a number of observations in particle physics, astrophysics and cosmology point to the existence of 
particles and interactions beyond the SM. In particular, the SM neutrinos are massless, but tiny masses are required for the explanation of the observed neutrino flavour oscillations.  Moreover, the SM lacks a particle that could explain the non-baryonic dark matter inferred from astrophysics and cosmology.\footnote{There exist some ideas attempting to explain dark matter within the SM, although with severe caveats. For instance the formation of primordial black holes might be possible, but is poorly understood.}  Last, but not least, the SM does not have enough CP violation and its dynamics in the early Universe does not allow for enough departure from equilibrium to explain the matter--anti-matter asymmetry.  

These three problems can be elegantly solved  in a minimal extension of the SM, {without introducing any new physics above the electroweak scale,}  by three sterile (SM-singlet) right-handed neutrinos $N_i$, $i=1,2,3$, having 
Dirac mass terms arising from Yukawa couplings with the Higgs and leptons, as well as explicit Majorana mass terms \cite{Asaka:2005an,Asaka:2005pn}.  Assuming that the Majorana masses are below the electroweak scale, it is possible to explain the tiny masses and the mixing of the (active) SM neutrinos by small Yukawa couplings of Dirac mass terms. The role of dark matter is played by the lightest of the sterile neutrinos (if it has a mass in the keV range  \cite{Dodelson:1993je}). The generation of the observed matter-antimatter asymmetry occurs via a variant of leptogenesis  exploiting neutrino oscillations  \cite{Akhmedov:1998qx}. 
See e.g.\ \cite{Canetti:2012kh} for a detailed description of this model, called $\nu$MSM, and \cite{Schneider:2016uqi,Perez:2016tcq} for the most recent and very stringent constraints on keV mass sterile neutrino dark matter.

It has been pointed out that the SM Higgs could drive inflation in the $\nu$MSM if it is coupled directly to the curvature scalar $R$ with a large coupling $\xi_H\sim 10^4$ \cite{Bezrukov:2007ep}. However, such a large value of $\xi_H$ implies that perturbative unitarity breaks down below the scale of inflation \cite{Barbon:2009ya,Burgess:2009ea}, 
 making the inflationary predictions unreliable. Moreover, Higgs inflation cannot be realised at all if the Higgs quartic coupling $\lambda_H$ runs negative at large (Planckian) field values. 
Although, given the current experimental uncertainties, a definite conclusion cannot yet be drawn, see e.g.~\cite{Alekhin:2012py,Degrassi:2012ry,Bezrukov:2012sa,Buttazzo:2013uya,Bednyakov:2015sca}, the currently favoured central values of the strong gauge coupling and the Higgs and top quark masses imply that $\lambda_H$ becomes negative at a field value corresponding to an energy scale $\Lambda_I \sim 10^{11}$ GeV, much lower than what is required for Higgs inflation, and is thus inconsistent with it.

In addition to these difficulties to account for inflation, the $\nu$MSM also has other drawbacks. In particular, it cannot address the strong CP problem and it requires very small right-handed neutrino Yukawa couplings (though they remain small under radiative corrections). These issues beg the question of whether a slightly less minimal model, with inflation embedded in a consistent way and able to solve the strong CP problem, can achieve a complete picture of particle physics and cosmology up to Planckian energies. 

The fundamental obstructions to have successful inflation within the $\nu$MSM can be surmounted by introducing a new scalar field, which drives inflation in the primordial Universe and, as we will see, also stabilises the potential. This field can be a ``hidden'' complex SM gauge singlet scalar field, $\sigma$,  that is charged under a new global $U(1)$ symmetry which is spontaneously broken. The role of the inflaton can indeed be played by the modulus $|\sigma|\equiv\rho/\sqrt{2}$, or by a combination of it with the modulus of the Higgs. Denoting by $\lambda_\sigma$ the quartic self-coupling of the new hidden scalar, the non-minimal coupling of this field to $R$ that is required to fit the amplitude of the Cosmic Microwave Background (CMB) temperature anisotropies is $\xi_\sigma\sim 10^5\sqrt{\lambda_\sigma}$, which can be of order unity if $\lambda_\sigma\sim 10^{-10}$. 
This eliminates the need for a large $\xi_H$, effectively raising  the unitarity breakdown scale to Planckian values.\footnote{Although $\xi_\sigma\sim 1$ forces the model upon a small parameter, we will see that the rest of them can take much more moderate values, and all of them are radiatively stable.}  Remarkably, the hidden scalar helps to stabilise the effective potential through a (completely generic) Higgs portal interaction with coupling $\lambda_{H\sigma}$. At the electroweak vacuum, where the value of $\rho$ is $v_\sigma$, its mass $m_\rho^2\sim\lambda_\sigma v_\sigma^2$ is large but smaller than $\Lambda_I$. When $\rho$ is integrated out at low energies, {$\lambda_{H\sigma}$} gives a negative contribution to the Higgs quartic coupling. Therefore,  what we measure in the SM as the Higgs quartic is the difference  $\lambda_H-\lambda_{H\sigma}^2/\lambda_\sigma$. At energies above $\sim m_\rho$, the true (and larger!) value of $\lambda_H$ is revealed by integrating $\rho$ in. This threshold  stabilisation mechanism can make the potential of the model absolutely stable up to the Planck mass~\cite{Lebedev:2012zw,EliasMiro:2012ay} and is especially effective for small $\lambda_\sigma$, which is precisely the situation we will be interested in. Given $\lambda_\sigma\sim 10^{-10}$, the required value of the ratio for threshold stabilisation, $\lambda_{H\sigma}^2/\lambda_\sigma\sim 10^{-2}$,  points to a relatively small value of the portal 
coupling, $\lambda_{H\sigma}\sim 10^{-6}$, which is nevertheless sufficient to reheat the Universe, as we will show. As a secondary stabilisation effect,  the portal coupling gives an extra positive contribution to the beta function of the Higgs quartic coupling, see e.g.\ \cite{Gonderinger:2009jp}, which can also aid the threshold mechanism we just discussed, despite the small value of $\lambda_{H\sigma}$. 

A new intermediate scale $v_\sigma$ (between the electroweak scale $v=1/\sqrt{\sqrt{2}G_F}=246$\,GeV and $M_P=1/\sqrt{8\pi G}=2.44\times 10^{18}$\,GeV), of order $v_\sigma\sim\Lambda_I \sim 10^{11}$ GeV, is extremely suggestive. First, {it could act as a} seesaw scale. 
Thus, coupling right-handed neutrinos to $\sigma$ with Yukawa interactions, we give them large Majorana masses $M_i\propto v_\sigma$. This allows to explain the smallness of the SM neutrino masses via the seesaw formula $m_i \propto v^2/v_\sigma$~\cite{Minkowski:1977sc,GellMann:1980vs,Yanagida:1979as,Mohapatra:1979ia,Chikashige:1980ui,Schechter:1981cv,Gelmini:1980re}, without requiring tiny values of the Yukawa couplings.   Moreover, in this framework, the out-of-equilibrium decays of the lightest right-handed neutrino can easily generate the baryon asymmetry of the Universe in the simplest viable leptogenesis scenario~\cite{Fukugita:1986hr}.  

For this choice of Majorana mass scale ($M_i\propto v_\sigma$),  the lightest sterile neutrino is too unstable to be a proper dark matter candidate. However, quite remarkably, the angular part of $\sigma$ allows to turn this potential issue from being a burden into a virtue in a simple way, which has the great added value of solving as well the strong CP problem \cite{Dias:2014osa}. Global lepton number is spontaneously broken by the vacuum expectation value (VEV) of  $\sigma$, and its angular part becomes a Nambu-Goldstone boson (NGB), the majoron (which turns out to be safe from any dangerous phenomenological consequences). To complete the model we just have to introduce a new chiral pair of colour triplets $Q_L, Q_R$ with opposite lepton number and couple them to $\sigma$ with Yukawa interactions. In this way, the lepton number $U(1)$ symmetry becomes colour-anomalous, effectively promoted to a Peccei-Quinn symmetry, and the majoron becomes the QCD axion. The Peccei-Quinn mechanism~\cite{Peccei:1977hh} thus solves the strong CP problem~\cite{Kim:1979if,Shifman:1979if} and gives a small mass to the axion $m_A\simeq 57\times({\rm 10^{11} GeV}/f_A)\, \mu{\rm eV}$~\cite{Weinberg:1977ma,Wilczek:1977pj,diCortona:2015ldu} where the role of the axion decay constant is played here by the VEV of $\sigma$, i.e.\ $f_A = v_\sigma$.
Best of all, for $f_A=v_\sigma\sim 10^{11}$ GeV,  the axion is a perfect cold dark matter 
candidate~\cite{Preskill:1982cy,Abbott:1982af,Dine:1982ah}!

Much of the appeal of this Standard Model - Axion - seesaw - Higgs portal inflation (SMASH) model lies also in its predictivity. 
Demanding hidden scalar  (or a mixture of Higgs and hidden scalar) inflation with values of $\xi_\sigma,\xi_H\lesssim 1$ 
sets the inflationary scale, an upper limit for $\lambda_\sigma$ and a lower limit for the {CMB} tensor to scalar ratio around 0.01. Requiring also absolute stability of the effective potential, implies a minimum value for $\lambda_{H\sigma}\sim 10^{-6}$. {This can lead to efficient reheating, ensuring that} the lepton-Peccei-Quinn symmetry is restored by thermal effects after inflation.  In this case, axion dark matter production happens in the most predictive scenario, where axions are produced by an average vacuum realignment mechanism and the decay of axionic strings and domain walls. In principle, if the efficiency of these mechanisms would be perfectly known, it would lead to a concrete prediction for $f_A$ and thus the axion mass, which is the key for the detection of axion dark matter. 
In fact, the vacuum realignment contribution to the axion dark matter density can be predicted to a remarkable accuracy  \cite{Borsanyi:2016ksw}. Unfortunately, the current  uncertainties in the prediction of the contribution arising form the decay of topological 
defects  are much larger \cite{Kawasaki:2014sqa,Fleury:2016xrz,Moore:2016itg,Fleury:2015aca} and thus prevent us from an accurate axion mass prediction, but we are optimistic that this will improve in the future. 

It is well known that this type of solution of the strong CP problem is catastrophic if there is more than just one  extra quark $Q$ (notice that $Q_L,Q_R$ form a Dirac fermion with mass $m_Q\propto v_\sigma$) contributing to the colour anomaly of the Peccei-Quinn symmetry~\cite{Sikivie:1982qv}. For each $Q$ there is a physically distinct CP-conserving minimum separated by a very energetic domain wall and all of them would be populated during reheating, leading to a Universe dominated by the energy of the walls. Thus, our construction requires to consider one  and only one  extra quark $Q$. However, note that the cosmology of this new particle could still be very problematic. If $Q$ is an electroweak singlet without hypercharge, it has no decay channels and is thus cosmologically stable~\cite{Nardi:1990ku}. A thermal bath of $Q,\bar Q$'s would then be produced after reheating in thermal equilibrium with the SM through the strong interactions, but at temperatures below the $Q$ mass,  the annihilation $\bar Q Q\to gg$ freezes-out, leaving a substantial relic abundance of these coloured particles. At the QCD confining phase transition, the $Q$'s  hadronise with SM quarks and become heavy hadrons with electric charge, which are strongly constrained by searches of charged massive stable particles (CHAMPS) \cite{Perl:2001xi,Perl:2004qc,Perl:2009zz,Chuzhoy:2008zy}. 
The most minimal way to avoid this issue is to allow the $Q$'s to mix with SM quarks, making possible their co-annihilation and decay. 
This can only be achieved if the $Q$'s have hypercharge $-1/3$ or $+2/3$ and can mix with down or up quarks respectively, the only other colour triplet weak singlets available. In this way, predictivity strikes again, because the hypercharge of the new quark sets the strength of the axion-photon coupling, which is the most relevant parameter for axion dark matter detection.

Some parts of our SMASH model have been considered separately, e.g.\ the possible relation between the seesaw mechanism and either the axion \cite{Kim:1981jw,Mohapatra:1982tc,Shafi:1984ek,Langacker:1986rj,Shin:1987xc,He:1988dm,Dias:2005dn,Dias:2014osa,Celis:2014iua,Celis:2014jua,
Bertolini:2014aia,Ng:2015eia,Bertolini:2015boa,Salvio:2015cja,Carvajal:2015dxa,Barenboim:2015cqa,Clarke:2015bea,Ahn:2015pia,Berezhiani:1983hm,Berezhiani:1985in}, or Higgs (Hidden Scalar) Inflation \cite{Boucenna:2014uma,Okada:2015zfa,Budhi:2015sha}. A relation between the latter and scalar \cite{Lerner:2009xg,Lerner:2011ge,Khoze:2013uia,Kahlhoefer:2015jma} or fermionic  \cite{Aravind:2015xst}   WIMP dark matter (plus right-handed (RH) neutrinos \cite{Haba:2014zda,Haba:2014zja}) or axion dark matter \cite{Fairbairn:2014zta} has also been studied. Leptogenesis and dark matter have not only been linked in the $\nu$MSM, but also in works such as \cite{Khoze:2016zfi}. Higgs stability has been related to scalar  \cite{Gonderinger:2009jp} and scalar plus vector dark matter \cite{Khoze:2014xha}.\footnote{A superficially similar extension of the SM to the one we propose appeared in \cite{Salvio:2015cja}.  There, in analogy to the $\nu$MSM, the right-handed neutrinos do not carry PQ charges and, consequently,  their masses are independent of $v_\sigma$. Another key difference from SMASH is that the inflaton of \cite{Salvio:2015cja} is identified with the (non-minimally coupled) Higgs. Therefore, the proposal of \cite{Salvio:2015cja} suffers from the same unitarity problem as any other embedding of Higgs Inflation. An earlier  extension of the SM --even more different from SMASH-- but also inspired by minimality  was put forward in \cite{Davoudiasl:2004be}. In that work two ad-hoc singlets are added to account for dark matter and inflation, whereas the SM neutrino masses and baryogenesis originate from RH neutrinos. Like the $\nu$MSM, this model did not address the strong CP problem.} We will show that putting these puzzle pieces together in the appropriate way, a strikingly simple and testable picture of particle physics --up to the Planck scale--  and cosmology --from inflation to the present epoch-- arises.\footnote{A cosmological constant is needed to account for the current accelerated expansion of the Universe.} In essence, we will demonstrate that by solving the strong CP problem in a minimal way and linking this mechanism to the generation of small neutrino masses, one can automatically get baryogenesis, dark matter and successful primordial inflation. Remarkably, this is achieved by introducing a single new physics scale: $f_A = v_\sigma\sim 10^{11}$ GeV.

In this paper, we quantify the region of SMASH parameter space in which five fundamental problems of particle physics and cosmology are solved in one stroke: 
\begin{itemize}
\item[(i)] the origin of neutrino masses and their  flavour mixing,  
\item[(ii)]  the strong CP problem,
\item[(iii)] the nature of the inflaton, 
\item[(iv)] the generation of the matter-antimatter asymmetry in the Universe, and 
\item[(v)] the identity of dark matter.
 \end{itemize}

{The introduction of the new scale, $f_A$, which is very nicely motivated by the phenomenology we address, implies a hierarchy with respect to the electroweak scale. Given the benchmark values of the parameters that we have mentioned, the Higgs mass parameter of the Lagrangian needs to be adjusted to reproduce the Higgs mass. Notice however that a hierarchy is also present implicitly in the SM due to the existence of the Planck scale. Successful leptogenesis, together with requirements coming from stability and dark matter, enforce a small level of degeneracy ($\sim4\%$) between some of the right-handed neutrino masses. We accept these hierarchies, as we do with a small cosmological constant to account for the present dark energy density. }

The reader that is avid for knowing our main results and conclusions may now move directly to Section \ref{summary} or consult our Letter \cite{Ballesteros:2016euj}. Instead, for a gradual presentation of our findings, the outline of the paper is the following:  In Section \ref{model} we summarise the particle content of the SMASH model following Ref.\ \cite{Dias:2014osa} and we remind how it solves the problems (i) and (ii).  In Section \ref{inflation} we explain how inflation  (problem (iii)) is realised in SMASH.
We show that the inflationary predictions of SMASH are in perfect agreement 
with the current cosmic microwave background (CMB) observations. Importantly, we demonstrate that 
Hidden Scalar Inflation is possible for a non-minimal coupling of order unity, {and  in Section \ref{Sunit} we discuss} how this solves the predictivity issue of standard Higgs Inflation (that originates from the breakdown of perturbative unitarity).  In Section \ref{stability} we determine the parameter region in which the effective potential of SMASH (including radiative corrections) is positive up to Planckian  values of the scalars. 
This is crucial to ensure that successful inflation is a generic property of the model. We develop a semi-analytic understanding of the stability region in parameter space. In Section 
\ref{sec:reheating}, we present a comprehensive investigation of reheating in SMASH. Importantly, unlike in many other models of inflation, here the inflaton's couplings ($\lambda_\sigma$, $\lambda_{H\sigma}$, $\xi_\sigma$) are specified and well constrained by stability and unitarity.
Therefore, solid estimates of the reheating temperature can be derived.
We show that the Peccei-Quinn symmetry is effectively restored during reheating as long as $f_A\lesssim 10^{16}$\,GeV. Furthermore, the RH neutrinos are thermalised at a maximum temperature of order $10^{10}$\,GeV.
These predictions from reheating are key inputs for baryogenesis (problem (iv)) and for axion dark matter (problem (v)), as will be discussed in detail in  
Sections \ref{baryogenesis} and \ref{dark_matter}, respectively.
Apart from an executive summary, Section \ref{summary} contains
a discussion of possible variants of SMASH and gives 
a perspective for possible experimental tests of SMASH  and further theoretical in-depth studies. Several appendices are also included to make 
the paper self-contained. {There we discuss the effective potential at finite temperature (\ref{finiteT}), the beta functions at two-loops (\ref{betas}),} {  the amount of degeneracy required for the right-handed neutrinos (\ref{app:CPsources}), and finally, the level of tuning in the model in relation to the hierarchy {between $f_A$ and the electroweak scale} (\ref{tune}).}

\section{\label{model}The SMASH model}
The SMASH model, first proposed in~\cite{Dias:2014osa}, has the following boson and fermion (Weyl spinor notation) representations beyond the SM: 
\begin{itemize}
\item In the scalar sector there is a new complex singlet $\sigma$ (``hidden scalar field"). 
\item 
Three SM-singlet neutrinos $N_i$, with $i=1,2,3$. 
\item 
$Q$ ($\tilde Q$) in the (anti)-fundamental of $SU(3)_c$, with charge $-1/3$ ($+1/3$) under $U(1)_Y$.
\end{itemize}
 These hypercharge assignments\footnote{It is also possible to make $Q (\tilde Q)$ hypercharge assignments of $+2/3(-2/3)$. This variant of SMASH leads to almost identical phenomenology. In the rest of this work we will focus exclusively on the choice $-1/3$ ($+1/3$) under $U(1)_Y$. We will comment on the other alternative in Section~\ref{summary}.}
ensure that the new {quark can mix with the right-handed SM down-type  quarks,} allowing its decay to the latter, thereby evading overabundance problems~\cite{Nardi:1990ku,Berezhiani:1992rk}.
 
All these new fields are charged under a global $U(1)$ symmetry, which acts as {Peccei-Quinn (PQ) and lepton number (L) symmetry.\footnote{This symmetry may be an accidental symmetry of the low energy effective field theory emerging from an exact discrete symmetry of a more fundamental theory which includes quantum gravity. For 
explicit examples, see \cite{Dias:2014osa,Ringwald:2015dsf}.}} The charges under this symmetry are vector-like for SM particles and RH neutrinos and axial for the new quark $Q$, see Table \ref{tab:smash_charges}.  
\begin{table}[t]
\centering
\begin{equation}{\nonumber
\begin{array}{|c|c|c|c|c|c|c|c|c|}
\hline
  q & u & d & L & N & E    & Q &\tilde Q & \sigma  \\
\hline
 1/2 & -1/2 & -1/2 & 1/2 & -1/2 & -1/2   & -1/2 & -1/2 &1 
\\[.5ex]
\hline
\end{array}
}
\end{equation}
\caption{\small Charge assignments of the fields in SMASH under the new $U(1)$ PQ and L symmetry. 
The remaining SM fields have no charge under this new symmetry. 
\label{tab:smash_charges}}
\end{table}

The most general Yukawa couplings compatible with the new U(1) symmetry  are
\begin{equation}
\begin{aligned}
  {\cal L}\supset&-\Bigg[{Y_u}_{ij}q_i\epsilon H u_j+{Y_d}_{ij}q_i H^\dagger d_j+G_{ij} L_i H^\dagger E_j + F_{ij}L_i\epsilon H N_j+\frac{1}{2}Y_{ij}\sigma N_i  N_j \\
 &
+y\, \tilde Q \sigma Q+\,{y_{Q_d}}_{i}\sigma Q d_i+h.c.\Bigg]\,,
\label{lyukseesaw}
\end{aligned}
\end{equation}
where we have included also the Higgs field $H$. The fields $L_i$ and $E_i$ represent, respectively, the left-handed lepton doublets of the SM and the left-handed fields related to the conjugates of the usual right-handed leptons. The two last terms in the first row of \eqref{lyukseesaw} give rise
to a neutrino mass matrix of the form
\begin{equation}
M_\nu = \left( \begin{matrix} 0&M_D\\ M_D^T& M_M \end{matrix}\right)
= \frac{1}{\sqrt{2}} \left( \begin{matrix} 0& F v\\ F^T v & Y\, v_{\sigma} \end{matrix}\right).
\end{equation}
This realises the seesaw mechanism \cite{Minkowski:1977sc,GellMann:1980vs,Yanagida:1979as,Mohapatra:1979ia}, explaining the smallness of the masses of the left-handed SM active neutrinos by the hierarchy between $v$ and $v_\sigma$:
\begin{equation}
\label{seesaw}
m_{\nu} = - M_D M_M^{-1} M_D^T = -  \frac{F\,Y^{-1}\,F^T}{\sqrt{2}}\ \frac{v ^2}{v_{\sigma}}
= 0.04\,{\rm eV}  \left( \frac{10^{11}\,{\rm GeV}}{v_{\sigma}} \right)
\left( \frac{-  F\,Y^{-1}\,F^T}{10^{-4}}\right)
\,.
\end{equation}

The scalar potential of SMASH is
\begin{equation}
\label{scalar_potential}
V(H,\sigma )= \lambda_H \left( H^\dagger H - \frac{v^2}{2}\right)^2
+\lambda_\sigma \left( |\sigma |^2 - \frac{v_{\sigma}^2}{2}\right)^2+
2\lambda_{H\sigma} \left( H^\dagger H - \frac{v^2}{2}\right) \left( |\sigma |^2 - \frac{v_{\sigma}^2}{2}\right)\,.
\end{equation}
The self-couplings in the scalar potential are assumed to satisfy $\lambda_H, \lambda_\sigma >0$ and 
$\lambda_{H\sigma}^2 <  \lambda_H \lambda_\sigma$, in order to ensure that the minimum of the scalar potential
is attained at  the VEVs
\begin{equation}
\aver{H^\dagger H} = v^2/2, \hspace{6ex}
\aver{|\sigma |^2}=v_{\sigma}^2/2\,,
\end{equation}
where $v=246$\,GeV. Instead, $v_\sigma$ corresponds to a much higher scale; roughly $v_\sigma\sim 10^{11}$ GeV although we will explore all possible working values.  
The hidden scalar $\rho$ can be defined as a fluctuation around the VEV $v_\sigma$:  
\begin{equation}
\label{sigma:}
\sigma (x) =\frac{1}{\sqrt{2}}\big[v_{\sigma}+\rho (x)\big]e^{iA(x)/v_{\sigma}}
\,,
\end{equation}
as it gets a mass from spontaneous symmetry breaking. The same happens for the other new fields, $N_i$ and $Q$, whose masses are mostly determined by the VEV of $\sigma$:
\begin{equation}
\label{eq:masses}
M_{ij} = \frac{Y_{ij}}{\sqrt{2}} v_\sigma + \mathcal{O}\left(  \frac{v}{v_{\sigma}}\right), \hspace{3ex}
m_\rho =   \sqrt{2\,\lambda_\sigma}\, v_{\sigma} + \mathcal{O}\left(  \frac{v}{v_{\sigma}}\right), \hspace{3ex}
m_Q = \frac{y}{\sqrt{2}}\, v_{\sigma} + \mathcal{O}\left(  \frac{v}{v_{\sigma}}\right). 
\end{equation}
As long as the dimensionless couplings $Y_{ij}$, $\lambda_\sigma$, and $y$ are sizeable, all these masses will be large. Therefore, as far as physics around the electroweak scale or below is concerned, these heavy particles can be integrated out. The corresponding low-energy Lagrangian only contains a new field beyond the SM: the axion $A$ introduced in \eq{sigma:}. 

This field plays the role of a KSVZ-type \cite{Kim:1979if,Shifman:1979if} axion \cite{Weinberg:1977ma,Wilczek:1977pj} and of the majoron, the NGB of spontaneous breaking of global lepton 
number \cite{Chikashige:1980ui,Gelmini:1980re,Schechter:1981cv}, which is usually called $J$. 
At energies above the scale of strong interactions, $\Lambda_{\rm QCD}$, but below electroweak symmetry breaking scale $v$, 
the low-energy effective Lagrangian of the field $A/J$ --the NGB of SMASH-- reads \cite{Kim:1979if,Shifman:1979if,Chikashige:1980ui,Gelmini:1980re,Schechter:1981cv,Shin:1987xc,Dias:2014osa}
\begin{equation}
{\mathcal L}_A = \frac{1}{2} \partial_\mu A \partial^\mu A 
- \frac{\alpha_s}{8\pi}\,\frac{A}{f_A}\,G_{\mu\nu}^c {\tilde G}^{c,\mu\nu}  
- \frac{\alpha}{8\pi}\,\frac{2}{3}\frac{A}{f_A}\,F_{\mu\nu} {\tilde F}^{\mu\nu}
{- \frac{1}{4}  \frac{\partial_\mu A}{f_A} \ \overline\nu_i \gamma^\mu\gamma_5 \nu_i}
,
\label{axion_leff}
\end{equation}
where $\alpha_s$ ($\alpha$) is the strong (electromagnetic) fine structure constant, $G_{\mu\nu}^c$, $c=1,\ldots 8$, 
($F_{\mu\nu}$) is the field strength of $SU(3)_c$ ($U(1)_{\rm em}$) and $\tilde G^{c,\mu\nu}$ ($\tilde F^{\mu\nu}$)
its dual. In this expression, $\nu_i$ are {Majorana fermions corresponding to the} active neutrinos of the SM in the mass basis and the $A/J$ decay constant is
\begin{equation}
\label{F1}
f_{A}=v_{\sigma}. 
\end{equation} 

The second term on the right hand side of equation \eqref{axion_leff} will ensure that the 
strong CP problem is solved. 
Essentially, the theta parameter in QCD gets replaced by a dynamical quantity, 
\begin{equation}
\theta (x)= \frac{A(x)}{f_A}, 
\end{equation} 
which spontaneously relaxes to zero, $\langle \theta\rangle =0$, due to QCD effects -- thereby explaining the 
non-observation of strong CP violation \cite{Peccei:1977hh}. The fourth term of \eq{axion_leff}
describes low energy effective interactions between the axion and the active neutrinos induced by the fundamental interactions of $\sigma$ with the 
heavy right-handed singlets $N_i$ in equation \eqref{lyukseesaw}. The SMASH axion has no tree-level interactions with SM quarks and charged 
leptons. The loop induced interactions with the SM charged leptons can be larger than in the pure KSVZ
axion model  \cite{Shin:1987xc}.

At energies below $\Lambda_{\rm QCD}$,  the interactions of the SMASH axion with photons, nucleons, 
$\psi_N=p,n$, and neutrinos are described by \cite{Kaplan:1985dv,Srednicki:1985xd}
\begin{equation}
{\mathcal L}_A = \frac{1}{2} \partial_\mu A \partial^\mu A - V(A) 
- \frac{\alpha}{8\pi}\,C_{A\gamma}\frac{A}{f_A}\,F_{\mu\nu} {\tilde F}^{\mu\nu}
+ \frac{1}{2}  C_{AN}\frac{\partial_\mu A}{f_A} \ \overline\psi_N \gamma^\mu\gamma_5 \psi_N
- \frac{1}{4}  \frac{\partial_\mu A}{f_A} \ \overline\nu_i \gamma^\mu\gamma_5 \nu_i
,
\label{axion_leff_app}
\end{equation}
where a recent analysis \cite{diCortona:2015ldu} of the ${\cal O}(1)$ couplings gives $C_{A\gamma} =    
\tilde C_{A\gamma} - 1.92(4)$, being\footnote{In the case of hypercharge assignments $Q(\tilde Q)\rightarrow +2/3(-2/3)$, one gets $\tilde C_{A\gamma}={8}/{3}$.}  $\tilde C_{A\gamma}={2}/{3}$ in SMASH, $C_{Ap}=-0.47(3)$ and $C_{An}=-0.02(3)$, where the uncertainty of the axion-nucleon couplings arises from the 
uncertainty in the quark mass ratio $z=m_u/m_d=0.48(3)$ and higher order corrections~\cite{diCortona:2015ldu}.
The axion potential $V(A)$ {is determined by the $\theta$-dependence} of the vacuum energy density in QCD, 
\begin{eqnarray}
V(A) \equiv - \frac{1}{\mathcal V} \ln \frac{Z (\theta)}{Z (0)}\Bigg{|}_{\theta = A/f_A}
,
\label{axion_potential_free_energy_density}
\end{eqnarray}
where $\mathcal{V}$ is the Euclidean space-time volume. It has 
an absolute minimum at $\theta = A/f_A = 0$. In fact, this is the reason why in this extension of the SM there is no strong CP problem \cite{Peccei:1977hh}. The curvature around this minimum determines the axion mass, which is inversely proportional to the axion decay constant:
\begin{equation}
m_A^2 \equiv \frac{\partial^2 V(A)}{\partial A^2}\Bigg{|}_{A=0}   = \frac{\chi}{f_A^2}.
\end{equation}
The proportionality constant is given by the so-called topological susceptibility, i.e.  
the variance of the $\theta =0$ topological charge distribution,
\begin{equation} \label{topos}
\chi  \equiv \int d^4x \langle q (x) q (0)\rangle|_{\theta =0}
= \lim_{{\mathcal V}\to \infty}  \frac{\langle Q^2\rangle|_{\theta =0}}{\mathcal V}\,,
\end{equation}
where $Q\equiv\int d^4x\, q(x)$, with $q(x) = \alpha_s\,G_{\mu\nu}^c(x) {\tilde G}^{c,\mu\nu}(x)/{8\pi}$ the topological
charge density. 

The topological susceptibility and thus the axion mass can be estimated using chiral perturbation theory, giving $m_A\sim m_\pi f_\pi/f_A$~\cite{Crewther:1977ce,DiVecchia:1980ve,Leutwyler:1992yt}.  
The latest results include NLO corrections~\cite{diCortona:2015ldu}, revealing that at zero temperature 
\begin{equation}
\label{top}
\chi_0 \equiv m_A^2 f_A^2 = [75.5(5) {\rm MeV}]^4 , 
\end{equation}
which agrees beautifully with the latest lattice QCD calculation, $\chi_0 = [75.6(1.8)(0.9) {\rm MeV}]^4$ \cite{Borsanyi:2016ksw}. 
Equation \eqref{top} gives a zero-temperature axion mass, 
\begin{equation}
\label{zeroTma}
m_A= 
{57.0(7)\,   \left(\frac{10^{11}\rm GeV}{f_A}\right)\mu \text{eV}. }
\end{equation}

Currently, the most stringent lower bound on $f_A$ for SMASH arises from the non-observation of an anomalously short neutrino pulse of the supernova SN1987A, which would be attributable to the fast core cooling due to nucleon bremsstrahlung initiated axion emission~\cite{Raffelt:2006cw,Fischer:2016cyd}. This translates into
\begin{equation}
f_A \gtrsim   4\times  10^8\ {\rm GeV}\,,
\end{equation} 
which is just approximate due to the uncertainty on the axion emission rates. This corresponds to an upper bound on the mass of the axion:
\begin{equation}
m_A \lesssim 14.3\, {\rm meV} .
\end{equation} 

The interactions of the $A/J$ with neutrinos have been studied in laboratory experiments, in particular neutrino-less double beta decay, in astrophysics, and cosmology. However, on account of the on-shell vacuum relation 
\begin{equation}
\label{double_suppression}
{\mathcal L}_A\supset -\frac{1}{4}  \frac{\partial_\mu A}{f_A} \ \overline\nu_i \gamma^\mu\gamma_5 \nu_i
= \frac{i}{2}\,\frac{m_i}{f_A}\, A\, \overline\nu_i \gamma_5 \nu_i ,
\end{equation}
the $A/J$ interactions with neutrinos are suppressed by the small ratio $m_i/f_A$ and therefore the bound on $f_A$ from 
them is much weaker than the ones arising from the consideration of the photon and nucleon interactions of the $A/J$. Indeed, the best limit on the neutrino interactions arising from 
cosmic microwave background (CMB) observations 
is \cite{Hannestad:2005ex}
\begin{equation}
{m_i} \lesssim 10^{-7}{f_A}, 
\end{equation}
which translates into
\begin{equation}
f_A \gtrsim {\rm MeV}\, \left( \frac{m_i}{\rm 0.1\ eV}\right) .
\end{equation}
A comparable limit is obtained from energy loss considerations of the supernova SN1987A \cite{Kachelriess:2000qc,Tomas:2001dh,Farzan:2002wx}.


\section{\label{inflation}Inflation}

In this section we discuss inflation in SMASH, which in principle may  occur with the Higgs (Higgs Inflation, HI), the hidden scalar (HSI) or a mixture of both (HHSI) playing the role of the inflaton. We will later see in Section \ref{Sunit} that HI should be discarded in favour of HSI and HHSI due to the violation of perturbative unitarity at large field values in HI, as we already anticipated in the Introduction. Throughout this section we assume that the potential is absolutely stable and we will find the parameters of the model that can fit the CMB observations. In Section \ref{stability} we will investigate the stability of the potential.

\subsection{\label{subsec:2field}Two-field inflation with non-minimal couplings to $R$}
Our analysis builds upon  Higgs Inflation \cite{Bezrukov:2007ep}, realising (in a particularly well motivated model) the ideas of e.g.\ \cite{Kaiser:2010ps,Kaiser:2010yu,Giudice:2010ka,Lebedev:2011aq,Kaiser:2012ak,Greenwood:2012aj,Kaiser:2013sna,Schutz:2013fua} for two fields non-minimally coupled to gravity. Including gravity, the most general SMASH action at operator dimension four\footnote{Notice, however, that once the graviton is properly normalised by giving it dimensions of mass, the operators $\xi_H\, H^\dagger H\, R$ and $\xi_\sigma\, \sigma^* \sigma\ R$ have dimension five by power-counting.} is completed (in the Jordan frame) by including a term
\begin{equation}
  \label{Lmain}
  S\supset - \int d^4x\sqrt{- g}\,\left[
     \frac{M^2}{2}  + \xi_H\, H^\dagger H+\xi_\sigma\, \sigma^* \sigma  
  \right] R
  \,,
\end{equation}
where $\xi_H$ and $\xi_\sigma$ are dimensionless non-minimal couplings to the curvature scalar $R$, and the mass scale $M$ is related to the actual reduced Planck mass, $M_P^2=1/(8\pi\, G)$, by 
\begin{align}
\label{eq:MMP}
M^2_P=M^2+\xi_H v^2+\xi_\sigma v^2_\sigma.
\end{align}
We recall that these non-minimal couplings are generated radiatively, even if they are set to zero at some scale, and therefore they should be included in a general analysis. {{Note that, since 
the} {action is non-renormalisable,} {in principle one has to consider higher dimensional operators suppressed by powers of $M_P$ in \eqref{Lmain}. We are assuming that in the infrared the} {action} {flows
to \eqref{Lmain}, with the coefficients of other operators becoming negligible. Non-zero values will be generated at higher scales, but it is possible to argue that, for backgrounds involving
energy scales that remain suppressed with respect to $M_P$, such corrections will remain small,} {see, e.g. \cite{Burgess:2009ea}.} These energy scales include those associated with derivatives of the background fields, the effective potential and its derivatives, the masses of all particles in the background, { and the dimensionful combinations $(\xi_\sigma \sigma^*\sigma)^{1/2},(\xi_H H^\dagger H)^{1/2}$.} For the preferred parameter space in SMASH, it turns out that these energy scales remain below $M_P$ even in the inflationary background, so that it is consistent to restrict ourselves to the interactions in \eqref{Lmain}.
In the following, we will assume that $\xi_H,\xi_\sigma$ are positive. We will also require that $\lambda_{H\sigma}>-\sqrt{\lambda_H\lambda_\sigma}$, which is needed for tree-level absolute stability of the potential. The non-minimal couplings do not affect our considerations on the stability of the potential, which we discuss in the next section. 

 As far as the tree-level dynamics is concerned, it is sufficient to consider the 
Higgs in the unitary gauge and the modulus of $\sigma$. It is useful to organise them as the two components of a field $\phi$:  
\beq \label{choiceg}
H(x)=\frac{1}{\sqrt{2}}{0 \choose h(x)}\,,\quad 
|\sigma(x)|=\frac{\rho (x)}{\sqrt{2}} \,,\quad \phi(x)= (h(x),\rho(x))\, .
\eeq
Performing a Weyl transformation of the metric:
\begin{equation} \label{weyl}
\tilde g_{\mu \nu}(x) = \Omega^{2}(h(x),\rho(x))\, g_{\mu \nu} (x) ,
\end{equation}
where $\Omega^2$
 is defined as
\begin{equation}
\Omega^2= 1+\frac{\xi_H (h^2-v^2) +\xi_{\sigma}(\rho^2-v^2_\sigma)}{M_P^2} \,,
\label{conf_fac}
\end{equation}
we get that the relevant part of the action  in the Einstein frame reads
\begin{align} \label{Eact}
S_{\rm SMASH}^{\rm (E)}\supset {\int} d^4x\sqrt{-\tilde{g}}\left[-\frac{M_P^2}{2}\tilde{R}+\frac{1}{2}\sum_{i,j}^{1,2}\mathcal{G}_{ij}\tilde g^{\mu\nu}\partial_\mu\phi_i\partial_\nu\phi_j-\tilde{V}\right]\,,
\end{align}
where $\mathcal{G}_{ij}$ is an induced metric in field space, given by 
\begin{align} \label{imet}
\mathcal{G}_{ij}= \frac{\delta_{ij}}{\Omega^{2}}+\frac{3 M_P^2}{2}\frac{\partial\log\Omega^2}{\partial {\phi_i}}\frac{\partial\log\Omega^2}{\partial {\phi_j}}\,.
\end{align}
The potential in this (Einstein) frame is the same as the original one of SMASH, but rescaled by an appropriate power of $\Omega$: 
\begin{align}
\label{VEpot}
 \tilde{V} (h,\rho) 
= \frac{1}{\Omega^4(h,\rho)} 
\left[
\frac{\lambda_H}{4} \left(  h^2- v^2\right)^2
+\frac{\lambda_\sigma}{4} \left( \rho^2 - v_{\sigma}^2\right)^2+
\frac{\lambda_{H\sigma}}{2} \left(  h^2- v^2\right) \left( \rho^2 - v_{\sigma}^2\right)\right].
\end{align}
Notice that the change of frame makes explicit the presence of high-dimensional operators suppressed by $M_P/\xi_i$, $M_P/(\xi_H \xi_\sigma)^{1/2}$ or $M_P/\xi_i^{1/2}$ (neglecting the VEVs of the fields). The smallest of these scales is the ultraviolet cut-off of the theory, which will be shown to have important implications in Section~\ref{Sunit}.

The factor $\Omega^{-4}$ factor in the Einstein frame potential that is responsible for its flattening at large field values. For instance, if we assume that the Higgs field is the relevant one at large field values, we see that $\Omega^{4}$ grows as $h^4$, which asymptotically compensates the term $h^4$ in the numerator of \eq{VEpot} coming from the original (Jordan frame) potential, thus  leading to a constant at large $h$ values. The same kind of behaviour is also possible along other directions in the two-dimensional field space of $h$ and $\rho$, as we will now discuss.

Since the potential \eq{VEpot} is symmetric under $h \to - h$ and $\rho \to - \rho$, we can restrict our attention to $h > 0$ and $\rho > 0$.
It is convenient to use polar field-space coordinates,\footnote{Analogous results can be achieved with a slightly different choice of variables, see \cite{Lebedev:2011aq}, where the fields $\tan \varphi$ and $\log(\xi_h h^2+\xi_\sigma \rho^2)$ are used.} $\phi$ and $\theta$, defined as follows: 
\beq
h = \phi \sin \varphi~,~\rho = \phi \cos \varphi\,.
\eeq
For large values of the radial field $\phi$, neglecting the VEVs of $h$ and $\rho$ and assuming also that $\Omega^2 \gg 1$, which means that $M_P^2/\phi^2\ll \xi_H \cos^2\theta+\xi_\sigma \sin^2\theta$, the potential can be approximated by a function of $\varphi$ alone:
\beq \label{vlim}
\tilde V \simeq   \frac{\lambda_H \sin^4 \varphi + \lambda_\sigma \cos^4 \varphi +2 \lambda_{H\sigma} \cos^2 \varphi  \sin^2 \varphi}{4\left(\xi_H  \sin^2 \varphi +\xi_\sigma  \cos^2 \varphi\right)^2} M_P^4\,.
\eeq
The function \eq{vlim} has extrema for $\varphi=0$, $\varphi=\pi/2$ and also for
\beq \label{threethetas}
\sin^2\varphi=  \frac{ {\kappa_\sigma}}{{\kappa_H + \kappa_\sigma }}\,,
\eeq
with    
\begin{equation}
\kappa_H  \equiv \lambda_{H\sigma} \xi_H - \lambda_H \xi_\sigma\,,\quad
\kappa_\sigma  \equiv \lambda_{H\sigma} \xi_\sigma - \lambda_\sigma \xi_H.
\label{kappadef}
\end{equation}
Whether the directions $h$, $\rho$ and \eq{threethetas} correspond to approximate asymptotic valleys or ridges of the potential depends on the signs of $\kappa_H$ and $\kappa_\sigma$, which determine the sign of the second derivative of \eq{vlim} on its extrema. {We are interested in valleys because they are attractors of inflationary trajectories. Concretely, if $\kappa_H>0$ the direction $\varphi=\pi/2$ is a valley (which can support Higgs inflation (HI)), 
cf. Fig. \ref{fig:pot_einstein} (left), 
  and, vice versa, if $\kappa_H<0$ it is a ridge. Similarly, if $\kappa_\sigma>0$ the direction $\varphi=0$ is a valley, along which Hidden Scalar Inflation (HSI) may occur, 
cf. Fig. \ref{fig:pot_einstein} (middle), 
and it is instead a ridge if $\kappa_\sigma<0$. The sign of the second derivative of \eq{vlim} along the direction defined by the angle \eq{threethetas} is equal to the sign of $-\kappa_H\kappa_\sigma(\kappa_H\xi_\sigma+\kappa_\sigma\xi_H)$. Therefore, if both $\kappa_H$ and $\kappa_\sigma$ are positive, \eq{threethetas} is a ridge. Instead, if both $\kappa_\sigma$ and $\kappa_H$ are negative, \eq{threethetas} is a valley, which can support mixed Higgs-Hidden Scalar inflation (HHSI),
cf. Fig. \ref{fig:pot_einstein} (right).  
Finally, if only one among $\kappa_\sigma$ and $\kappa_H$ is positive, the direction \eq{threethetas} does not extremise the expression \eq{vlim} because the equation \eq{threethetas} has no real solution for $\varphi$. In that case, this direction plays no special role for inflation. The different directions that, a priori, may support successful inflation within SMASH are summarised in Table \ref{tab:summa}. 
\begin{table}[h]
  \footnotesize
  \small
  \renewcommand\arraystretch{1.3}
  \centering
  \begin{tabular}{|c|c|c|}
    \hline
    {$\text{sign}(\kappa_H)$} &{$\text{sign}(\kappa_\sigma)$} & Inflation   \\ 
    \hline
    \hline
    +&  + &    HI {\it or} HSI      \\ \hline 
    +&  $-$ &    HI      \\ \hline 
    $-$&  + &     HSI      \\ \hline 
    $-$&  $-$ &    HHSI    \\ \hline 
       \end{tabular}
  \caption{\label{tab:summa} \small  The inflaton may in principle be the neutral component $h$ of the Higgs (HI), the modulus $\rho$ of the Hidden Scalar (HSI) or a combination of both (HHSI) depending on the signs of the parameters $\kappa_H$ and $\kappa_\sigma$ defined in \eq{kappadef}. }
\end{table}

\begin{figure}[t]
\begin{center}
\includegraphics[width=0.3\textwidth]{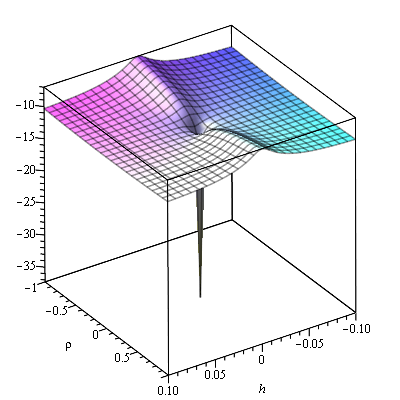}
\includegraphics[width=0.3\textwidth]{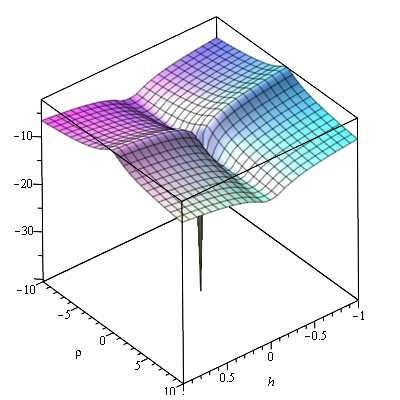}
\includegraphics[width=0.3\textwidth]{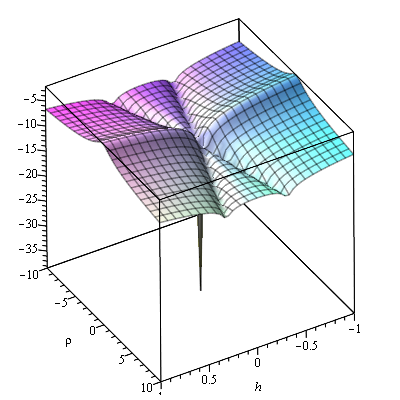}
\caption{ \small   Decadic log of the SMASH scalar potential in the Einstein frame, as a function of $h$ and $\rho$, all in units of $M_P$,
for parameters yielding pure Higgs Inflation ($\kappa_H>0$, $\kappa_\sigma < 0$) 
({\em left panel}), pure Hidden Scalar Inflation ($\kappa_H<0$, $\kappa_\sigma >0$)
({\em middle panel}),
and mixed Higgs-Hidden-Scalar Inflation ($\kappa_H<0$, $\kappa_\sigma <0$)
({\em right panel}), respectively. 
The couplings have been chosen such that the amplitude of primordial scalar perturbation is properly normalised.
}
\label{fig:pot_einstein}       
\end{center}
\end{figure}

Generically, the direction orthogonal to the inflaton gets an effective mass squared $\sim 4\lambda_{H\sigma}\,M_{P}^2/\xi_\sigma$, that is heavy compared to the Hubble expansion rate --see also \cite{Giudice:2010ka,Lebedev:2011aq}-- and prevents uphill motion transverse to the rolling of the fields \cite{Burgess:2012dz}. Generically, in all the cases that we have discussed (HI, HSI and HHSI) inflation proceeds towards a single-field attractor along the valleys in field space for a large spectrum of initial conditions. If the initial conditions are chosen so that the fields are set atop a ridge with some velocities, they will eventually roll into a neighbouring valley. For moderate initial velocities, the last stages of inflation (which are those relevant for the observations) will take place along a valley for which the single field slow-roll approximation can be applied \cite{Kaiser:2013sna,Schutz:2013fua}. Motion in field space in the direction orthogonal to the valley gets damped away (thanks to Hubble friction). {For large initial velocities, Hubble friction is still active as the time derivatives of the fields contribute to the energy density; thus, the velocities will be damped and slow-roll conditions
may again be achieved. }

In order to study the properties of inflation along the three possible directions in field space, it is convenient to work with redefined fields $\chi_\rho$ and $\chi_h$ defined as follows, see for instance \cite{Lerner:2009xg}:
\begin{align} 
\label{chi_h}
\Omega^2\frac{\partial\chi_h}{\partial h} = \sqrt{{\Omega^2 + 
6\,\xi_H^2\frac{ h^2}{M_P^2}}}\,,\quad
\Omega^2\frac{\partial \chi_\rho}{\partial \rho} = \sqrt{{\Omega^2 + 
6\,\xi_\sigma^2\frac{\rho^2}{M_P^2}}}.
\end{align}  
In terms of them, the action \eq{Eact} is
\begin{equation}  \label{Einstein_frame_lagrangian_canonical}
\begin{aligned}
S_{\rm SMASH}^{\rm (E)} \supset  
 \int d^4x\sqrt{-\tilde{g}}  {\Bigg [} -\frac{1}{2} M_P^2\tilde{R} 
& +\frac{1}{2} \,
\tilde{g}^{\mu\nu}\partial_\mu \chi_h\partial_\nu \chi_h + \frac{1}{2} \,
\tilde{g}^{\mu\nu}\partial_\mu \chi_\rho \partial_\nu \chi_\rho  \\
&+\frac{1}{2} \, K(\chi_h,\chi_\rho )\, \tilde{g}^{\mu\nu}\partial_\mu \chi_h \partial_\nu \chi_\rho  
- \tilde V(\chi_h,\chi_\rho )
\Bigg{]}\,,
\end{aligned}
\end{equation}
where the kinetic mixing is 
\beq
\label{kinmix}
K(\chi_h,\chi_\rho ) = 6\frac{\xi_H\xi_\sigma} {\Omega^4}
\frac{\partial h}{\partial \chi_h} \frac{\partial \rho}{\partial\chi_\rho} 
\frac{h \rho}{M_P^2} \,,
\eeq
and the potential is just given by \eq{VEpot}, but expressed in terms of the new fields.

Although it is not possible in general to make a field redefinition for which the kinetic mixing vanishes, it is possible to simplify the action further by considering the  directions in field space that we identified above. In the limit of $\Omega\gg 1$, we obtain that $K$ vanishes  along the  $\cos\varphi =0$ (HI) and $\sin\varphi = 0$ (HSI) directions. 
In these cases $\Omega^2$ only depends on one field. We can obtain analytical expressions for the (now canonically normalised fields) $\chi_h$ or $\chi_\rho$ by integrating \eqref{chi_h}. 
In the HI case ($\rho=0$), we have   
\bea
\label{chi_h_H_inf_sol}
\sqrt{\xi_H}\frac{\chi_h}{M_P} = \sqrt{1+6{\xi_H}}\,\text{arcsinh}\left(\sqrt{1+6\xi_H}u(h)\right)
-\sqrt{6\xi_H}\,\text{arcsinh}\left(\sqrt{6\xi_H}\frac{u(h)}{\sqrt{1+u(h)^{2}}}\right)\,,
\eea
where $u(h)\equiv \sqrt{\xi_H}\,{h}/{M_P}$. An analogous result is obtained for HSI ($h=0$) after exchanging $h$ and $\rho$. 

The HHSI case is more involved because the kinetic mixing does not {in principle} vanish. {However, for reasons related with unitarity and stability, which will become clear in Sections \ref{Sunit} and \ref{stability},} we will only be interested in the limit in which inflation in the HHSI case describes a trajectory that is parametrically close to the HSI case. A first approximation to this limit can be obtained by neglecting $\xi_H$ in the action. Doing so, {the mixing can be ignored and,} expressing $h$ as a function of $\rho$ using \eq{threethetas}, we can  normalise canonically the resulting kinetic term by defining a new field $\tilde\chi_\rho$ which is given by equations similar to those of \eq{chi_h}:
\begin{align} 
\label{chi_hs}
\Omega^2\frac{d\tilde\chi_{\rho}}{d\rho} = \sqrt{b\,\Omega^2 + 
6\,\xi_\sigma^2\frac{ \rho^2}{M_P^2}}\,,
\end{align}
where now $\Omega^2\simeq 1+\xi_\sigma\rho^2/M_P^2$ and we introduce the parameter
\begin{align} \label{adef}
b\equiv 1+\left|{\lambda_{H\sigma}}/{\lambda_H}\right|\,,
\end{align}
which determines the angle $\varphi$ of \eq{threethetas} in the limit $\xi_H=0$ through the relation $\sin^2\varphi= 1-b^{-1}$. Notice that in this limit 
\beq \label{tra}
\frac{\rho}{h} = \sqrt{-\frac{\lambda_H}{\lambda_{H\sigma}}} + \mathcal O \left( \frac{\xi_H}{\xi_\sigma}\right)\,.
\eeq
This leads to a result analogous to \eq{chi_h_H_inf_sol}:
\bea
\label{gens}
\sqrt{\frac{\xi_\sigma}{b}}\frac{\tilde\chi_\rho}{M_P} = \sqrt{1+6\frac{\xi_\sigma}{b}}{\rm arcsinh} \left(\sqrt{1 +6 \frac{ \xi _{\sigma }}{b}}u(\rho)\right)-\sqrt{6 \frac{\xi_\sigma}{b}}\, {\rm arcsinh}\left(\sqrt{6\frac{\xi_\rho}{b}}\frac{u(\rho)}{\sqrt{1+u(\rho)^{2}}}\right)\,,
\eea 
where now $u(\rho) \equiv  \sqrt{\xi_\sigma}\,{\rho}/{M_P}$ and  the deviation from HSI is controlled by the parameter $b$. {For $|\lambda_H/\lambda_{H\sigma}|\gg1$, one has $b\simeq1$ in equation \eqref{adef}, and the inflationary trayectory \eqref{tra} is indeed parametrically close to $h=0$, as in HSI. Such choice is motivated by unitarity and stability, as will be further elaborated in Sections \ref{Sunit} and \ref{stability}.}


\subsection{Confronting CMB observations}
The inflationary potential in the Einstein frame along the field directions that are interesting for us is approximately  
\be
\label{genpotential}
\tilde V(\chi) = \frac{\lambda }{4}\phi(\chi)^4\(1+\xi\frac{\phi(\chi)^2}{M_P^2}\)^{-2}\,,
\ee 
where, depending on the  specific direction, the canonically normalised inflaton $\chi$, the quartic coupling, $\lambda$, the non-canonical inflaton $\phi$ and the non-minimal coupling $\xi$ must be understood as:
\begin{align}
\text{HI :}\,  \chi_h, \lambda_H, h, \xi_H\,,\quad\quad\quad\quad
\text{HSI :}\,  \chi_\rho, \lambda_\sigma, \rho, \xi_\sigma\,,\quad\quad\quad\quad
\text{HHSI :}\, \tilde \chi_\rho, \tilde \lambda_\sigma, \rho, \xi_\sigma\,;
\end{align}
and the effective quartic coupling for HHSI is
\begin{align} \label{tildes}
\tilde\lambda_\sigma= \lambda_\sigma-\frac{\lambda_{H\sigma}^2}{\lambda_H}\,,
\end{align}
as it can be directly seen from \eq{VEpot}. We recall that in HHSI, the angle between $h$ and $\rho$ is described by $b$, defined in \eqref{adef}, and {as mentioned before} we will be interested in $b\simeq 1$. The expression \eq{genpotential} clearly shows that at large $\phi$ values (which correspond to large $\chi$ and that are those relevant for inflation) the potential becomes asymptotically flat, tending to\footnote{As with any asymptotically flat potential one could wonder what is the probability of having a field value that leads to slow-roll inflation fitting the CMB data. It is straightforward to estimate that eternal inflation happens for field values above $\sim 40 M_P$, which are considerably higher than the $\sim 5 M_P$ that we need for the initial value of the inflation. Computing the relative probability of occurrence between these two field values goes beyond the scope of this work.}
\begin{align} \label{limitlf}
\tilde V\sim \frac{\gamma}{4} M_P^4\,,
\end{align}
where the coupling $\gamma$ is
\begin{align} \label{gammap}
\text{HI :}\,  \lambda_H/\xi_H^2\,,\quad\quad\quad\quad
\text{HSI :}\,  \lambda_\sigma/\xi_\sigma^2\,,\quad\quad\quad\quad
\text{HHSI :}\, \tilde\lambda_\sigma/\xi_\sigma^2\,.\quad\quad\quad\quad
\end{align}
The CMB measurements from the Planck \cite{Ade:2013zuv,Planck:2013jfk,Ade:2015xua,Ade:2015lrj} and BICEP2/Keck \cite{Array:2015xqh} collaborations show a preference for plateau-like inflationary potentials. See \cite{Martin:2013nzq} for a Bayesian approach to inflationary model comparison in agreement with this conclusion.\footnote{Other plateau-like models not included in \cite{Martin:2013nzq}, such as the general formulation of \cite{Ballesteros:2015noa} for renormalisable potentials, are also able to fit well the data, supporting further this conclusion.} In SMASH,
the way in which the potential approaches the plateau only depends on the size of the relevant non-minimal coupling $\xi$ in each case. Therefore, the shape of the spectra of primordial perturbations and the amount of inflation in the slow-roll approximation are determined by the values of $\chi$ and $\xi$ at any given number of e-folds before the end of inflation,\footnote{The precise number of e-folds that is needed to solve the horizon problem is determined by the dynamics of reheating after inflation, which depends on whether inflation is HI, HSI or HHSI, see Sections \ref{sec:efolds} and \ref{sec:reheating}.} whereas the amplitude of scalar perturbations is determined by the combination $\gamma$, defined in \eq{gammap} for each case. 

Indeed, as it is well-known, the primordial scalar and tensor spectra, $P_s(k)$ and $P_t(k)$, can be computed in the slow-roll approximation from the (potential) slow-roll parameters: $\epsilon=M_P^{2}(\tilde V'/\tilde V)$, $\eta=M_P^2 \tilde V''/\tilde V$, $\zeta=M_P^4 \tilde V' \tilde V'''/\tilde V^2$, \ldots, where the primes denote derivatives with respect to $\chi$. In terms of them, the scalar and tensor spectra around a fiducial scale $k_*$ are
\begin{align}
P_s(k)=A_s\left(k/k_*\right)^{n_s-1+{1/2\,\alpha}\log (k/k_*)+\cdots}\,,\quad P_t(k)= A_t\left(k/k_*\right)^{n_t+\cdots},
\end{align}
where the amplitude of scalar perturbations $A_s$, their spectral index $n_s$, its running $\alpha$, the tensor-to-scalar ratio  $r=A_t/A_s$, and the tensor spectral index $n_t=-r/8$ (assuming single-field slow-roll inflation), are given by
\be
A_s = \frac{1}{24 \pi^2 \epsilon}\frac{\tilde V}{M_P^4} \, , \quad 
n_s = 1-6\epsilon+2\eta \, , \quad \alpha = -2\zeta +16\epsilon\,\eta-24 \epsilon^2\,,\quad 
r = 16 \epsilon\,,
\ee
evaluated at the field value corresponding to the time when the scale $k_*$ exits the horizon.  Besides, the number of e-folds from some initial field value $\phi_I$ until the end of inflation at $\phi_{\rm end}$ can also be approximately computed analytically {(see also \cite{Kaiser:1994vs})}: 
\be
\label{eq:efoldssimple}
N(\chi_{\rm I},\chi_{\rm end})\simeq \frac{1}{M_P}\int_{\chi_{\rm end}}^{\chi_{\rm I}}\frac{d\chi}{\sqrt{2\epsilon}}=
\frac{b+6\xi}{8 {M_P^2}}\(\phi_I^2-\phi_{\rm end}^2\)-\frac{3}{4}\ln\(\frac{M_P^2+\xi \phi_I^2}{M_P^2+\xi \phi_{\rm end}^2}\)\,,
\ee
where we have used \eq{chi_hs}. Actually, $N(\chi_{\rm I},\chi_{\rm end})$ can be obtained exactly by solving the dynamics of the inflaton as a function of the number of e-folds itself, which is given by equation \cite{Ballesteros:2014yva}:
\begin{align} \label{eq:efolds}
\frac{d^2\chi}{dN^2}+3\,\frac{d\chi}{dN}-\frac{1}{2M_P^2}\left(\frac{d\chi}{dN}\right)^3+\left(3M_P-\frac{1}{2M_P}\left(\frac{d\chi}{d N}\right)^{2}\right)\sqrt{2\epsilon}=0\,,
\end{align}
and using the condition $\epsilon_H\equiv-\dot{\mathcal{H}}/\mathcal{H}^2=1$, where {$\mathcal{H}=\dot a/a$ is the Hubble expansion rate (in terms of the Friedmann-Lema\^itre-Robertson-Walker (FLRW) metric's scale factor $a(t)$),} to determine the value of $\chi$ at the end of inflation. 

\begin{figure}[t]
\begin{center}
\includegraphics[width=0.9\textwidth]{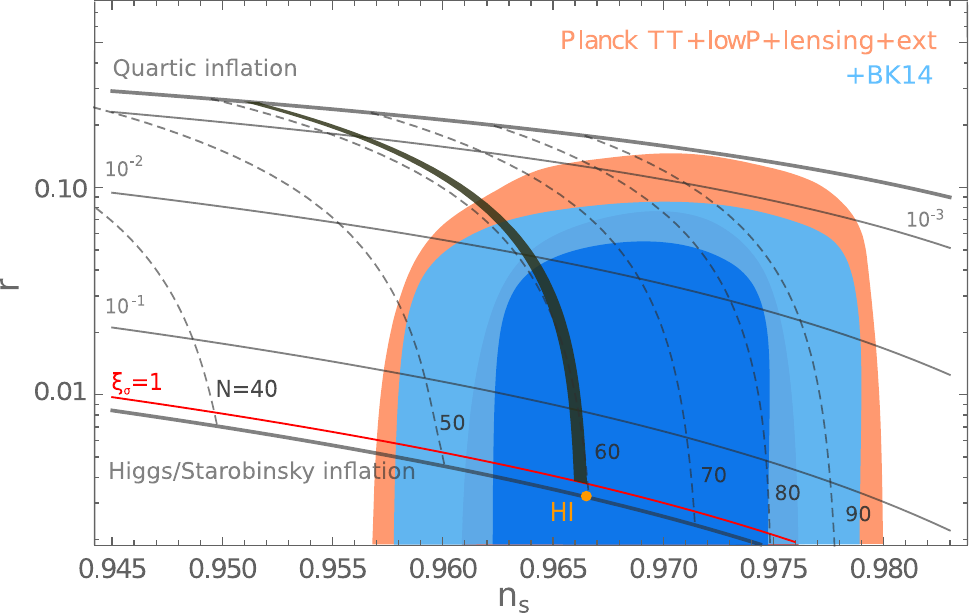}
\caption{\label{fig:r_vs_ns}  \small 
{Inflationary predictions for the potential of equation \eqref{genpotential}
in the $r$ vs $n_{s}$ plane at 0.002 Mpc$^{-1}$. Black solid lines show isocontours of the non-minimal coupling, $\xi_\sigma$. 
The thick black line is the SMASH prediction, which takes into account the fact after inflation the Universe enters directly into the radiation-dominated era, see Sec.~\ref{sec:efolds}.  For comparison, the thin dotted lines are isocontours of the number of e-folds for the potential \eq{genpotential}, without taking into account any information about the reheating process after inflation. 
We also show the 68\% and 95\% C.L.  regions at 0.002 Mpc$^{-1}$ taken from \cite{Array:2015xqh}. The most stringent of these (in blue) include CMB polarisation data from Planck (lowP) and BICEP2/Keck, as well as BAO, SN and lensing reconstruction, see \cite{Array:2015xqh} and \cite{Ade:2015xua}.  
The line indicated as ``Quartic inflation" shows the prediction for a purely quartic monomial potential (corresponding to the limit $\xi_\sigma \rightarrow 0$), which is ruled out by the data. We show as well the (black continuous) line corresponding to the limit $\xi_\sigma\rightarrow \infty$, for which the potential \eqref{genpotential} is analogous to that of Starobinsky \cite{Starobinsky:1980te} and Higgs inflation \cite{Bezrukov:2007ep}. 
{The HI result of \cite{Gorbunov:2012ns} is indicated as a point on this line. We stress that this point ignores the instability of the Higgs potential and the effects of new physics needed to restore unitarity at inflationary scales, see Section~\ref{Sunit}.}
}}
\end{center}
     \end{figure}

The numerical predictions in the $n_s$ -- $r$ plane using the expression $\eq{gens}$ are {shown\footnote{See also Fig. 3 in Ref. \cite{Fairbairn:2014zta}.}} in Figure \ref{fig:r_vs_ns}. {Note that {as anticipated before,} we take $b\simeq 1$ so that HSI and HHSI give the same results for inflation. The thin solid lines indicate the values of the non-minimal coupling,  bounded by the quartic inflation limit ($\xi_\sigma=0$) and the $\xi_\sigma\rightarrow \infty$ limit, for which the potential becomes that of Starobinsky inflation \cite{Starobinsky:1980te}. The dashed lines show  the number of e-folds till the end of inflation, computed using \eq{eq:efolds}. The number of e-folds can be constrained by using the post-inflationary evolution of the Universe (which is determined by the matter content of the model) to match the scales of current CMB perturbations to their values at horizon crossing during inflation, as discussed in Sections \ref{sec:efolds} and \ref{sec:reheating}. This narrows the SMASH predictions to the thick black line in Figure \ref{fig:r_vs_ns}. The figure also shows the 68\% and 95\% C.L.  regions  at 0.002 Mpc$^{-1}$ from the combined data analysis of Planck and BICEP/KECK \cite{Array:2015xqh}, assuming zero running of the scalar spectral index.
If instead $\alpha$ is included in the fit, the values of $n_s$ and $\alpha$ as determined by the Planck collaboration at the scale $k_*=0.05$ Mpc$^{-1}$  are 
$ n_s=0.9644\pm 0.0049$  and $\alpha=-0.0085\pm 0.0076$ both at $68\%$  C.L. \cite{Ade:2015lrj}. 
Therefore, whereas $n_s$ is already very well constrained, $\alpha$ is not and it is compatible with zero. In fact, all the spectral parameters other than $n_s$ and $A_s$ are compatible with zero. In particular, the tensor-to-scalar ratio is only bounded from above (see Figure \ref{fig:r_vs_ns}). The current strongest bound on this parameter (which assumes $\alpha=0$) is (again at $k_*=0.05$ Mpc$^{-1}$) 
\be
\label{cmb1}
r< 0.07 \quad    (68\%\rm C.L.)\  , 
\ee
and is obtained including data from the BICEP/Keck array \cite{Array:2015xqh}. 
Since primordial gravitational waves have not been detected, the constraints on $n_t$ are very weak, and the consistency relation, $r=-8 n_t$, has not been tested. To satisfy the current CMB constraints, any model of inflation must predict the right value of $n_s$, {
\be
\label{cmb2}
n_s=0.969 \pm 0.004\quad    (68\%\rm C.L.)\  ,  
\ee
(see Fig. \ref{fig:r_vs_ns}), }
and a normalisation of the spectrum in agreement with 
\be
\label{cmb3}
A_s=(2.207\pm 0.103)\times 10^{-9} \quad (68\%\rm  C.L.)\ , 
\ee
both at the scale $k_*=0.05$ Mpc$^{-1}$ \cite{Ade:2015lrj}. From our previous discussion on the parameter dependence of our model of inflation \eq{genpotential}, we see that given a certain number of e-folds from $k_*$ to the end of inflation, the value of the non-minimal coupling $\xi$ is determined by $n_s$ at that scale. Once this is fixed, $\gamma$  is then determined by $A_s$, and so is $\lambda$. For instance, the Figure \ref{fig:r_vs_ns} shows that $\xi_\sigma\sim 10^{-2}$ is perfectly compatible with current bounds on $n_s$ and $r$ and it also gives an adequate number of e-folds. As the non-minimal coupling is decreased, the predictions of the model get closer to those of a standard quartic monomial potential (see Figure \ref{fig:r_vs_ns}), which is ruled out by the upper bound on $r$. Besides, the model generically predicts a small negative running of the scalar spectral index, of the order of $|\alpha|\sim 10^{-4} - 10^{-3}$, which may be probed in the future by 21cm line observations { \cite{Adshead:2010mc,Shimabukuro:2014ava}.}

For further reference, in Figs. \ref{fig:chi2} and \ref{fig:nechi2} we show the 95\% C.L. contours arising from \eqref{cmb1}, \eqref{cmb2}, and \eqref{cmb3} for the various parameters and observables of SMASH at the pivot scale of $0.05$ Mpc$^{-1}$. 
\begin{figure}[htbp]
\begin{center}
\includegraphics[width=6.7cm]{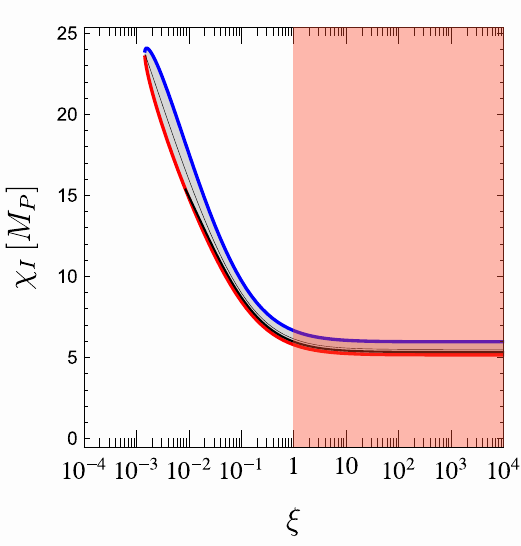} \includegraphics[width=7cm]{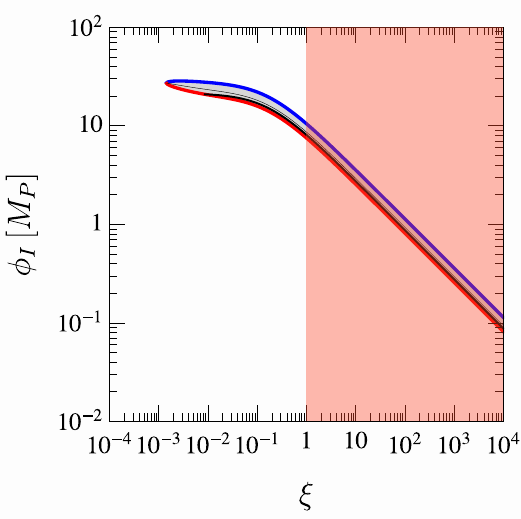}
\includegraphics[width=7.cm]{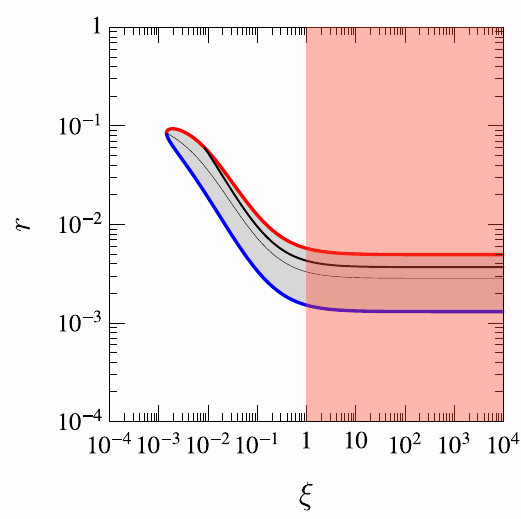} \includegraphics[width=7.1cm]{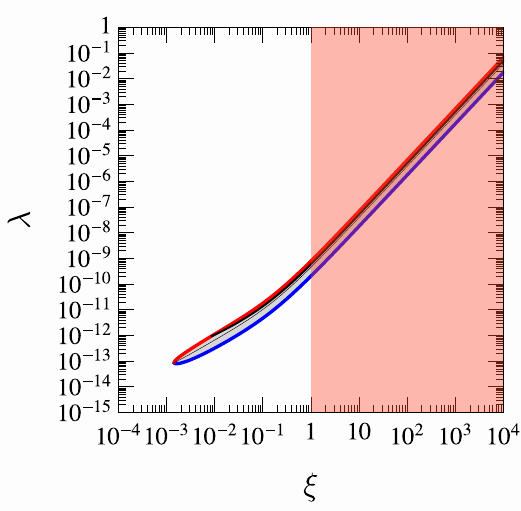}
\caption{\small {95\% C. L. contours for the parameters of our non-minimally coupled potential \eqref{genpotential} giving inflation as constrained by cosmological observations, all at the pivot scale $0.05$ Mpc$^{-1}$. Shown are: the canonical inflaton value $\chi_I$ (up left) and its non-canonically normalised $\phi_I$ (up right), the tensor-to-scalar ratio (down left) and the value of the quartic coupling (down right), as a function of the non-minimal coupling parameter $\xi$. The thin black line corresponds to the best fit for a given $\xi$, while the red and blue curves correspond to minimum and maximum values of $n_s$, i.e. to a redder or bluer primordial spectrum of curvature perturbations. The thicker black line corresponds to the predictions when taking into account the SMASH HHSI prediction of radiation domination immediately after inflation, see Sec.~\ref{sec:efolds}.} The shaded regions to the right of $\xi>1$ indicate, approximately, the region where the predictiveness of inflation starts to be threatened by the breakdown of peturbative unitarity.}
\label{fig:chi2}
\end{center}
\end{figure}

\begin{figure}[t]
\begin{center}
\includegraphics[width=7.2cm]{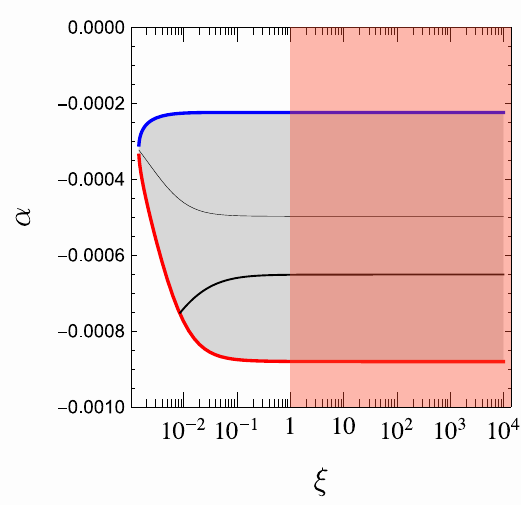} \includegraphics[width=7cm]{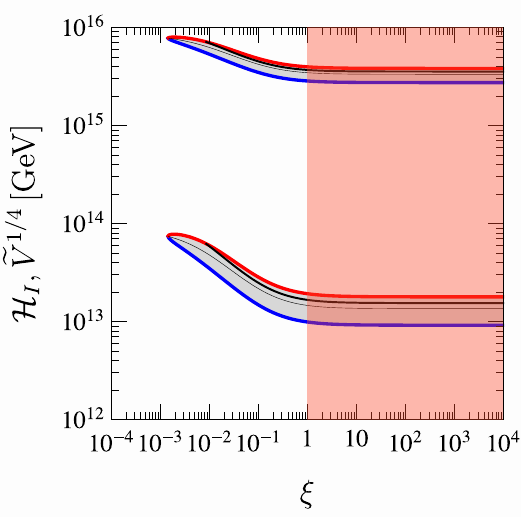}
\caption{\small  {95\% C. L. contours for the running of the spectral index (left) and Hubble scale and energy scale during inflation $H_I,\tilde V^{1/4}(\phi_I)$ (right) - ${\cal H}_I$ is the lower band. {Colour codes are like in Fig.~\ref{fig:chi2}.}}}
\label{fig:nechi2} 
\end{center}
\end{figure}

We point out that our conclusions of this section have been obtained using only the tree-level form of the action for inflation. A more detailed analysis should in principle include radiative corrections from matter and graviton loops. The generic form of these corrections for matter (which are the least suppressed) has been studied in \cite{Barvinsky:2008ia,Barvinsky:2009ii} and their estimated numerical value does not change significantly the results. 

\subsection{{Number of e-folds}} \label{sec:efolds}

As we will see in Sec.~\ref{sec:reheating}, the Universe expansion history in SMASH is remarkably simple, which allows to make a powerful connection between inflation and the CMB observations. In HHSI, inflation is followed by preheating and reheating, both of which happen with the Universe expanding as during radiation domination, i.e.\ there is no intermediate period of matter domination. 
In HSI, there is a small period of matter domination during a few e-folds before reheating, which we will neglect here. 
This allows to compute the number of e-folds between horizon crossing and the end of inflation in a simple way. 
For a mode with comoving momentum $k$, we have  
\begin{equation}
\label{eq:Reheatquart}
 N_e(k)\simeq \log\frac{a_{\rm eq}\, {\cal H}_{\rm eq}}{a_0\, {\cal H}_0}-\frac{1}{4}\log\frac{3\,{\cal H}_{\rm eq}^2}{M_P^2}-\log\frac{k}{a_0\,{\cal H}_0}+\frac{1}{2}\log\frac{V_k}{M_P^4}+\frac{1}{4}\log\frac{M_P^4}{V_{\rm end}}\,,
\end{equation}
where $V_k$ and $V_{\rm end}$  denote, respectively, the energy density at the time of the mode's horizon crossing and at the end of inflation. The subscripts ``$\rm eq$'' and ``0'' refer to the time of matter-radiation equality and today, see e.g.\ \cite{Liddle:2003as}. 
{By equating this expression with the result of integrating \eqref{eq:efolds} (or the simplest but less accurate expression \eqref{eq:efoldssimple}) and using $A_s$ to fit $\lambda$ we can compute the value of the inflaton field when a given scale $k$ exited the horizon (for a given $\xi$) and give a definite prediction for $n_s$ and $r$.  
We have used this expression to draw the prediction ({thick black} line) of Figure \ref{fig:r_vs_ns} and the thick black lines of Figs.~\ref{fig:chi2} and \ref{fig:nechi2}. }
}

\subsection{Analytical approximations}

In order to gain further insight on the properties of inflation driven by the field $\chi$ of \eq{gens} with the potential \eq{genpotential}, it is useful to obtain analytical expressions in the limits of large and small values of the non-minimal coupling $\xi$. We will see that simple analytical expressions can be obtained for $\xi$ larger than $\sim 5$ and smaller than $\sim 10^{-2}$. The intermediate region, see Figure \ref{fig:r_vs_ns}, has to be described by the equations \eq{gens} and \eq{genpotential}. {Below we obtain useful expressions in the large and small $\xi$ limits. Expressions for the spectral parameters in the strict limits of $\xi\rightarrow 0$ and $\xi\rightarrow\infty$ and an approximate interpolation between them can also be found in \cite{Bezrukov:2013fca}.}

\subsubsection{Large non-minimal coupling limit}

For large values of the fields, the relation \eq{chi_h_H_inf_sol} (or \eq{gens} for $b\simeq 1$) can be approximated by  
\begin{eqnarray}
\chi  \simeq  \sqrt{\frac{3}{2}} \,M_P\,\log\left[1+\xi\frac{\phi^2}{M_P^2}\right]\,,
\end{eqnarray}
and the potential \eq{genpotential} can be written as follows:
\begin{align}
\tilde V(\chi) \simeq \frac{\gamma}{4}M_P^4\left[1-\exp\left({-\sqrt{\frac{2}{3}}\frac{\chi}{M_P}}\right)\right]^2, 
\label{VgenL}
\end{align}
where $\chi$ is the (canonically normalised) inflaton and $\gamma$ is the effective coupling, defined in \eq{gammap}, that depends on the specific direction for which this limit applies. Concretely, whenever one of the two field directions ($h$ or $\rho$) has $\xi_i\gg 1$, and for  field values satisfying $M_P^2\ll\xi_i\phi_i^2$ and $\phi_i^2 \gg \xi_j^2\phi_j^2$ (with $j\neq i$), the expression \eq{VgenL} provides a good description of inflation. Numerically, we find that for values of the non-minimal coupling larger than $\xi\sim 5$, the error committed by using the potential \eq{VgenL} instead of \eq{genpotential} and \eq{gens} is negligible. Thus, the expression \eqref{VgenL} is appropriate for Higgs inflation. This potential is the same as that of Starobinsky's $R^2$ inflation \cite{Starobinsky:1980te}.

 Defining $\log x\equiv \sqrt{2/3}\,\chi/M_P$, the spectral parameters derived from the potential \eq{VgenL} in the slow-roll approximation are:
\begin{align}
\label{eq:slowrollxigg1}
A_s\simeq \frac{\gamma}{128\pi^2}\frac{(x-1)^4}{x^2}\,,\quad
n_s \simeq 1-\frac{8}{3}\frac{x+1}{(x-1)^2}\,,\quad \alpha\simeq-\frac{32}{9}\frac{x(x+3)}{(x-1)^4}\,,\quad r\simeq \frac{64}{3(x-1)^2}\,,
\end{align}
from which we obtain the relations
\begin{align} \label{relax}
\frac{n_s-1}{r}\simeq \frac{1+x}{8}\,,\quad\frac{(n_s-1)^2}{\alpha} \simeq -2\left(1+\frac{1}{x}\right)\,.
\end{align}
Planned probes of B-modes could reach a sensitivity of $\Delta r \simeq 10^{-3}$ \cite{ Matsumura:2013aja} or even $\Delta r \simeq 5 \times 10^{-4}$  \cite{Andre:2013afa}. Also, according to \cite{Mao:2008ug}, the running of the scalar spectral index might be measured with an absolute error of $10^{-4}$ with 21 cm tomography. Since the value of $x$ can be determined with an accurate prediction for the reheating process (and hence of the number of e-folds), this gives an extra handle to use the conditions \eq{relax}  to distinguish the model (in the large $\xi$ limit) from other models of inflation. 

Notice that these results are independent of $\gamma$, whose value can only be determined from the amplitude of the primordial spectrum $A_s\simeq 2\times10^{-9}$, giving
\begin{align} \label{ct}
\gamma\sim 10^{-10}\,.
\end{align}
Depending on the specific realisation of inflation that occurs among the various possibilities of SMASH listed in Table \ref{tab:summa}, this relation constrains the relative values of different couplings of the actual potential of the model \eq{VEpot}, see \eq{gammap}. In the case in which inflation proceeds along the Higgs direction (HI), the relevant value of $\lambda_H$ is the one evaluated at the field value of the Higgs during inflation, which is uncertain. With the current central values of the top and Higgs mass measurements, $\lambda_H$ is actually negative above a field value $\Lambda_I\sim 10^{11}$ GeV (see Fig. \ref{fig:LambdaI}), which  makes HI inconsistent.  In Section \ref{stability} we will explain how this problem can be overcome thanks to the hidden scalar $\sigma$. On the other hand, if $m_t$ is in reality close to the experimental lower limit, the coupling $\lambda_H$ remains positive in the SM at the scale relevant for inflation. Either way, $\lambda_H(\sim M_p)\sim {\cal O}(0.01)$ is a good estimate for $\lambda_H$ at such scales. This implies that
 \begin{equation}
\xi_H\sim 10^5 \sqrt{\lambda_H}\sim 10^4,
\label{ampl_constr_HI}
\end{equation} 
which is the usual result for standard Higgs {inflation. We see from \eq{ampl_constr_HI} that in order to have $\xi_H\sim 1$, the value of $\lambda_H$ (at $\sim M_P$) has to be $\sim 10^{-8}$. This small value of $\lambda_H$ already suggests that keeping it of that order for a range of $h$ sufficient for inflation would require a delicate tuning between the Higgs and top masses. As a matter of fact, in the SM and for $m_h=125.09$ GeV, the minimum possible value of $\lambda_H$ that can be reached at $M_P$ without running into an instability is $\lambda_H \sim 10^{-5}$,  which corresponds to a plateau with $m_t\sim 171.75$ GeV \cite{Ballesteros:2015noa}. Using \eq{ampl_constr_HI}, this implies $\xi_H\sim 3\times 10^2\gg 1$. Avoiding this constraint would necessarily require adding extra matter to the SM.}

If instead inflation is HSI or HHSI, the non-minimal coupling can be much smaller than $10^{4}$ while still having inflation in the large $\xi$ limit, see \eq{gammap}. The coupling $\lambda_\sigma$ can in principle take values much smaller than $\lambda_H$ at the scales relevant for inflation and the same holds for $\tilde\lambda_\sigma$ of \eq{tildes}. Indeed, notice that in the case of HHSI, $\xi$ could be much smaller than in standard Higgs Inflation without having to require a very small $\lambda_\sigma$, provided that $\lambda_{H\sigma}^2\sim \lambda_H\lambda_\sigma$.  For instance, $\xi\sim 1$ requires $\lambda_\sigma,\tilde \lambda_\sigma \sim 10^{-10}$.  It is worth recalling that in this discussion the couplings are implicitly evaluated around $M_P$, which is the order of magnitude relevant for the canonically normalised field during inflation. 

\subsubsection{Small non-minimal coupling limit} \label{smallnmc}

For small values of the non-minimal coupling, the flattening of the potential at large field values still occurs, but the predictions for inflation are different from those of standard Higgs Inflation because the potential approaches the constant value \eq{limitlf} with a different functional form. In this case, the second term of the expression \eq{chi_h_H_inf_sol} is negligible and therefore: 
\begin{align} \label{aq1}
\sqrt{\frac{\xi}{b}}\frac{\chi}{M_P} \simeq \text{arcsinh}\left(\sqrt{\xi}\frac{\phi}{M_P}\right)\,,
\end{align} 
where we recall that $b$ was defined in \eq{adef}.  The corresponding potential in this limit is
\begin{equation} \label{aq2}
\tilde V(\chi)\simeq\frac{\gamma}{4}M_P^4\,\left[{\rm tanh}\left(\sqrt{\frac{\xi}{b}}\frac{\chi}{M_P}\right)\right]^4\,.
\end{equation}
If one the two field directions ($h$ or $\rho$) satisfies $\xi_i\lesssim 1$ and if $M_P^2\ll\xi_i\phi_i^2$ and $\xi_i^2\phi_i^2 \gg \xi_j^2\phi_j^2$, then the mixing matrix $\mathcal{G}_{ij}$ of \eq{imet} becomes approximately diagonal and, after canonical normalisation, the potential \eq{aq2} is obtained. In practice, we find that the error committed by using $\eq{aq2}$ instead of \eq{genpotential} and \eq{gens} is negligible for $\xi$ smaller than $\sim 10^{-2}$. 

Unlike in the large non-minimal coupling limit, the inflationary predictions for $n_s$, $\alpha$, and 
$r$ now depend  explicitly on $\xi$. This can be seen in Figure \ref{fig:r_vs_ns}, where curves of constant $\xi$ get flatter as this parameter is increased. Using the standard slow-roll formulae and defining $u\equiv\sqrt{\xi/b}\,\chi$, we obtain:
\begin{align}
\begin{aligned}
\label{eq:slowrollxill1}
n_s \simeq\, 1+ \frac{8\xi}{b} \left( {\rm sech}^2u-3 \,{\rm csch}^2u\right)\,,\quad\quad  \alpha\simeq\, \frac{64\,\xi^2}{b^2} \left( {\rm sech}^4u-3 \,{\rm csch}^4u\right)\,,\\ \quad r\simeq\, \frac{ 512\,\xi}{b}\,{\rm csch}^2(2u),\,\quad\quad
A_s\simeq\, \frac{b\,\gamma}{768\pi^2\xi}\,{\rm sinh}^4u\,{\rm tanh}^2u\,.\quad\quad\quad \quad
\end{aligned}
\end{align}
As shown in Figure \ref{fig:r_vs_ns}, values of $\xi$ smaller than $\sim 10^{-3}$ are clearly excluded by the data, due to the upper bound on the tensor-to-scalar ratio. Therefore, the formulas provided above for the small non-minimal coupling limit only apply to a relatively narrow band of the allowed range for $\xi$. As in the large non-minimal coupling limit, the relations between the spectral parameters, $\xi$ and $N_e$ will allow to test inflation in this regime.

\section{The unitarity problem of Higgs Inflation as a motivation for SMASH} \label{Sunit}

{As argued in Section \ref{subsec:2field}, in the Einstein frame the theory has higher-dimensional operators, with an associated cutoff which, for small VEVs of the fields, goes as $\Lambda_U\sim {\rm min}\{M_P/\xi_i,M_P/\sqrt{\xi_\sigma \xi_H},M_P/\sqrt{\xi_i}\}$. In standard Higgs inflation, the large non-minimal coupling $\xi_H\sim 10^4$ required due to  condition \eq{ct}, or equivalently \eq{ampl_constr_HI}, lowers the perturbative unitarity cut-off down to}
\begin{align} \label{cutU}
\Lambda_U=\frac{M_P}{\xi_H}\ \sim 10^{14}\,{\rm GeV}\,,
\end{align}
which is below the Higgs field values $\sim M_P/\sqrt{\xi_H}\sim 10^{16}$\,GeV needed for inflation and comparable to the scale set by the height of the inflationary potential. In order to restore unitarity, physics beyond the SM is needed at or before $\Lambda_U$, presumably altering the inflationary dynamics. This implies that standard Higgs Inflation suffers from a serious problem of lack of actual predictive power \cite{Barbon:2009ya,Burgess:2009ea}. 

Of course, the same problem appears for inflation in the Higgs direction (HI) in SMASH. However, we saw in the previous section that the potential \eq{VEpot} may also provide inflation along the direction of $\rho$ (HSI) or along a mixed $h$--$\rho$ direction (HHSI) with $\xi_\sigma\ll 10^4$ (and keeping also $\xi_H$ small, even zero). In fact, we have seen that $\xi_\sigma$ can be of order 1 (or even smaller) and produce viable inflation. If $\xi_H$ is also small, the unitarity cut-off is then of the order of $M_P$, which is much higher than \eq{cutU}, effectively solving the predictivity problem of standard Higgs Inflation. 

This raises the obvious question of whether HSI and HHSI are compatible in SMASH with the non-inflationary (e.g.\ stability, baryogenesis and dark matter) constraints on the parameters of the model and, if so, for which values $\xi_\sigma$. We will address this question in later sections of the paper.  For the time being, we devote the rest of this section to discuss in more detail the unitarity problem and how it is circumvented in SMASH. The take home message of this section is twofold. First, we will show that the lack of predictive power of Higgs Inflation is a very strong motivation for adding the scalar field $\sigma$ of SMASH. Without such a scalar, a minimal extension of the SM such as the $\nu$MSM is incomplete, in the sense that it cannot really account for successful (and predictive) inflation. Moreover, we will show that preserving unitarity selects $\xi_\sigma\lesssim 1$. 

\subsection{The unitarity problem of Higgs Inflation}

In the SM, the Higgs field plays two roles. First, through the mechanism of spontaneous symmetry breaking, it gives mass to the other elementary particles. And second, it ensures unitarity in the scattering of gauge bosons. This second role is the fundamental reason why the Higgs (or another trick of nature) had to be expected around the electroweak scale. The idea of Higgs Inflation (with a large non-minimal coupling) entrusts the Higgs field with yet another task, which turns out to be too heavy a burden. As we have already discussed, in order to allow the Higgs to drive inflation compatible with the CMB data, it was proposed  that it should be coupled to the curvature scalar $R$ through a term in the action of the form \cite{Bezrukov:2007ep}
\begin{align} \label{hiterm}
\xi_H \int d^4x \sqrt{-g}\, R\,\,H^\dagger H\,,
\end{align}
where the coupling $\xi_H$ must be large: $\xi_H\sim 10^4$. It was shown in \cite{Burgess:2009ea,Barbon:2009ya} that such a coupling introduces the scale \eq{cutU}, at which perturbative unitarity is broken. For a large value of $\xi_H$, such as that required for inflation in the SM, the scale $\Lambda_U$ is significantly lower than $M_P$, where unitarity would anyway be expected to be lost. The early breakdown of unitarity implies that the physics must change at $\Lambda_U$, in order to ensure the consistency of the (completed) model as a quantum field theory. Presumably, some new particles  must become relevant at that scale (or even at a lower one).\footnote{A more exotic possibility would be that the model becomes strongly coupled at $\Lambda_U$ without having to introduce new degrees of freedom. Either way, unitarity must be restored and the predictiveness of the model for inflation above that scale is lost.}  At energies below the scale $\Lambda_U$ these particles must manifest as non-renormalisable operators, which appear suppressed by $\Lambda_U$ itself. 

As we anticipated, the consistency problem of Higgs Inflation stems from the fact that the Higgs field values required for inflation are of the order of $M_P/\sqrt{\xi_H}$, which is well above $\Lambda_U$. Since the preservation of unitarity requires that the action has to be modified at energies of the order of $\Lambda_U$, any prediction for field values $\sim M_P/\sqrt{\xi_H}\gg \Lambda_U$  should not be trusted, because it is out of the domain of validity of the effective field theory. 

This issue can be exemplified by modelling the feedback into the potential of the physics reinstating univarity in terms of non-renormalisable operators with cutoff $\Lambda_U$. Generically, the action below $\Lambda_U$ will contain an infinite series of operators that modify the Higgs potential. These operators are of the form $(H^\dagger H)^{N+2}/\Lambda_U^{2N}$, with $N\geq 1$, and their importance becomes larger as $H^\dagger H$ approaches $\Lambda_U^2$ from below. The predictive power of the model is gone at $H^\dagger H \sim M_P^2/\xi_H$,  since the action above the scale $\Lambda_U$ is not known. This is heralded by the hierarchy of non-renormalisable operators getting out of control.\footnote{If these operators were abnormally suppressed by very small coefficients, the problem could be milder. However, there is no reason to expect that the physics restoring unitarity should have such property.} The problem is particularly acute for inflation, whose predictions for the CMB critically depend on the shape of the potential (as well as on the form of the rest of the action). 

Another argument that shows the issue comes from noticing that the Hubble rate during inflation is precisely of the order of $\Lambda_U$, which means (via Friedmann equations) that changes in the effective action at that scale will be directly imprinted in the dynamics of inflation.\footnote{It was suggested in \cite{Bezrukov:2010jz} that an approximate shift symmetry might keep unitarity safe at $\Lambda_U$ while keeping the predictions for inflation unchanged. However, unless a UV completion with such a property is provided, the apparent asymptotic shift symmetry in the Einstein frame (for a sufficiently small variations of the Higgs) cannot be considered other than an accident. Such a UV completion has not been proposed so far.}

In summary, introducing a large coupling $\xi_H$ to extract primordial inflation out of the Higgs, lowers the unitarity cut-off to a value $\Lambda_U$  that hinders the predictivity of the model and threatens its self-consistency. One of our main goals in this paper is to construct an extension of the SM that allows to embed primordial inflation into it. In consequence, we must check whether our (SMASH) model solves the unitarity and predictivity problem that we have just explained. 

\subsection{Solving the unitarity problem in SMASH}

We can focus directly on the action \eq{Lmain} with the potential of equation \eq{VEpot}, with the fields written as in \eq{choiceg}.  As we have discussed, inflation with this potential can proceed along the Higgs direction $h$ (HI), the direction of the field $\rho$ (HSI), or along a combination of the two (HHSI); depending on the parameters of the model, see Table \ref{tab:summa}. Inflation along the direction $h$ suffers from the problem we have just described and so we must determine if the other possibilities (HSI and HHSI) are free from this issue. 

{A possible  way out of the unitarity problem of Higgs inflation was put forward in \cite{Giudice:2010ka} (see also \cite{Barbon:2015fla} for another approach). In the proposal of \cite{Giudice:2010ka},} inflation takes place along a mixed direction which runs almost parallel to $\rho$, with a very small contribution from $h$. This direction is characterised by our expressions \eq{threethetas} and \eq{tra}, in the limit of $\xi_H\ll \xi_\sigma$, and belongs to the class we call HHSI. In the scenario of \cite{Giudice:2010ka}, the unitarity cut-off is thus set above \eq{cutU} by freeing the Higgs from the task of providing inflation. A very large $v_\sigma^2$ in \eq{VEpot} gives an important contribution (of the order of $\xi_\sigma v_\sigma^2$) to the actual value of the observed Planck mass squared $M_P^2$, given by \eq{eq:MMP}. This mechanism allows to raise the unitarity cut-off up to $\sim M_P$,  provided that $M_P^2\sim \xi_\sigma v_\sigma^2$.  An explicit expression for the unitarity breaking scale  as a function of the non-minimal coupling $\xi_\sigma$ is given by  \cite{Bezrukov:2010jz,EliasMiro:2012ay}:
\begin{align}\label{UB}
\Lambda_U=\frac{M^2+\xi_\sigma\,v_\sigma^2+6\xi_\sigma^2\, v_\sigma^2}{\xi_\sigma\sqrt{M^2+\xi_\sigma\, v_\sigma^2}}\,,
\end{align}
where $M_P^2\simeq M^2+\xi_\sigma\, v_\sigma^2$, using \eq{eq:MMP} and neglecting $\xi_H\,v^2$. {As mentioned before, this result can be easily read from the kinetic part of the action in the Einstein frame,   which controls the scattering of scalars around the vacuum of the theory.} For $\xi_\sigma\gg 1$ and $\xi_\sigma\,v_\sigma^2\sim M_P^2/6$, the unitarity breaking scale \eq{UB} is of the order of $M_P$, thus solving the issue of Higgs Inflation. An interesting feature of this ``unitarisation'' mechanism is that since inflation takes place in the large $\xi_\sigma$ limit, the predictions for inflation are essentially analogous to those of Higgs Inflation. 

However, the regime of \cite{Giudice:2010ka} is not consistent in SMASH because it requires a too large value of $f_A\sim v_\sigma$. Indeed, for $\xi_\sigma\,v_\sigma^2\sim M_P^2/6$ and $\lambda_\sigma\sim 4\pi$ (at the limit of perturbativity), the constraint \eq{ct} implies $f_A\gtrsim 10^{15}$ GeV, which is well above the values allowed for SMASH. We recall that the value of the Hubble scale during inflation for SMASH is in the range $\mathcal{H}\sim (10^{13}, 10^{14})$ GeV, see Figure \ref{fig:nechi2}.  This means that such a large $f_A$ is ruled out by the bounds on axion isocurvature fluctuations, which we discuss  in Section \ref{dark_matter}. Moreover, as we will explain in Section \ref{dark_matter}, in order to account for all the dark matter with the axion, the value of $f_A$ cannot be larger than {$\sim 1.2\times 10^{11}$ GeV,} which sets an even stronger upper bound. Notice that values of  $\lambda_\sigma$ smaller than $4\pi$ (such as those expected for inflation) only make $f_A$ larger than $10^{15}$ GeV and so more incompatible. In this argument we have assumed that inflation happens in the limit HSI, in which $h$ does not participate in the dynamics. Using \eq{tildes}, it is straightforward to check that HHSI requires even higher values of $f_A$ than HSI. 

Fortunately, there is however a different possibility to approach the unitarity problem within SMASH. It consists simply in  assuming small values for both $\xi_H$ and $\xi_\sigma$, but respecting the hierarchy $\xi_H\ll \xi_\sigma$. This is the small non-minimal coupling that we discussed in subsection \ref{smallnmc}. The inflationary potential in this limit, given by \eq{aq2}  is different from \eq{VgenL} and therefore the predictions for the primordial parameters and the number of e-folds deviate from those of the large non-minimal coupling limit; see Figure \ref{fig:r_vs_ns}. The expressions \eq{eq:MMP} and \eq{UB} reveal that now, for $\xi_\sigma\lesssim 1$ and $v_\sigma\ll M_P$, the Planck mass  is $M_P\sim M$ and $\Lambda_U$ goes formally above $M_P$; effectively pushing the unitarity breaking scale up to the Planck scale. This gets rid of the issue of loss of predictiveness by unitarity breaking in standard Higgs Inflation and is a very strong argument for the introduction of the extra scalar in SMASH.

The field values during inflation in the small non-minimal coupling limit are Planckian, both for $\chi$ and $\phi(\chi)$, like in any other model of large field inflation. Whether this is an issue (and if so, finding a solution) depends on knowing in detail how gravity behaves at such scales, which goes beyond the scope of our work.  Having said this, we stress that the improvement with respect to Higgs Inflation --which also requires Planckian scales, well above the breaking of unitarity at $\sim 10^{-4} M_P$ in that case-- is very significant. Besides, most large field models of inflation lack a detailed particle physics implementation, which we provide here with SMASH in a very well motivated scenario. Therefore, the consistency and predictive power of inflation in SMASH is considerably better than in any other minimal extension of the SM in which the inflaton is identified with the Higgs, as for e.g.\ in \cite{Asaka:2005an,Asaka:2005pn} or \cite{Salvio:2015cja} supplemented by Higgs Inflation \cite{Bezrukov:2007ep}. 

To summarise: in order the solve the predictivity problem of standard HI, inflation in SMASH must be of the types HSI or HHSI, with $\xi_\sigma\lesssim 1$. Then, the constraint \eq{gammap} from the requirement of fitting the amplitude of primordial perturbations implies that the corresponding quartic coupling has to be small $\lambda\equiv (\lambda_\sigma\, \text {or}\,\tilde \lambda_\sigma) \lesssim 5 \times 10^{-10}$. Taking also into account the current upper limit on primordial tensor modes, we read from Figure \ref{fig:chi2} the following ranges for the effective quartic {coupling} during inflation, {compatible with the Planck/BICEP results:}
\be
\label{inflationconstraint}
6\times 10^{-3} \lesssim \xi_\sigma \lesssim 1 \quad \text{which implies}\quad
5\times10^{-10} \gtrsim \lambda\gtrsim  5\times10^{-13},
\ee
{where the limits} {come from} {the predictions accounting for the post-inflationary evolution in SMASH, given by the thick black lines in Figures \ref{fig:r_vs_ns}, 
\ref{fig:chi2} and \ref{fig:nechi2}. Along these lines, the windows of allowed values for the cosmological parameters $r$, $n_s$ and $\alpha$ are:
\be
\label{eq:cosmoranges}
\begin{aligned}
 &0.960\lesssim n_s\lesssim0.964,\quad 0.0038\lesssim r\lesssim0.07,\quad -6.8\times10^{-4}\lesssim\alpha\lesssim-5.8\times10^{-4}, \quad \text{at}\quad k=0.05 \,\,{\rm Mpc}^{-1},\\
 &0.962\lesssim n_s\lesssim0.966,\quad 0.0042\lesssim r\lesssim0.07,\quad -7.5\times10^{-4}\lesssim\alpha\lesssim-6.5\times10^{-4}, \quad \text{at}\quad k=0.002 \,\,{\rm Mpc}^{-1}.
\end{aligned}
\ee
}
{In the next section we will see that the range \eqref{inflationconstraint}, when combined with  constraints from stability and perturbativity, implies $\lambda_H/|\lambda_{H\sigma}|\gg1$, justifying the choice $b\simeq1$ in Section \ref{inflation} (see equation \eqref{adef}), and implying that HHSI inflationary trajectories remain close to HSI ones, as follows from equation \eqref{tra}.}


\section{\label{stability}Vacuum stability}

It is well known that in the SM the top Yukawa coupling can drive the Higgs quartic coupling to negative values. Indeed, with the current central values of the Higgs and top quark masses~\cite{Aad:2015zhl,ATLAS:2014wva}, the effective SM potential becomes negative at large Higgs values, see e.g.\ \cite{Degrassi:2012ry,Buttazzo:2013uya,Bednyakov:2015sca}. We show in Figure~\ref{fig:LambdaI} the instability scale $\Lambda_I$ at which the SM Higgs quartic coupling crosses zero, as a function of the top mass, for fixed $m_h=125.09$ GeV. We show in blue the  uncertainty bands due to the dominant experimental error in the measurement of $\alpha_s(m_Z)=0.1185\pm0.0006$ \cite{Agashe:2014kda}. Our results are compatible within these bands with the state-of-the art calculation of \cite{Bednyakov:2015sca}, which includes further two-loop electroweak corrections in the determination of the top Yukawa coupling from the experimental measurements of the top mass, $m_t$. The two-sigma band due to the effect of varying $m_h$ within its experimental error, although not shown in Figure \ref{fig:LambdaI}, roughly overlaps with the one-sigma contour associated to $\alpha_s(m_Z)$.

\begin{figure}
\begin{center}
\includegraphics[width=0.6\textwidth]{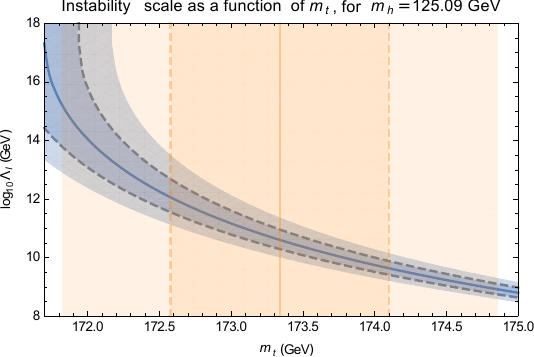}
\caption{  \small The instability scale $\Lambda_I$ at which the SM Higgs quartic coupling becomes negative as a function of the top mass, for $m_h=125.09$ GeV. We show the deviations corresponding to varying $\alpha_s(m_Z)$ within its one (darker blue shade) and two-sigma (lighter blue shade) experimental uncertainties. The orange shades correspond to the one and two-sigma uncertainties of the combined  measurement of the top quark's mass in reference \cite{ATLAS:2014wva}. The curves have been obtained using the tree-level potential, improved by two-loop RGEs.}  
\label{fig:LambdaI}       
\end{center}
\end{figure}

This instability of the SM effective potential threatens very severely the possibility of having inflation by minimally coupling the Higgs to $R$, because this kind of inflation requires Planckian field values, that are generically above or around the instability scale of the potential. This is especially important for particle physics models such as the $\nu$MSM, which postulates a vast desert of scales in between the Higgs mass and the Planck mass. If more refined measurements and calculations with increased precision strengthen the evidence for the SM instability, the idea of Higgs Inflation (HI) without extra higher energy degrees of freedom should definitely be abandoned.\footnote{An idea to make compatible the instability of the effective potential of the SM with Higgs Inflation was suggested in \cite{Bezrukov:2014ipa}, which assumes (among other things) the existence of an unknown UV completion with very specific asymptotic properties. }   This provides further motivation (alongside the unitarity issue of HI) in favour of HSI and HHSI in SMASH, but we must check that these scenarios do not run into a similar instability problem. Therefore, in this section we determine the ranges of parameters of SMASH for which the effective potential is positive up to very large values ($\sim 30\, M_P $) of the fields, in order to guarantee successful inflation. 

For this aim, it is sufficient to focus exclusively on the behaviour of the (Jordan frame) effective potential of SMASH at large field values, since the non-minimal couplings $\xi_\sigma$ and $\xi_H$  will not affect our considerations on the stability.\footnote{They change the depth of the tree-level minima, but they do not determine whether unstable vacua (that may originate due to the quartic couplings running negative) exist or not.  However, if the potential had vacua with negative energy, the non-minimal couplings would affect the probability of tunneling into them. For $\xi_i<0$, the field receives a positive contribution to its effective
mass, which makes it less likely to fluctuate during inflation. Conversely, if $\xi_i>0$, the effective mass is diminished, making  the tunneling more likely.} We recall the tree-level potential of SMASH, equation \eq{scalar_potential}:
\begin{equation}
\label{scalar_potential2}
V(H,\sigma )= \lambda_H \left( H^\dagger H - \frac{v^2}{2}\right)^2
+\lambda_\sigma \left( |\sigma |^2 - \frac{v_{\sigma}^2}{2}\right)^2+
2\lambda_{H\sigma} \left( H^\dagger H - \frac{v^2}{2}\right) \left( |\sigma |^2 - \frac{v_{\sigma}^2}{2}\right)\,.
\end{equation}
A minimum with $V=0$ is attained at  $h=v,\rho=v_\sigma$ provided that 
\begin{align}
\lambda_H(\mu),\lambda_\sigma(\mu)>0\quad \quad \text{and}\quad \quad\lambda_{H\sigma}^2(\mu)<\lambda_H(\mu)\lambda_\sigma(\mu)\,\quad\text{ for }\quad\mu\lesssim m_\rho. \label{cond1}
\end{align}
The choice of scale $\mu\lesssim m_\rho$ reflects the fact that the constraint comes from considering the potential near the vacuum, which is  more accurately described by choosing a renormalisation scale of the order of the largest of the particle masses in that field region, which is typically $m_\rho$. As will be seen later, the second identity in \eqref{cond1} can be violated at higher scales --corresponding to the potential evaluated at larger values of the fields-- while still maintaining stability and a minimum at  $h=v,\rho=v_\sigma$.
 In the previous section we learnt that solid predictions in  HSI require the upper bound $\lambda_\sigma(\mu\sim M_P)\lesssim 10^{-9}$ (for $\xi_\sigma\lesssim 1$).  
At this level, the condition \eq{cond1} implies $\lambda_{H\sigma} \lesssim 10^{-5}$ at low scales (using $\lambda_H\sim 0.1$ near the Higgs vacuum, and assuming that $\lambda_\sigma$ does not run much between inflationary and low scales, as does indeed happen for small $\lambda_{H\sigma}$ and small enough Yukawas).
In the HHSI case, the role of $\lambda_\sigma$ is played by $\tilde \lambda_\sigma=\lambda_\sigma-\lambda_{H\sigma}^2/\lambda_H$, which allows larger values of $\lambda_\sigma$, and thus of $\lambda_{H\sigma}$,  if there is some degree of cancellation. Notice that a small value of $\lambda_{H\sigma}$ is consistent with the choice of $b\simeq 1$ --defined in \eq{adef}-- that we made when we studied the inflationary parameter space of HHSI leading to Figures \ref{fig:chi2} and \ref{fig:nechi2}.

We will be interested in {\it absolute stability} of the full effective potential (including radiative corrections), which in our case is the requirement that the SMASH potential has to be positive everywhere away from its $V=0$ absolute minimum. Absolute stability ensures ``by brute force'' that no tunnelling into unstable regions of the potential will occur during inflation. In general, tunnelling is an important concern, since fields sufficiently lighter than the Hubble scale, $\mathcal{H}$, acquire large quantum fluctuations (proportional to $\mathcal{H}$) during inflation. This has been studied for the Higgs in various works \cite{Espinosa:2007qp,Degrassi:2012ry,Lebedev:2012sy,Kobakhidze:2013tn,Fairbairn:2014zia,Enqvist:2014bua,Hook:2014uia,
Herranen:2014cua,Shkerin:2015exa,Kearney:2015vba,Espinosa:2015qea,East:2016anr}.  Nevertheless, it is worth stressing that  if the potential were unstable, successful inflation could in principle be possible along specific directions away from the instability, as long as inflation ends in a long-lived stable vacuum, with quantum fluctuations into unstable regions suppressed both during and after inflation. This would further open the allowed parameter region of SMASH. For instance, sufficiently large non-minimal couplings to gravity can protect from such instabilities \cite{Espinosa:2007qp,Espinosa:2015qea,Herranen:2014cua}, allowing the construction of models in which large-field instabilities do not spoil the cosmological evolution of the scalar fields. However, after inflation, thermal fluctuations could drive the fields into unstable regions if the reheating temperature is high enough.\footnote{Using the finite-temperature effective potential (see Appendix \ref{finiteT}), we have calculated thermal tunneling rates towards the Higgs instability, concluding that they can only be problematic for large top masses and for temperatures above the typical reheating temperatures produced in  Higgs or hidden scalar inflation (see Section \ref{sec:reheating}). We find results similar to those of \cite{Rose:2015lna} for the SM: for a Higgs mass of 125.09 GeV, thermal tunneling requires $m_t>174$ GeV, and for this limiting value the temperature needs to be above $10^{17}$ GeV.} For simplicity, we restrict our analysis in this paper to the context of absolute stability, proving that successful inflation is viable in SMASH, but it should be kept in mind that this may not cover all the viable possibilities.

\subsection{Stability in the singlet direction and its interplay with leptogenesis and axion dark matter}

Instabilities in the effective potential of SMASH can arise not only from fermionic quantum corrections to $\lambda_H$, but also to $\lambda_\sigma$, i.e.\ both in the Higgs and $\rho$ directions. 
As we just mentioned, in the Higgs direction, an instability can be triggered mainly by the top quark Yukawa; whereas in the $\rho$ direction the Yukawas of the RH neutrinos, $Y_{ii}$, and the extra quark $Q,\tilde Q$, denoted $y$, are the main negative contributions to the beta function of the quartic {coupling $\lambda_\sigma$.} The $\rho$ direction may actually be critical in SMASH because of the small values of $\lambda_\sigma$ required for inflation and the role of this parameter in leptogenesis and axion dark matter, which we study in Sections \ref{baryogenesis} and \ref{dark_matter}, respectively.

At one loop, the most relevant contributions to the running of $\lambda_\sigma$ are\footnote{See Appendix \ref{betas} for the full two-loop beta functions.}
\be \label{b1}
\frac{d \lambda_\sigma}{d \ln \mu} = \frac{1}{16\pi^2}\(-\sum_iY_{ii}^4- 6 y^4+ 20\lambda_\sigma^2  +8\lambda_{H\sigma}^2 + ... \) .   
\ee
Since $y$ can be chosen very small without running into trouble and since $\lambda_\sigma$ must be very small due to the inflationary constraint \eq{inflationconstraint}, we can neglect their effects on \eq{b1} for the rest of the argument. Therefore, we only have the competing effects of $Y_{ii}$'s and $\lambda_{H\sigma}$, which tend to drive $\lambda_\sigma$ in opposite directions. Using \eq{b1} and assuming that the Yukawas run slowly, we can easily estimate the minimum value of $\sum_iY_{ii}^4$ that would make the one-loop corrections to $\lambda_\sigma$ at inflationary field values (say, $\mu\sim 30\, M_P$) comparable to its upper bound of $10^{-9}$, see \eq{inflationconstraint}. This condition is simply
\be
\label{naivestability}
\lambda_\sigma(m_\rho)+\left.\Delta \lambda_\sigma\right|_Y 			\sim \lambda_\sigma(m_\rho)-\frac{\sum_iY_{ii}^4}{16\pi^2} \ln \(\frac{\mu}{m_\rho}\)>0\,.
\ee

As reviewed in Section \ref{baryogenesis}, vanilla leptogenesis requires the lightest RH neutrino mass to satisfy $M_1 = Y_{11}v_\sigma/\sqrt{2}>5\times 10^8$~GeV,  and it also demands a sufficient hierarchy, for instance $Y_{33}=Y_{22}=3Y_{11}$ \cite{Buchmuller:2002rq}. 
With this hierarchy, $\sum_iY_{ii}^4=163\ Y_{11}^4$ and \eqref{naivestability} implies 
\begin{align}
\label{Y11stability}
Y_{11}\lesssim 2\times 10^{-3} \({\lambda_\sigma}\times{10^{9}}\)^{1/4}\,,
\end{align}
where $\lambda_\sigma$ is evaluated at $\mu \sim 30\, M_P$. 
But the leptogenesis constraint implies a lower limit, 
\be
\label{vanillaY11}
Y_{11}\gtrsim 3.5\times 10^{-2}\, \frac{10^{11}\, \rm GeV}{f_A} \,.
\ee
Therefore, once $f_A$ is determined by axion dark matter considerations, see Section \ref{dark_matter}, vanilla leptogenesis requires a minimum value of the lightest RH neutrino Yukawa that may easily enter in conflict --depending on $f_A$-- with the requirements on the parameter $\lambda_\sigma$ for (stable and predictive) HSI inflation.  

The bound \eqref{vanillaY11} could be evaded if the positive contribution to the running of $\lambda_\sigma$ from the portal coupling, $\lambda_{H\sigma}$, were sufficiently large. This would require $|\lambda_{H\sigma}|\gtrsim 10^{-5}\sqrt{\lambda_\sigma\times 10^{9}}$, which contradicts the low-scale condition  of equation \eq{cond1}: $|\lambda_{H\sigma}|<\sqrt{\lambda_\sigma\lambda_H}\sim 10^{-5}\sqrt{\lambda_\sigma\times 10^{9}}$ (for the largest allowed value of $\lambda_H(m_h)=0.13$). Thus, at first glance, it seems that the effects of $\lambda_{H\sigma}$ in the running of $\lambda_\sigma$ cannot help stabilising the latter while maintaining at the same time the desired spontaneous symmetry breaking at low scales.

Notice that with inflation of HHSI type, there is more freedom to play with the parameter space than in HSI, given the fact that the effective quartic coupling relevant for inflation, $\tilde\lambda_\sigma$, differs from $\lambda_\sigma$, see \eq{tildes}. This allows predictive (i.e.\ free from unitarity issues) inflation with larger values of $\lambda_\sigma$, while still satisfying the constraints from vanilla leptogenesis and stability. In any case, as will be elaborated in Section \ref{baryogenesis}, the vanilla leptogenesis bound of equation \eqref{vanillaY11} can always be evaded by accommodating some degree of resonant enhancement of leptogenesis \cite{Pilaftsis:1997jf,Pilaftsis:2003gt}. This requires some degeneracy among the  Yukawas for the RH neutrinos, which may take values smaller than the bound in \eqref{vanillaY11}.

The simple estimates of this subsection indicate that a thorough analysis of the stability conditions in SMASH is necessary to clarify whether the model can easily be compatible with successful and predictive inflation, axion dark matter and leptogenesis; and if so, whether the latter can be vanilla-like or with some resonant enhancement. We will present detailed numerical analyses in Section \ref{subsec:numerics}, but first we will move on to consider stability in the Higgs direction and characterise the restrictions in the SMASH parameter space imposed by the requirement of stability.

\subsection{Stability in the Higgs direction: threshold  mechanism}

The (apparent) stability in the Higgs direction in the SM can be improved in the presence of an additional scalar \cite{Lebedev:2012zw,EliasMiro:2012ay}, such as the $\sigma$ field in SMASH. Aside from the obvious positive contribution of the Higgs portal coupling $\lambda_{H\sigma}$ to the beta functions of the Higgs quartic coupling,\footnote{See e.g.\ \cite{Gonderinger:2009jp} for a discussion of this effect in the context of scalar singlet dark matter.} which is analogous to \eq{b1}, there is a more powerful effect that occurs if the additional scalar is heavy and acquires a large VEV. In this situation, the SM has to be understood as a low energy effective theory arising when the heavy field is integrated out.\footnote{We recall that in SMASH the axion also survives at scales below the heavy fermions and scalar modulus $\rho$, but it does not play any role in what follows.} The VEV of this  heavy scalar can make the Higgs quartic coupling at high energies larger than in the SM, which may be enough to guarantee positivity of the potential in that direction. This is known as the threshold stabilisation mechanism \cite{EliasMiro:2012ay,Lebedev:2012zw}. Notice that in SMASH the smallness of $\lambda_{H\sigma}\lesssim 10^{-5}$ in comparison to $\lambda_H$ make stabilisation impossible from the change in the beta function alone, and the threshold effect is needed. Since dark matter constraints require a large $f_A=v_\sigma\sim 10^{11}$ GeV (see  Section \ref{dark_matter}), SMASH is ideally suited for this mechanism, which we review in this section. 

We will first study the generic properties of the sign of the potential for large and small field values. That analysis will serve us to characterise precisely the stability conditions under the threshold mechanism. 

\subsubsection{Effective potential in the large-field and small-field regimes}
At energies below $m_\rho$, the  field $\rho$ can be integrated out by using its equation of motion at zero momentum \be
\label{sline1}
\frac{\partial V}{\partial \rho} =0 \quad \implies \quad \rho^2 - v_\sigma^2= \frac{\lambda_{H\sigma}}{\lambda_\sigma}(v^2-h^2) , 
\ee 
which we substitute into the potential \eqref{scalar_potential2}  {to get an effective potential which should be matched with that in the SM:}
\be
\label{scalar_potential_lambdabar}
V(h) = \frac{\overline\lambda_H}{4}\left( h^2 - v^2\right)^2,
\ee
where
\begin{align}
\label{eq:treematching}
\overline\lambda_H=\lambda_H -\frac{\lambda_{H\sigma}^2}{\lambda_\sigma}.
\end{align}
{Similarly, defining quadratic parameters $m_\sigma^2\equiv -\lambda_\sigma v_\sigma^2-\lambda_{H\sigma}v^2$  and $m^2_H\equiv -\lambda_H v^2-\lambda_{H\sigma}v_\sigma^2 $, the SM matching imposes
the following relation with the SM quadratic Higgs coupling $\bar m^2_H$ (of the order of the physical Higgs mass squared \footnote{{In SMASH, the physical} Higgs mass arises as a tree-level cancellation of two large quantities. This is a generic property of all Higgs portal models with a large VEV. For a brief discussion of the degree of tuning associated to this expression see Appendix \ref{tune}.}),}
\begin{align}
\label{eq:massmatching0}
 \bar m^2_H=m^2_H-\frac{\lambda_{H\sigma}}{\lambda_\sigma}m^2_\sigma.
\end{align}

The essence of the threshold stabilisation mechanism lies in the fact that the low energy parameter characterising the Higgs mass measured at the LHC is not $\lambda_H(m_h)$ but the combination $\overline\lambda_H(m_h)=\left. \lambda_H-\lambda_{H\sigma}^2/\lambda_\sigma \right|_{\mu=m_h}$. 
The quartic coupling $\lambda_H$, gets revealed only at sufficiently high energies, integrating $\rho$ in.

Notice that $\lambda_H$ is larger than $\overline\lambda_H $ (provided that $\lambda_\sigma$ is positive) and can stay positive up to very high energies if the tree-level threshold correction 
\be \label{THc}
\delta \equiv \frac{\lambda^2_{H\sigma}}{\lambda_\sigma} , 
\ee
is sufficiently large. Of course, a necessary requirement for the threshold mechanism to work is that the matching scale $m_\rho$ must be smaller than the SM instability scale $\Lambda_I$.

Let us now discuss in detail the conditions for stability. For large values of the fields, we can neglect the VEVs in the SMASH potential \eqref{scalar_potential2} to find, 
\be
V\simeq \frac{\lambda_H}{4}h^4+\frac{\lambda_\sigma }{4}\rho^4+\frac{\lambda_{H\sigma}}{2}\rho^2h^2, 
\ee
which is always positive if
\be
\label{crit_largefield}
\lambda_H>0 , \quad \lambda_\sigma>0 \quad \text{and} \quad  \lambda_{H\sigma} > -\sqrt{\lambda_H\lambda_\sigma} \,  .
\ee

In the low-field regime, $v$ and $v_\sigma$ cannot be neglected and quadratic interactions become important. Since we need $\lambda_H>0$ and $\lambda_\sigma>0$ at large values of the fields and they both typically grow at lower scales, new instabilities can only come from the portal coupling. For convenience, we rewrite the potential for $h$ and $\rho$ as  
\bea
\label{scalar_potential3}
V &=& \frac{\lambda_\sigma v_\sigma^4}{4}\( \Theta\,p_1^2 +p_2^2-2 \Theta\,p_1\, p_2\) \\
\label{scalar_potential4}
&=& \frac{\lambda_\sigma v_\sigma^4}{4}\[ (\Theta\, p_1 -p_2)^2 + p_1^2\,\Theta\,(1-\Theta)\] \, ,   \\
\label{eq:ab}
{\rm with} &&p_1 =  \frac{h^2-v^2}{\Lambda_h^2} \, , \quad p_2 = 1-\frac{\rho^2}{v_\sigma^2},  \\
{\rm and} && \Lambda_h^2 = \frac{\lambda_{H\sigma}}{\lambda_H}v_\sigma^2 , 
\eea
where in the last equation the values of the couplings have to be taken at the scale $\Lambda_h$ and the equation solved self-consistently. 
Most importantly, we have defined the most relevant parameter for the analysis of stability,  
\be
\Theta = \frac{\lambda_{H\sigma}^2}{\lambda_H\lambda_\sigma}=\frac{\delta}{\lambda_H}\,, 
\ee
where $\delta$ was introduced in \eq{THc}.

With positive quartic couplings, $\lambda_H>0,\lambda_\sigma>0$, the last term of the potential \eqref{scalar_potential4}  is the only one that can be negative, which happens if $\Theta>1$. The other term is always non-negative (and exactly zero along $\Theta\, p_1 - p_2 =0$). The closer are the fields to the curve $\Theta\, p_1 - p_2 =0$, the more relevant will be the destabilising effect of a $\Theta >1$. Clearly,
\bea
\Theta <1\quad \text{and}\quad \lambda_H>0 \quad &\Longrightarrow& 
\quad \tilde \lambda_H =  \lambda_H - \frac{\lambda^2_{H\sigma}}{\lambda_\sigma}>0 \, ,    \\
\Theta <1\quad \text{and}\quad \lambda_\sigma >0 \quad&\Longrightarrow& 
\quad \tilde \lambda_\sigma =  \lambda_\sigma - \frac{\lambda^2_{H\sigma}}{\lambda_H}>0 \, ,    
\eea
where $\tilde\lambda_\sigma$ was already introduced in \eq{tildes}. Importantly, $\tilde \lambda_H$ runs with the couplings in the full SMASH model, and so it is different from the previously defined $\overline \lambda_H$, see \eq{eq:treematching}, which runs {(by definition)} only with the SM field content. Armed with these results, we can now easily determine the stability conditions by considering two cases, depending on the sign of $\lambda_{H\sigma}$.\footnote{The sign of $\lambda_{H\sigma}$ in SMASH can change under the renormalisation flow due to the box diagram involving RH neutrinos and leptons, which contributes with a term $-2{\rm Tr}\[Y^\dagger Y F^\dagger F\]$ to the beta function of $\lambda_{H\sigma}$, see Appendix \ref{betas}. This term is comparable in size to the RH neutrino contribution to $\beta_{\lambda_\sigma}$, but in SMASH $\lambda_{H\sigma}$ is much larger than $\lambda_\sigma$, so if we solve the $\lambda_\sigma$ instability we can be quite sure that this effect is small and the sign will not change.}

\subsubsection{Case $\lambda_{H\sigma}>0$}
The portal term of \eqref{scalar_potential2} can only become negative in one of two separate regions: 1) $\{h>v\,,\rho<v_\sigma\}$ and 2) $\{h<v\,,\rho>v_\sigma\}$. Let us focus now on the first one. This region is away from the HSI and HHSI inflationary paths, but as we have explained, our aim is to stabilise the potential in the whole field space. In this region, the parameters $p_1$ and $p_2$ of equation \eqref{eq:ab} satisfy $p_1> 0$ and $p_2\in (0,1]$. The second term in the potential \eqref{scalar_potential3} is always positive (for $\lambda_\sigma>0$ and $\lambda_H>0$) and minimum for $p_2=0$, while the last term is always negative and minimum for $p_2=1$.  Setting the second term to zero and the third to its most negative value, we have the inequality 
\be
V(h,\rho) > \lambda_\sigma v_\sigma^4 \Theta p_1 ( p_1 -2) 
=\lambda_H (h^2-v^2)\(h^2-v^2-2\Lambda_h^2\).
\ee
We see that the potential can be negative only for $p_1<2$, which is equivalent (neglecting $v$ in comparison with $h$) to $h<\sqrt{2}\Lambda_h$. Therefore we need to impose $\Theta<1$ in that region. The scale $\sqrt{2}\Lambda_h$ constitutes then the practical boundary between the large and small $h$ field regimes.\footnote{
Equation \eqref{scalar_potential4} allows to understand in a different way the stability criteria derived in \cite{Ballesteros:2015iua}. For intermediate values, this reference identified regions where the quartic and quadratic terms of the potential compete, and it was argued that stability should be imposed there, since those regions are more prone to develop negative values of the potential. The curve $\Theta a-b=0$ corresponds to the ``s-line" of \cite{Ballesteros:2015iua} and the criterion $\lambda_H-\lambda_{H\sigma}^2/\lambda_\sigma>0$ is $\Theta<1$. The second region (``h-line") of \cite{Ballesteros:2015iua} can extend much further in $h$ than $h=\sqrt{2}\Lambda_h$, and $\Theta<1$ was considered as well a necessary condition there. Here we have shown that assuming $\lambda_H,\lambda_\sigma>0$, then $\Theta<1$ for $h<2\Lambda_h$ is a sufficient condition for stability in the whole region, so that an instability in the h-line for $h>\sqrt{2}\Lambda_h$ can only come from a violation of the positivity of the quartics. 
}
We can easily get an idea of the hierarchy of scales in this case. 
The mass of the hidden scalar is $m_\rho=\sqrt{2 \lambda_\sigma}v_\sigma$, so that
\be
m^2_\rho = 2\Lambda^2_h \frac{\lambda_\sigma \lambda_H}{\lambda_{H\sigma}} 
= 2\Lambda^2_h \frac{\lambda_{H\sigma}}{\Theta}\ll 2\Lambda^2_h = 
 \frac{2\lambda_{H\sigma}}{\lambda_H} v^2_\sigma \ll v_\sigma^2,
\ee
where we have used that $\Theta$ and $\lambda_H$ cannot be as small as $\lambda_{H\sigma}\lesssim 10^{-5}$, because otherwise the threshold mechanism would hardly work.

Analogous arguments apply to the $\{h<v\,,\rho>v_\sigma\}$ region, which contains the HSI inflationary direction. We assume again that the quartic self-couplings are positive there.  In this region, $-v^2/\Lambda_h^2<p_1<0$ and $p_2<0$; and we can write
\be
V  > \lambda_\sigma v_\sigma^4 \(p_2^2 -2 \Theta \frac{v^2}{\Lambda_h^2}|p_2|\) , 
\ee
and thus the safe region for stability is $|p_2|>2 \Theta v^2/\Lambda_h^2$, which corresponds to 
$\rho^2-v_\sigma^2>2 v^2 \lambda_{H}/\lambda_\sigma$. So, we see that in this case  the potential can only be negative in a tiny region (tiny because $\lambda_H\gg\lambda_\sigma$) above $\rho=v_\sigma$, where we then have to impose $\Theta<1$ to guarantee stability. In this region, $\rho-v_\sigma\sim (v^2/v_\sigma)\lambda_{H}/\lambda_\sigma $ and $h-v$ are both of order $v$. If we make sure that $\Theta<1$ for $\mu<\sqrt{2}\Lambda_h$ this region will be stable.

\subsubsection{Case $\lambda_{H\sigma}<0$}
This case is the relevant one for HHSI. There are again two regions in which the potential can become negative due to the portal coupling, the low-fields region $\{h<v, \rho<v_\sigma\}$ and the high-fields region 2) $\{h>v,\rho>v_\sigma\}$. 
The equations for the potential \eqref{scalar_potential3} or \eqref{scalar_potential4} are still valid, but keeping in mind that now $\Lambda_h^2<0$. 
The curve $\Theta\,p_1-p_2=0$ goes from the region 1) to 2)  tending asymptotically to $\rho^2/h^2=|\lambda_{H\sigma}|/\lambda_\sigma$, see \eq{tra}. Therefore, $\Theta<1$ needs to be enforced in both regions to ensure stability and so we have to demand $\Theta<1$ up to the inflationary scales.

\subsubsection{Summary}
In order to have an absolutely stable potential in SMASH, from the previous analysis we conclude that one needs to impose the following sufficient conditions: 
\begin{itemize}
\item For $\lambda_{H\sigma}>0$:
\bea
\label{stabilitycondition1}
\left\{\begin{array}{cc}  
\tilde \lambda_H, \tilde \lambda_\sigma>0, \quad\text{for}\quad h < \sqrt{2}\Lambda_h \\
         \lambda_H,         \lambda_\sigma>0, \quad\text{for}\quad h > \sqrt{2}\Lambda_h \\
\end{array}\right.
\eea
\item For $\lambda_{H\sigma}<0$:
\bea
\label{stabilitycondition2}
   \quad \quad \,
\tilde \lambda_H, \tilde \lambda_\sigma>0\quad \text{for all}\quad  h\,.
\eea
\end{itemize}
We recall that HSI and HI require $\lambda_{H\sigma}>0$ while HHSI requires $\lambda_{H\sigma}<0$. 

As was argued before, the coupling $\tilde\lambda_{H}$ must match the SM Higgs quartic $\overline\lambda_H$ at small scales. If the running of $\tilde\lambda_H$ is also 
SM-like (meaning that the effects of $\lambda_{H\sigma}$ and other high energy parameters are negligible), then for $\lambda_{H\sigma}>0$ the stability condition $\tilde \lambda_H>0$ for $h<\sqrt{2}\Lambda_h$ can be satisfied
whenever
\begin{align} \label{stability0}
\Lambda_I>\sqrt{2}\Lambda_h\,.
\end{align}
We stress that {this} requirement is only valid whenever the coupling $\tilde\lambda_H$ does not deviate from the SM coupling $\overline\lambda_H$  at large scales. As we will see, such deviations can indeed happen, and in fact are the only way to obtain stable HHSI models with $\lambda_{H\sigma}<0$. This follows from the fact that in this case stability demands $\tilde\lambda_H>0$ for all scales, which cannot happen if $\tilde\lambda_H$  runs as in the SM.

\subsection{Stable SMASH models}

With the previous results, we can already sketch how to obtain stable SMASH models starting from values of $m_h$ and $m_t$ for which the Higgs quartic coupling turns negative at $\Lambda_I$ in the SM.  

Choosing a value of $f_A$ and a value $\lambda_\sigma(m_\rho)$, the first thing that has to be checked is that the latter coupling remains positive up to scales larger than the field values required for inflation, for instance  $30 M_P$. Neglecting the effects of $y$, which is reasonable provided that $y\lesssim {\rm min}\{Y_{ii}/2\}$, this imposes a limit on the RH neutrino Yukawa couplings, 
\be
\label{SSMY33}
Y_{33}^4+Y_{22}^4+Y_{11}^4\lesssim \frac{16\pi^2 \lambda_\sigma}{\log\(\frac{30 M_P}{\sqrt{2\lambda_\sigma}f_A}\)} . 
\ee 
In particular, for $Y_{33}=Y_{22}=3Y_{11}$ and $f_A\sim 10^{11}$ GeV, this yields \eqref{Y11stability}. Neglecting the running of the $Y's$ for this estimate is a good approximation. 
 
To ensure stability for $\lambda_{H\sigma}>0$, we also need to satisfy \eq{stabilitycondition1}. In SMASH examples which solve the unitarity problem of Higgs inflation, the beta function of $\lambda_H$ is SM-like so that the evolution of $\lambda_H$ only depends on $m_t$, $m_h$ and the value of the threshold $\delta$ appearing from the matching with measurements of the Higgs mass: $\bar \lambda_{H}(m_h)=0.132 = \lambda_{H}(m_h)-\delta(m_h)$, where we have chosen $m_h=125.09$ GeV. In particular, the beta function of $\lambda_H$ is  
\be \label{likeSM}
\frac{d \lambda_H}{d\ln \mu}=\frac{1}{16\pi^2}\(12 y_t^2 \lambda_H+24 \lambda_H^2-6 y_t^4 \) + ... \,,
\ee
like in the SM. Solving the SM RGEs with an initial value $\lambda_{H}(m_h)=0.132+\delta$, we obtain the minimum threshold correction needed to have $\lambda_H$ positive up to large scales. In particular, requiring positivity of $\lambda_H$ up to $\sim 30\, M_P$, the minimum $\delta$ that is needed as a function of $m_t$ is shown in Figure~\ref{fig:mindelta} (left), obtained with one-loop RGEs. This sets a lower limit on $\lambda_{H\sigma}$, 
\be
\label{condi1}
 \delta_{\rm min}\lesssim\frac{\lambda_{H\sigma}^2}{\lambda_\sigma}\Big|_{m_h}\,.
\ee
Whenever $\tilde\lambda_H$, which matches the SM quartic at low scales, also runs as in the SM, stability in the $h$ direction also requires $\Lambda_I> \sqrt{2} \Lambda_h$, see \eq{stability0}. This leads to an upper limit on $\lambda_{H\sigma}$,
\be 
\label{condi2}
\lambda_{H\sigma}(\Lambda_h) \lesssim {\frac{\lambda_H(\Lambda_h)}{2}}\(\frac{\Lambda_I}{f_A}\)^2 . 
\ee
In SMASH, $\lambda_{H\sigma}$ runs multiplicatively and typically increases with the RG scale $\mu$, with contributions coming from $\lambda_H$, the top quark Yukawa, the gauge couplings. Combining \eq{condi1} and \eq{condi2}:
\be
\label{condi3}
\lambda_\sigma(m_h) \lesssim \frac{\lambda_H^2(\Lambda_h)}{4\,\delta_{\rm min}}\(\frac{\Lambda_I}{f_A}\)^4 .  
\ee
This constraint can be easily satisfied for $f_A$ giving axion dark matter in the post-inflation PQ symmetry restoration scenario, i.e. $3\times 10^{10}<f_A/{\rm GeV}<5\times 10^{11}$ (see Section \ref{dark_matter}), for $171<m_t/{\rm GeV}<175$ and  $\lambda_\sigma$ in the range required for predictive HSI, see \eq{inflationconstraint}. This ensures that there are always values of $\lambda_{H\sigma}$ such that the $h$-direction is stabilised by the tree-level threshold mechanism. Larger values of $f_A$ may become incompatible with large $m_t$ and $\lambda_\sigma$, but such values are not phenomenologically interesting for  SMASH. 

\begin{figure}[t]
\begin{center}
\includegraphics[width=0.46\textwidth]{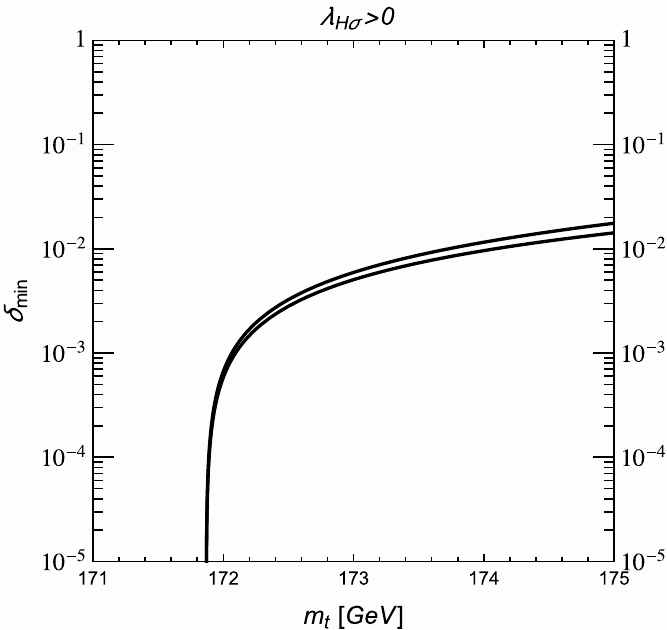}
\includegraphics[width=0.46\textwidth]{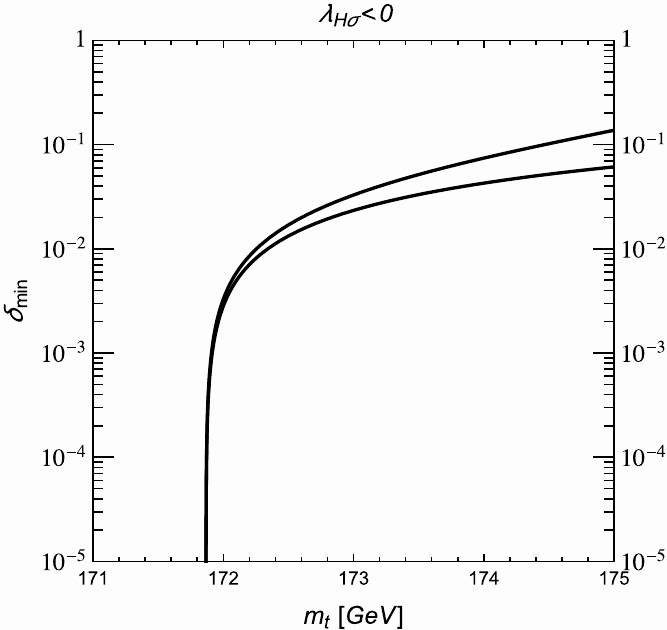}
\caption{\small  Minimum values of the threshold correction $\delta$, defined in \eq{THc}, (as a function of the top quark mass) for threshold stabilisation with $\lambda_{H\sigma}>0$ (left) and $\lambda_{H\sigma}<0$ (right) to ensure absolute stability in SMASH. The lower and upper lines correspond to $\delta$ at the low energy matching  scale $m_\rho$ and at $\sim 30 M_P$. }  
\label{fig:mindelta}       
\end{center}
\end{figure}

The bound of equation \eqref{condi3} can be avoided if there exist additional one-loop contributions to the tree-level matching relation of equation \eqref{eq:treematching} (as will be seen in our numerical scans), or if the running of $\tilde\lambda_H$ deviates from that of the SM Higgs quartic. The latter can ensure stability in the case $\lambda_{H\sigma}<0$, which is of particular interest for HHSI inflation. The stability condition in this case is \eq{stabilitycondition2}, and we need to consider the running of both $\lambda_H$ and $\delta$. As said before, for the small values of $\lambda_{H\sigma}$ of interest for SMASH the beta function of $\lambda_H$ can be taken as SM-model like. To estimate the running of $\delta$, one can neglect the running of $\lambda_\sigma$, given the smallness of  $\lambda_\sigma$, $\lambda_{H\sigma}$ in scenarios of interest. Then as long as the $Y's$ are not fine tuned to the limit \eqref{SSMY33}, we can neglect their running as well, and  compute $\lambda_{H\sigma}$ approximately from SM-like parameters only:
\be
\frac{\lambda_{H\sigma}(\mu')}{\lambda_{H\sigma}(m_h)}=\exp\[\int_{
\log m_h}^{\log\mu'} \frac{d\log\mu}{16\pi^2}\(12 \lambda_H^2+6 y_t^2-\frac{9}{2}g_2^2-\frac{9}{10}g_1^2\) \] . 
\ee
Given the positive contribution from the top Yukawa, $\lambda_{H\sigma}(\mu)$ typically grows with the scale, making $\delta$ larger. Thus, the stabilisation of $\tilde\lambda_H=\lambda_H-\delta$ can only come from the running of $\lambda_H$. How can this happen with a SM-like beta function, as in equation \eqref{likeSM}? In the SM and for a Higgs mass of $m_h=125.09$ GeV, $\bar \lambda_H(m_h)\simeq 0.132$ is sufficiently small that the negative top contribution dominates. In SMASH, $\lambda_H(m_h) = \bar \lambda_H(m_h) + \delta(m_h)\sim 0.132+\delta(m_h)$  is larger, and for a sufficiently large $\delta$ the positive top-Higgs or Higgs self coupling terms will dominate.
In summary, the threshold contribution to $\lambda_H$ increases its positive contribution to the SM-like beta function,  such that running effects can overcome $-\delta$ and stabilise $\tilde \lambda_H>0$. The minimum value of $\delta(m_h)$ or $\delta(30 M_P)$ that ensures $\tilde \lambda_H>0$ up to Planckian scales is displayed in Figure \ref{fig:mindelta} (right). 

An upper limit for the values of the threshold correction $\delta$ can be obtained imposing perturbativity along the RG flow. The dangerous coupling is $\lambda_H$, which in scenarios with a large threshold contribution can be larger at Planck scales than at low energies. 
At large values of $\delta(30 M_P)$ we have $\lambda_H(30 M_P)\sim \delta(30 M_P)$, so keeping $\delta(30 M_P)\lesssim 1$ keeps perturbativity under control. At low energies, this corresponds to 
\begin{align}
\label{eq:deltapert}
\delta\lesssim 0.1\,.
\end{align}
This result will be confirmed in the next section with detailed numerical scans, see Figure \ref{fig:scan1_2}.  Notice that this regime of large $\delta$ is at the edge of violating the stability conditions. For $\lambda_H\simeq \delta$ (at large field values), we have $\tilde \lambda_H/\lambda_H\ll 1$, which is close to the boundary of stability $\Theta\lesssim 1$.}

\begin{figure}[h!]
\begin{center}
\includegraphics[width=0.32\textwidth]{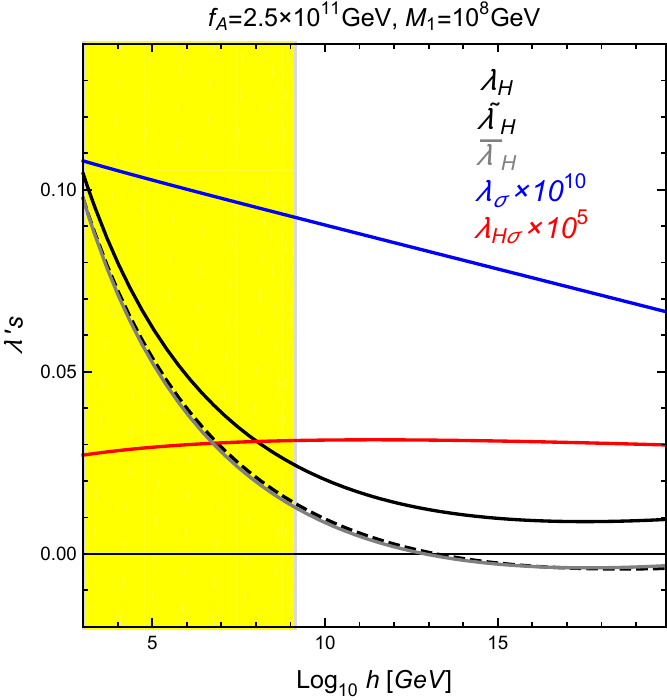}
\includegraphics[width=0.32\textwidth]{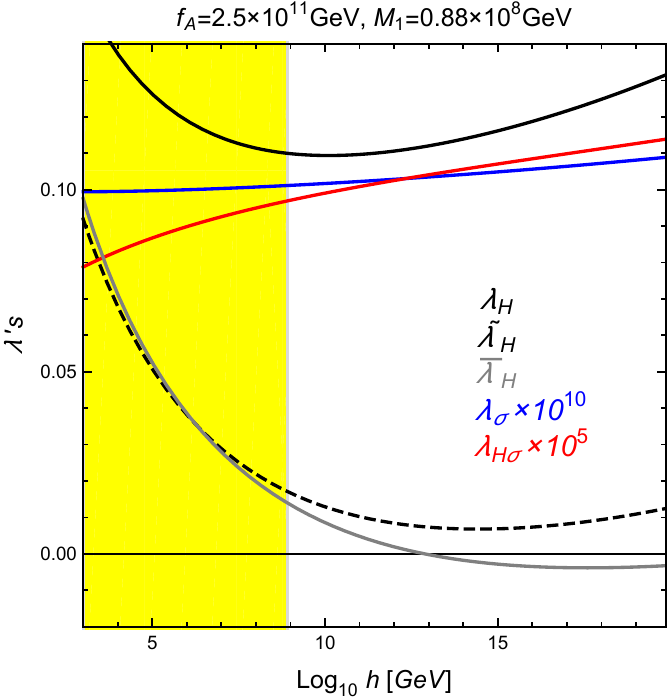}
\includegraphics[width=0.32\textwidth]{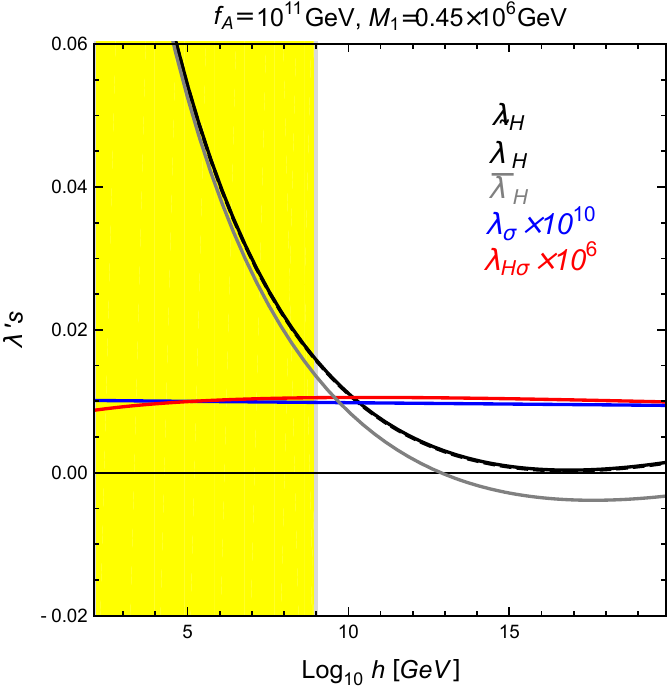}
\caption{\small  Examples of running couplings $\lambda_H$ (black), $\tilde\lambda_H$ (black dashed), the SM quartic coupling $\bar\lambda_H$ (grey),  $\lambda_\sigma$ (blue) and $\lambda_{H\sigma}$ (red), as functions of the RG scale $\mu=h$, for $m_t=172.38$ GeV and $y=Y_{11}$, on three stabilised examples with $\lambda_{H\sigma}>0$. The shaded regions correspond to $h<\Lambda_h$. 
In the center panel, $\tilde\lambda_H$ stays positive below $30 M_P$ thanks to the large $\delta$ which increases the positive running of $\lambda_H$. The right plot shows an example of one-loop threshold corrections splitting very slightly $\lambda_H$ and $\bar \lambda_H$ at low scales (with a small $\delta\sim 10^{-4}$) which are further split and stabilised by SM running terms in $\lambda_H$. 
}  
\label{fig:lambdarunning}       
\end{center}
\end{figure}

Stable SMASH models are then to be built as follows. Choose first $f_A$  from axion dark matter considerations, and $m_t$ from measurement. For an HSI model, choose $\lambda_\sigma$ at the inflationary scale, choose small enough Yukawas not to have a large running down to the $m_\rho$ scale. If \eqref{condi3} is satisfied, choose a delta close to $\delta_{\rm min}$ of Fig. \ref{fig:mindelta} (left). If the condition is not satisfied or we want larger values of $\delta$, choose a value larger than the minima of Fig. \ref{fig:mindelta} (right). 
For an HHSI model, choose the required $\tilde \lambda_{\sigma}$ at the inflationary scale and a value of $\delta$ larger than the minimum of Fig. \ref{fig:mindelta} (right). The value of $\lambda_\sigma$ at the inflationary scale is given by 
\be
\lambda_\sigma = \frac{\tilde \lambda_\sigma}{1-\frac{\delta}{\lambda_H}} \, , 
\ee 
which can be used to set bounds on the $Y's$ so as not to have too strong a running of $\lambda_\sigma$. 

{The previous results on perturbativity and stability  justify the choice $b\simeq1$ made in Section \ref{inflation} for the parameter defined in equation \eqref{adef}, and imply as well as that 
predictive HHSI trajectories remain very close to HSI ones. Indeed, we have $|\lambda_{H\sigma}|/\lambda_H=\sqrt{\delta\lambda_\sigma}/\lambda_H$. Predictive inflation demands $\lambda_\sigma$ to be in the window of \eqref{inflationconstraint}, while stability and perturbativity require $\delta\lesssim0.1$. Thus  $|\lambda_{H\sigma}|/\lambda_H\lesssim10^{-5}/\lambda_H$. The quartic $\lambda_H$ is larger than in the SM, as follows from \eqref{eq:treematching}, and unless the model is just barely stabilised, at scales relevant for inflation it takes values $\lambda_H(M_P)\gg10^{-5}$. Thus $\lambda_{H\sigma}|/\lambda_H\ll1$, giving $b\sim1$ in equation \eqref{adef} and $\rho/h\gg1$ for the HHSI inflationary trajectory of equation \eqref{tra}.}

Before moving onto the numerical scans, we give three different examples of running couplings in Figure~\ref{fig:lambdarunning} for different choices of $\lambda_{H\sigma}>0$. On the left one, $\delta\sim10^{-2}$ and $\tilde\lambda_H$ has a SM-like running, becoming negative  for $h>\Lambda_I$, while $\lambda_H$ stays positive thanks to the threshold correction. The plot in the center has $\delta\sim0.1$, and the coupling $\tilde\lambda_H$ remains positive at all scales, deviating from the SM running thanks to the sizable value of $\delta$. The plot on the right is an example of stabilisation despite a very small value of $\delta\sim10^{-4}$ which, according to the results of Figure \ref{fig:mindelta}, would naively seem to make threshold stabilisation impossible. However, in this case loop effects in the matching become important and ensure that $\lambda_H$ can be sufficiently larger than the SM coupling $\overline\lambda_H$ at the matching scale and beyond.

\subsection{\label{subsec:numerics}Scanning the parameter space}

To investigate the feasibility of the threshold stabilisation mechanism in SMASH and to check the validity of the estimates of the previous subsection, we have performed numerical scans looking for models in which the above stability conditions are satisfied. Each realisation of SMASH in these scans was constructed by starting with a choice of the axion scale $f_A$, accompanied by boundary conditions at the renormalisation scale $\mu=m_\rho$ for the couplings $\lambda_\sigma,\lambda_{H\sigma}$ in~\eqref{scalar_potential}, as well as for $y,Y_{ij}$  and  ${y_{Qd}}_i$ in~\eqref{lyukseesaw}. In the following, whenever we give numerical values for  couplings in SMASH without specifying a renormalisation scale, it should be understood that they are evaluated at $\mu=m_\rho$. We choose ${y_{Qd}}_i=0$ for simplicity and impose a diagonal ansatz for $Y_{ij}$, requiring $Y_{33}=Y_{22}=3Y_{11}$ to accommodate vanilla leptogenesis requirements (see Section~\ref{baryogenesis}). Once the $Y_{ij}$ are fixed, the Yukawas $F_{ij}$ can be determined requiring neutrino masses compatible with observations. For this we use the relation between $F_{ij}$ and the light and heavy neutrino masses obtained in \cite{Casas:2001sr}:
\begin{align}
\label{eq:Fboundary}
 F=\frac{1}{v}\,U^*_\nu D_{\sqrt{m}} O D_{\sqrt{M}}.
\end{align}
The previous equation has to be understood as a $3\times 3$ matrix identity for the couplings $F_{ij}$. The matrix $U_\nu$ is the PMNS neutrino mixing matrix, $O$ is a $3\times3$ orthogonal matrix, and $D_{\sqrt{m}}, D_{\sqrt{M}}$ are diagonal matrices containing the light and heavy neutrino masses, respectively. For the light neutrinos we set the lightest mass to zero, and fix the other masses and the mixing matrix $U_\nu$ from the fits of reference~\cite{Forero:2014bxa}, choosing either a normal or inverted hierarchy. The matrix $O$ can be set to the identity in our case, while the heavy masses are approximated by $Y_{ii}f_A/\sqrt{2}$.\footnote{We are applying equation~\eqref{eq:Fboundary} at the scale $f_A$, while using low-energy neutrino data. Strictly, the latter should be run up to the scale $f_A$, but since we find that the results we obtain for stability do not end up depending on the details of the neutrino spectrum --e.g. normal versus inverted hierarchy-- the effect of this running can be neglected.}

\begin{figure}[h!]
\begin{center}
\includegraphics[width=0.49\textwidth]{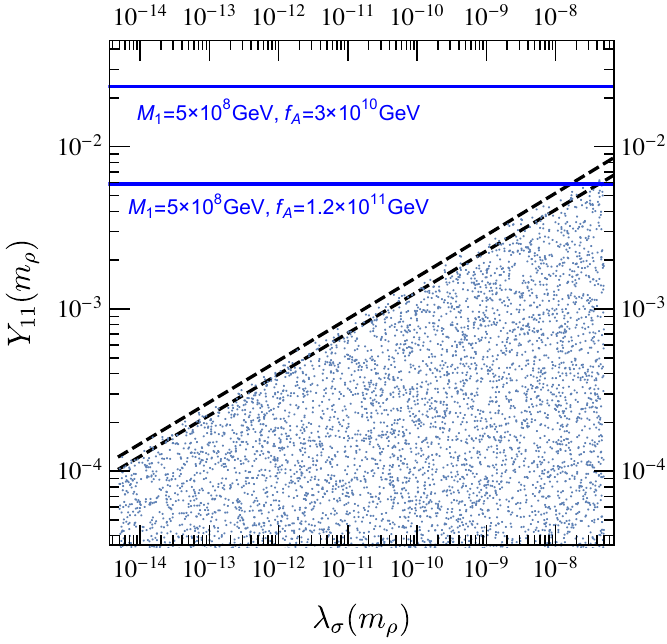}
\includegraphics[width=0.49\textwidth]{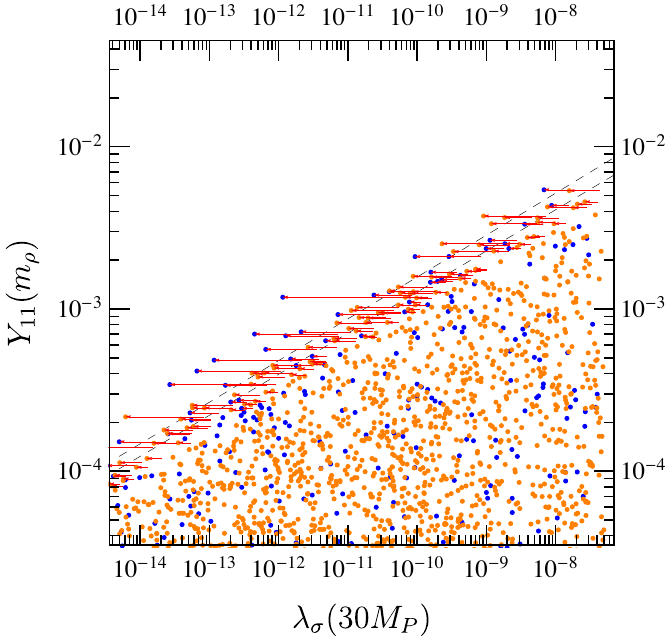}
\caption{\small  Left: Four-parameter scan in the SMASH model, represented in the $(Y_{11},\lambda_\sigma)$ plane, with randomised $f_A,\lambda_{H\sigma}$. Each point corresponds to a choice of parameters with a stable $\rho$ direction. The horizontal lines represent values of $Y_{11}$ associated with different choices of $M_{1}$ and $f_A$. 
Right: Same scan as the left plot, represented now as a function of $\lambda_\sigma$ evaluated at $30 M_P$.  
All the points have stable $\rho$ and $h$ directions, with the orange points having $\tilde\lambda_H>0$ for all scales. 
Arrows indicate the running from $\lambda_\sigma(m_\rho)$ to the $\lambda_\sigma(30 M_P)$ values when it is larger than 30\%.}  
\label{fig:scan1}       
\end{center}
\end{figure}

The remaining couplings are determined by matching relations with the SM at the scale $f_A$. On the SM side, we impose $m_h=125.09$ GeV, following from the latest ATLAS and CMS combination \cite{Aad:2015zhl}. For the top mass, we first focus on two representative values: $m_t=173.34$ GeV, taken from 
a joint Tevatron and LHC combination \cite{ATLAS:2014wva}, or $m_t=172.38$ from the latest combined CMS analysis \cite{CMS:2014hta} and consider only later a broader range. The gauge and Yukawa couplings are fixed using the latest Particle Data Group listings \cite{Agashe:2014kda}. In the determination of the top Yukawa from the top mass we include one-loop electroweak corrections \cite{Hempfling:1994ar} as well as three-loop strong coupling corrections at the scale $m_t$ \cite{Chetyrkin:1999qi,Melnikov:2000qh}.  
The Higgs VEV is determined from the Fermi constant using the one-loop electroweak matching
 relations of reference~\cite{Hempfling:1994ar}, while the Higgs quartic and quadratic couplings are determined by using the two-loop effective potential, requiring it to be minimised at the Higgs VEV,\footnote{This is equivalent to setting the Higgs tadpole diagrams to zero.} and demanding that one gets the correct pole Higgs mass from the self-energy. The latter is obtained from the second derivative of the effective potential supplemented by a one-loop finite momentum correction, as discussed for example in  reference~\cite{Degrassi:2012ry}.

\begin{figure}[h!]
\begin{center}
\includegraphics[width=0.4\textwidth]{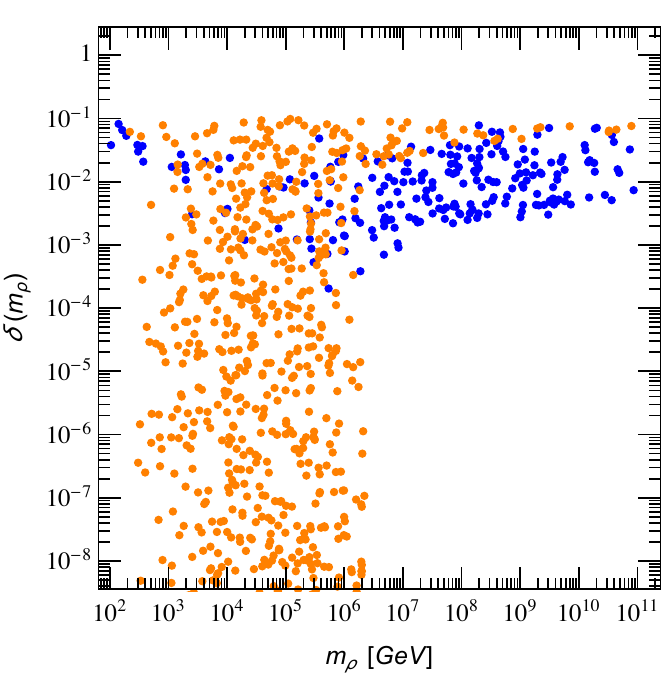}\includegraphics[width=0.4\textwidth]{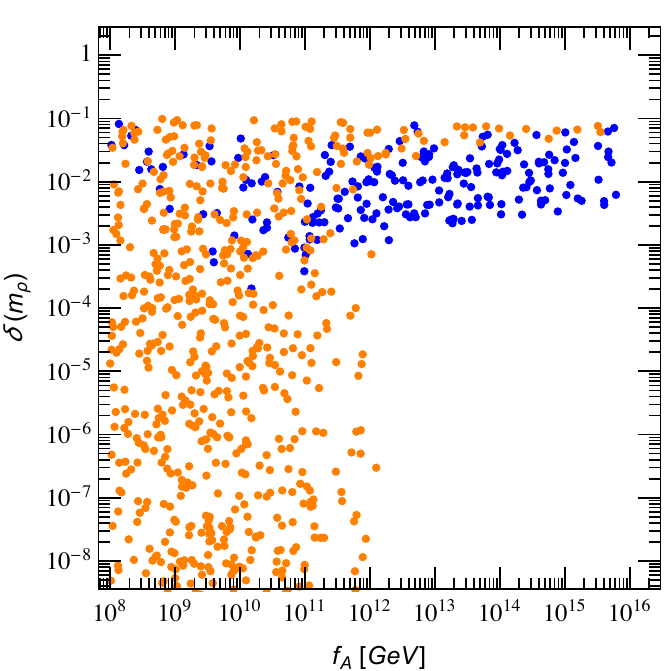}
\includegraphics[width=0.4\textwidth]{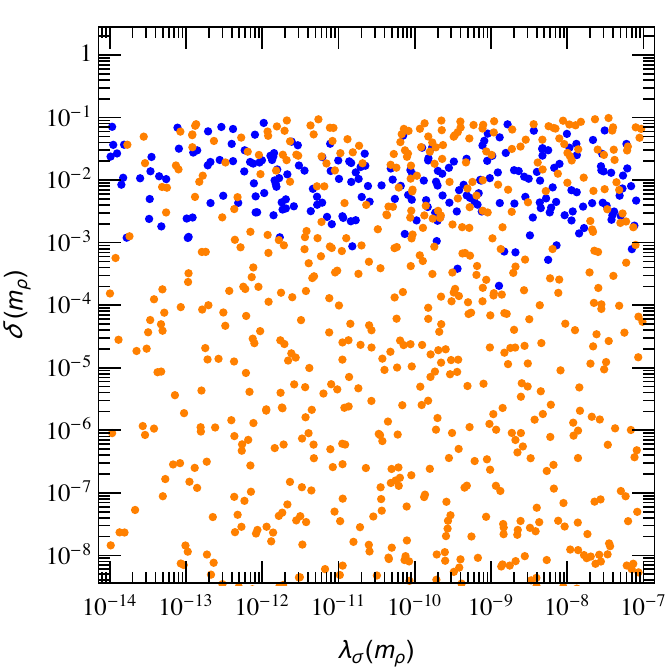}\includegraphics[width=0.4\textwidth]{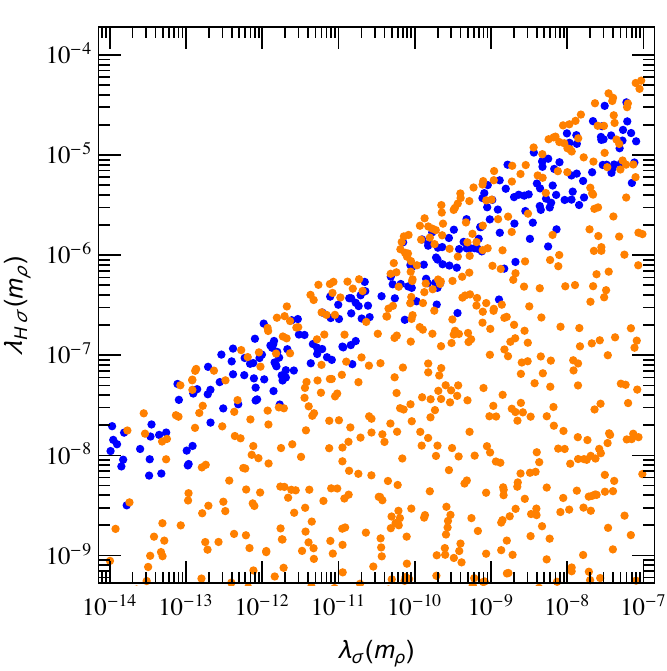}
\caption{\small  Same scan as the one in Figure \ref{fig:scan1}, represented in different planes of the parameter space. All the points have stable $\rho$ and $h$ directions, with the orange points having $\tilde\lambda_H>0$ for all scales. The quantity $\delta$ is defined in \eq{THc}.  }  
\label{fig:scan1_2}       
\end{center}
\end{figure}
 
Once the SM parameters are known, they are run with two-loop RG equations up to the scale $m_\rho=\sqrt{2\lambda_\sigma} f_A$. At that scale, the couplings shared with the SMASH model are matched across the threshold. We go beyond the matching of equation~\eqref{eq:treematching} by doing the matching at one-loop. This is achieved by identifying  the derivatives of the SM effective potential with those of the SMASH effective potential evaluated at the one-loop corrected version of \eqref{sline1}. 

Once the SMASH couplings are determined at the scale $f_A$, the RG equations are solved with two-loop precision, including full flavour-mixing and complex phases for the neutrino couplings $F_{ij},Y_{ij}$. The solution to the RG flow allows to check the stability conditions \eqref{stabilitycondition1} and \eqref{stabilitycondition2} at all scales. 

For simplicity, we performed first a four-parameter scan, varying $(\lambda_\sigma, \lambda_{H\sigma}, Y_{11}, f_A)$ with $\lambda_{H\sigma}>0$ setting $y=Y_{11}=Y_{33}/3=Y_{22}/3$ (and fixing the rest of the parameters as described above), such that all the (dangerous) negative Yukawa contributions are encoded in the single parameter $Y_{11}$. We choose a normal hierarchy for the light neutrinos, and checked that the resulting figures are essentially identical if an inverted hierarchy is chosen instead.  

The plots in Figure \ref{fig:scan1} show the regions in the plane $(Y_{11},\lambda_\sigma)$  corresponding to stability in the $\rho$ direction,  for $m_t=172.38$ GeV. In the left plot, points with a stable $\rho$ direction are shown as dots. We also show the minimum values of the Yukawa $Y_{11}$ needed to satisfy the vanilla leptogenesis requirement $M_1=f_A Y_{11}/\sqrt{2} \geq 5\times10^8$ GeV (see Section \ref{baryogenesis}) for two extreme values of $f_A$ that can, within uncertainties, provide enough axion dark matter in the scenario with PQ symmetry restoration after inflation: $f_A=3 \times10^{10}$ GeV and $f_A=1.2\times 10^{11}$ GeV at the upper end of the range, see \eqref{farangePQrestoration} in Section \ref{dark_matter}.
As we anticipated,  stability in the $\rho$ direction enforces a minimum value of $\lambda_\sigma$ for a given $Y_{11}$, which is clearly seen in the figure.  The dashed lines are derived by demanding positivity of $\lambda_\sigma$ using the one-loop contribution of the $Y_{ij}$ Yukawas to its beta function, as in equation \eq{naivestability}. The two lines reflect the mild logarithmic dependence of the scanned range $f_A\in(10^8,2\times 10^{17})$ GeV. We do not find any stable point that violates our approximate stability criterion \eq{Y11stability}. Moreover, we do not find any stable model in which the positive contribution of $\lambda_{H\sigma}$ helps overcoming the negative effect of $Y_{11}$, which would show up as violations of the criterion. 
In the right plot, we show in the abscissa the values of $\lambda_\sigma(30 M_P)$ relevant for inflation, displaying only models that are also stable in the $h$ direction (discussed below). We notice that the sharp boundary that divides stable and unstable models is now slightly diffused. Models close to the upper boundary have large $Y_{11}$ (and negative running of $\lambda_\sigma$) and thus a sizeable decrease of $\lambda_\sigma$ from $m_\rho$ to $30 M_p$. We have chosen to show models stable up to $30 M_P$, which is always above our requirements for inflation, see Figure \ref{fig:chi2} upper right. 
Models for which with  $\lambda_\sigma$ runs by more than 30\% between these scales of more than are indicated by an arrow which starts at the value $\lambda_\sigma(m_\rho)$ (and points to the left). We see that only models with the largest Yukawas compatible with stability can run so much.

The plots in Figure \ref{fig:scan1_2} show several aspects of the stability requirement in the $h$ direction in four different planes of the parameter space. Again, for the purpose of illustration, the top quark mass $m_t$ was fixed at $172.38$ GeV, which corresponds to $\Lambda_I\simeq 6.5\times 10^{12}$ GeV. Similar figures are obtained for higher values of $m_t$. All the dots in these plots represent stable models. First, orange dots have $\tilde\lambda_H>0$ for scales up to $30 M_P$. Instead, blue points have $\tilde\lambda_H$ becoming negative for scales larger than $\sim \Lambda_h$, but  with $\lambda_H$ remaining positive up to $30 M_P$. We can further distinguish two groups of orange  points: those corresponding to large values of $\delta \simeq 0.01\sim 0.1$ and those with $m_\rho<10^6\, {\rm GeV}$ and $\delta\lesssim 10^{-3}$, see Figure \ref{fig:scan1_2} upper left. The stabilisation for the first group of orange points (those with ``large'' $\delta$), with $\tilde\lambda_H>0$ up to Planckian scales, is due to the effect discussed before in which, despite a SM-like beta function of $\lambda_H$ (as corresponds to the small values of $\lambda_{H\sigma}$ in Figure \ref{fig:scan1_2}),  the threshold contribution to $\lambda_H$ enhances the positive contributions to the beta function of $\tilde\lambda_H$, which can remain positive up to large scales, as in the example in the center panel of Figure \ref{fig:lambdarunning}. Regarding the orange points with small  values of $\delta$, they seem to be in conflict with the results of Figure \ref{fig:mindelta} (left), which suggests  a minimum value of $\delta\lesssim10^{-3}$ to ensure stability. The resolution of this apparent contradiction  is that, as mentioned earlier, one-loop  threshold effects can become important, and the true threshold is given by the tree-level $\delta$ of equation \eqref{THc} supplemented by additional corrections. This effect is also present (to a lesser extent) at the overlapping region between the two groups of orange points. Notice that this stabilisation by one-loop threshold effects for very small values of $\delta$ (in general associated with very small values of $\lambda_{H\sigma}$) is only important for low values of $f_A$ and the singlet mass $m_\rho$, as reflected in Figure \ref{fig:scan1_2}. This can be understood from the decoupling of virtual excitations in the $\rho$ direction for large values of $m_\rho$. An example of running couplings in the group of orange points with small delta is illustrated by the right panel in Figure \ref{fig:lambdarunning}.

\begin{figure}[h!]
\begin{center}
\includegraphics[width=0.49\textwidth]{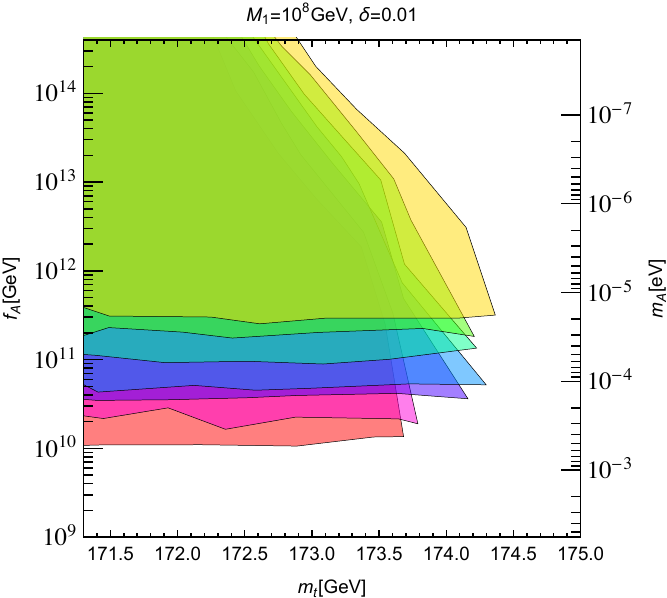}
\includegraphics[width=0.49\textwidth]{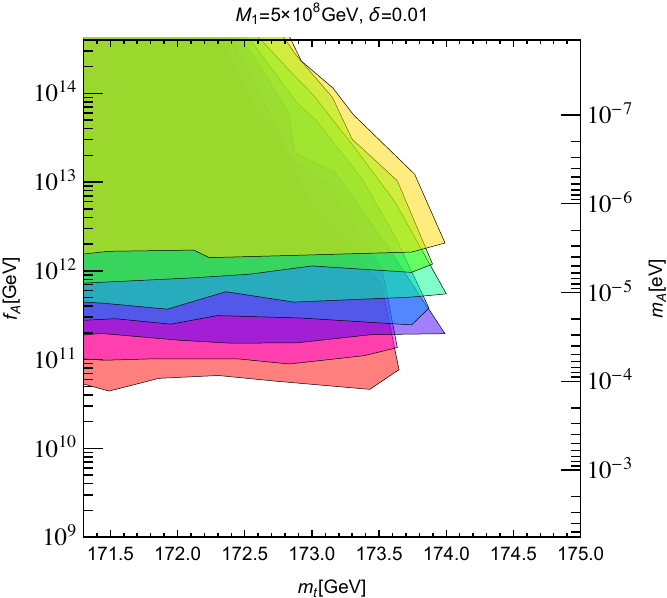}
\caption{\small  Regions with stable examples in the $(m_t,f_A)$ plane, colour-coded according to the value of $\lambda_\sigma(m_\rho)$: yellow ($10^{-13}-10^{-12}$), green ($10^{-12}-10^{-11}$), cyan  ($10^{-11}-10^{-10}$), blue ($10^{-10}-10^{-9}$), purple ($10^{-9}-10^{-8}$), pink ($10^{-8}-10^{-7}$) and red  ($10^{-7}-10^{-6}$). The threshold parameter $\delta$ was fixed to $10^{-2}$. On the left panel, $M_1=10^8$~GeV, whereas on the right $M_1=5\times10^8$ GeV.}
\label{fig:famt}       
\end{center}
\end{figure}

Figure \ref{fig:famt} illustrates the minimum values of $\lambda_\sigma$ attainable in stable examples of SMASH for given choices of $M_1$, $\delta$, $f_A$ and $m_t$. The plots show $\lambda_\sigma$-dependent  upper and lower bounds in $f_A$ for fixed $\delta,M_1$ and $m_t$. The upper bounds can be understood from  equation \eqref{condi3}, valid for scenarios with a SM-like running of $\tilde\lambda_H$; this is the case for the choice of $\delta=0.01$ in the figure. The higher the value of $m_t$, the lower the value of the instability scale $\Lambda_I$ (see Figure \ref{fig:LambdaI}) and thus the bound in $f_A$ is reduced.
The lower bound for $f_A$ follows from the stability condition in the $\rho$ direction, equation \eqref{Y11stability}, once the relation $M_{ii}=Y_{ii}f_A/\sqrt{2}$ is substituted. Again, values of $\lambda_\sigma\lesssim 10^{-10}$, of interest for unitarity-safe HSI inflation, are possible with values of $f_A\sim10^{11}$\,GeV compatible with dark matter constraints, studied in Section \ref{dark_matter}. For these HSI scenarios there is marginal compatibility with the  vanilla leptogenesis constraint of equation \eqref{leptogenesis}, which can be circumvented with a {very} mild resonant enhancement of leptogenesis. The latter allows to lower the value of $M_1$, which in turn permits reaching larger values of $\lambda_\sigma$ for lower values of $f_A$. In the case of unitarity-safe HHSI scenarios, one can allow for lower values of $\lambda_\sigma$ and still have vanilla leptogenesis with the correct dark matter abundance.

\begin{figure}[t]
\begin{center}
\includegraphics[width=0.49\textwidth]{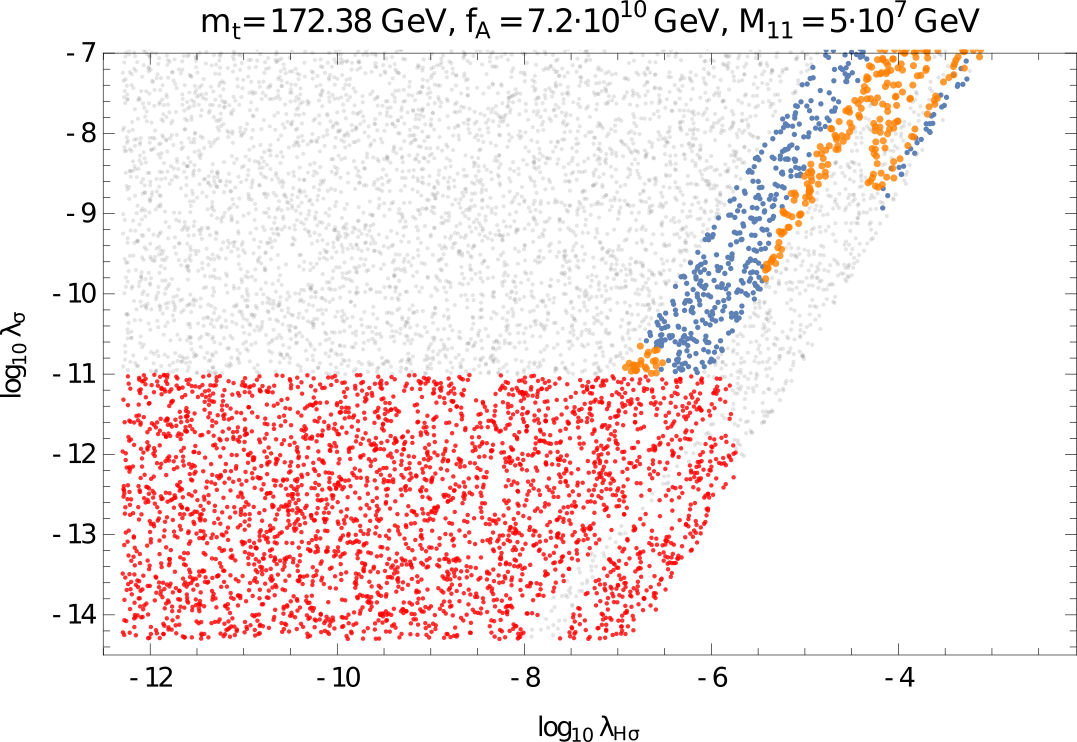}
\includegraphics[width=0.49\textwidth]{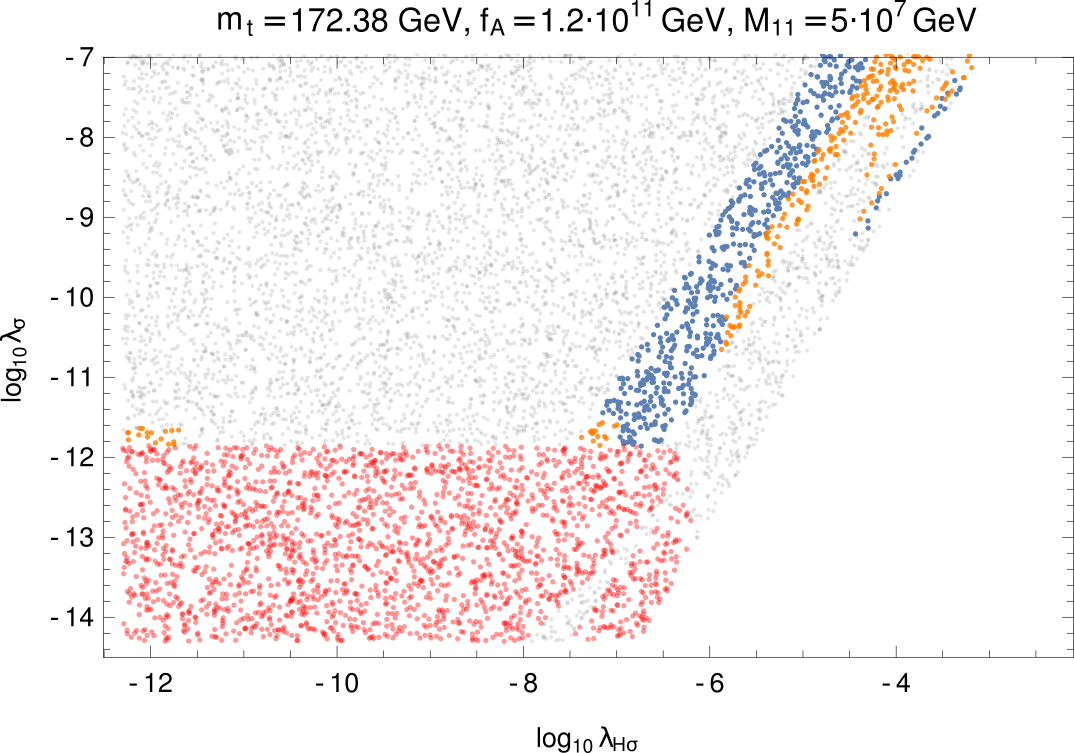}
\vskip0.2cm
\includegraphics[width=0.49\textwidth]{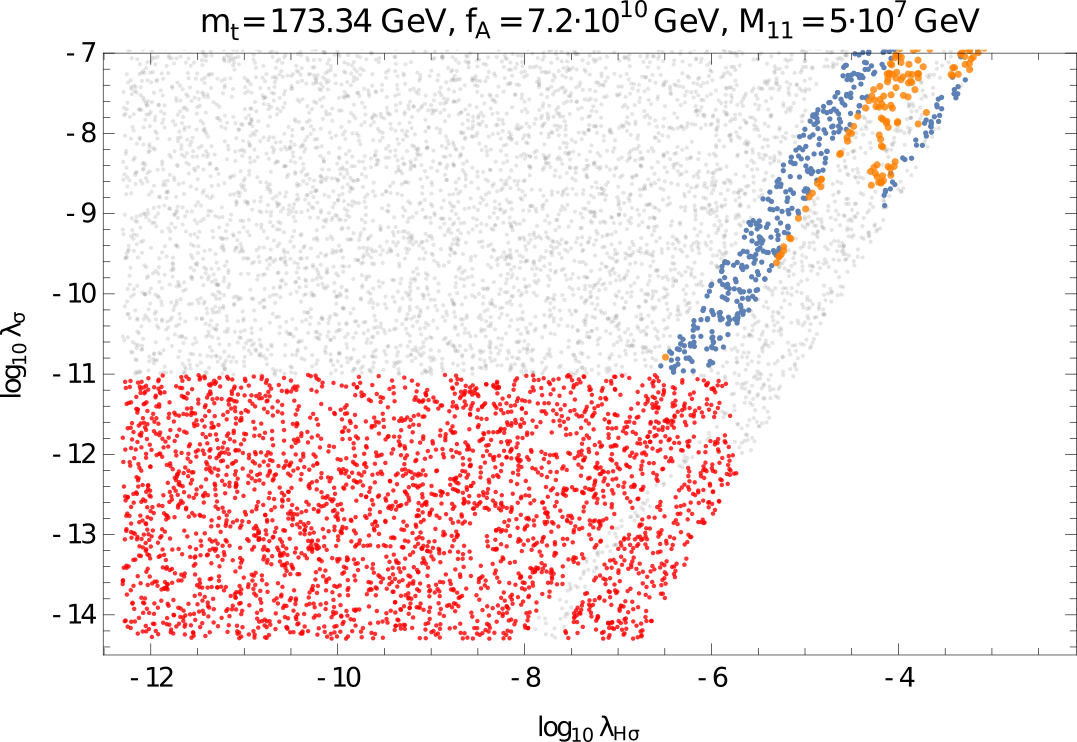}
\includegraphics[width=0.49\textwidth]{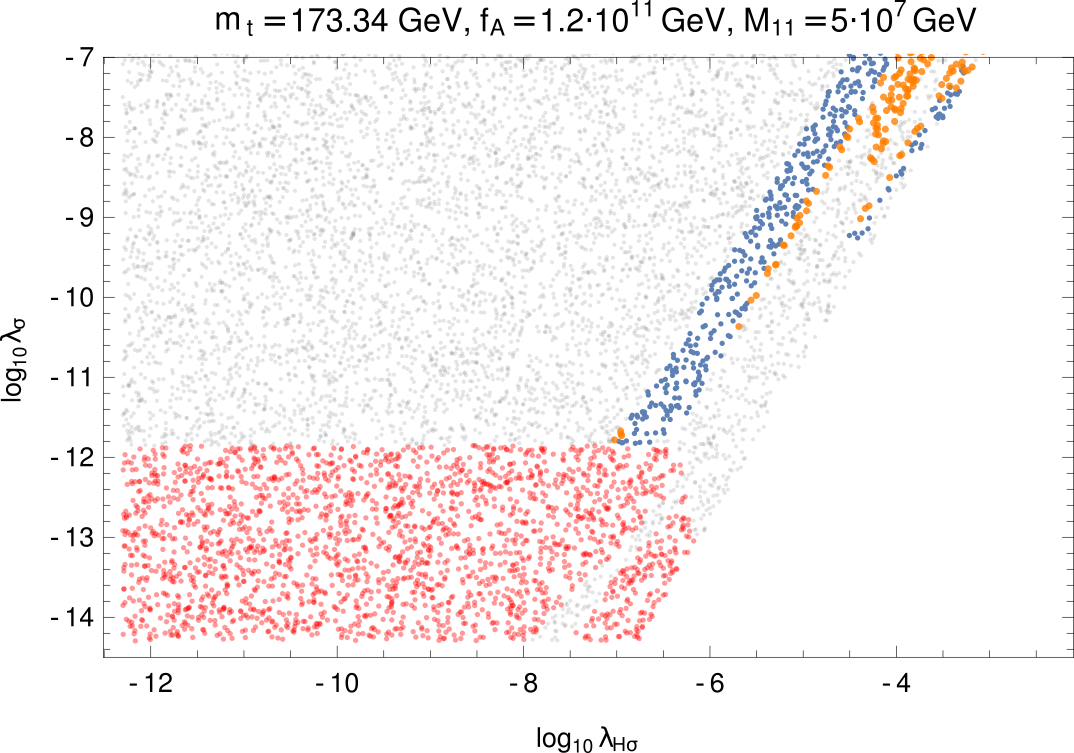}
\caption{\small  Two-parameter scans in the  $(\lambda_{\sigma},\lambda_{H\sigma})$ plane, for $M_1=5\times10^{7}$ GeV, $m_t=172.38$ GeV (top) and  $m_t=173.34$ GeV (bottom), and with $f_A=7.2\times10^{10}$ GeV (left) and  {$f_A=1.2\times10^{11}$} GeV (right). Red points are unstable in the $\rho$ direction, and grey points have an instability along $h$. Blue and orange points are stable, with the orange points having $\tilde\lambda_H>0$ for all scales. }  
\label{fig:scan2}       
\end{center}
\end{figure}

\begin{figure}[t]
\begin{center}
\includegraphics[width=0.49\textwidth]{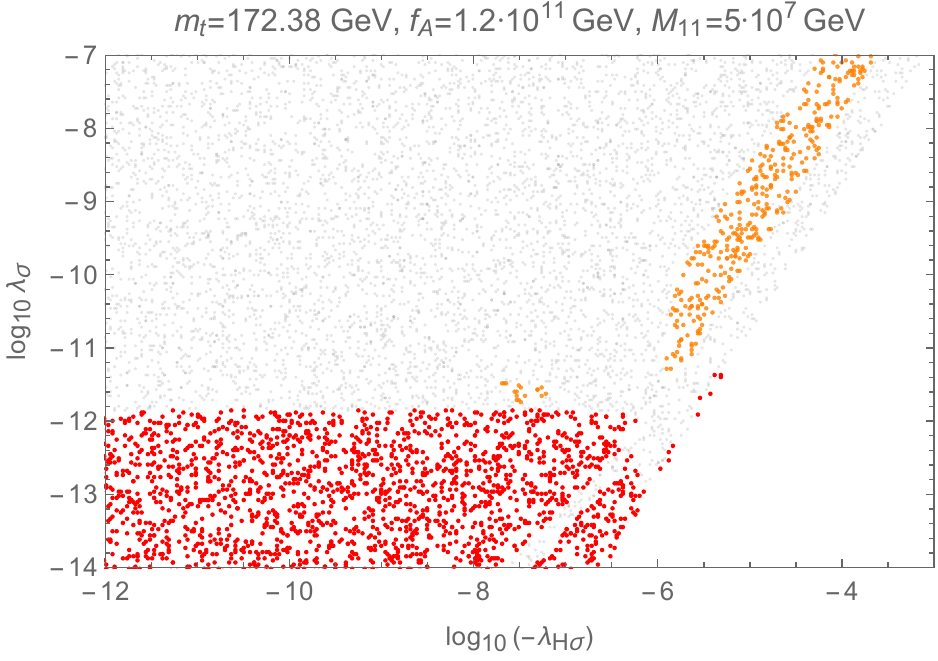}
\includegraphics[width=0.49\textwidth]{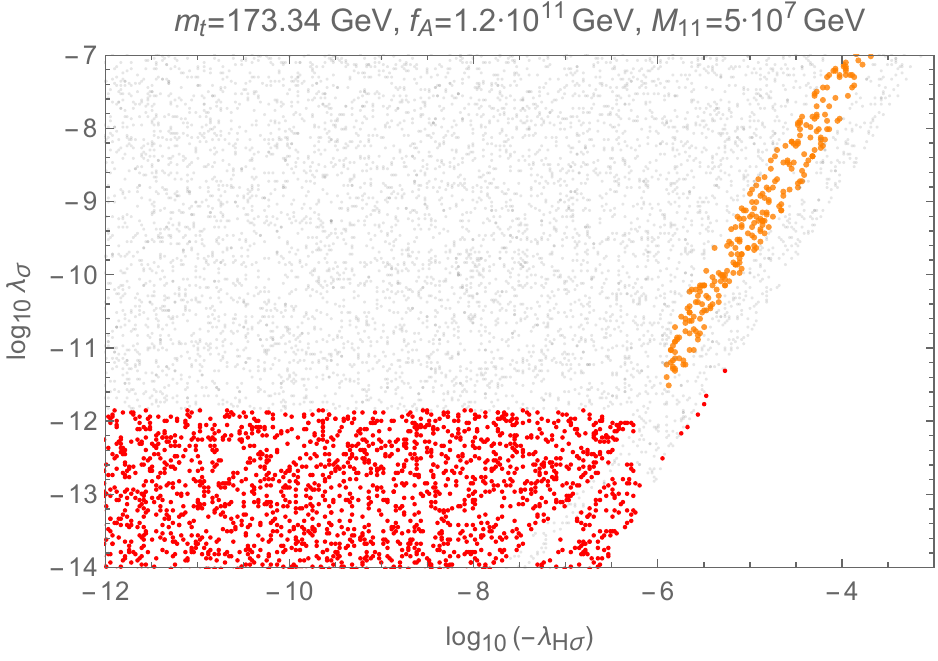}
\caption{\small  Two-parameter scans in the  $(\lambda_{\sigma},\lambda_{H\sigma})$ plane for two different values of the top quark mass:  $m_t=172.38$ GeV (left) and  $m_t=173.34$ GeV (right). The other parameters are $\lambda_{H\sigma}<0$, $M_1=5\times10^{8}$ GeV and {$f_A=1.2\times10^{11}$} GeV for both figures. The red points have an unstable $\rho$ direction, and the grey points have an instability along $h$. The orange points have $\tilde\lambda_H>0$ for all scales. }
\label{fig:scan3}       
\end{center}
\end{figure}

The different stability regions in the $\lambda_\sigma,\lambda_{H\sigma}$ plane can be more clearly seen with two-parameter scans with $f_A, m_t$ fixed to different values,  and $M_{1}$ fixed at a value  $5\times10^{7}$ GeV, which would correspond to a mild resonant enhancement of leptogenesis. This is shown in Figure \ref{fig:scan2}. The red and grey points have instabilities in the $\rho$ and $h$ directions, respectively, while the blue and orange points represent stable scenarios with the same color coding as in the previous plots. Notice how, as expected from the results of Figure \ref{fig:mindelta}, scenarios with $\tilde\lambda_H>0$ up to Planckian scales typically require larger values of $\delta$. The diagonal boundary beyond which there are no coloured points is due to the fact that the scan was restricted to scenarios in which $\delta=\lambda^2_{H\sigma}/\lambda_\sigma$ remains perturbative ($<4\pi$). For completeness, Figure \ref{fig:scan3} shows analogous scans with  $\lambda_{H\sigma}<0$.

\begin{figure}[h!] 
\begin{center}
\includegraphics[width=0.49\textwidth]{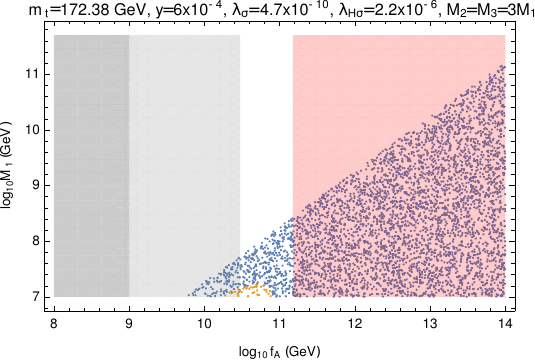}
\hfill
\includegraphics[width=0.49\textwidth]{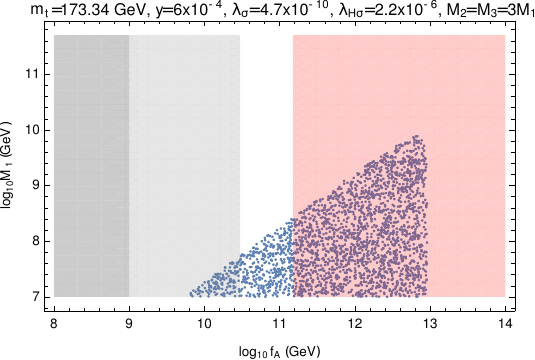}
\caption{  \small Parameter scans in the SMASH model for $m_t=172.38$ GeV (left) and $m_t=173.34$ GeV (right), with varying $M_1$ and $f_A$, {choosing a small value of $\lambda_\sigma$ compatible with inflation with a non-minimal gravitational coupling of order 1}. All coloured points in these plots correspond to stability up to energies above the Planck scale, with the orange points having $\tilde\lambda_H>0$ for all scales. 
{The light grey coloured band is excluded by insufficient dark-matter relic abundance and the dark grey region by the SN1987A energy loss argument. In the red region on the right, the correct amount of axion dark matter can only be produced if the PQ symmetry is not restored after inflation. In the white region, it can be produced  in both the PQ  restoration and non-restoration scenarios. }
}  
\label{fig:SMASHscans}       
\end{center}
\end{figure}

To finish illustrating the parameter space corresponding to stable SMASH scenarios, Figure~\ref{fig:SMASHscans} displays the results of further scans in the $(M_1,f_A)$ plane, for  $m_t=172.38$ GeV and $m_t=173.34$ GeV, and choosing {$y=6\times 10^{-4},\lambda_\sigma=4.7\times10^{-10},\lambda_{H\sigma}=2.2\times10^{-6}$.} The orange and blue points have stability up to the Planck scale, with the same colour coding as before. The shaded bands have been chosen to indicate different areas of the relevant SMASH parameter space. First, the left vertical dark-grey band covers the region $f_A<4\times 10^8$ GeV discarded by supernovae constraints. 
The vertical light-grey band with {$f_A<3\times 10^{10}$} GeV  represents the region of parameter space in which the axion relic abundance is insufficient to account for all dark matter {(see \eqref{farangePQrestoration} in Section \ref{dark_matter}).} The right red band is the region {$f_A>1.2\times10^{11}$} GeV, for which the right axion dark matter abundance can happen only if the PQ symmetry is not restored after inflation.\footnote{Points in this region have to satisfy specific isocurvature constraints that depend on the Hubble scale during inflation.}
The dependence with the axion scale $f_A$  of the maximum value of $M_1$ allowing for stability {reflects the bound of equation \eqref{Y11stability}, while the upper bound in $f_A$ can be explained from equation \eqref{condi3}.}

\section{\label{sec:reheating}Reheating}

A remarkable feature of SMASH is that the mechanism of reheating {can be well described} and therefore {the temperature}  at the start of the radiation dominated stage, $T_R$, can be estimated. This is due to the fact that {the SMASH model is complete and} the couplings {of the new particles (including the inflaton)}  to the Standard Model are either known or well-constrained. For the purposes of this paper, there are two fundamental questions that a study of reheating must answer. First, is the PQ symmetry restored after inflation? 
If so, there is a one-to-one relation between the axion dark matter relic abundance and the axion decay constant. Fitting the former,  implies $f_A\sim 10^{11}$ GeV, see \ Section~\ref{dark_matter}. If not, {both smaller and larger values of $f_A$ can be allowed}, but {axion isocurvature constraints must be respected.} Second, is the reheating temperature, {$T_R$,} large enough to ensure a thermal bath of RH neutrinos that produces the adequate lepton asymmetry?
It turns out that both {questions} can be answered  in SMASH,  and for both the answer can be positive {(in the appropriate regions of parameter space).}

{Let us} consider first the dynamics of the background {fields} after inflation. For $\lambda_{H\sigma}>0$ the inflaton {is} $\rho$, while for $\lambda_{H\sigma}<0$ it {has} a Higgs component, with inflation going along the direction $h/\rho\sim(|\lambda_{H\sigma}|/\lambda_H)^{1/2}\ll1$. {In both cases,} when the slow-roll period {(and inflation)} ends, $\epsilon \sim 1$, the inflaton field, which will be denoted as $\phi$, has a value
\be
\frac{\phi_{\rm end}^2}{M_P^2} \simeq \frac{\sqrt{1+32\xi+192\xi^2}-1}{2\xi (1+6\xi)}, 
\ee
which changes from $8$ to $1$ in the {$\xi$-range} $10^{-3}-1$. 
These are still {large values of $\phi$,} but not enough for the non-minimal coupling to affect sizeably the potential because $\xi \phi_{\rm end}^2/M_P^2 \lesssim 1$ and $\Omega^2\sim 1$. Therefore, it is justified to neglect the effects of the non-minimal coupling from here on, to study the reheating {process.\footnote{For $\xi_\sigma\sim 1$, one gets $\Omega^2\sim 2$. This value of $\Omega^2$  is sufficiently low to neglect the non-minimal coupling for reheating, because after a single oscillation the potential seen by the inflaton is essentially quartic.}} Once freed from the large friction term induced by the accelerated expansion of the Universe, the inflaton {rolls} towards the origin, accumulating kinetic energy and {then} oscillating around $\phi\sim 0$ with decreasing amplitude. When the amplitude of oscillations is {sufficiently} large, we may neglect {the effect of $v_\sigma=f_A$} and effectively consider that the inflaton rolls in a quartic potential, with an equation (neglecting fluctuations of other fields):
\begin{equation}
 \ddot\phi+3{\cal H}\dot\phi+\lambda\phi^3=0.
\end{equation}
As seen in Section \ref{inflation}, the effective coupling $\lambda$ is given by $\lambda_\sigma$ in the case $\lambda_{H\sigma}>0$ (HSI), or by equation \eqref{tildes} in the case $\lambda_{H\sigma}<0$ (HHSI).
Given the absence of any relevant dimensionful couplings, it is possible to define rescaled field and space-time variables such that the {Hubble} expansion of the Universe is {effectively} factored out of the dynamical equations.  This {is} achieved by defining a rescaled, dimensionless field $F$ and using a rescaled conformal {time $\tau$ \cite{Greene:1997fu},\footnote{Here $t$ represents the standard cosmic time of a metric $ds^2=dt^2-a(t)^2 dx^2$.}}
\begin{align}
\label{eq:F}
 F=\frac{a \phi}{\phi_{\rm end}},\quad
 a d\tau = dt \sqrt{\lambda}\phi_{\rm end},
\end{align}
where $a$ is the FLRW metric's scale factor. {We choose to indicate the scale factor at the end of inflation by} $a_{\rm end}=a(0)=1$ so that $F(0)=1$. Denoting the derivatives with respect to $\tau$ with primes, the equation for the inflaton becomes
\be
F''+F^3=0 ,
\ee
where we have neglected ${\cal H}''$ terms, irrelevant after a few oscillations. 
The solution, with initial condition $F(0)=1, F'(0)\simeq 0$, is described by an oscillating Jacobi sine,
\be
\label{eq:bgosc}
F(\tau)= {\rm sn}\(\frac{\tau}{\sqrt{2}}+K(-1),-1\)\sim\cos(c \tau), \quad\text{ with}\quad K(-1)=1.31103,\ c=0.8472.
\ee
Evaluating Friedmann's equation, one sees that the conformally oscillating inflaton makes the Universe expand as during radiation domination, with 
\begin{align}
\label{eq:atau}
 a(\tau)=1+\frac{1}{2\sqrt{3}}\frac{\phi_{\rm end}\tau}{M_P}, \quad \tau=2\cdot(3\lambda)^{1/4}\sqrt{tM_P}. 
\end{align}
In the original coordinates, the inflaton's oscillations have a decaying amplitude $\phi_0(t)= \phi_{\rm end}/a$, and a frequency proportional to the latter \cite{Shtanov:1994ce},
\begin{align}
\label{eq:oscill}
 \phi(t)=\phi_0(t)\cos\(2c\sqrt{\lambda}\phi_0(t)t\).
\end{align}
The approximation of neglecting $v_\sigma=f_A$ in the evolution is valid until the scale factor has increased such that $\phi_0(\tau)\sim f_A$. {After that, the expansion ceases} to be conformal. This happens at a time $\tau_{\rm PQ}\sim 2\sqrt{3}M_P/f_A$, when the scale factor is $a_{\rm PQ}\sim \phi_{\rm end}/f_A$; {and the PQ symmetry (which had been non-thermally restored in the process), now gets broken, as we will  see later in detail.}

Reheating {proceeds} as the oscillating background loses its energy through decays {and/or} annihilations. This will cause a feedback in the oscillations, affecting the evolution of their amplitude $\phi_0(t)$. In conformal rescaled units, we may generalise  \eqref{eq:bgosc} to the ansatz
\be
\label{eq:bgosc2}
F(\tau)\sim F_p(\tau)\cos\(c F_p(\tau) \tau\).
\ee

Reheating in models with non-minimal gravitational couplings has been studied in a variety of papers \cite{Bezrukov:2008ut,GarciaBellido:2008ab,Lerner:2009xg,Lerner:2011ge}; however, these articles focused 
on the $\xi\gg 1$ regime (either for the Higgs or another scalar), for which the inflaton oscillates in a potential which is approximately quadratic. 
Thus, our treatment {(and results!)} will be somewhat different. Another peculiarity of SMASH that influences the sequence of events after inflation is that, in the preferred regions of parameter space, the self-coupling $\lambda$ of the inflaton is much smaller than its couplings to other fields, such as the Higgs,  right-handed neutrinos and vector quarks, as well as vector bosons in the HHSI case. This has 
the effect that the fields that couple to the inflaton get time-dependent masses which, most of the time, remain larger than the oscillation frequency of the background, {$\sim\sqrt{\lambda}\phi_0(t)$, see equation \eq{eq:oscill}.} For $\lambda_{H\sigma}>0$, neglecting again {all the dimensionful couplings,} the tree-level inflaton-induced masses for the fluctuations are:
\be
\label{eq:HSImasses}
\begin{aligned}
&\lambda_{H\sigma}>0:\\
&
m^2_{\sigma_1} =3{\lambda_\sigma} \phi^2 \quad;\quad m^2_{\sigma_2} = \lambda_{\sigma}\phi^2 \quad ; \quad
m^2_h = \lambda_{H\sigma}\phi^2 \quad ; \quad
m_{Y_i} = \frac{Y_{i}}{\sqrt{2}}\phi \quad ; \quad
m_Q = \frac{y}{\sqrt{2}}\phi .
\end{aligned}
\ee

In the equations above,  $\sigma_1$ and  $\sigma_2$ denote the two components of the complex field $\sigma$. We {use} $\sigma_1$ to {denote} the inflaton direction (the one having a VEV during inflation) and $\sigma_2$ its orthogonal direction.\footnote{Note that $\sigma_1$ does not have to be aligned with $\theta=0$, i.e. the CP conserving minimum of the axion potential.} For convenience, in this section (and only in this section) we have chosen to  define $\rho=|\sigma|$ reabsorbing a factor of $\sqrt{2}$. 

For $\lambda_{H\sigma}<0$ the inflaton is a linear combination of $\sigma_1$ and the {Higgs, with a small} mixing angle {$|\lambda_{H\sigma}|/\lambda_H$.} Ignoring dimensionful parameters and going to a basis that diagonalises the mass matrix (corresponding to the parallel and orthogonal directions to the bottom of the potential energy valley along which the background oscillates), the relevant tree-level induced {(and oscillating)} masses {are,} to lowest order in $|\lambda_{H\sigma}|/\lambda_H$,
\be
\label{eq:HHSImasses}\begin{aligned}
&\lambda_{H\sigma}<0:\\
&
m^2_{||} =3\widetilde \lambda_\sigma \phi^2 \quad;\quad m^2_{\sigma_2} = \widetilde \lambda_{\sigma}\phi^2 \quad ; \quad
m^2_{\perp} = 2|\lambda_{H\sigma}|\phi^2 \quad ; \quad
m_{Y_i} = \frac{Y_{i}}{\sqrt{2}}\phi \quad ; \quad
m_Q = \frac{y}{\sqrt{2}}\phi,\\
&m^2_W=\frac{g^2}{4}\frac{|\lambda_{H\sigma}|}{\lambda_H}\phi^2 \quad ; \quad m^2_Z=\frac{g^2+{g'}^2}{4}\frac{|\lambda_{H\sigma}|}{\lambda_H}\phi^2 \quad ; \quad m_t=\frac{y_t}{\sqrt{2}}\sqrt{\frac{|\lambda_{H\sigma}|}{\lambda_H}}\phi.
\end{aligned}
\ee
Note how in this case the  Higgs component of the inflaton induces a mass for the gauge bosons and the top quark. {The symbol} $m^2_{||}$ corresponds to the mass {squared} of {(longitudinal)} fluctuations parallel to the inflaton (mostly aligned with $\sigma_1$) while $m^2_{\perp}$ denotes the mass of excitations orthogonal to the inflaton (mostly aligned with the Higgs).

The oscillating background  can be understood as a condensate of {inflaton} excitations with {average} energies equal to the oscillation frequency. {When this frequency $\omega \sim\sqrt{\lambda}\phi_0(t)$} stays below the masses of {the} other particles, decays or annihilations of the {condensate}  are forbidden. This hampers the energy transfer from the inflaton to the SM particles that couple to it, and the usual perturbative estimates of reheating through fermionic or scalar couplings (such as the perturbative results {of} \cite{Shtanov:1994ce}, as well as the {usual} parametric resonance estimates in the adiabatic regime), are not valid as they do not take into account {these} kinematic blocking effects. 

{Contrary to what happens for Higgs particles,} right-handed neutrinos and $Q,\tilde Q$ fermions (and gauge bosons and top quarks in the HHSI case), {the excitations of } $\sigma$ (i.e. non-zero modes of $\sigma_1$ and $\sigma_2$) {are} more easily produced, {which prompts two successive questions.}
First, whether an efficient growth of the inflaton perturbations may restore the PQ symmetry and/or destroy the coherence
of the background. Such loss of coherence could end up opening the decays towards SM particles. If the answer {to this first question} is negative,  the ensuing question is whether the oscillating background can still lose energy by producing  SM particles, and fast enough so that these may reheat {the Universe.} This production is typically more efficient for bosonic fields coupling to the inflaton, such the Higgs in HSI and HHSI, as well as the gauge bosons in HHSI.

{As will be seen {in the next subsection,} the efficient production of $\sigma$ excitations {does lead} to a {non-thermal} restoration of the PQ symmetry. Although an oscillating zero mode survives {the growth of these fluctuations,} its initial energy {becomes} shared with {them,} which grow up to the point of stopping the oscillations of the {background (and hence of the induced Higgs mass). In HSI, the} {resulting (large)} Higgs mass prevents {the} production of {Higgs excitations at this point,} which makes reheating in HSI inefficient until {the time $\tau_{\rm PQ}$ at which the Peccei-Quinn symmetry is spontaneously broken (after inflation has ended).} In the HHSI case, the induced masses of the gauge bosons are not affected by $\sigma$ fluctuations and keep oscillating, so that particle production can be maintained and reheating temperatures $\sim 10^{10}$ GeV {(which are larger than in HSI, where they are $\sim 10^7$ GeV)} can be achieved. In the following, we analyse in turn the preheating {process} into $\sigma$ excitations, reheating in HSI, and reheating in HHSI. }

\subsection{Preheating into $\sigma$ excitations}

{As we already mentioned above, the} production of $\sigma$ excitations {is less kinematically blocked than that of the rest of the particles} and  could {in principle} be efficient {enough} {to threaten} the coherence of the background. 
Even though the {self-interactions} of the {two components of $\sigma$,} determined by $\lambda_\sigma$, {are} tiny, nonperturbative processes {could} give rise to resonant production. Such effects, which  effectively resum many-body processes $\phi+\phi+\phi+... \to ...$, can be described in a semiclassical way. To simplify the treatment, here we will consider the HSI case. Since in HHSI the inflaton is mostly aligned with the $\sigma$ field, we expect similar results {regarding the production of} $\sigma$ excitations in both cases.

 Following the standard reference \cite{Greene:1997fu}, resonant production  can be understood   by expanding the fluctuations in Fourier modes. 
In {analogy} to equation \eqref{eq:F}, we {define} $F_i(x)={a \sigma_i}(x)/{\phi_{\rm end}}$ and their {Fourier} components. The zero mode $F_{1,0}=F$ corresponds to the homogeneous inflaton, which evolves according to \eqref{eq:bgosc}, while non-zero {$\kappa$-modes} (fluctuations) satisfy
\bea
\label{perturbations}
&F_{1,\kappa}'' +(\kappa^2+ 3F^2(\tau))F_{1,\kappa} = 0, \\
&F_{2,\kappa}'' +(\kappa^2+ F^2(\tau))F_{2,\kappa} = 0, 
\eea
at first order in perturbations. We label the fluctuations with a rescaled comoving momentum $\kappa=k/(\sqrt{\lambda_\sigma}\phi_{\rm end})$. Owing to the time dependent frequency {$W_{i}(\kappa,\tau)=\kappa^2+b_iF^2(\tau)$} ($b_1=3,b_2=1$), modes in certain frequency bands are unstable and grow exponentially: $F_{i,k}(\tau)\sim F_{i,k}(0) e^{\mu_i \tau}$, where {$\mu_i$} is the corresponding {Floquet} index. This {exponential growth} corresponds to resonant particle production. For $b_1=3$ the instability band is very narrow $\kappa\in(1.5,\sqrt{3})$ and the maximum Floquet index inside the band is small $\mu_1=0.03598$. 
Fluctuations in the orthogonal direction $F_2$ are unstable in a larger band $\kappa\in(0,0.5)$ and have a larger maximum Floquet index $\mu_2=0.1470$ \cite{Greene:1997fu}. With  initial conditions fixed by quantum uncertainties,
the fluctuations {reach} $\langle F^2_{2}(x)\rangle \sim {\cal O}(1)$ around $\tau\sim 100$. This corresponds to $\sim 14$ full oscillations and a Universe expansion factor of {$a(100)/a(0)\sim 30 \phi_{\rm end}/M_P$ from the end of inflation. 
Once these fluctuations grow significantly,} rescattering effects, $F_{2,\kappa}F_{2,\kappa'}F_{1,0}\to F_{1,\kappa+\kappa'}$, that we have neglected in \eqref{perturbations} accelerate very much the growth of $F_1$ perturbations, which would otherwise take much longer to grow \cite{Tkachev:1998dc}, so that they also become ${\cal O}(1)$ at $\tau \sim 100$. 

In order to confirm this picture and to check whether the PQ symmetry is indeed restored, we have performed lattice numerical simulations of the preheating stage following the work of \cite{Tkachev:1998dc}. We have solved numerically the equations of motion of $\sigma_1,\sigma_2$, in their rescaled version
\bea
\label{eq:fullsigma}
F_1''-\nabla^2F_1+F_1(F_1^2+F_2^2)=0 \\ \nonumber
F_2''-\nabla^2F_2+F_2(F_1^2+F_2^2)=0 
\eea
with initial conditions given by a homogeneous zero mode $F_{1,0}(0)=1$ and non-zero modes populated statistically according to the quantum uncertainty principle, see \cite{Tkachev:1998dc}. 
Our results confirm the findings of \cite{Tkachev:1998dc}. The main messages are conveyed in Figs.~\ref{fig:osc} and   \ref{fig:SSB}.  {There is a first period of time,} so-called preheating, {in which} fluctuations in the $\sigma_1$ and $\sigma_1$ direction grow exponentially until they reach values comparable to the {background amplitude, $\langle \delta\sigma^2_1\rangle\sim \langle \delta\sigma^2_2\rangle \sim\, \phi_0^2$.    
Although one cannot speak of temperature yet, the PQ symmetry becomes effectively restored as the value of $\theta=\arctan\(\sigma_2/\sigma_1\)$ takes now random values between $-\pi$ and  $\pi$ in different regions of the Universe. Such behaviour is shown in Fig.~\ref{fig:SSB}, which shows values of $\theta$ in {a} 2D slice at the end of our simulation ($\tau=400$), although times after $\tau\sim 100$ show similar behaviour.  
We observe that after $\tau\sim 100$, the fluctuations of the modulus of $\langle |F|^2 \rangle$ tend to a constant value of $0.5$, with small oscillations that decrease in size very slowly. 
This is the process of turbulent thermalisation described in \cite{Micha:2002ey,Micha:2004bv}, with the {difference} that in our case the inflaton is a complex field. Once the fluctuations are {of} ${\cal O}(1)$, the real and complex scalar field cases are very similar and \cite{Micha:2002ey,Micha:2004bv} provide a good guidance to the thermalisation of the zero mode, i.e. the slow transfer of energy from the zero mode to the fluctuations that we start to see from $\tau=100$ to $\tau=400$ in Fig. \ref{fig:osc}. After $\tau\sim 100$ the {spectra of $\sigma_1$ and $\sigma_2$ fluctuations become power laws with an exponential cutoff that grows in time, increasing slowly as energy is drained from the zero and {lowest} momentum modes with $\kappa\sim 1$ towards {higher momentum modes, with the system evolving towards a thermal distribution.} 

\begin{figure}
\begin{center}
\includegraphics[width=0.83\textwidth]{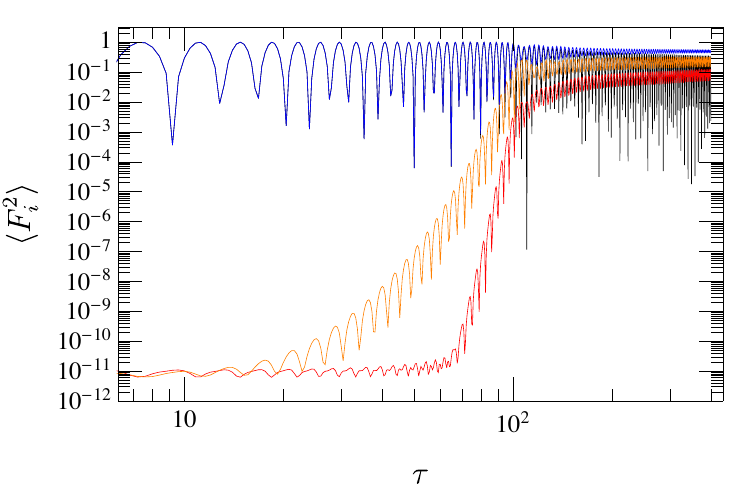}
\includegraphics[width=0.83\textwidth]{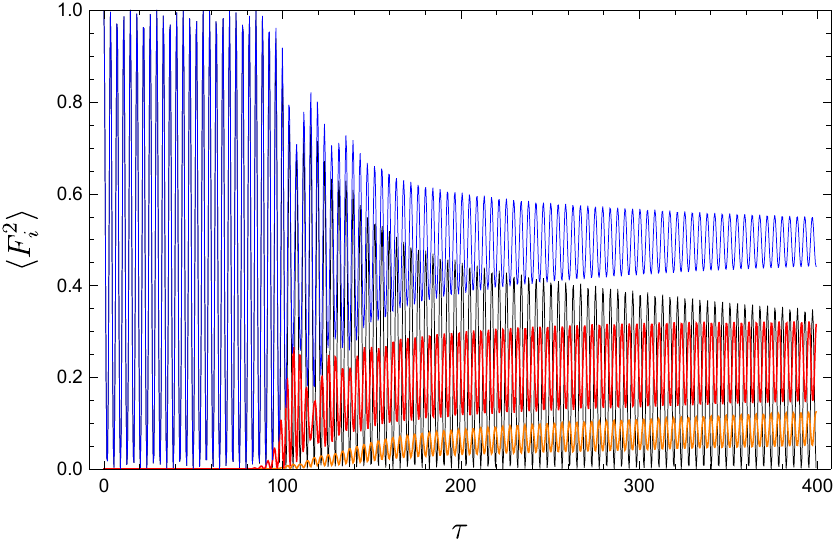}
\caption{\small  {Time-evolution of spatial averages of the  perturbations squared: inflaton zero mode $F_{1,0}^2$ (black), fluctuations in the direction of the inflaton $\langle(F_1-F_{1,0})^2\rangle$ (orange),  in the orthogonal direction $\langle F_2^2\rangle$ (red) and the total sum $\langle F_1^2+F_2^2\rangle$ (blue). Note  the distribution of energy between the background zero mode and the inflaton excitations for $\tau>100$. All quantities are in conformal rescaled units.}}  
\label{fig:osc}       
\end{center}
\end{figure}

\begin{figure}[h!]
\begin{center}
\includegraphics[width=0.8\textwidth]{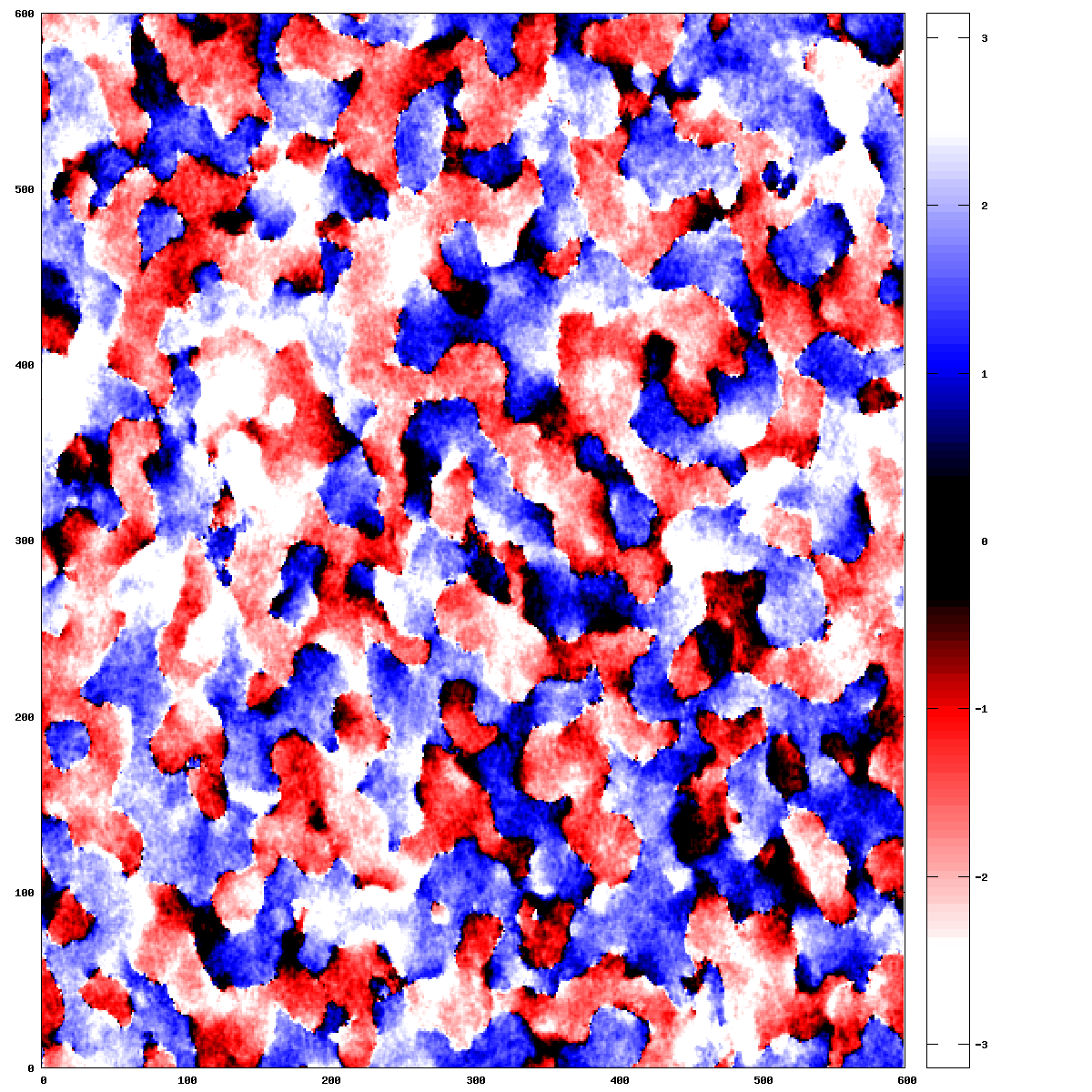}
\caption{\small  2D slice of our 3D simulations of the growth of perturbations of the inflaton during preheating at $\tau=400$ showing patches of the Universe with different values of $\theta$. The length in the abscissa  is given in comoving units $1/(\sqrt{\lambda}\phi_{\rm end})$.}  
\label{fig:SSB}       
\end{center}
\end{figure}

We recall that our conclusion is based on neglecting $f_A$ in the inflaton potential and effects of the other fields. 
The first assumption is justified if around $\tau\sim 100$ the typical amplitude of the oscillations {$\langle |\phi(\tau=100)|^2\rangle\gg f_A^2$.} Using {$\langle |\phi(\tau=100)|^2\rangle\sim 0.5 \phi_{\rm end}^2/a^2$} and \eqref{eq:atau} we conclude that 
for $f_A \lesssim 4\times 10^{16}$ GeV the PQ symmetry gets restored. For larger values of {$f_A=v_\sigma$,  the field} $\sigma_1$ will get trapped around the true minimum $\rho=f_A$ before the fluctuations grow large enough to homogeneise {the angle} $\theta$. 
We comment later on on this scenario {of large $f_A$,} which {turns out to be} excluded by {either} the dark matter abundance or isocurvature perturbations in the CMB. {For the time being} we concentrate in the small $f_A$ scenario. 

Let us now {explain} why other fields are not relevant during preheating. Fluctuations in the Higgs field are more strongly coupled to the inflaton {than} {those of} $\sigma_1,\sigma_2$ because {$|\lambda_{H\sigma}|\gg \lambda_\sigma$} in SMASH. Their exponential growth is faster, but it is quenched very early due to {the} large {Higgs} induced mass, which drives the fluctuations away from their resonance band; see \cite{Greene:1997fu}. Thus, the Higgs plays no significant role in the dynamics of the growth $\sigma$ fluctuations. This conclusion {holds} independently of the sign of  $\lambda_{H\sigma}$, yet it should be noted that for $\lambda_{H\sigma}>0$ parametric resonance is further hindered because Higgses with a large inflaton-induced mass can decay very quickly into massless tops or gauge bosons. {Instead, for} $\lambda_{H\sigma}<0$, the background induces a mass on the tops, as in equation \eqref{eq:HHSImasses}, and this decay is typically closed.

{

\subsection{Reheating in HSI}

In HSI, the oscillating background can in principle decay or annihilate at tree-level into Higgses, quarks $\tilde Q, Q$ and right-handed neutrinos. The decays and annihilations are open as long as the momentum of the background condensate is larger than the sum of the  induced masses of the product particles. As noted before, the couplings determining the induced masses, ($\sqrt{\lambda_{H\sigma}}, y$ and $Y_{ii}$, respectively) are typically much larger than $\sqrt{\lambda_\sigma}$, which sets the frequency of oscillation of the background, $\omega\sim\sqrt{\lambda_{\sigma}}\phi_0(t)$.
Hence for an oscillating  $\phi=\sigma$, which induces oscillating masses, the decays or annihilations can only happen when $\sigma$ crosses the origin and the induced masses approach zero. 
This is possible during preheating, i.e. $\tau\lesssim 100$, but not after because effective masses are actually proportional to $\langle |\sigma|^2\rangle $ which soon tends to $0.5 \phi_{\rm end}^2/a^2$ (and will decrease very slowly but without further crossing zero), see Fig.\ \ref{fig:osc}. This implies that the masses of the particles coupling directly to the inflaton set on values much above the frequency of oscillation $\omega$. This closes the particle production channels from the background condensate. The decays of the excitations themselves are also  closed at this stage, as their typical momentum is of the order of $\omega$.

Thus, reheating is quenched until our assumption of neglecting $f_A$ is not valid anymore. 
When the amplitudes become of the order of $f_A$, the PQ symmetry becomes broken. 
This happens when quadratic terms in the potential become relevant, and the $\sigma$ fluctuations end up confined in potential wells around $|\sigma|=f_A/\sqrt{2}$. After  {the PQ} symmetry breaking, the physics is {described more easily using} a massive modulus excitation $\rho$, with mass $m_\rho=\sqrt{2\lambda_\sigma}f_A$ and a {massless axion; see equations \eq{sigma:} and \eqref{eq:masses}).} When the amplitude of {$\rho$ becomes} small enough, {it} can decay into Higgses. When the decay opens, the decay rate quickly overcomes the Hubble rate, so that the $\rho$ excitations transfer all their energy into Higgses, which {then} thermalise with other particles. 

The reheating temperature of the {resulting} particle bath can be estimated from the energy stored in $\rho$ excitations at the time at which the {decay opens.} In order to calculate the reheating temperature, we may start by finding out the time of spontaneous breaking of the PQ symmetry. This happens when $\langle |\sigma|^2\rangle=\phi_{\rm end}^2 (a_{\rm end}/a)^2\langle F_1^2+F_2^2\rangle\sim f_A^2$. 
The $\sigma$ fluctuations in rescaled conformal coordinates evolve approximately as $\langle F_1^2+F_2^2\rangle \sim 0.5 (100/\tau)^{2/7}$ for $\tau>100$ \cite{Micha:2004bv}. Using Equation \eqref{eq:atau} for the evolution of the scale parameter, we get the following estimate for the time $\tau_{\rm PQ}$ of PQ breaking,
\be
\tau_{\rm PQ}\sim 4\left(\frac{M_P}{f_A}\right)^{7/8}\sim  10^7 \(\frac{10^{11}\rm GeV}{f_A}\)^{7/8}.
\ee
After the PQ breaking, the quadratic terms cannot be neglected, and as said before the appropriate description of the physics {is better done using the} modulus {$|\sigma|=(\rho+v_\sigma)/\sqrt{2}$}  and the axion, rather than the fluctuations $\sigma_i$. 
The $\rho$ particles (fluctuations) cannot decay immediately into Higgses because their mass is\footnote{In addition, there are non-thermal corrections $\propto \langle \rho^2\rangle$ that are not necessary for this discussion.}  $m_{\rho}(\tau_{\rm dec})=\sqrt{2\lambda_\sigma} f_A$ 
and the Higgs mass is $m_h= \sqrt{\lambda_{H\sigma}\langle\rho^2\rangle}$, parametrically larger because $\lambda_{H\sigma}\gg \lambda_\sigma$ and $\langle\rho^2\rangle\sim f_A^2$ at the time of the PQ {phase} transition. 
The decays will be open at a {later} time, $\tau_{\rm dec}$ when $m^2_{\rho}(\tau_{\rm dec})=2\lambda_\sigma f^2_A = 4m^2_h=4 \lambda_{H\sigma}^2\langle\rho^2(\tau_{\rm dec})\rangle$. 

Thus we need to estimate the time evolution of $\langle\rho^2\rangle$ during the PQ phase transition and after {it,} to compute the time of the decay. 
Naively, we expect that during the PQ phase transition the energy stored in the {fluctuations of $\sigma_1$ and $\sigma_2$} is transferred democratically to {fluctuations} of $\rho$ and {the} axion $A$. However we {do not know} of any numerical calculation {that confirm} this intuition. We have thus performed a numerical simulation of the PQ phase transition to answer this question. 
{We solved the classical evolution of $\sigma$ with the full $\sigma$ potential, i.e. including $f_A$. In conformal version \eqref{eq:fullsigma}, this amounts to change $F_1^2+F_2^2\to F_1^2+F_2^2-f_A^2a^2/\phi_{\rm end}^2$ in the last term. Again, we start with initial conditions including the inflaton as a zero-mode plus quantum fluctuations.} We used the value $f_A=5.4\times 10^{15}$ GeV, which gives $\tau_{\rm PQ}\sim 800$, to separate as much as possible the PQ phase transition from the dynamics at $\tau\sim100$ allowing a relatively large degree of turbulent thermalisation without compromising a lengthy simulation. 
We wrote the energy density as a function of $\rho$ and $A$ and plot their evolution in Fig.~\ref{fig:PQtransi}. 
We can see that before $\tau\sim 100$ all the energy is in the radial mode, which of course is the inflaton, but after that time it is shared between $\rho$ and $A$, as we would expect from {the} symmetry restoration {that we observed before {while} studying the fluctuations of $\sigma_1$ and $\sigma_2$.} Up to $\tau_{\rm PQ}\sim 800$ the graph is not particularly revealing because $\rho$ and $A$ are not the adequate degrees of freedom. After $\tau_{\rm PQ}$, the PQ symmetry breaks down and, relatively fast, the energy becomes equipartitioned between $\rho$ and $A$, confirming our intuition. 
The typical momentum of {the fluctuations of $\sigma_1$ and $\sigma_2$} before the PQ phase transition {is} $\kappa_{ 0}(\tau)\sim \kappa_{100}(\tau/100)^{1/7}$, where we introduce the shorthand notation 
$\kappa_{100}=\kappa_{ 0}(\tau=100)$. With typical values {$\kappa_{100}\sim 1$,} the momentum is ${\cal O}(1)$ larger than their effective mass so they are mildly relativistic. 
{The typical momenta are also larger than the corresponding masses for $\rho$ and of course with $A$ (which is massless)} {after the phase transition.}

\begin{figure}[t]
\begin{center}
\includegraphics[width=0.8\textwidth]{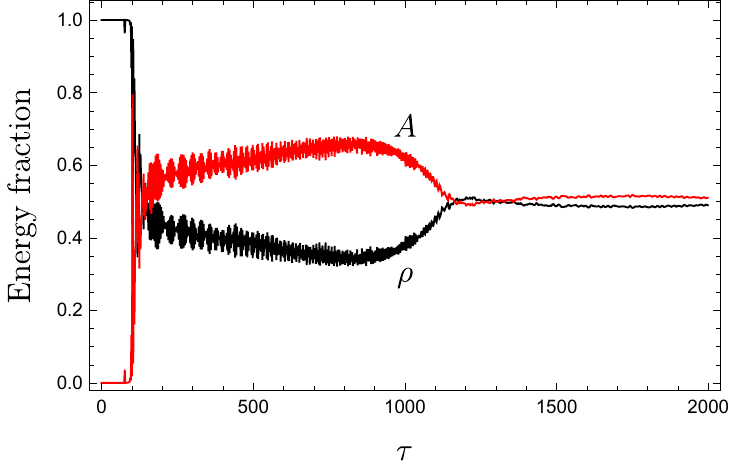}
\caption{\small  Evolution of the energy density in $\rho$ and the $A$ fields across the phase transition at $\tau_{\rm PQ}\sim 800$. After the phase transition the energy density is equipartitioned. }  
\label{fig:PQtransi}       
\end{center}
\end{figure}
 
After the PQ phase transition, we expect the $\rho$ modes to evolve in such a way that $\omega_k \langle \rho^2\rangle a^3$, with $\omega_k=\sqrt{(k/a)^2+m^2}$ the mode's energy, remains constant. {In other words, the} conservation of the comoving number of quanta {should be satisfied, as it follows} from the WKB approximation. With this, we can estimate the scale factor at $\tau_{\rm dec}$ {(when the decays open)}. Using that at $\tau=\tau_{\rm PQ}$ one has $\langle \rho^2(\tau_{\rm PQ})\rangle\sim f_A^2$, we find
\begin{align}
 \frac{a(\tau_{\rm dec})}{a(\tau_{\rm PQ})}\sim 78\left(\frac{\kappa_{100}^3\, \delta^2_3 }{{f_{11}}  \lambda^2_{10}}\right)^{1/12}, 
\end{align}
where we have defined the following SMASH dimensionless parameters for future convenience:
\be
\label{eq:SMASHpars}
\lambda_{10}=\frac{\lambda_\sigma}{10^{-10}}  \quad ; \quad \delta_{3}=\frac{\delta}{0.03 }\quad ; \quad f_{11}=\frac{f_A}{10^{11}{\rm GeV}}.
\ee
As momenta redshift with the scale factor, and {the $\rho$ particles} are only barely relativistic at the PQ phase transition, {they} are non-relativistic when they decay. This implies that the Universe will be matter dominated for a little while before $\rho$ particles decay and reheat the Universe. {Note however, that matter domination lasts typically  $\sim 1$ e-fold so the the effect can be neglected in the prediction of the number of e-folds, which we discuss in Section \ref{sec:efolds}.}

Finally, the reheating temperature may be estimated as follows. The preheating simulations showed an equipartition of the inflaton's energy so that half the total energy goes into {$\sigma$} excitations. The particle number $n_\rho(\tau_{\rm PQ})$ just before PQ breaking can be estimated as the total energy over the typical momentum $k_{\rm 0}(\tau_{\rm PQ})$. {After  PQ breaking, the $\rho$ particles get a mass} $m_\rho$ and their total energy when the decay opens can be estimated as $m_\rho n_\rho(\tau_{\rm dec})=m_\rho n_\rho(\tau_{\rm PQ}) (a(\tau_{\rm PQ})/a(\tau_{\rm dec}))^3$. After decay, all this energy is transferred into Higgses that subsequently thermalise with other particles. Assuming instant thermalisation with a relativistic thermal bath containing all the SM degrees of freedom, {we obtain} the following estimate of the reheating temperature:
\be
\label{eq:HSITR}
T_R\sim 10^7 \frac{f_{11}\lambda_{10}^{3/8}}{\kappa_{100}^{1/4}\, \delta_3^{1/8}}\quad {\rm GeV}.
\ee

The above estimate is based on the energy equipartition between $\rho$ and axion excitations. This means that there is a relativistic gas of axion particles that can contribute to the effective number of neutrino species, $N_{\rm eff}$. In order to estimate the present energy density of axions, we note that equipartition and the fact that $\rho$ and axions are relativistic with the same typical momentum implies that there is an equal number of axion and $\rho$ particles. At the time $\tau_{\rm dec}$, the energy of the axions will be of the order of $k_{\rm 0}(\tau_{\rm PQ}) a(\tau_{\rm PQ})/a(\tau_{\rm dec})$, while that of the $\rho$ particles will be given by $m_\rho$. All the $\rho$ particles decay into SM degrees of freedom, which reheat to a temperature which is not enough to guarantee thermal equilibrium between the SM and axion radiation baths, see Sec.~\ref{sec:axionradiation}.
Therefore the SM and axion radiation remain decoupled, and  the ratio of energies of the SM and axion radiation bath at $\tau=\tau_{\rm dec}$ is simply
\be
\frac{\rho_{\rm SM}(\tau_{\rm dec})}{\rho_A(\tau_{\rm dec})}=
\frac{m_\rho}{k_{0}(\tau_{\rm PQ})}\frac{a(\tau_{\rm dec})}{ a(\tau_{\rm PQ})}\sim  
2.8 \left(\frac{\delta_3 f_{11}}{\lambda_{10}\,\kappa_{100}^4}\right)^{1/6}.
\ee
The ratio at the present time can be obtained by accounting for the change of relativistic degrees of freedom in the SM bath as particles decouple from it and inject entropy into the SM radiation, reheating it. 
Conservation of comoving entropy between reheating and today gives 
\be
\left.\frac{\rho_{\rm SM}}{\rho_A}\right|_{\rm today}=\frac{\rho_{\rm SM}(\tau_{\rm dec})}{\rho_A(\tau_{\rm dec})}
\frac{g_{*,\rm today}}{g_{*,\rm dec}}\(\frac{g_{*S,\rm dec}}{g_{*S,\rm today}}\)^{4/3}=
\frac{\rho_{\rm SM}(\tau_{\rm dec})}{\rho_A(\tau_{\rm dec})}
\frac{3.36}{106.75}\(\frac{106.75}{3.9}\)^{4/3}=
7.3 \left(\frac{\delta_3 f_{11}}{\lambda_{10}\,\kappa_{100}^4 }\right)^{1/6}. 
\ee
With the SM radiation density fixed by the measured CMB temperature, the ratio $\rho_{\rm SM}/\rho_A$ allows to estimate the axion radiation density today {and} its contribution to the current effective number of relativistic neutrinos,  
\be
\label{eq:HSINeff}
\Delta N_{\rm eff}=
\frac{\rho_A}{\rho_{1\nu}}= 
\frac{\rho_A/\rho_{\rm SM}}{\rho_{1\nu}/\rho_{\rm SM}}=
\frac{\rho_A/\rho_{\rm SM}}{7.4}\simeq
\left(\frac{\lambda_{10}\,\kappa_{100}^4}{\delta_3 f_{11}}\right)^{1/6}.
\ee
where we have used that the density of one standard neutrino species is $\rho_{1\nu}= \frac{\pi^2}{30}\frac{7}{4}T_0^4\(\frac{4}{11}\)^{4/3}$ and $\rho_{\rm SM}=3.36 \frac{\pi^2}{30}T_0^4$. 

Using $\kappa_{100}\sim 1, \delta_3\sim 1, f_{11}\sim 0.6$ (see {Sec. \ref{dark_matter}),} $\Delta N_{\rm eff}$ lies in the interval $0.35-1.6$ in the SMASH range $\lambda_{10}\in(10^{-3},10)$. This range is mostly ruled out by the latest measurements from baryon acoustic oscillations and the CMB, which give the constraint $N_\nu^{\rm eff} = 3.04 \pm 0.18$ at 68\% CL  \cite{Ade:2015xua}. 
Thus, although it is borderline for $\lambda\sim 10^{-13}$ (corresponding $\xi_\sigma\sim 2\times 10^{-3}$ and the largest {possible} tensor to scalar ratio {$r\sim 0.07$ at 0.05 Mpc$^{-1}$}), we conclude that HSI inflation is in conflict with dark radiation constraints.
{In the next subsection we will analyse the HHSI case which turns out to evade these constraints.}

\subsection{Reheating in HHSI}

Reheating in HHSI, i.e. for $\lambda_{H\sigma}<0$, is qualitatively different from its HSI counterpart. The reason is that in {HHSI} the inflaton has a Higgs component, which, although small,
allows for the production of gauge bosons from the background. Moreover, our preheating simulations showed that there is no efficient production of Higgs fluctuations, and thus $h^2$ will be dominated
by the oscillating contribution from the zero mode. {Therefore,} the gauge boson masses will keep oscillating in time (in contrast to the Higgs masses, as seen before), crossing zero when the inflaton's zero mode approaches {the} origin, and allowing for sustained particle production at the times near the crossings in which the annihilation of the zero-mode condensate into gauge bosons remains open.

It should be {recalled} that {the requirement of stability of the effective potential} disfavour HHSI realisations which are small deformations of {HSI}. This is because {stabilisation by means of the tree-level threshold effect} enforces a minimum of $|\lambda_{H\sigma}|$ {for a given $\lambda_\sigma$} (see {Figure} \ref{fig:mindelta}). {Therefore,} {typical} stable HHSI models lead to a post-inflationary history that is substantially different from that in {HSI.}{\footnote{Possible exceptions involve models which are stabilised by one-loop threshold contributions, which can happen for 
very small $\lambda_{H\sigma}$, see Fig.~\ref{fig:scan1_2}.}}

The mass of the gauge bosons in the zero mode background (see equation \eqref{eq:HHSImasses}) is determined by an effective coupling $\tilde g\equiv g^2|\lambda_{H\sigma}|/(4\lambda_{H})\gg \lambda_\sigma$,
so that {the production of gauge bosons from the condensate of particles} is kinematically blocked except when the inflaton is near the origin. Although gauge bosons are not produced away from the crossings,  this does not mean that the energy loss of the inflaton stops. Once gauge bosons are created
after a crossing, their induced masses start to grow as soon as the inflaton moves away of the {origin. As the inflaton} reaches its maximum and starts to move back, causing a decrease in the  induced masses, the energy ceded to the created particles would in principle be transferred back to the condensate. However, in the time in between crossings, when the gauge bosons are very massive, they can have enhanced decays into light quarks and leptons, which couple very weakly to the inflaton and thus remain light. In this way, the energy transferred by the inflaton to the emitted gauge bosons is dumped into light particle radiation, and this process can successfully reheat the SM plasma. The energy loss of the inflaton  can be further enhanced by thermal effects once the light particles thermalise, since the production of gauge bosons when they become light at the crossings will not only receive contributions due to annihilations or decays of the condensate, but also from annihilations of the thermally excited particles in the plasma. Again, the produced gauge bosons will decay  into radiation when their mass increases, and thus the rate of  energy loss of the inflaton will feed on itself. This thermal enhancement 
will be more efficient while the temperature of the light particle bath remains below the maximum value of the inflaton-induced gauge boson masses, because otherwise the produced gauge bosons will reach thermal equilibrium with the plasma, and their inverse  decays will stop being suppressed away from the crossings. Although at the crossings the inflaton will also produce RH neutrinos and vector quarks $Q,\tilde Q$, the latter have decay rates to lighter particles that are suppressed with respect to those of the $W$ and $Z$, and so they will play a lesser role in draining energy from the inflaton and will be ignored. {Similarly, the Higgs component of the inflaton can produce top quarks. However, their effective mass in the background is larger than that of the gauge bosons, and thus the nonthermal production rate of tops at the crossings will be smaller than that of $W,Z$ (note that the number of top final states is the same as that of $W^\pm$ and $Z$). Away from the crossings the tops can decay into $W$s and $b$ quarks, and the plasma can in turn contribute  tops at the crossings. However, the latter thermal production is suppressed with respect to that of $W,Z$, because gauge bosons are not in equilibrium with the thermal bath. Therefore, top quarks will be created from the plasma predominantly through annihilation processes, with the corresponding phase-space suppression with respect to inverse decays. This means that the thermal enhancement of top production is subdominant with respect to the case of gauge bosons, and so we will base our estimates on the latter.}

{It is interesting to note that in  HHSI, the Higgs is ultimately responsible for reheating the Universe. The dynamics, though, is different from that in Higgs Inflation or in the scenarios of reference \cite{Figueroa:2016dsc}. In HI the oscillating regime of the inflaton is different, and parametric production of gauge bosons is less hindered by kinematic blocking. On the other hand, reference \cite{Figueroa:2016dsc} considers a  Higgs condensate arising as a result of large perturbations induced by an external inflationary background, as opposed to the situation in HHSI in which the Higgs condensate is part of the inflationary background, and its fluctuations are suppressed due to the large induced mass.}

Next we summarise the three main processes taking place in this post-preheating evolution:  gauge boson production at the crossings, energy transfer to a light particle radiation bath away from the crossings, and the thermal enhancement of the energy loss of the inflaton. 

\subsubsection{\label{subsubsec:gbproduction}Nonthermal gauge boson production}

As was mentioned before, the production of gauge bosons remains kinematically blocked most of the time, and is only efficient near the times at which the inflaton crosses the origin.
Let us first estimate the production of particles during a single crossing, ignoring resonant effects. For simplicity we will start ignoring gauge-group and polarisation indices, and {model each gauge field as a scalar boson with an oscillating mass, and simply adding an overall} factor of 3 for the final rate to account for the three polarisations of a massive boson. We will also forego the distinction between $W$ and $Z$ bosons and consider a mass given by $m_W= \tilde g \phi^2$; again, we will approximate the final rate of $W$ and $Z$ production as twice the estimate for a single gauge boson. In analogy with the rescaled inflaton field $F=a \phi/\phi_{\rm end}$, we may express the bosonic fluctuation $w$ in terms of a dimensionless, conformally rescaled field $W$ with $w=a W/\phi_{\rm end}$. At first order $W$ satisifies the equation
\be
W_{k}'' + \Omega^2_k W_{k}  = 0 ,  \quad \Omega_k=\sqrt{\kappa^2+\frac{\tilde g}{\lambda}F^2},
\ee
where $\kappa$ is the momentum in rescaled conformal units, $\kappa=k/E$, with $E=\sqrt{\lambda}\phi_{\rm end} a_{\rm end}/a$.
One can study particle production by following the occupation number of the fluctuations. Following a semiclasical reasoning, the occupation number {for a mode of momentum $\kappa$} is defined as \cite{Greene:1997fu}
\be
f^{(W)}_\kappa = \frac{\Omega_k}{2 \lambda}\(|W_k|^2+\frac{|W'_k|^2}{\Omega_k^2}\)-\frac{1}{2}.
\ee
For initial boundary conditions motivated by quantum uncertainty,  $f^{(W)}_\kappa$ takes an initial value close to $f^{(W)}_\kappa \simeq   0$. 
The evolution of the fluctuations is mostly adiabatic in the occupation number whenever $\Omega'_k(\tau)<\Omega^2_\kappa(\tau)$. This condition is violated when the inflaton 
is close to the origin and for a small region of momenta, where  particle number change may occur. {This} can be calculated by expanding the modes 
$\Omega_k$ in negative and positive frequencies, with Bogoliubov coefficients. Away from the crossings, the coefficients remain approximately constant, while their  change after each crossing can be computed in the limit $\tilde g/\lambda\gg1$ ({as it is the case} for SMASH) by relating the problem to that of a quantum mechanical wave being transmitted and reflected in a parabolic potential \cite{Greene:1997fu}. We find that after a single crossing with an initial particle number of zero, the comoving gauge boson number density is, accounting for a factor of 6 coming from the 3 polarisations and  the
two types of gauge bosons,
\be
\label{nH1}
\bar n^c_W=6\int\frac{d^3\kappa}{(2\pi)^3} f^{(H)}_\kappa=6 \int\frac{d^3\kappa}{(2\pi)^3}  e^{-(\kappa/\kappa_c)^2} = \frac{3\kappa_c^3}{4\pi^{3/2}},
\ee
where we have defined
\be
\kappa_c= \frac{\tilde g^{1/4}}{\pi^{1/2} (2\lambda)^{1/4}}.
\ee
Using the parameters of equation \eqref{eq:SMASHpars}, and similarly defining $\tilde \delta_3\equiv \tilde g/0.03$, we can express $\bar n^c_W$ as
\be
\bar n^c_W= 21.6 \(\frac{\tilde\delta_3}{\lambda_{10}}\)^{3/8}.
\ee
We are denoting comoving quantities in our dimensionless units (rescaled by factors of $1/E=a/(\sqrt{\lambda_\sigma}\phi_{\rm end})$) with bars.
As will be seen in the next subsection, the gauge bosons produced during a crossing will decay almost completely into light quarks and leptons between crossings, so that for each successive crossing one starts, {in practice,} with a zero abundance of gauge bosons. Thus the production in the first crossing given above is the key quantity to compute the light particle abundance which will lead to reheating.


\subsubsection{Inelastic production of light quarks and leptons}

The gauge bosons produced immediately after a crossing have a small mass, but  since the latter  depends on the inflaton value, $\bar m_W = \sqrt{\tilde g/\lambda}F$, it grows quite large as the inflaton amplitude emerges from $F=0$. This opens up the decays into light quarks and leptons. Due to their small Yukawa couplings, the induced masses of the latter particles are much smaller than those of the gauge bosons, and so the produced light particles will be relativistic. We may refer to this process as ``inelastic light fermion production'', given the fact that the fermions are produced with an energy much larger than the inflaton's oscillation frequency.

Using the decay rate of $W$ bosons into light fermions in conformal rescaled units,
\be
\bar\Gamma_{W} = \frac{3 g^2}{16\pi } \bar m_W, \quad  \bar m_W =\frac{m_W}{a\sqrt{\lambda_\sigma} \phi_{\rm end}},
\ee 
we have checked that the Higgs population can decrease very efficiently in between crossings. For the preferred SMASH parameters, the survival probability after half an oscillation
is typically less than one percent, scaling as
\be
1-\exp\(-\int_0^{T/2} d\tau  {\bar \Gamma}_{W} \)\equiv\exp\(-2 \gamma_c \),\quad  \gamma_c = 2.3\(\frac{g}{0.5}\)^2\, \tilde\delta_3^{1/4}\lambda_{10}^{-1/4}.
\ee

This means that essentially all gauge bosons decay in between crossings, so that the occupation number cannot grow like it does in ordinary parametric resonance. However, the energy density in light quarks and leptons accumulates and will grow linearly with conformal time. If we have a number $\bar n^c_W$ of gauge bosons after a crossing and each gives its energy --given essentially by its mass-- to light fermions, the comoving energy gained by the fermion bath before the next crossing ($\tau = T/2$) is 
\bea
\Delta \bar \rho_f &=& \int_0^{T/2}  d\tau \bar \Gamma_{W}  \bar m_W \bar n^c_W e^{-\int_0^\tau d\tau' \bar \Gamma_{W}}.
\eea
At first, gauge boson production proceeds in bursts that decay into fermions, and the energy density of the latter will grow on average as 
\bea
\label{eq:rhortau}
\bar \rho_f &=& \frac{2\tau}{T} \Delta \bar \rho_f  \sim  128\, \tau\, F_p^2Y(F_p,\gamma_c) \tilde\delta_3^{5/8}\lambda_{10}^{-5/8},
\eea
where we used the ansatz \eqref{eq:bgosc2} for the background, neglected the effect of the backreaction in the period of oscillation, and $Y(F_p,\gamma_c)$ is given by the following integral,
\bea
 Y(F_p,\gamma_c)=\int d\tau
c \gamma_c F_p \sin^2(c F_p\tau)
 \exp\[-2\gamma_c\sin^2\(\frac{c F_p \tau}{2}\)\].
\eea
For $F_p=1$, when the backreaction of produced particles is not important, $Y(1,\gamma_c)$ takes a maximum value around 0.66 for $\gamma_c=1.5$ and then decreases as $1/\sqrt{\gamma_c}$. This reflects that for a large gauge boson decay rate, and hence for large $\gamma_c$, the decays are so fast that most gauge bosons decay before having reached the maximum mass, and thus transfer less energy into fermions. 

If this linear growth of the energy density of the produced fermions is maintained, one can estimate a reheating temperature by finding the moment
at which the energies of the fermion bath and the inflaton background plus excitations coincide. The contribution of the inflaton background plus excitations to the total energy can be estimated in a first approximation by considering that, if the backreaction of $W$ production is neglected, the total energy of the inflaton background plus fluctuations in comoving units is conserved, and so must be equal to the inflaton's energy at the beginning of the oscillation phase, that is  $\bar \rho_\phi={1}/{(4\lambda})$.  Assuming that the light particle bath thermalises with $g_*$ degrees of freedom, then the temperature of the light particle bath when $\rho_f=\rho_\phi$ is obtained by imposing
\be
\frac{\pi^2}{30} g_* T_{R}^4 = \bar\rho_f\(\frac{\sqrt{\lambda}\phi_{\rm end}}{a}\)^4.
\ee
Taking $g_*=124.5$ --corresponding to all SMASH particles in thermal equilibrium and behaving as a relativistic plasma-- and again neglecting backreaction effects (setting $F_p=1$) leads  to an estimate
\bea
T_{R} &\sim&   10^9\, {\rm GeV}\,  \lambda_{10}^{5/8}\,\tilde\delta_3^{5/8} Y(1,\gamma_c).
\eea

For typical SMASH parameters one would get a reheating temperature near $10^9$ GeV, which would be borderline compatible with vanilla leptogenesis, which requires
RH neutrinos with thermal abundances and masses above $5\times10^8$ GeV, cf. Section \ref{baryogenesis}.  Moreover, the previous estimate neglected backreaction effects, which slow down the production
of gauge bosons as the amplitude and the frequency of the background decrease. However, the assumption of a maintained linear growth with conformal time of the light fermion energy density turns out to be too conservative, as it ignores the fact that the large population of fermions can thermalise and enhance the production of gauge bosons at the crossings. This is explored in the next subsection, in which we will also take into account the feedback of the radiation bath in the evolution of the background.

\subsubsection{Thermal enhancement of gauge boson production}

As the abundance of light fermions increases, scatterings can reorganise their momenta, eventually leading to a thermal distribution. The thermalisation time can be estimated by studying the scattering rates by gluon exchange {between any pair of coloured particles \cite{Davidson:2000er}, e.g. $\Gamma_{{\rm gluon}}=n_q \sigma_{q_i\bar q_i q_j\bar q_j}$, where $n_q$ is the total number of light quarks --all except the top, which can get a large mass in the inflationary background-- and $q_i,q_j$ represent any choice of light quark flavours (with all of them being nearly massless, the different scattering rates are approximately equal).   Thermalisation is expected to be quickly achieved when the interaction rate becomes faster than the rate of expansion of the Universe, i.e. $\Gamma_{\rm gluon}>{\cal H}$. Imposing this requirement, and estimating a ratio of $\sim2$ from the number of light quarks to leptons originated in the decays of $W^\pm,Z$, (following from a simple counting of possible final states), we estimate that thermalisation of the light particle bath is achieved for 
\be
\label{eq:taueq}
\tau_{\rm eq} \sim \frac{2\times 10^3}{F_{p}^{1/2}}\(\frac{\tilde\delta_3}{\lambda_{10}}\)^{1/16}.
\ee}
Around the same {time} the quark and gluon interactions thermalise as well, and we expect the formation of a thermal bath with all the light SM particles, except for the weak gauge bosons and Higgses, whose masses are affected by the inflaton background and fluctuations. Thus we will assume that after $\tau_{\rm eq}$ one ends up with a light particle bath with $g_\star=96.75$ relativistic degrees of freedom. Matching the energy density of the light particles at this time with that of the thermal bath gives a temperature 
\be
T_{\rm eq} =  3 \times 10^{11}\, {\rm GeV}\,\times Y^{1/4}F_{p}^{5/8} \tilde\delta_3^{7/64}{\lambda_{10}}^{25/64} . 
\ee

Given this, for $\tau>\tau_{\rm eq}$ the gauge bosons become coupled to a thermal bath, which can affect their abundance. When the inflaton crosses the origin and the mass of the gauge bosons decreases, inverse fermion-antifermion decays can produce gauge bosons. This represents a new source complementing the nonthermal production from coherent scatterings of the inflaton. Away from the origin, the inverse decays of the fermions become Boltzmann suppressed due to the large gauge boson mass, and the heavy bosons decay out equilibrium into light fermions. As before, the bosons decay when their energy is greater than that with which they were produced, and so the thermally produced particles allow to drain more energy from the background. Since the rate of thermal gauge boson production is proportional to the equilibrium abundance of fermions, which grows with the energy density of the fermion bath, we  expect that this ``thermal feedback'' will cause an enhancement in the growth of $\rho_f(\tau)$ after the thermalisation time $\tau_{\rm eq}$. This growth beyond the linear $\rho_f(\tau)\sim \tau$ regime of the previous section  will lead to a higher reheating temperature. The enhancement will wear down as the temperature grows, because when the gauge bosons are massive the inverse decays will become less suppressed,  and not all the thermally produced bosons will decay back to light fermions. At some point the gauge bosons
will thermalise with the radiation bath of the light particles and the thermal feedback will be further suppressed.

\begin{figure}[h!]
\begin{center}
\includegraphics[width=.7\textwidth]{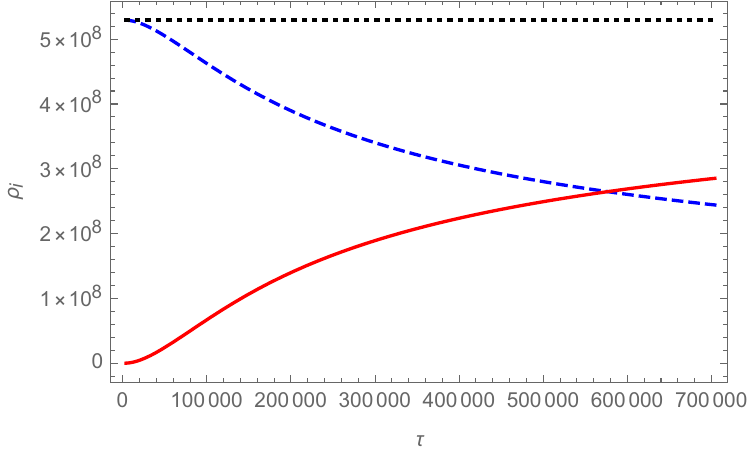}
\vskip.2cm
\caption{\small  Evolution of the rescaled comoving energy densities for the  background (blue, dashed)  radiation bath (solid red), and their sum (black dotted), for HHSI with  $\xi_\sigma=1$, $\lambda_{10}=4.7$, $\delta_3=1$. $\tau$ is the rescaled conformal time, with $\tau=0$ corresponding to the end of inflation.}  
\label{fig:rhos}       
\end{center}
\end{figure}

The previous dynamics of the gauge bosons coupled to a thermal bath is captured by a Boltzmann equation including the non-thermal source from the inflaton, as well as thermal decays and inverse decays. 
\be
\label{boltzman}
\bar n_W' = -\bar \Gamma_{W}(\bar n_W- \bar n_{eq}) + \sum_{F(\tau_i)=0} \bar n^c_W\delta(\tau-\tau_i), 
\ee
where $\bar n_{eq}$ is the equilibrium gauge boson density, which depends on the slowly oscillating gauge boson mass $\bar n_{eq}(T,m_W(\tau))$. The last term accounts for the non-adiabatic production of gauge bosons when $F(\tau_i)=0$, here taken to be instantaneous. 
On the other hand, the energy transfer to the fermion bath is modelled by
\be
\label{eq:rhorp}
\bar \rho_f' =\bar m_W \bar \Gamma_{W}(\bar n_W-\bar n_{eq}) . 
\ee 
Finally, the feedback on the inflaton background can be taken into account by using covariant conservation of the energy momentum tensor, or equivalently 
the fact that the total energy density must scale as $a^{-4}$. For our rescaled comoving densities, this means that the inflaton's comoving energy density $\bar\rho$ satisfies
\begin{align}
\label{eq:rhoinf}
\bar  \rho_f'+\bar\rho'=0.
\end{align}
This can be traded for an equation for the amplitude $F_p$ of the inflaton's oscillations (see \eqref{eq:bgosc2}), when writing $\bar\rho$ in terms of $F_p$. For this we may start from the 
potential \eqref{scalar_potential}, go to a unitary gauge in which the only physical component of $H$ is the Higgs field $h$, and separate the modulus $\rho$ and $h$ into zero mode and excitations. For the field components 
$h_0,\rho_0$ of the zero
mode $\phi_0$ we may write, using $|\lambda_{H\sigma}/\lambda_H|\ll1$,
\begin{align*}
 h\sim \sqrt{\frac{\lambda_{H\sigma}}{\lambda_H}}\phi_0,\quad \rho_0\sim \phi_0.
\end{align*}
Then, motivated by our lattice simulations, we may assume that the efficient parametric production of inflaton excitations yields fluctuations of the modulus $\rho$ which are of the order of {$\sim \phi_{\rm end}/(2a)$}
(see Figure \ref{fig:osc}), while fluctuations in $h$ are negligible. The total energy of the inflaton background plus excitations can be then estimated in conformal rescaled units as
\begin{align*}
 \bar\rho_\phi=\frac{1}{64\tilde\lambda_\sigma^2}\left[F_p^2 \left(16 \lambda _{\sigma }+8 \tilde\lambda_\sigma \right)+8 F_p \lambda _{\sigma }+16 \tilde\lambda_\sigma  F_p^4+32 \tilde\lambda_\sigma  F_p^3+\lambda _{\sigma }\right],
\end{align*}
where $F_p$ is the amplitude of the oscillations of the background, as in equation \eqref{eq:bgosc2}.

 Solving equations \eqref{boltzman}, \eqref{eq:rhorp}, \eqref{eq:rhoinf} numerically   for conformal times beyond the fermion thermalisation time $\tau_{\rm eq}$,
with an initial abundance of $\bar n_W$ set to zero at the first maximum of the inflaton, we can have more accurate estimates for the reheating temperature by again looking for the moment in which $\rho_\phi=\rho_f$. Figure \ref{fig:rhos} shows the resulting evolution of the energy densities for $\lambda_{10}=4.7$, $\delta_3=1$, corresponding to $\xi_\sigma\sim 1$. Equality of radiation densities is achieved for $\tau\sim 6\times10^5$, corresponding to a reheating temperature $T_R\sim 10^{10}{\rm GeV}.$

\begin{figure}
\begin{center}
\includegraphics[width=.65\textwidth]{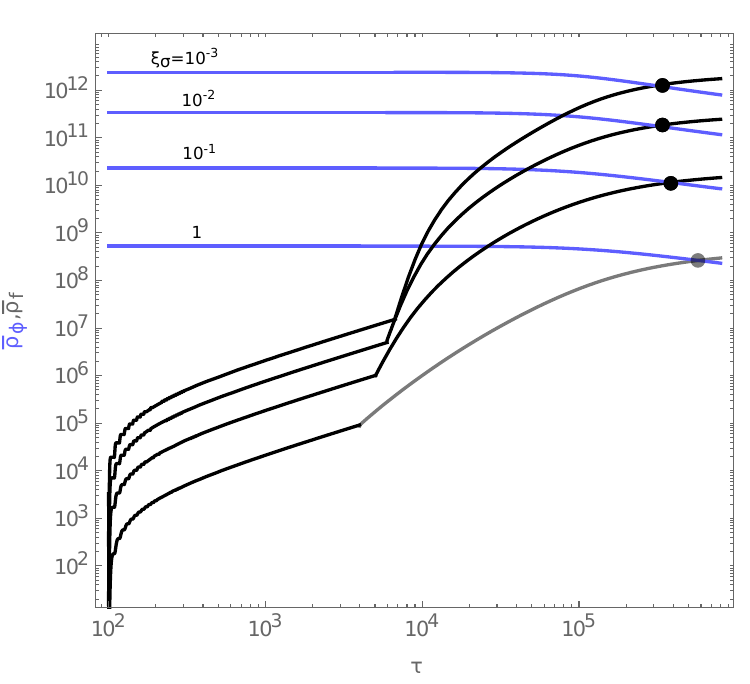}
\includegraphics[width=.65\textwidth]{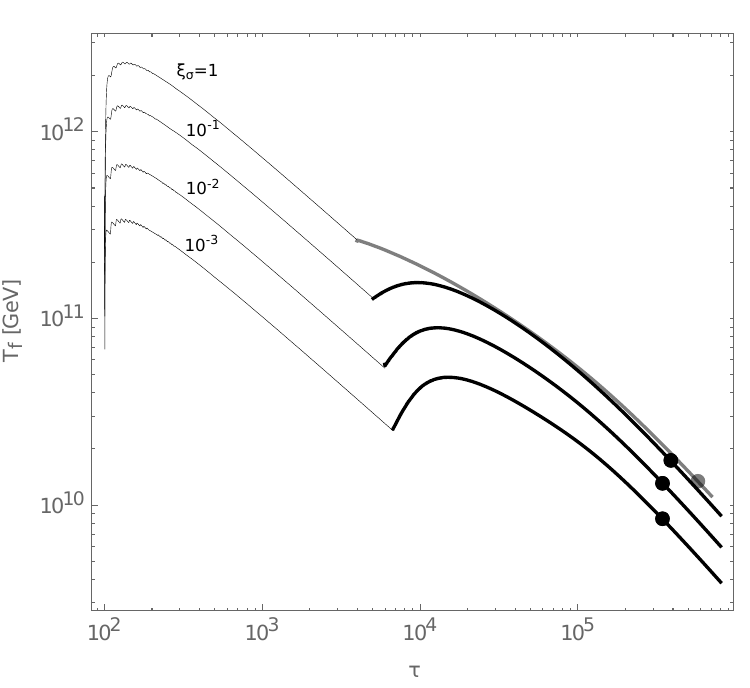}
\caption{\small  {Top: Evolution of the rescaled comoving energy densities of the inflaton background (blue) and the radiation bath (black), for 
different values of $\xi_\sigma$ and fixed $\delta_3=1$, as a function of the rescaled conformal time $\tau$. Note the early linear growth of the radiation density and the enhanced growth after thermalisation.
Bottom: Evolution of the temperature of the radiation bath (thick lines), or, for $\tau<\tau_{eq}$, the equivalent temperature of a thermal bath with the same energy density (thin lines). We fixed again $\delta_3=1$.}
}  
\label{fig:reheating}       
\end{center}
\end{figure}

We have also solved equations \eqref{boltzman}, \eqref{eq:rhorp}, \eqref{eq:rhoinf} for conformal times before $\tau_{\rm eq}$, by dropping the thermal source proportional to the equilibrium number density in \eqref{boltzman}. We show the corresponding results for the  evolution of the energy densities 
and radiation temperature for different values of $\xi_\sigma$ (keeping $\delta_3=1$) in Figs. \ref{fig:reheating}. In Fig. \ref{fig:reheating} {(top),} the blue and black lines represent the comoving energy density of the background and the light particles, respectively. Note how for low values of $\tau$ the growth of $\rho_\tau$ is linear in $\tau$, with our numerics confirming the analytic estimate of equation \eqref{eq:rhortau}. The initial wiggles clearly follow the first half-oscillations of the inflaton, with  $T/2=\pi/c=3.7$, and show particles being created from gauge boson decays in between crossings. At the corresponding thermalisation time $\tau_{\rm eq}$, given by equation \eqref{eq:taueq}, the thermal source term was activated in \eqref{boltzman}, with the resulting  thermal feedback enhancing the growth of the radiation density and tapering off when the gauge bosons approach thermal equilibrium. The black points mark the times of equality between the energy density stored in the inflaton background plus excitations and the energy density of the light fermion radiation. The evolution of the temperature of the radiation bath is shown for the same scenarios in Fig. \ref{fig:reheating} {(bottom).} For $\tau<\tau_{\rm eq}$, for which the bath of light particles has not reached equilibrium, the thinner lines show the temperature that would correspond to a thermal bath with the same energy density. Estimates for the reheating temperature can be read directly from the black points. It should be kept in mind that for different models, same values of the rescaled conformal time correspond to different physical times, given that the relation between $\tau$ and $t$ depends not only on the evolution of the scale factor of the Universe, but also the energy scale at the end of inflation and the coupling $\lambda$ (see equation \eqref{eq:F}).

At the typical reheating temperatures for predictive SMASH models, the amplitude of the inflaton's oscillations is such that the maximum values of the inflaton-induced masses for the RH neutrinos (assuming $Y_{ij}\sim 10^{-3}$) are precisely of the order of $T_R$, so that these particles will thermalise shortly after. The reheating temperatures allow for a thermal restoration of the PQ symmetry, since they exceed the value of the critical {temperature, cf. equation \eqref{eq:TC} in Appendix \ref{finiteT},} for quartic couplings in the window of predictive inflation in equation \eqref{inflationconstraint}, and within the stability window of $10^{-3}\lesssim\delta\lesssim10^{-1}$ (see Figure \ref{fig:mindelta}).  Moreover, the temperatures also allow for thermal abundances for heavy right-handed neutrinos, as required for leptogenesis scenarios.

\section{\label{dark_matter}Axion dark matter and radiation}

As it is well known, for large values of $f_A\gtrsim 10^{9}$ GeV, the axion may contribute substantially to the amount of dark matter in the {Universe; for reviews see  \cite{Sikivie:2006ni,Ringwald:2012hr,Kawasaki:2013ae}}. The amount of axion dark matter produced in the big bang is non-thermal and thus sensitive to the axion initial conditions, i.e. to the early stages of the big bang. 
In particular, it depends on whether the $U(1)$ PQ symmetry is restored or not after inflation. 
Further dependencies can arise if the heavy particles in the axion model ($\rho,Q,N_i$ {in our case}) are light or long-lived. 

We have already seen that in SMASH, the PQ symmetry is {first broken and then} restored non-thermally during preheating for 
$f_A\lesssim 4\times 10^{16}$ GeV. For larger $f_A$ values, the inflaton {would settle} to $\rho\sim f_A$  after a few oscillations in the quartic regime and there is no time for PQ breaking. We have not studied this {second} case in detail {because,} as we will see, it is not viable. Even if the PQ symmetry were restored afterwards in the later stages of reheating, {this case} is killed by overabundance of DM, and if it were not, it would nevertheless be ruled out} by producing too large isocurvature perturbations in the CMB {(which is a generic prediction for axions with such values of $f_A$)}. Let us then review generically, both cases: PQ restoration and non-restoration after inflation.

\subsection{Dark matter in the restored PQ scenario}

If the PQ symmetry is restored after inflation (the case in SMASH for $f_A\lesssim 4\times 10^{16}$ GeV), there is a unique relation between the axion scale $f_A$ and the amount of axion cold dark matter, which is produced through the vacuum re-alignment mechanism and the decay of topological defects. 

Below the critical temperature of the PQ phase transition, $T_c$, cf. equation \eqref{eq:TC} in Appendix \ref{finiteT}, the field $\rho$ takes its VEV $v_\sigma=f_A$ and the axion field $A$, which is essentially massless, gets random initial values at distances beyond the causal horizon $\mathcal{H}^{-1}$. 
At the same time, a network of axionic strings forms with tiny cores where the $U(1)$ symmetry is still preserved (i.e. with $\sigma=0$). As the temperature of the Universe drops, the axion inhomogeneities decay as radiation as soon as they enter the causal horizon and, at the same time, they are replenished by radiation from oscillating strings and the collapse of string loops. The energy density in the network of global strings reaches a scaling behaviour \cite{Kibble:1976sj}, widely discussed in the literature, which redshifts as radiation in the radiation dominated Universe. 

Eventually, at a temperature that is slightly above the QCD confining phase transition, $T_{\rm QCD} \sim 157$ MeV, the axion potential induced by QCD can no longer be ignored. At $T\gg T_{\rm QCD}$, the potential features a cosine shape and the height of the potential is determined by the topological susceptibility --See equation \eq{topos}-- as a function of temperature,  
\be
\label{axionmasschi}
V(A) = \chi(T)\(1-\cos\(\frac{A}{f_A}\)\)\,.
\ee

Very recently, lattice QCD calculations of the topological susceptibility have been made available~\cite{Petreczky:2016vrs,Borsanyi:2016ksw} and there is no need to rely on phenomenological models like the interacting instanton liquid model (IILM)~\cite{Wantz:2009mi,Wantz:2009it} or the dilute-instanton-gas-approximation (DIGA)~\cite{Borsanyi:2015cka}. 
These results for the topological susceptibility agree well with previous analytical calculations~\cite{Buchoff:2013nra,diCortona:2015ldu} at low temperature, where they lie between the predictions of IILM and the DIGA of~\cite{Borsanyi:2015cka}, and follow the DIGA slope at high temperatures (as observed for quenched QCD in~\cite{Borsanyi:2015cka}). A previous lattice calculation, which predicted a much softer slope~\cite{Bonati:2015vqz}, was most probably suffering from strong cut-off effects \cite{Petreczky:2016vrs,Borsanyi:2016ksw}. All these results are shown in Fig. \ref{suscep}. 

\begin{figure}[t]
\begin{center}
\includegraphics[width=0.7\textwidth]{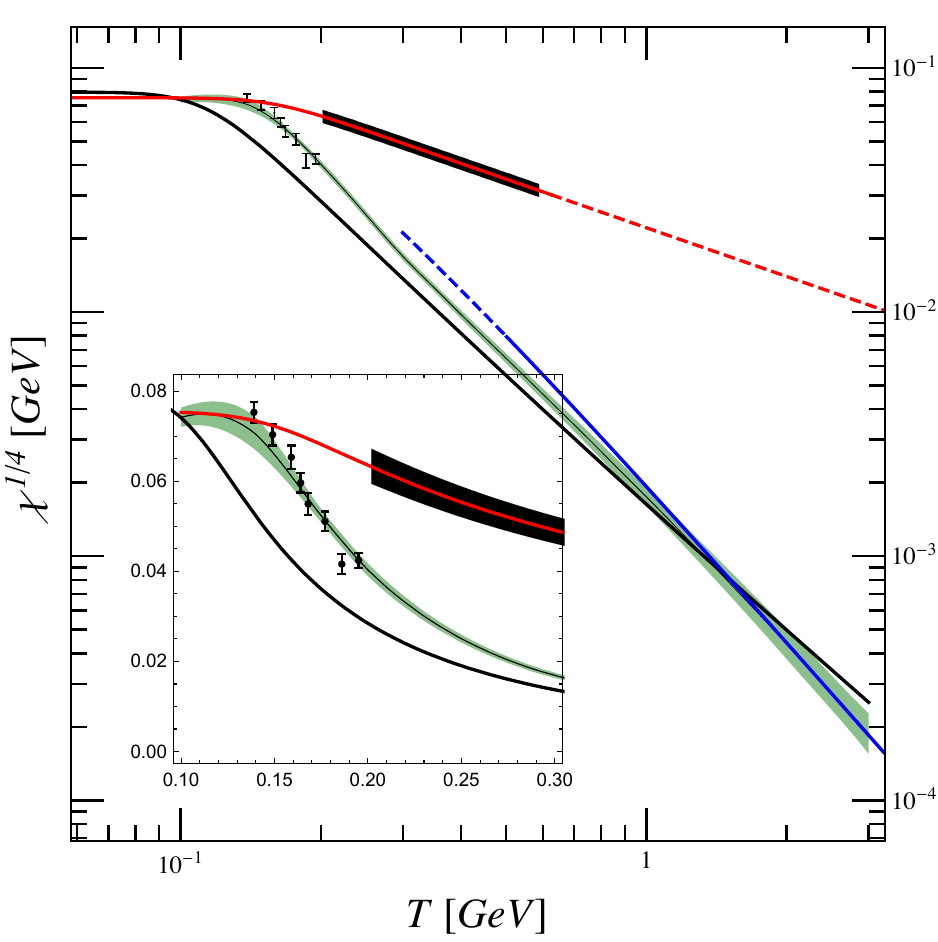}
\caption{Topological susceptibility from the latest lattice QCD results: Borsanyi et al. \cite{Borsanyi:2016ksw} (green band)\exclude{, Brookhaven \cite{Petreczky:2016vrs} {()}} and Bonati et al \cite{Bonati:2015vqz} (red line and black band). We also show the predictions from the IILM model~\cite{Wantz:2009mi,Wantz:2009it} (black) and the 2-loop DIGA~\cite{Borsanyi:2015cka} (blue) with schematic extrapolations in dashed lines. 
The points are from the lattice results of~\cite{Buchoff:2013nra}, computed at $m_\pi\simeq 200$ MeV rescaled by $(135/ 200\,{\rm MeV})^4$.  }
\label{suscep}
\end{center}
\end{figure}

The topological susceptibility $\chi$ sets the dynamical energy scale for the axion field to respond to the QCD potential, which is encoded in a temperature dependent axion mass $m_A=\sqrt{\chi(T)}/f_A$. The axion field starts to oscillate classically around the CP conserving minimum $\theta=0$, when the temperature decreases sufficiently so that the Hubble friction is {comparable} to the effect of the potential, i.e. $3 {\cal H}(T_1) \simeq m_A(T_1)$. Solving this equation gives an estimate of the temperature {$T_1$} at which the axions become non-relativistic and thus dark matter:
\bea
T_1 &\simeq&\left\{
\begin{array}{ll}
2.09\, {\rm GeV} \(\frac{f_A}{10^{10}\rm GeV}\)^{-0.1655} & (f_A \lesssim 3 \times 10^{17}\, \rm GeV) ,\\
126\, {\rm MeV} \(\frac{f_A}{2\times 10^{17}\rm GeV}\)^{-0.47} & (f_A \gtrsim  3\times 10^{17}\, \rm GeV )\,,
\end{array}
\right.
\eea
where we have used the equation of state of QCD (i.e.\ the effective number of radiation degrees of freedom) provided also in \cite{Borsanyi:2016ksw}. For $f_A \lesssim 3 \times 10^{17}\, \rm GeV$, the axions start to oscillate when $\chi$ is suppressed by thermal effects, while for $f_A \gtrsim 3 \times 10^{17}\, \rm GeV$, the axions start to oscillate when the temperature effects are negligible and their mass is the vacuum mass \eqref{zeroTma}.

The amount of DM can be artificially divided into a re-alignment contribution and a contribution from radiation from axionic strings and domain walls. 
{To compute the re-alignment contribution, one considers relatively homogeneous patches of the Universe with constant initial condition $\theta=\theta_I$,} follows their evolution around $T_1$, 
\be
\label{evo0mode}
\ddot \theta +3 {\cal H} \dot \theta + m_A^2(T)\sin\theta = 0, 
\ee
computes the number density at some late time, when it adiabatically scales as decoupled dark matter $n_{A,\rm real} \propto a^{-3}$, and extrapolates to the current size of the Universe assuming constant comoving entropy $\propto g_S(T)T^3a^3$.  
If one models $\chi$ as a power law $\chi = \chi_0\({T_{\rm QCD}}/{T} \)^n$ with $T_{\rm QCD}\simeq 157$ MeV and $n\sim 8.16$ the result is approximately 
\bea
\label{misa1}
\rho_{A,\rm real} &=& \chi_0 \frac{T^3_0 g_S(T_0)}{T_{\rm QCD}^3 g_S(T_1)}\(\frac{f_A}{M_P}\frac{T^2_{\rm QCD}}{\sqrt{\chi_0}}\sqrt{\frac{2\pi^2 g_*(T_1)}{45}}\)^\frac{6+n}{4+n} A(n) ,
\eea
where $A(n)\simeq 7.673+0.531(n-8)$ accounts for the average over initial random values of $\theta_I\in(-\pi,\pi)$.  
A convenient fit to the full numerical results using the latest data for $\chi(T)$ gives 
\bea
\label{misa2}
\Omega_{A,\rm real}h^2 &=& (0.12\pm 0.02)\(\frac{f_A}{1.92 \times 10^{11}\rm GeV}\)^{1.165} , 
\eea
where the errors come from the topological susceptibility.\footnote{Interestingly, direct numerical simulations do not confirm this estimate~\cite{Fleury:2015aca} in the sense that they give less energy for the sum of re-alignment and string radiated axions than the value derived here only from the misalignment.  
The reason is probably that at any time there is typically one cosmic string per horizon together with all the axions radiated previously. Since the axion field takes values from 0 to $2\pi$ around a string, this means that the homogeneous regions cannot be larger than the horizon at $T_1$ and thus the gradient term that we dropped in \eqref{evo0mode}, $-\frac{1}{R^2}\nabla^2\theta$, is relevant. Moreover, a large part of the energy comes from regions where $\theta_I\simeq \pi$ because the cosine potential is quite flat there and the axion field takes longer time to start rolling down its potential (would be $\infty$ at $\theta_I$), storing the QCD energy for a longer time. But these regions are not really homogeneous. Shorter wavelength axions produced by the strings have non-linear interactions with the long wavelength mode through the $\sin\theta$ potential and could destabilise the field making it roll before. These two effects go in the direction of decreasing the re-alignment estimate probably by a factor around $\sim 2$. We will include this factor in the uncertainties but the reader should keep in mind that the artificial division of the DM yield into misalignment and strings has its systematic uncertainties.}

{The axions radiated from strings and domain walls are subject to even more uncertainties. The string network consists of string loops that emit axions while tightening, oscillating, reconnecting and collapsing during the Universe expansion.} Numerical simulations support the reach of a scaling solution for the energy density stored in strings (mostly in their associated Goldstone axion field) 
\be
\label{scaling}
\rho_{s} = \zeta \frac{\mu_s}{t^2}\,,\quad \quad  \mu_s = \pi f_A^2 \ln\(\frac{f_A t \sqrt{\lambda_\sigma }}{\sqrt{\zeta}}\)
\ee 
with $\zeta\sim O(1)$~\cite{Kawasaki:2014sqa}.  
Numerics can only be performed on fluffy strings with $\ln ( \sqrt{\lambda_\sigma} f_A t )$ much smaller than the realistic value and thus, they require an extrapolation that has been recently challenged in \cite{Fleury:2015aca,Fleury:2016xrz}. A new numerical method has been proposed in \cite{Fleury:2016xrz,Moore:2016itg} to do these estimates, with current trends pointing towards less axion dark matter for the same $f_A$ and thus a larger $f_A$ required for accounting for all the DM.  

Equation \eqref{scaling} implies an instantaneous axion energy production rate~\cite{Kawasaki:2014sqa},  
\be
\frac{d\rho_{A,s}}{dt} = \pi\,\zeta  \frac{f_A^2}{t^3}\[\ln\(\frac{f_A t \sqrt{\lambda_\sigma }}{\sqrt{\zeta}}\)-1\]\,,
\ee 
which, {appropriately weighted,} can be integrated in time to give an estimate of the comoving number of axions radiated from strings at time $t_s$
\be
\label{Nastring}
{N_{A,s}(t_s)} = \int^{t_s}dt \frac{a^3(t)}{\langle\omega\rangle (t)}\frac{d\rho_{A,s}}{dt} 
\simeq
\frac{a^3(t_s)}{\langle\omega\rangle (t_s)}\frac{2 \pi f_A^2\zeta}{t_s^2}\[\ln\(\frac{f_A t_s \sqrt{\lambda_\sigma }}{\sqrt{\zeta}}\)-3\]
\ee
where $\langle\omega\rangle(t)$ is a proper average energy for the radiated axions, which is close to the Hubble scale (momenta close to the causal cut-off)   {$\langle\omega\rangle(t)\equiv \epsilon_A 2\pi /t$.}
The integral is mildly dominated by times close to the upper limit of integration $t_s$, so that early produced axions are not as relevant as those produced close to the QCD confining transition when axions become more massive. {Around $t_1$,} {which is the time associated to the temperature $T_1$, for which the axions become non-relativistic, the  axion mass starts becoming relevant. However $\langle\omega\rangle(t_1)$ is still larger than $m_a(t_1)$ so axions can still be radiated.} {We define the cut-off time $t_{co}$ as the time when $\langle\omega\rangle(t_{co})=2\pi\epsilon_A/t_{co}=m_a(t_{co})$. After that time, the axions radiated have to be necessarily much more energetic because of their larger mass. Moreover, since the mass increases very fast with time} {$m_a\propto t^{n/2}\sim t^4$} {the cut-off is extremely sharp.  
The contribution to the axion dark matter density today is thus given by the number density at $t_{co}$ diluted by the expansion multiplied by the mass that each acquires from QCD effects, $\rho_{A,s}=m_A N_A(t_{co})/a_0^3$. }

It is interesting to do the math with the power law approximation for $\chi(T)$. {One} gets a very similar formula to the average re-alignment, {where the role of $t_1$ is played by $t_{co}$}
\bea
\label{stringDM}
\rho_{A,s} &\sim& \chi_0 \frac{T^3_0 g_s(T_0)}{T_{\rm QCD}^3 g_s(T_{co})}\(\frac{f_A}{M_P}\frac{T^2_{\rm QCD}}{\sqrt{\chi_0}}\sqrt{\frac{2\pi^2 g_*(T_{co})}{45}}\)^\frac{6+n}{4+n} \[\frac{\zeta}{\epsilon_A}\(2\pi \epsilon_A\)^\frac{2}{4+n}\ln\(\frac{f_A t_{co} \sqrt{\lambda_\sigma }}{\sqrt{\zeta}}\)\] .
\eea
The first three {factors are the same as those of \eq{misa1} and give the scaling with $f_A$.} Thus, the relative importance between the two production mechanisms is given by the last {factors,} the $A(n)\sim 7.6$ factor of \eq{misa1} {for} the re-alignment {mechanism} vs the bracketed log enhancement of the strings.

Using the $\chi(T)$ results from \cite{Borsanyi:2016ksw}  we find,  
\be
\label{aDMstringy}
\Omega_{a,s}h^2 \sim 0.37^{+0.3}_{-0.2} \(\frac{f_A}{1.92\times 10^{11}}\)^{1.165} \frac{\ln\({f_A t_{co}\sqrt{\lambda_\sigma /\zeta}}\)}{50},  
\ee
where we have used $\epsilon_A \simeq 4\pm 0.7 $ and $\zeta = 1\pm 0.5$ as suggested by the simulations of \cite{Kawasaki:2014sqa}. The errors correspond to the maximum values, i.e. have not been added in quadrature. Note that they correspond roughly to a factor 2 up and down the central value. 

{Around $t_{co}$}  {the remaining strings collapse very fast, accelerated by the fastly growing energy per area of the domain walls attached to them. Here the energy density is emitted in the form of axions, but since their mass is by then relatively large, the} {increase in the} {axion number density cannot be significant. Indeed, reference~\cite{Kawasaki:2014sqa} finds that the contribution of the final cataclysm is $\sim 1/2$ of the string radiations by} {the time} {$t_1$~\cite{Kawasaki:2014sqa}. This is approximately the factor that we gain by extending the integration to $t_{co}$} {which confirms that we  can neglect any further contribution beyond the time $t_{co}$, as we anticipated above.}  

In SMASH, we have $\lambda_\sigma\sim 10^{-13}-10^{-9}$, {see \eq{inflationconstraint},} and axions from string radiation  are $\sim 3-4$ more abundant than those created by mere re-alignment. The {range of} $\lambda_\sigma$ gives an ${\cal O}(10\%)$ model variability from the string log in \eqref{aDMstringy}, which is not included in \eqref{aDMstringy}.

Considering {finally} the sum of {the contributions of} re-alignment and string radiation, axion dark matter fits the  observed value $\Omega_{A,\rm real}h^2+\Omega_{A,s}h^2 {\simeq} 0.12$ in the range,\footnote{We have scaled the re-alignment {contribution} with an uncertainty factor $0.5-1$ and used $\lambda_\sigma=10^{-11}$ {for} the string contribution.} 
\be
\label{farangePQrestoration}
3\times 10^{10}\, {\rm GeV}\lesssim f_A \lesssim 1.2 \times 10^{11}\, {\rm GeV}
\quad ; \quad 
50\, \mu{\rm eV} \lesssim m_A \lesssim 200\, \mu{\rm eV} ,
\ee
which will be narrowed {once the theoretical and computational} uncertainties in {the amount of  axions} radiated from strings and walls are under control.  
Let us emphasise that the systematic errors are quite tied up to our splitting of the DM yield in {re-alignment} and strings and the particular approximations we have made to compute them. Although we have done our best to provide {an accurate} estimate, we {cannot fully dismiss the possibility} that a full numerical calculation {accounting for both effects simultaneously} will not change the yield beyond them {and therefore it is possible that the estimated range \eq{farangePQrestoration} could change in the future.}

\subsection{Dark matter in the non-restored PQ scenario}
If the PQ symmetry is broken during inflation and never restored afterwards, {which is the case in SMASH for $f_A\gtrsim 4\times 10^{16}$ GeV,} only re-alignment contributes to the dark matter density, {since strings are not formed after the end of inflation.} Any preexisting strings are typically diluted by inflation and we will assume that we are not in one of the few {Hubble-sized patches of the Universe that are} crossed by one {of them.} 

{For those values of $f_A$ in SMASH,\footnote{And in general for any axion scenario with non restoration of PQ after inflation.} inflation}
homogenises the axion field such that {each Hubble patch  at the end of inflation} has {a certain} initial value of $\theta_I$ plus some small fluctuations. 
Since {a} large $f_A$ implies $\Omega_{A}h^2\gg 1$ for $\theta_I\sim {\cal O}(1)$, the initial {conditions have} to be relatively close to $\theta_I$ {so as not to} overclose the Universe. Therefore, we can solve \eqref{evo0mode} {assuming that $\theta \sim \sin\theta $.} The equation is {then linear and homogeneous in the angle $\theta$} and the initial condition $\theta_I$ factors out leading to 
\be
\Omega_{A,\rm real} h^2 \sim  0.35 \(\frac{\theta_I}{0.001}\)^2 \times
\left\{
\begin{array}{ll}
 \(\frac{f_A}{3\times 10^{17}\rm GeV}\)^{1.17} & (f_A \lesssim 3\times 10^{17}\, \rm GeV) ,\\
  \(\frac{f_A}{3\times 10^{17}\rm GeV}\)^{1.54} & (f_A \gtrsim  3\times 10^{17}\, \rm GeV )\,,
\end{array}
\right.
\ee
{where the splitting comes from the different temperatures $T_1$ at which the axions become non-relativistic, as a function of $f_A$; see \eq{aDMstringy}.}

{Therefore, for $f_A\gtrsim 4\times 10^{16}$ GeV in SMASH, the correct relic density could be reproduced for} a wide range of $f_A$ if $\theta_I$ {is} appropriately chosen as initial condition in our Universe. We denote by $\theta_{I,c}$ such value and compute it by inverting $\Omega_{a,\rm real}h^2(\theta_{I,c})=0.12$. We obtain, 
\be
\label{thetaIc}
\theta_{I,c}  \sim 0.0006 \times
\left\{
\begin{array}{ll}
 \(\frac{f_A}{3\times 10^{17}\rm GeV}\)^{-0.504} & (f_A \lesssim 3\times 10^{17}\, \rm GeV) ,\\
 \(\frac{f_A}{3\times 10^{17}\rm GeV}\)^{-0.77} & (f_A \gtrsim  3\times 10^{17}\, \rm GeV ) .
\end{array}
\right.
\ee

However, {in this case} the quantum  fluctuations of the axion acquired during inflation are not erased afterwards and would be imprinted as isocurvature fluctuations in the anisotropies of the temperature of the CMB~\cite{Turner:1990uz,Fox:2004kb,Beltran:2006sq,Hertzberg:2008wr,Hamann:2009yf,Ade:2015lrj}, which are {strongly} constrained by Planck, {as we discuss in the next subsection.}

\subsubsection{Constraints from axion dark matter isocurvature fluctuations}

During HSI or HHSI {--See Section \ref{inflation}--} the {value of} $\rho$ breaks spontaneously the PQ symmetry and the corresponding orthogonal direction in field space (the axion) is an exactly flat direction that experiences quantum fluctuations {as any other light field would do in de Sitter space. It is convenient to write the kinetic term of $\sigma$ as follows 
\begin{equation}
{\cal G}_{\rho\rho}(\partial_\mu \sigma^\dagger )(\partial^\mu \sigma) =
\frac{1}{2}{\cal G}_{\rho\rho}(\partial_\mu \rho )(\partial^\mu \rho)+
\frac{1}{2}{\cal G}_{\rho\rho}\rho^2(\partial_\mu \theta) (\partial^\mu \theta)\,,
\end{equation}
from which we can define a canonically normalised axion $\bar A$ satisfying $\partial_\mu \bar A = \sqrt{{\cal G}_{\rho\rho}}\,\rho\, \partial_\mu \theta$. The variance of the fluctuations of this field has the usual expression for massless fields,
\begin{equation}
\langle |\delta \bar A|^2\rangle \simeq
\left(\frac{{\cal H}}{2\pi}\right)^2\,,
\end{equation}
where $\mathcal{H}$ is the Hubble parameter during inflation. Since ${\cal G}_{\rho\rho}$ is approximately constant during inflation, we can define an effective Einstein frame through {a rescaling involving $\chi_{I,\rm iso} \equiv  \sqrt{{\cal G}_{\rho\rho}}\rho_I$.} It is important to observe that {$\chi_{I,\rm iso}$} is not the same as the canonically normalised inflaton field, $\chi_\rho$ (defined by the differential equation \eqref{chi_hs}) because ${\cal G}_{\rho\rho}$ depends on $\rho$ but not on the orthogonal direction, $\theta$. The values of $\chi_{I,\rm iso}$ and $\chi_\rho$ are shown in Fig.~\ref{fig:chi2_2} for the SMASH parameters of HSI, HHSI within 95\% C.L. of the current bounds on the CMB spectra.} Note that the uncertainty in {the scalar spectral index $n_s$} cancels out considerably in the {allowed range for} $\chi_{I,\rm iso}$.

\begin{figure}[t]
\begin{center}
\includegraphics[width=7cm]{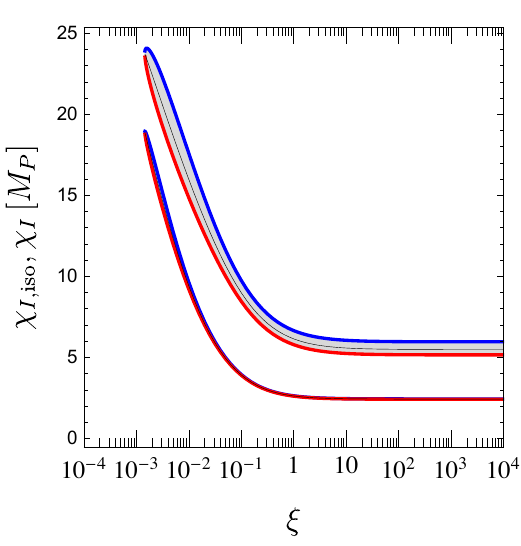} 
\includegraphics[width=7cm]{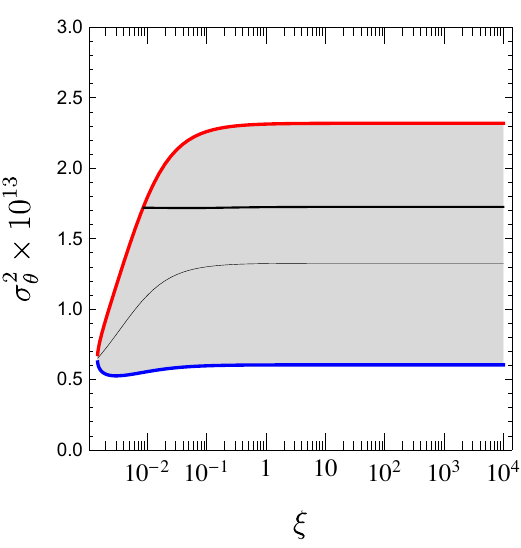} 
\caption{\label{fig:chi2_2} Left: \small 95\% C.L. contours for the {effective} Einstein frame normalisation scale of axion fluctuations $\chi_{I,\rm iso}$ (lower band). We also show from Fig. \ref{fig:chi2} the ranges of the canonically normalised inflaton field {$\chi_I=\chi_\rho$} for comparison (upper band). Right: \small 95\% C.L. contours for the variance of $\theta$ fluctuations in SMASH {(for HSI and HHSI)}. }
\end{center}
\end{figure}

The {variance of the} corresponding fluctuations of the angular field {$\theta = \bar A / \chi_{I,\rm iso} $ is
\begin{equation}
\sigma_{\theta}^2\equiv \langle |\delta\theta|^2\rangle \simeq 
\left(\frac{{\cal H}}{2\pi  \chi_{I,\rm iso} }\right)^2\,,
\end{equation}}
which is shown in Fig.~\ref{fig:chi2_2} (right) as a function of the non-minimal coupling $\xi=\xi_\sigma$.}
Note that the dependence on $\xi$ is weak but the experimental value of $n_s$ (difference between the red and blue lines) induces a large uncertainty.    {The perturbations of {$\theta$ lead to isocurvature modes} and manifest  as fluctuations in the number density of axions, $\delta (n_A/s)\neq 0$, {where $s$ is the total entropy density, see} \cite{Turner:1990uz,Fox:2004kb,Beltran:2006sq,Hertzberg:2008wr,Hamann:2009yf}. If the {PQ} symmetry is not restored after inflation, these perturbations remain until the QCD phase transition, when axions get a mass from QCD temperature-dependent effects, and get thus imprinted into temperature anisotropies of the CMB (cf. Ref. \cite{Beltran:2006sq} and references therein). Axion fluctuations are completely uncorrelated to the inflaton fluctuations and show up as distortions from the pure adiabatic spectrum, which are not observed and indeed severely constrained by CMB anisotropy measurements, in particular by Planck. 

Assuming {that all the dark matter is made of axions,} the isocurvature power spectrum due to axion fluctuations with {comoving} number $k$ is given by 
\begin{equation}
{\mathcal P}_i(k)  
 = \left\langle \left\vert \frac{\delta (n_A/s)}{n_A/s} \right\vert^2\right\rangle = 
\frac{2  \sigma_{\theta}^2
\left(  2\theta_I^2 +\sigma_{\theta}^2\right)}{\left(  \theta_I^2 +\sigma_{\theta}^2\right)^2}\,,
\end{equation}
{where $\theta_I$ is the initial misalignment angle.}

Planck constraints the fraction of isocurvature {perturbations} in the power spectrum  \cite{Ade:2015lrj}, 
\begin{equation}
r_i \equiv   
\frac{{\mathcal P}_i(k)}{{\mathcal P}_s(k)+{\mathcal P}_i(k)}\bigg|_{k=k_0} 
 < 0.037, \hspace{3ex} {\rm at\ } 95\%\  {\rm CL\  and\  a\ scale\ } k_0 = 0.05\  {\rm Mpc}^{-1}\,,
\label{def:isocurvature_fraction}
\end{equation}
where ${\mathcal P}_s(k)$ is the power spectrum of {scalar and adiabatic} fluctuations.  For $r_i\ll 1$, the {isocurvature} fraction can be expressed as  
\begin{equation}
r_i (k_0)\simeq   
{A_s^{-1}}
\left\{\begin{array}{ll}
 {4{\sigma_{\theta}^2}/{\theta_I^2}}
,&\text{for  $\theta_I^2 \gg \sigma_{\theta}^2$,}\\[1ex]
2,&\text{for  $\theta_I^2 \ll \sigma_{\theta}^2$.}
\end{array}
\right. ,
\label{smallri}
\end{equation}
where $A_s$ is the {amplitude} of the scalar power spectrum {\eqref{cmb3}.}
Clearly, the parameter region with 
$\theta_I^2 \ll \sigma_{\theta}^2$ is ruled out by the isocurvature fraction \eqref{def:isocurvature_fraction}.  

To assess the implications {for} $\theta_I^2 \gg \sigma_{\theta}^2$, we exploit the prediction of the initial vacuum angle required for $\Omega_{\rm CDM}h^2=0.12$ in terms of the axion decay constant $f_A$, cf.~\eqref{thetaIc}. Setting $\theta_I = \theta_{I,c}$, 
we arrive at the following prediction {for} the isocurvature fraction in SMASH {(for HSI and HHSI types of inflation.)}
\be
r_i \simeq 500 \frac{\sigma_\theta^2}{10^{-13}}\left\{
\begin{array}{ll}
 \(\frac{f_A}{3\times 10^{17}\rm GeV}\)^{1.084} & (f_A \lesssim 3\times 10^{17}\, \rm GeV) ,\\
 \(\frac{f_A}{3\times 10^{17}\rm GeV}\)^{1.54} & (f_A \gtrsim  3\times 10^{17}\, \rm GeV ) .
\end{array}
\right.
\ee
{Since $\sigma^2_\theta>0.6\times 10^{-13}$, as it} follows from fitting the observed $n_s$, {see} Fig.~\ref{fig:chi2_2}, the isocurvature constraint \eqref{def:isocurvature_fraction} implies   
{(see also Ref. \cite{Fairbairn:2014zta})}
\be \label{axionisoc}
f_A < 1.4\times 10^{14}\, {\rm GeV}.     
\ee
We emphasise that this is a conservative 95\% C.L.\ {limit} that corresponds to the largest values {of $n_s$.}  {Importantly,} this rules out {entirely the possibility of having the PQ symmetry not restored} after inflation {(which would require $f_A\gtrsim 4\times 10^{16}$ GeV). Therefore, the axion isocurvature constraint \eq{axionisoc} on $f_A$ has a strong implication in SMASH for the mechanisms that participate in the generation of axion dark matter. In SMASH dark matter thus receives contributions both from the re-alignment mechanism and from the decay of cosmic strings.}

\subsection{Cosmic axion background radiation}\label{sec:axionradiation}

{Relativistic axions equilibrate with the SM bath for temperatures larger than}
\beq
T_A^{\rm dec} \sim 1.7\times 10^9\ {\rm GeV} \left( \frac{f_A}{10^{11}\ {\rm GeV}}\right)^{2.246}\,,
\eeq 
{{as long as $T_A^{\rm dec}<T_c$ (since otherwise the symmetry is restored and there is no Goldstone), see \cite{Masso:2002np,Graf:2010tv,Salvio:2013iaa}.} In SMASH {with} HSI, the reheating temperature \eqref{eq:HSITR} is {smaller than $T_A^{\rm dec}$. This means that the axions produced during reheating and the PQ symmetry restoration never thermalise and therefore lead to an amount of dark radiation \eqref{eq:HSINeff} which is typically in conflict with observations.}

On the other hand, in HHSI the reheating temperature is {above $T_A^{\rm dec} $ so that the axions do thermalise in this case with the rest of the SM,} attaining the equilibrium abundance 
\beq
Y_A^{\rm eq} (T) = \frac{n_A^{\rm eq}}{s}= \frac{45\,\zeta(3)}{2\pi^4 g_*(T) } \simeq 2.6\times 10^{-3}
\left( \frac{427/4}{g_{*s}(T)} \right),
\eeq
where we have used the SM value, $g_{*s}(T)=427/4$, as a benchmark\footnote{This value is slightly larger in SMASH 
if $Q$ and  $N_i$ have a mass below $T_A^{\rm dec}$. In this case, $Q$ ($N_i$) adds to 
$g_{*s}(T)$ a term $21/2$ ($7/4$).} for the effective number of entropy degrees of freedom  at $T\gtrsim T_A^{\rm dec}$. 
Entropy conservation then implies that today they have a temperature 
\beq
T_A^0 = T_0 \left( \frac{g_{*s}(T_0 )}{g_{*s}(T_A^{\rm dec} )} \right)^{1/3}
\simeq 0.907 \ {\rm K}\ \left( \frac{427/4}{g_{*s}(T_A^{\rm dec})} \right)^{1/3},
\eeq
where $T_0=2.73$\,K is the present temperature of the CMB photons and $g_{*s}(T_0 )=43/11$. SMASH thus predicts 
a Cosmic Axion Background Radiation (CABR)
corresponding to an excess effective number of neutrinos,
\beq 
\Delta N_\nu^{\rm eff} = \frac{4}{7} \left( \frac{T_A^0}{T_\nu^0}\right)^4 \simeq 0.0268\ \left( \frac{427/4}{g_{*s}(T_A^{\rm dec})} \right)^{4/3},
\label{eq:delta_neff}
\eeq 
where $T_\nu^0 = T_0\, (4/11)^{1/3}$ is the present temperature of the relic active neutrinos.
Future CMB polarisation experiments may be sensitive to this prediction \cite{Abazajian:2013oma,Errard:2015cxa}.

\section{\label{baryogenesis}Solving the matter--anti-matter asymmetry problem}

For sufficiently large Majorana neutrino masses in a type-I seesaw model and a comparable or even larger 
reheating temperature, 
the baryon asymmetry of the Universe can be easily generated by thermal leptogenesis \cite{Fukugita:1986hr}.
In fact, 
in vanilla leptogenesis, corresponding to the following simplifying assumptions,
\begin{itemize}
\item 
a hierarchical neutrino mass spectrum of the $N_i$, with  $M_3 = M_{2} \gtrsim 3\, M_{1}$, 
\item 
negligible flavour effects,  in particular excluding fine tuned-cancellations among the different 
terms contributing to the neutrino masses in the seesaw formula \cite{Davidson:2002qv,Hamaguchi:2001gw}, 
\end{itemize}
one finds that the observed baryon asymmetry is generated as long as \cite{Buchmuller:2002rq,Giudice:2003jh,Buchmuller:2004nz} 
\begin{equation}
M_{1}\gtrsim 5\times 10^8\ {\rm GeV}; \hspace{6ex}
 (M_D M_D^T)_{11}/M_1 \lesssim 10^{-3}\,  
 {\rm eV},
\label{leptogenesis}
\end{equation} 
for a thermal initial abundance of $N_1$. As follows from the results of the reheating section, thermal abundances of heavy RH neutrinos satisfying the previous bound are only guaranteed {in HHSI,} 
which are also free of dark radiation problems. The large reheating temperatures in HHSI ensure a thermal restoration of the PQ symmetry. {This means that RH neutrinos are massless at high energies. 
They can be efficiently created from the plasma through Higgs decays, as the Higgs gets a non-zero thermal mass  $m^2_H\sim y^2_t\, T^2/4$. This guarantees an initial relativistic equilibrium abundance of the $N_i$ unless the Yukawas $F_{ij}$ are extremely small; this is expected to happen as long as $\Gamma_{H\rightarrow NL}/{\cal H}>1$, which can be seen to require $F_{ij}\gtrsim10^{-8}$ (and correspondingly, $Y_{ij}\gtrsim10^{-12}$ in order to guarantee light neutrino masses below the cosmological bound of 0.23 eV \cite{Ade:2015xua}, see \eqref{seesaw}). Since stability in the $\sigma$ direction allows for $Y_{ij}\lesssim 10^{-3}$ (see \eqref{Y11stability}), we conclude that RH neutrinos are thermally produced in the vast majority of parameter space}.
Given that the Majorana neutrinos get a mass from the $\sigma$ field, leptogenesis can only happen once the PQ symmetry is broken,  and 
the usual interplay between CP-violating decays and washout reactions is complicated by the possibility of annihilation of the RH neutrinos through their interactions with $\sigma$.
After reheating and thermal PQ restoration, the RH neutrinos become massive at the PQ phase transition and 
can retain an equilibrium abundance if washout reactions are still active after $\sigma$ gets a VEV.  
This requires $M/T_c\lesssim1$, where $M$ designates the scale of the RH neutrino masses. It turns out
that the stability condition \eqref{SSMY33} guarantees $M_1/T_c<1$, as can be seen by using the lower  bounds for $T_c$ of equation \eqref{eq:TcfA}. Thus, at least the lightest RH neutrino will
have a thermal abundance after the PQ transition.
Moreover, the annihilation reactions of the RH neutrinos can be neglected with respect to the decays if the combinations of couplings  $Y^2,Y\lambda_\sigma,Y\lambda_{H\sigma}$ remain lower in 
absolute value than  $F_{ij}$. Given equation \eqref{eq:Fboundary}, annihilations will be subdominant if the following equality is satisfied,
\begin{equation}
 M^{3/2}<\frac{{f_A^2}\sqrt{m}}{2v},
\end{equation}
where $M$ and $m$ are typical scales for the heavy and light neutrinos, and $v$ is the Higgs VEV. For $f_A\sim10^{11}$ GeV and neutrino masses near the cosmological bound of 0.23 eV, this can be satisfied for $M<4\times 10^9$ GeV. Therefore, in principle SMASH allows to realise vanilla leptogenesis scenarios in which the lepton asymmetry arises from the decays of a single massive RH neutrino  with an initial thermal abundance. In this case the constraint \eqref{leptogenesis} would apply, which can be  in tension with predictive inflation and stability. This is due to the stability bound of equation \eqref{Y11stability}, which  applies for scenarios with hierarchical RH neutrinos.  When substituting $Y_{11}=\sqrt{2}M_1/f_A$ one gets $M_1<0.3 \lambda_\sigma^{1/4} f_A$, which for $f_A$ around $10^{11}$ GeV and $\lambda_\sigma$ in the window of \eqref{inflationconstraint} is in tension with the  vanilla leptogenesis requirement of equation \eqref{leptogenesis}.

The former tension can be relaxed in scenarios with less hierarchical neutrino spectra. In such cases the stability bound can ensure that all RH neutrinos can retain thermal abundances after the phase transition, and moreover leptogenesis can happen for smaller values of $M_1$ than in equation \eqref{leptogenesis} thanks to the possibility of a resonant enhancement of the CP sources  \cite{Pilaftsis:1997jf,Pilaftsis:2003gt}. This enhancement requires some degree of degeneracy between the RH neutrinos, but only a mild one is needed. For approximately degenerate RH neutrinos, the stability bound \eqref{SSMY33} becomes $M_1<0.8 \lambda_\sigma^{1/4} f_A$, which for $f_A$ {in the window of equation \eqref{farangePQrestoration}} and $\lambda_\sigma$ as in equation \eqref{inflationconstraint} can be satisfied for maximum values of $M_1$ in between $2\times 10^7$ GeV and the bound in \eqref{leptogenesis}. Thus, only a small enhancement of the CP sources is needed with respect to vanilla scenarios. {We estimate a degeneracy around 4\% for $M_1=10^7$ GeV, much less than the one part in $10^8$ in the low-scale resonant leptogenesis scenarios considered for example in \cite{Pilaftsis:1997jf}; more details are given in appendix \ref{app:CPsources}.}

Finally, it is worth mentioning that in SMASH, axionic strings can support zero modes of the right-handed neutrinos and the new quark $Q$. This allows lepton number to be trapped in strings in RH neutrinos, which are released when string loops collapse and decay out of equilibrium injecting new lepton number in the Universe. Numerical estimates show that the contribution to the baryon asymmetry is negligible for the relatively small values of $M_1$ in which we are interested~\cite{Sahu:2005vu,Jeannerot:2005ah,Sahu:2004ir}. {However,} the conclusions are based on poor knowledge about the evolution of the string network so an updated study might be worth{while} but is beyond the scope of this paper.  {We also note that, for very small values of $F_{ij},Y_{ij}$, for which decay leptogenesis is not viable, one can have leptogenesis from neutrino oscillations, as in the $\nu$MSM \cite{Asaka:2005pn}. This requires some washout reactions to never reach equilibrium, as can be seen to happen for GeV-scale RH neutrino masses. This corresponds to $Y_{ij}\sim10^{-11}$.}


\section{\label{summary}Summary and perspectives}

We have considered a minimal extension of the SM by
\begin{itemize}
\item[(i)] a new global $U(1)$ PQ {and L} symmetry which is spontaneously broken by the VEV of a SM-singlet {complex} scalar field,
$v_{\sigma}=\sqrt{2\aver{|\sigma |^2}}$,
\item[(ii)] three right-handed SM-singlet neutrinos $N_i$, charged under the new global $U(1)$ symmetry,
\item[(iii)] a vector-like color triplet $Q$, also charged under the $U(1)$,
\end{itemize}
{whose most general} {interactions,} {the Yukawa interactions, the scalar potential, and the non-minimal couplings to gravity, have been given in Eqs. \eqref{lyukseesaw},  \eqref{scalar_potential}, and   \eqref{Lmain}, respectively.
We have shown that, in a particular region of its parameter space, notably for an intermediate symmetry breaking scale, $3\times10^{10}\,{\rm GeV}\lesssim v_\sigma\lesssim 1.2\times10^{11}\,{\rm GeV}$, this simple UV completion of the SM -- dubbed SMASH -- provides a consistent and complete description of particle physics up to the Planck scale 
and of cosmology from inflation until today, cf. Fig.  \ref{fig:history}.}

\begin{figure}[h!]
\begin{center}
\includegraphics[width=0.9\textwidth]{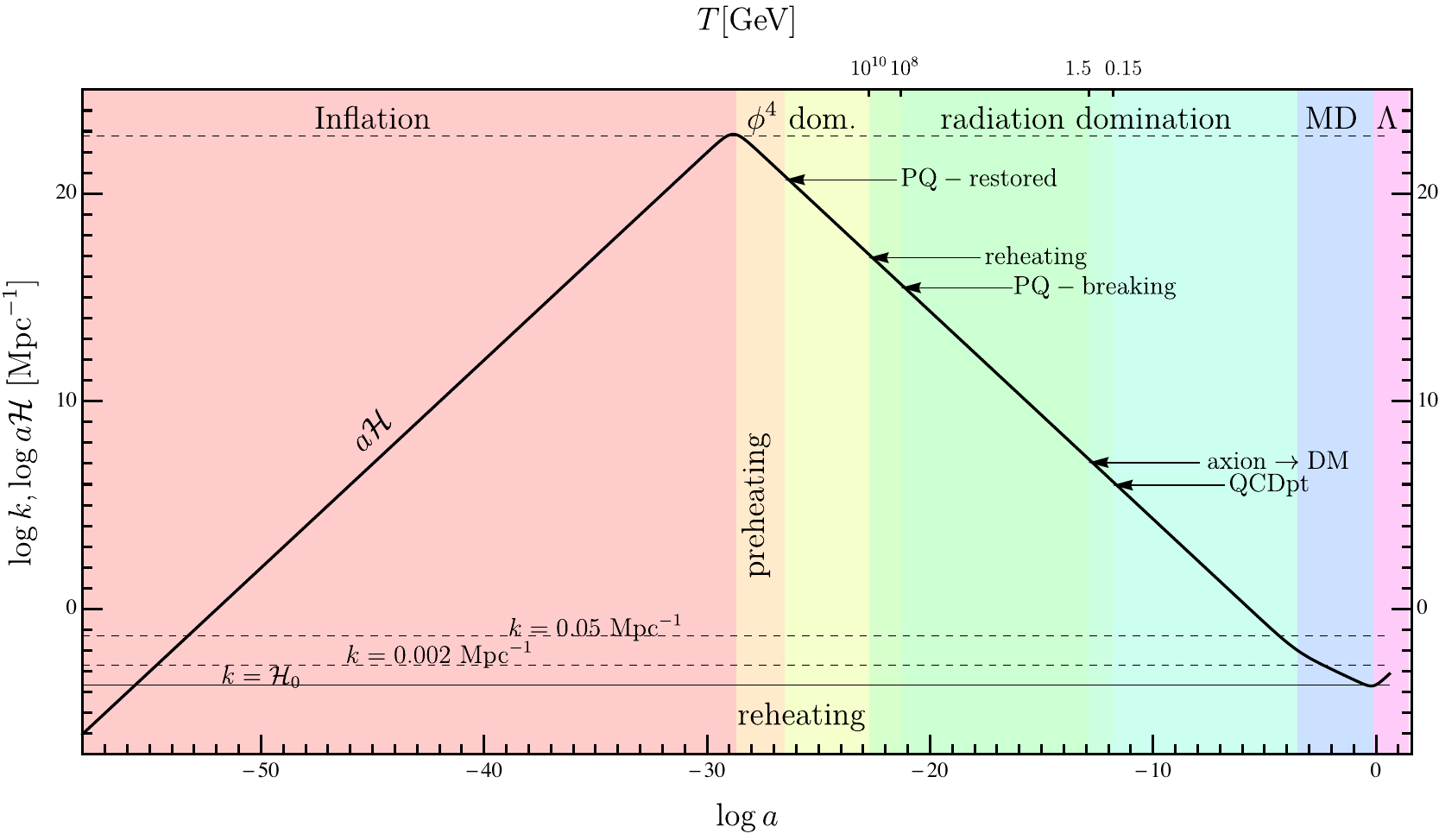}
\caption{\small The history of the Universe in SMASH HHSI, emphasising the transition from inflation to radiation-domination-like Universe expansion $a{\cal H}\propto 1/a$ before standard matter and cosmological constant domination epochs. The end of inflation, PQ-restoration, reheating and PQ breaking have ${\cal O}(1)$ SMASH model dependencies (mainly from $\lambda_\sigma$ or $\xi_\sigma$) not reflected in the big picture.  }  
\label{fig:history}       
\end{center}
\end{figure}

 {In fact,}
at low energies, SMASH reduces to the SM, augmented by type-I seesaw 
generated neutrino masses and mixing, plus the axion $A$ --the latter arising as a Nambu-Goldstone boson from the breaking of the $U(1)$ symmetry-- with a decay constant equal to the VEV of $\sigma$: $f_A\equiv v_\sigma$. The strong CP problem is automatically solved in this model. SMASH features small active neutrino masses, 
$m_i\propto v^2/f_A$, {cf. equation \eqref{seesaw},} 
and predicts also a tiny mass of the axion, $m_A\propto m_\pi f_\pi/f_A$, {cf. equation \eqref{top},} as well as very feeble 
couplings of the latter to SM particles, $g_{A\,{\rm SM}}\propto 1/f_A$, {cf. equation \eqref{axion_leff_app}.} 

The non-minimal couplings in SMASH {stretch} the scalar potential in the 
Einstein frame, which makes it convex and asymptotically flat at large field values (cf. Fig.\  \ref{fig:pot_einstein}). 
Depending on the signs of the parameters $\kappa_H  \equiv \lambda_{H\sigma} \xi_H - \lambda_H \xi_\sigma$ and $\kappa_\sigma  \equiv \lambda_{H\sigma} \xi_\sigma - \lambda_\sigma \xi_H$, involving the non-minimal couplings $\xi_H$ and $\xi_\sigma$,  the Higgs field, the hidden scalar field, or even a linear combination of both fields {could in principle play} the role of the inflaton (cf. Table  \ref{tab:summa}) and the predictions of the CMB parameters such as the spectral 
index $n_s$, its running $\alpha$, and the 
tensor-to-scalar ratio $r$ are in perfect consistency with the current CMB observations
for a wide range {of SMASH} parameter space and provide a target for future precision cosmology, 
see {Figs.\ \ref{fig:r_vs_ns}, \ref{fig:chi2}, and \ref{fig:nechi2}, {and} {the windows of cosmological parameters \eqref{eq:cosmoranges}.}} {Indeed, the requirement of predictive inflation, free of unitarity problems, demands { $r \gtrsim 0.004$}, which could be probed by upcoming experiments such as LiteBIRD \cite{Matsumura:2013aja} and PRISM  \cite{Andre:2013afa}}. Analogously,     future measurements, in particular of the 21cm line of Hydrogen \cite{Adshead:2010mc,Shimabukuro:2014ava}, will improve the determination of the running of the scalar spectral {index.} {We have argued that the validity of these predictions requires that inflation is driven either by the new scalar, $\sigma$,  or by a mixture of it with the Higgs. The reason is that Higgs Inflation (HI) suffers from an intrinsic lack of predictive power due to its low unitarity breaking scale.}
{Indeed, if} the inflaton is the hidden scalar field (HSI) or a mixture between the Higgs and the hidden scalar (HHSI), the size of the 
required non-minimal coupling, $\xi_\sigma\sim 10^5{\sqrt{\lambda}\gtrsim} \xi_H$, is not necessarily large {(unlike in HI)},
cf. Figs.\ \ref{fig:r_vs_ns} and \ref{fig:chi2}, 
 since 
the effective scalar self-coupling  
\begin{equation} \label{recap}
{\lambda} \equiv 
\left\{
\begin{array}{ll} 
\lambda_\sigma
,  & \mathrm{for\ HSI},  \\ 
\lambda_\sigma \left( 1-\frac{\lambda_{H\sigma}^2}{\lambda_\sigma\lambda_H} \right)
,  & \mathrm{for\ HHSI},
\end{array}
\right.
\end{equation}
can be small, {$ 5\times10^{-13} \lesssim \lambda\lesssim 5\times10^{-10}$.}
In this region of parameter space, the perturbative predictivity of SMASH is guaranteed and superior to Higgs Inflation, which {operates} at large $\xi_H$. 

{
In order to ensure self-consistency of inflation in SMASH, 
we have determined the region of parameter space in which the SMASH scalar potential is positive all the way up to the Planck scale.  Importantly, the Higgs portal term $\propto \lambda_{H\sigma}$ in the scalar potential helps to ensure absolute stability in the Higgs direction, which is threatened by top-quark loops, via the threshold stabilisation mechanism. We have found that stability can be achieved if the threshold parameter $\delta = \lambda_{H\sigma}^2/\lambda_\sigma$ is between $10^{-3}$ and $10^{-1}$, see Fig. \ref{fig:mindelta}. Instabilities could also originate  in the $\rho$ direction due to quantum corrections from the RH neutrinos and the exotic quark $Q$. Stability in the 
$\sigma$ direction then requires the Yukawas to satisfy the bound $\sum Y^4_{ii}+6 y^4\lesssim 16\pi^2\lambda_\sigma/\log\(30M_P/\sqrt{2\lambda_\sigma}v_\sigma\)$, cf. equation \eqref{SSMY33}.}
{These analytical findings were backed} {by precise numerical scans, as displayed in Figs.\ \ref{fig:scan1}, \ref{fig:scan2}, \ref{fig:scan3}, and \ref{fig:SMASHscans}.
}

{We have studied reheating after inflation in Section \ref{sec:reheating}, with the aim to determine whether the PQ and L symmetry is restored after inflation, leading then to a very predictive axion dark matter scenario, and to estimate the reheating temperature to investigate under which conditions efficient baryogenesis can occur}. {Slow-roll inflation ends at a value of $\rho\sim \mathcal{O}(M_P)$, where the effect of
$\xi_\sigma$ is negligible and the inflaton starts to undergo Hubble-damped oscillations in a quartic potential, with the Universe expanding as in a radiation-dominated era, which lasts until reheating, cf. Fig. \ref{fig:history}. After the latter, radiation domination continues, though driven by a bath of relativistic particles. This fixes the thick { black} line in Fig.\ \ref{fig:r_vs_ns} as the prediction for $r$, $n_s$ and $N$ in SMASH.}

{Importantly, we have shown that HSI inflation generically leads to an excess of dark radiation (cf. eqs.~\eqref{eq:HSITR}), which is incompatible with current observational bounds. This is because the reheating temperature is low ($\sim 10^7$ GeV) and the axions are never thermalised. For this reason, inflation in SMASH must be of HHSI type and therefore the inflaton contains a (small) Higgs component. This component is not only essential for inflation itself, but also after it, since it is thanks to the Higgs that the Universe gets reheated efficiently. The reheating temperature in this case} is predicted to be around {$T_R\sim  10^{10}$\,GeV, cf. Fig.\ \ref{fig:reheating} (bottom),
 for $\delta\sim 0.05$ (see Fig.~\ref{fig:mindelta}) and $\lambda\sim10^{-10}$ (which satisfy the requirements for stability and inflation).
Such temperature ensures a thermal restoration of the PQ symmetry for the relevant region of parameter space,  since the critical temperature $T_c$ of the PQ phase transition goes as ${T_c}/{v_\sigma}\simeq{2\sqrt{6\lambda_\sigma}}/{\sqrt{8(\lambda_\sigma+\lambda_{H\sigma})+\sum_i Y^2_{ii}+6 y^2}}$, cf. equation  \eqref{eq:TC}. A thermal background of axions is produced at this stage and leads to dark radiation with a 
 corresponding increase in the effective number of relativistic neutrino species of $\triangle N_\nu^{\rm eff}\simeq 0.03$, cf. equation \eqref{eq:delta_neff}. 
Although this is beyond the current accuracy of global fits to cosmological data, it may however be reached in the next decade by a stage-IV CMB polarisation experiment \cite{Abazajian:2013oma,Errard:2015cxa}.}

The hot plasma after reheating contains {a nearly thermal} population of relativistic right-handed singlet neutrinos $N_i$, provided {that} their mass $M_i=Y_{ii} f_A/\sqrt{2}$ is below $T_R$. Their subsequent  out-of-equilibrium 
CP, $L$, and $B-L$ violating decays into leptons and Higgs bosons, at temperatures 
$T\sim M_1$, generate  an 
$L$ and a $B-L$ asymmetry, which is  finally turned (by non-perturbative electroweak 
$B+L$ violating processes) into the observed baryon asymmetry of the Universe. This thermal leptogenesis 
mechanism explaining the matter-antimatter asymmetry of the Universe  requires a right-handed neutrino
mass above $M_1\gtrsim (3\textendash 5)\times 10^{8}$\,GeV, {cf. equation \eqref{leptogenesis}.} This value is compatible with vacuum stability in
SMASH if $\lambda_\sigma\gtrsim 10^{-10}$, which is not far from the border of the stability region, as follows from equation \eqref{SSMY33} and can be seen in  Figs.\ \ref{fig:scan1} (left) and \ref{fig:SMASHscans}. Lower values of $M_1$ can be attained in models in which there is some degree of degeneracy between the RH neutrinos, which can enhance the lepton asymmetry. We note that
no large resonant enhancement {is} needed to ensure compatibility with stability.

A robust prediction of SMASH is that dark matter is comprised by axions, with a mass in the range
$50\,\mu\mathrm{eV}\lesssim m_A \lesssim 200 \,\mu\mathrm{eV}$, 
corresponding to a PQ symmetry breaking scale in the range $3\times 10^{10}\,\mathrm{GeV}\lesssim f_A \lesssim   1.2\times 10^{11}\,\mathrm{GeV}$, {cf. equation \eqref{farangePQrestoration}.} This window is enforced solely by dark-matter constraints, the uncertainty  coming mainly from the difficulty in predicting the relative importance of the two main mechanisms of production of axionic dark matter, i.e. the {re-alignment} mechanism and the  decay of topological strings.\footnote{We recall that in SMASH scenarios, the PQ symmetry is restored either nonthermally or thermally unless $f_A\gtrsim 10^{16}$ GeV, in which case the scenarios are ruled out by isocurvature constraints.}
{Fortunately, the axion dark matter mass window will be probed  in the upcoming decade by axion dark matter direct detection experiments such as CULTASK~\cite{CULTASK}, MADMAX~\cite{Majorovits:2016yvk,TheMADMAXWorkingGroup:2016hpc}, and ORPHEUS~\cite{Rybka:2014cya}, see also~\cite{Horns:2012jf,Borsanyi:2016ksw} and Fig.~\ref{fig:pro} for our estimates of their future sensitivity.}}

{
In this scenario, axion dark matter is highly inhomogeneous at length scales of the order of the causal horizon around the time when the axion mass turns on, around 1 comoving mpc. Axions in dense regions suffer gravitational collapse around matter-radiation equality forming the so-called axion miniclusters~\cite{Hogan:1988mp,Kolb:1993zz,Kolb:1993hw,Zurek:2006sy}. Miniclusters might have survived until today and could be detected by femtolensing~\cite{Kolb:1995bu,Zurek:2006sy}. Some speculations suggest that they might be behind the recently discovered fast radio bursts~\cite{Iwazaki:2014wka,Tkachev:2014dpa,Iwazaki:2014wta,Pshirkov:2016bjr}. An encounter with the Earth would boost the signal of direct detection experiments by many orders of magnitude but it seems extremely unlikely. 
The encounter with their tidal streams might give a more moderate enhancement but it could be as frequent as 1 every 10 years~\cite{Tinyakov:2015cgg}. }

\begin{figure}[t]
\begin{center}
\includegraphics[width=0.7\textwidth]{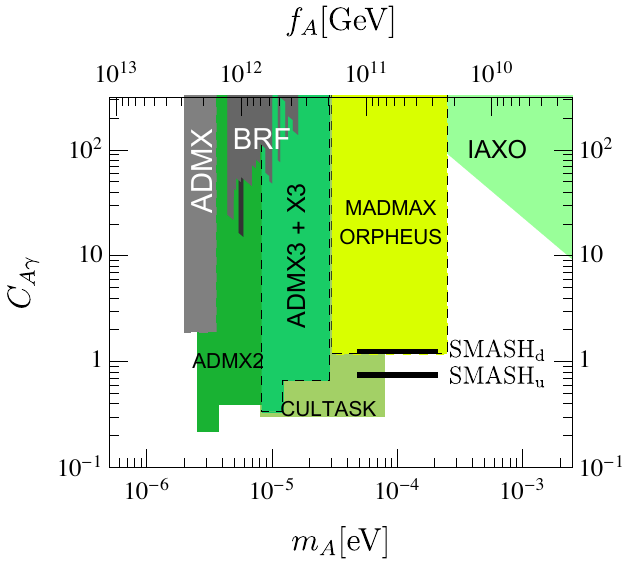}
\caption{\label{fig:pro} \small SMASH predictions for the axion-photon coupling (thick solid horizontal lines {labelled as SMASH$_{d(u)}$ predicted in the version exploiting a hypercharge assignment of $-1/3$ ($2/3$) for $Q$)}
with current bounds on axion DM (ADMX \cite{vanBibber:2013ssa},BRF) and prospects for next generation axion dark matter experiments, such as ADMX2(3)~\cite{ADMX}, CULTASK~\cite{CULTASK}, 
MADMAX~\cite{Majorovits:2016yvk,TheMADMAXWorkingGroup:2016hpc}, ORPHEUS~\cite{Rybka:2014cya}, X3~\cite{X3}, and the helioscope IAXO~\cite{Armengaud:2014gea}.}
\end{center}
\end{figure}

{An important} feature of SMASH is that the axion is at the same time the majoron. Therefore, it has a direct coupling to the active neutrinos, cf. Eq. \eqref{axion_leff}. Unfortunately, both the smallness of the neutrino mass and the largeness of the symmetry breaking scale suppresses its interaction strength so much, cf. equation \eqref{double_suppression}, that it seems impossible to test the majoron nature of the axion in the foreseeable future. 

We have discussed just one possible minimal extension of the SM exploiting the Axion, the {seesaw} and 
{Higgs portal inflation.} Clearly, one may consider similarly well motivated minimal extensions exploiting other axion or {seesaw}
models. It is then {an interesting question to determine} which of the predictions of SMASH are universal and which of them allow for a 
discrimination of different incarnations of the SMASH idea. 

{For instance, the HSI and HHSI predictions for the CMB observables} are universal for any axion variant of SMASH.  
Every PQ-like UV completion of the SM augmented by an axion features a hidden complex scalar field,  
whose potential has the form \eqref{scalar_potential} and whose 
modulus can play the role of the inflaton. The favored axion mass range, however, depends on the UV completion,
in particular on the domain wall number \cite{Kawasaki:2014sqa,Ringwald:2015dsf}. {The axion mass range gets extended} to larger masses, 
up to the meV range \cite{Kawasaki:2014sqa,Ringwald:2015dsf}, if one considers, as in Ref. \cite{Clarke:2015bea}, a variant of SMASH exploiting  two Higgs doublets, $H_u$ and $H_d$ (instead of the color-triplet $Q$) coupling to up and down like quarks, as in the DFSZ axion model \cite{Dine:1981rt,Zhitnitsky:1980tq}, which has domain wall number six.\footnote{{Importantly, in this case the PQ symmetry
is required to be an accidental rather than an exact symmetry
in order to avoid the overclosure of the Universe
due to domain walls [61].}}\footnote{In the KSVZ axion model, with the color-triplet $Q$, \cite{Kim:1979if,Shifman:1979if}, the domain wall number is one.}
Intriguingly, in this variant of SMASH, in which --in contrast to the KSVZ case-- the axion has a tree-level coupling to electrons, one may explain simultaneously the {recent} hints of anomalous excessive energy losses of red giants,  
white dwarfs and helium burning stars  \cite{Giannotti:2015kwo}. {The latter coupling may be probed in the  corresponding axion mass window by the axion dark matter experiment QUAX \cite{Barbieri:2016vwg}, the fifth force experiment ARIADNE \cite{Arvanitaki:2014dfa}, and the helioscope IAXO~\cite{Armengaud:2014gea}.}
Further variants of SMASH may invoke other implementations of the {seesaw} mechanism, such as in 
Refs.\ \cite{Bertolini:2014aia,Ahn:2015pia,Bertolini:2015boa}.

{Vacuum stability is an important requirement in SMASH and its variants. Therefore, it is extremely important 
to improve the experimental input to these investigations, in particular $\alpha_s$, $m_t$, and $m_h$. The main source of uncertainty is the top mass, which could be determined with a precision of the order of
$0.1$ GeV --mostly limited by theoretical uncertainties-- in linear colliders such as CLIC, CEPC, FCC-ee and ILC, see e.g.\ \cite{Seidel:2013sqa,Fan:2014vta}.

Direct tests of SMASH at present or near future accelerators are, however, only possible in corners of 
parameter space. The right-handed singlet neutrinos $N_i$ can be produced via Drell-Yan} {processes} {at the LHC, if their Yukawa couplings  
are sufficiently small, $Y_{ii}\sim M_i/f_A\lesssim 10^{-7}$, and then searched for in the same-sign dilepton sample \cite{Deppisch:2015qwa}. Similarly, the exotic quark $Q$ in SMASH is in reach of the 
LHC, if its Yukawa coupling is in the same ball-park, 
$y\sim m_Q/f_A\lesssim 10^{-7}$, cf.  \cite{Okada:2012gy}.  
Further SMASH opportunities at LHC arise 
if the DFSZ variant of SMASH is realised, because the latter features a full-fledged two-Higgs-doublet model
with all of its exciting electroweak-scale phenomenology \cite{Branco:2011iw}.
For even smaller Yukawa couplings of the right-handed neutrinos, 
$Y_{ii}\sim M_i/f_A\lesssim 10^{-11}$, the 
neutrino phenomenology of the $\nu$MSM \cite{Canetti:2012kh} is reproduced,\footnote{In this case, thermal leptogenesis, being based on decays of the $N_i$ into Higgses plus leptons, is kinematically forbidden, but the variant of leptogenesis based on coherent CP-asymmetric neutrino oscillations proposed in Ref. \cite{Akhmedov:1998qx} may lead to a sufficiently large matter-anti-matter asymmetry. Importantly, this SMASHy variant of the $\nu$MSM frees the lightest of the
$N_i$ from the burden to be 100\,\% of dark matter, thereby evading strong astrophysical bounds \cite{Schneider:2016uqi,Perez:2016tcq}. } which can be tested decisively at the proposed beam dump experiment SHiP \cite{Alekhin:2015byh}. 
}

{To conclude, it may be useful to recapitulate the values of the most significant new parameters added to the SM Lagrangian and some combinations of them in a standard SMASH scenario. The VEV of the new singlet, which is also the axion decay constant, has to be in the specific range $3\times 10^{10}\,{\rm GeV}<v_\sigma=f_A<5\times 10^{11}\,{\rm GeV}$ due to the requirement of the total dark matter abundance. The non-minimal couplings to $R$ of both the Higgs ($\xi_H$)  and the new singlet ($\xi_\sigma$) are of order 1 or smaller. The coupling $\xi_\sigma$ has also to be larger than approximately $6\times 10^{-3}$. These limits come from imposing perturbative unitarity up to $M_P$ and the current upper bound on the tensor-to-scalar ratio. The effective quartic coupling for inflation, defined above in Eq.\ \eq{recap} is then in the range $5\times10^{-10} \gtrsim \lambda\gtrsim  5\times10^{-13}$.  In a typical case in which $\lambda_\sigma\simeq \lambda$, this range together with stability imply that $2\times 10^{-8}\lesssim|\lambda_{H\sigma}|\lesssim 7\times 10^{-6}$. A specific combination of the Yukawa couplings of the right handed neutrinos to the new singlet ($Y$) and to the Higgs ($F$) are constrained to be of the order $FY^{-1}F^T\sim 10^{-4}$, for the seesaw mechanism. In order to ensure absolute stability of the effective potential, the Yukawa couplings $Y$ cannot be larger than approximately $10^{-3}$.} {A similar bound} {from stability applies to the Yukawa coupling $y$ of the new quark $Q$. Vanilla leptogenesis in the case $\lambda_\sigma\simeq \lambda$ would require that the weakest of the Yukawas $Y$ is also bound to be larger than approximately $3.5\times 10^{-2}$. This bound can be  significantly relaxed with a very mild resonant enhancement of leptogenesis. We have estimated that the required degeneracy in the masses of the right-handed neutrinos is very small, just about 4\% {for RH neutrino masses at $10^7$ GeV}. It is also important to remark that the allowed region of parameter space can be enlarged by relaxing the (conservative) assumption of absolute stability. This is an interesting theoretical direction to explore, which requires a detailed evaluation of the properties of the vacuum in that case, and outside the scope of this work.}

{{There are also other} appealing possibilities for {future theoretical} exploration. Even though SMASH is meant to be a full model of particle physics and cosmology, a string oriented theorist may be motivated by trying to find out how to get the SMASH field content and symmetries from a high energy perspective. Another interesting question is whether the structure of some SMASH variant could help to address the hierarchy problem of the Higgs mass.\footnote{In SMASH, the Higgs mass arises at tree-level from the difference of to large quantities. In Appendix \ref{tune} we discuss briefly the level of tuning associated.} Again, although SMASH is a complete model up to $M_P$, we do not provide a fundamental origin for the hierarchy between this scale, $v_\sigma$, and the electroweak one, which remains an open question.}

\section*{Acknowledgments}

We would like to thank Fedor Bezrukov, Alex G.\ Dias, Jose Ram\'on Espinosa, Daniel Figueroa, Juan Garc\'ia-Bellido, Lucien Heurtier, J\"org J\"ackel, Marieke Postma, Javier Rubio, Ken'ichi Saikawa, Mikhail Shaposhnikov, Brian Shuve, Sergey Sibiryakov, and Alexander Westphal for discussions. 
The work of G.B.\ is funded by the European Union's Horizon 2020 research and innovation programme under the Marie Sk\l{}odowska-Curie grant agreement number 656794 and was supported as well by the German Science Foundation (DFG) within the Collaborative Research Center SFB 676 “Particles, Strings and the Early Universe. G.B. thanks the DESY Theory Group and the CERN Theory Department for hospitality. G.B. and C.T.  are grateful to the Mainz Institute for Theoretical Physics (MITP) for its hospitality and its partial support during the completion of this work. C.T. acknowledges  partial support by the Munich Institute for Astro- and Particle Physics (MIAPP) of the DFG cluster of excellence ``Origin and Structure of the Universe". J.R. is supported by the Ramon y Cajal Fellowship 2012-10597 and by FPA2015- 65745-P (MINECO/FEDER).

\appendix
 
\section{\label{finiteT}Appendix: Effective potential at finite temperature}

In this appendix we briefly review the construction of the finite-temperature effective potential. At one-loop it is given by $V=V^0+V^T$ where $V^0$ is 
the zero-temperature potential (with one-loop Coleman-Weinberg corrections), and $V^T$ the finite-temperature contribution. $V^0$ takes the following form in the $\overline{\rm MS}$ scheme,
\begin{align}
 \label{eq:V0}V^0=V^{{\rm tree}}+\frac{1}{64\pi^2}&\left[\sum_V m^4_V(\phi_i)\left(\log\frac{m^2_V(\phi_i)}{\mu^2}-\frac{5}{6}\right)+\sum_S m^4_S(\phi_i)\left(\log\frac{m^2_S(\phi_i)}{\mu^2}-\frac{3}{2}\right)
 \right.\\
 \nonumber&\left.-\sum_F m^4_F(\phi_i)\left(\log\frac{m^2_F(\phi_i)}{\mu^2}-\frac{3}{2}\right)\right].
\end{align}
In the previous expression $V^{{\rm tree}}$ is the tree-level potential of equation \eqref{scalar_potential}, while $V, S, F$ denote gauge bosons, real scalars and Weyl fermions, respectively.\footnote{In equation \eqref{eq:V0} the sum over vector masses includes a sum over the three polarisations of massive gauge bosons, while the sum over fermions states includes a sum over the two spin/helicity states of each Weyl fermion.}  $m^2_{V/S/F}(\phi_i)$ are the field-dependent masses in the background of the scalar fields $\phi_i$. The one-loop
temperature corrections adopt the form
\begin{align}
 \label{eq:VT}
  V^T=\frac{T^4}{2\pi^2}\left[\sum_B J_B\left(\frac{m^2_B(\phi_i)}{T^2}\right)-\sum_F J_F\left(\frac{m^2_F(\phi_i)}{T^2}\right)\right],
 \end{align}
 where here $B$ denotes bosonic fields (gauge bosons and real scalars) and $F$ denotes Weyl fermions. The functions $J_B$ and $J_F$ are given by
 \begin{align}
  J_B(x)=\int_0^\infty dy\, y^2\log\left[1-\exp(-\sqrt{x^2+y^2})\right],\\
  \nonumber  J_F(x)=\int_0^\infty dy\, y^2\log\left[1+\exp(-\sqrt{x^2+y^2})\right].
 \end{align}
In our calculations we improve on the previous expressions by resumming zero-temperature effects for the scalars \cite{Martin:2014bca,Elias-Miro:2014pca}, and by performing a Daisy resummation in the finite-temperature potential \cite{Quiros:1999jp,Kapusta:2006pm}. These 
resummations increase the precision in the presence of light scalars, such as the Goldstone modes arising when scalar fields get VEVs. We implement the resummations by substituting the tree-level masses in equations 
\eqref{eq:V0} and \eqref{eq:VT} with expression including one-loop corrections at zero-temperature, as well as finite-temperature contributions in the large $T$ limit. The zero-temperature, one-loop two-point functions can be obtained from the general formulae of reference \cite{Martin:2003it}. The leading temperature-dependent corrections in SMASH for the scalar masses $\mu^2_H$, $\mu^2_\sigma$ and the longitudinal hypercharge gauge bosons $m^2_{B,L}$ are:\footnote{For the SU(2) gauge bosons, the corrections are in the Standard Model (the exotic quarks $Q$, $\tilde Q$ are charged under hypercharge but not under $SU(2)_L$)). Note that the corrections of equation \eqref{eq:deltaT} are given in the interaction basis; in order to implement equations \eqref{eq:V0} and \eqref{eq:VT} one has to compute the mass-matrix eigenvalues, including the zero-temperature mixing.}
\begin{align}
 \nonumber\Delta \mu^2_H=&\left(\frac{\lambda_H}{2}+\frac{\lambda_{H\sigma}}{6}+\frac{y_t^2}{4}+\frac{3g_2^2}{16}+\frac{3g^2_1}{80}\right)T^2,\\
 \label{eq:deltaT}\Delta \mu^2_\sigma=&\left(\frac{\lambda_\sigma}{3}+\frac{\lambda_{H\sigma}}{3}+\frac{y^2}{4}+\sum_i\frac{Y_{ii}^2}{24}\right)T^2,\\
 \nonumber\Delta m^2_{B,L}=\,&\frac{7 g^2_1 T^2}{6}.
\end{align}
 Using the one-loop, finite-temperature effective potential with the described improvements, we have estimated that the critical temperature for the PQ phase transition --that is, the temperature below which the potential restricted to the $\sigma$ axis develops a minimum-- has a lower bound in stabilised models given by
 \be
  \label{eq:TcfA}
 \begin{aligned}
  \frac{T_c}{f_A}&\,\gtrsim \,2\lambda_\sigma^{1/4},\quad \text{stable models,}\quad Y_{33}\sim Y_{22}\sim 3Y_{11}\sim3 y,\\
  \frac{T_c}{f_A}&\,\gtrsim \,\lambda_\sigma^{1/4},\quad \text{stable models,}\quad Y_{33}\sim Y_{22}\sim Y_{11}\sim  y.
 \end{aligned}
 \ee
  The transition turns out to be second order. The previous inequalities can in fact be reproduced analytically from the high temperature corrections to $\mu^2_\sigma$ in equation \eqref{eq:deltaT}, since the critical temperature for a second order transition can be estimated by requiring the effective mass to go to zero. In the high-T expansion, the critical temperature is given by
 \begin{align}
 \label{eq:TC}
  \frac{T_c}{f_A}=\frac{2\sqrt{6\lambda_\sigma}}{\sqrt{8(\lambda_\sigma+\lambda_{H\sigma})+\sum_i Y^2_{ii}+6 y^2}}.
 \end{align}
The value of $T_c/f_A$ will be minimal for the largest values of the Yukawas $Y_{ij}$ allowed by stability. These can be obtained from the bound of equation \eqref{SSMY33}; then neglecting the contributions from
the quartic couplings one gets the results of \eqref{eq:TcfA}.

   One may also use the finite-temperature potential to estimate the effect of thermal excitations on the decay rates towards unstable regions. For Higgs and top masses such as $m_h=125.09 $ GeV, $m_t=172.38$ GeV, and for small enough $\lambda_{H\sigma}$, the Higgs direction has an instability which cannot be cured by the threshold mechanism. In this case, thermal excitations after reheating can drive the fields towards the instability region, triggering a first-order phase transition. The  decay rate can be estimated by assuming that the thermal tunneling proceeds solely in the Higgs direction, and solving numerically the differential equation for the thermal bounce.  The temperature at which bubbles of the unstable phase nucleate and the Universe is taken over by the instability can be estimated from the requirement \cite{Quiros:1999jp}
 \begin{align}
  \frac{B(T)}{T}\lesssim 140,
 \end{align}
where $B(T)$ is the Euclidean action of the thermal bounce. We found results analogous to those obtained in reference \cite{Rose:2015lna} for the Standard Model, that is, that thermal tunneling towards 
regions affected by the Higgs instability is only possible for large top masses ($m_t > 174$ GeV for a Higgs mass of 125 GeV). Moreover, for the limiting $m_t$ values the nucleation temperature is above
$10^{17}$ GeV, which is above the typical reheating temperatures in SMASH. Therefore, in models in which the Higgs instability is not cured the dominant fluctuations towards unstable regions will be those sourced by the curvature
during inflation, which can be suppressed with small non-minimal gravitational couplings.


\section{\label{betas}Appendix: Two-loop beta functions}

Here we collect formulae for the two-loop RG equations in SMASH. The beta functions are valid in the $\overline{\rm MS}$ scheme, in the Landau gauge, and have been obtained from the results for general theories in references~\cite{Machacek:1983tz,Machacek:1983fi,Machacek:1984zw,Luo:2002ti}. In the formulae
we ignore mixing in the quark sector, but we take into account the full flavour structure in the leptonic sector, including phases. When the matrices of couplings  $F,G,Y$ are treated as spurions transforming under flavour symmetries, one gets selection rules that only allow flavour contractions compatible with the spurious symmetry. This allows to write the beta functions as linear combinations of the allowed flavour contractions. We have explicitly checked this by matching our results for the beta functions with such linear combinations, which required to solve systems of hundreds or thousands of overdetermined linear equations. The fact that that we found actual solutions is a nontrivial consistency check for our results. For the couplings
 that are already present in the SM, we give the beta functions in terms of the SM ones (for the SM beta functions at two-loops, see references~\cite{Machacek:1983tz,Machacek:1983fi,Machacek:1984zw,Luo:2002ti,Luo:2002ey}). In the formulae that follow, we identify the hypercharges of the extra quarks $Q,\tilde Q$ as $\pm q$. In the calculations of this  paper we take $q=-1/3$, which allows for the $y_{{Q_d}_i}$ couplings to the SM quarks which we however neglect in the expressions below.

\allowdisplaybreaks[1]
\begin{flalign*}
 \beta_{g_1}=&\beta^{SM}_{g_1}+\frac{12}{80\pi^2} g_1^3 q^2+\frac{1}{(16\pi^2)^2}\Big[\frac{108}{25} g_1^5 q^4+q^2 \left(\frac{48}{5} g_1^3 g_3^2-\frac{18}{5} g_1^3 y^2\right)-\frac{3g_1^3}{10}  \,\text{Tr}[FF^\dagger]\Big],
 \end{flalign*}
 \begin{flalign*}
 \beta_{g_2}=&\beta^{SM}_{g_2}-\frac{g_2^3}{2(16\pi^2)^2}\,\text{Tr}[FF^{\dagger }],
 \end{flalign*}
 \begin{flalign*}
 \beta_{g_3}=&\beta^{SM}_{g_3}+\frac{g_3^3}{24\pi^2} +\frac{1}{(16\pi^2)^2}\left[\frac{38 g_3^5}{3}-g_3^3 y^2+\frac{6}{5} g_1^2 g_3^3 q^2\right],
 \end{flalign*}
 \begin{flalign*}
 \beta_{y_t}=&\beta^{SM}_{y_t}+\frac{y_t}{(16\pi^2)}\text{Tr}[{FF}^{\dagger }] +\frac{1}{(16\pi^2)^2}\left[\frac{40}{9} g_3^4 y_t+2 
 \lambda _{H\sigma }^2 y_t+\frac{29}{25} g_1^4 q^2 y_t+\frac{1}{8}\text{Tr}[{FF}^{\dagger }] \left(10 y_b^2 y_t\right.\right.\\
 &\left.\left.+3 g_1^2 y_t+15 g_2^2 y_t-18 y_t^3\right)+\frac{\left.\text{Tr}[{GG}^{\dagger }{FF}^{\dagger }\right] y_t}{2}-\frac{9\text{Tr}\left[{FF}^{\dagger }{FF}^{\dagger }\right] y_t}{4}
 -\frac{3 \text{Tr}[Y^{\dagger }{YF}^{\dagger }F]y_t}{4}\right],
 \end{flalign*}
 \begin{flalign*}
 \beta_{y_b}=&\beta^{SM}_{y_b}+\frac{y_b}{16\pi^2}\text{Tr}[{FF}^{\dagger }] +\frac{1}{(16\pi^2)^2}\left[\frac{40}{9} g_3^4 y_b+2 y_b \lambda _{H\sigma }^2-\frac{1}{25} g_1^4 q^2 y_b+\frac{1}{8}\text{Tr}[{FF}^{\dagger }] \left(3 g_1^2 y_b+15 g_2^2 y_b\right.\right.\\
 &\left.\left.+10 y_b y_t^2-18 y_b^3\right)+\frac{\left.\text{Tr}[{GG}^{\dagger }{FF}^{\dagger }\right] y_b}{2}-\frac{9\text{Tr}\left[{FF}^{\dagger }{FF}^{\dagger }\right] y_b}{4}
  -\frac{3 \text{Tr}[Y^{\dagger }{YF}^{\dagger }F]y_b}{4}\right],
  \end{flalign*}
 \begin{flalign*}
 \beta_G=&\beta^{SM}_G+\frac{1}{16\pi^2}\left[{G}\text{Tr}[{FF}^{\dagger }]-\frac{3 {FF}^{\dagger }G}{2}\right]+\frac{1}{(16\pi^2)^2}\left[\frac{15 {FF}^{\dagger }G y_b^2}{4}+2 \text{G} \lambda _{H\sigma }^2+g_1^2 \left(\frac{3 {G}\text{Tr}[{FF}^{\dagger }]}{8}\right.\right.\\
 &\left.-\frac{27 {FF}^{\dagger }G}{16}\right)+g_2^2 \left(\frac{15 {G}\text{Tr}[{FF}^{\dagger }]}{8}+\frac{9 {FF}^{\dagger }G}{16}\right)-\frac{9 {G}\text{Tr}\left[{FF}^{\dagger }{FF}^{\dagger }\right]}{4}+\frac{{G}\text{Tr}\left[{FF}^{\dagger }{GG}^{\dagger }\right]}{2}\\
 &+\frac{15 {FF}^{\dagger }G y_t^2}{4}+\frac{5 {FF}^{\dagger }{G}\text{Tr}[{FF}^{\dagger }]}{4}+\frac{5 {FF}^{\dagger }{G}\text{Tr}[G^{\dagger }{G]}}{4}+\frac{11 {FF}^{\dagger }{FF}^{\dagger }G}{4}-{FF}^{\dagger }{GG}^{\dagger }G\\
 &\left.-\frac{9 \left.{GG}^{\dagger }{G}\text{Tr}[{FF}^{\dagger }\right]}{4}-\frac{{GG}^{\dagger }{FF}^{\dagger }G}{4}+\frac{7 {FY}^{\dagger }{YF}^{\dagger }G}{8}-\frac{3 G \text{Tr}[Y^{\dagger }{YF}^{\dagger }F]}{4}+\frac{99}{25} \text{G} g_1^4 q^2\right],
  \end{flalign*}
  \begin{flalign*}
 \beta_{F}=&\frac{1}{16\pi^2}\left[3 {F} y_b^2-\frac{9}{20}  {F} g_1^2-\frac{9 {F} g_2^2}{4}+3 {F} y_t^2+F\text{Tr}[{FF}^{\dagger }]+F\text{Tr}[{GG}^{\dagger }]+\frac{3 {FF}^{\dagger }F}{2}\right.\\
 &\left.-\frac{3 {GG}^{\dagger }F}{2}+\frac{{FY}^{\dagger }Y}{2}\right]+\frac{1}{(16\pi^2)^2}\left[g_1^2 \left(\frac{5 {F} y_b^2}{8}-\frac{27 }{20} {F} g_2^2+\frac{17 {F} y_t^2}{8}+\frac{3 F\text{Tr}[{FF}^{\dagger }]}{8}\right.\right.\\
 &\left.+\frac{15 F\text{Tr}[{GG}^{\dagger }]}{8}+\frac{279 {FF}^{\dagger }F}{80}-\frac{243 {GG}^{\dagger }F}{80}\right)+g_2^2 \left(\frac{45 {F} y_b^2}{8}+\frac{45 {F} y_t^2}{8}+\frac{15 F\text{Tr}[{FF}^{\dagger }]}{8}\right.\\
 &\left.+\frac{15 F\text{Tr}[{GG}^{\dagger }]}{8}+\frac{135 {FF}^{\dagger }F}{16}+\frac{9 {GG}^{\dagger }F}{16}\right)+g_3^2 \left(20 {F} y_b^2+20 {F} y_t^2\right)+y_t^2 \left(\frac{3 {F} y_b^2}{2}-\frac{27 {FF}^{\dagger }F}{4}\right.\\
 &\left.+\frac{15 {GG}^{\dagger }F}{4}\right)-\frac{27 {F} y_b^4}{4}+y_b^2 \left(\frac{15 {GG}^{\dagger }F}{4}-\frac{27 {FF}^{\dagger }F}{4}\right)+\frac{21 {F} g_1^4}{40}-\frac{23 {F} g_2^4}{4}+6 {F} \lambda _H^2+2 {F} \lambda _{H\sigma }^2\\
 &-\frac{27 {F} y_t^4}{4}-\frac{9 F\text{Tr}\left[{FF}^{\dagger }{FF}^{\dagger }\right]}{4}+\frac{F\text{Tr}\left[{FF}^{\dagger }{GG}^{\dagger }\right]}{2}-\frac{9 F\text{Tr}\left[{GG}^{\dagger }{GG}^{\dagger }\right]}{4}-12 {FF}^{\dagger }F \lambda _H\\
 &-\frac{9 {FF}^{\dagger }F\text{Tr}[{FF}^{\dagger }]}{4}-\frac{9 {FF}^{\dagger }F\text{Tr}[G^{\dagger }\text{G]}}{4}+\frac{3 {FF}^{\dagger }{FF}^{\dagger }F}{2}-\frac{{FF}^{\dagger }{GG}^{\dagger }F}{4}\\
 &+\frac{5{GG}^{\dagger }F\text{Tr}[{FF}^{\dagger }]}{4}+\frac{5 {GG}^{\dagger }F\text{Tr}[G^{\dagger }\text{G]}}{4}-{GG}^{\dagger }{FF}^{\dagger }F+\frac{11 {GG}^{\dagger }{GG}^{\dagger }F}{4}-\frac{3 F \text{Tr}[F^{\dagger }{FY}^{\dagger }Y]}{4}\\
 &-4 {FY}^{\dagger }Y \lambda _{H\sigma }+\frac{7 {FY}^{\dagger }F^tY^*F}{4}-\frac{{FY}^{\dagger }{YF}^{\dagger }F}{8}-\frac{{FY}^{\dagger }{YY}^{\dagger }Y}{8}-\frac{9}{4}  {FY}^{\dagger }Y y^2\\
 &\left.-\frac{3 \left.{FY}^{\dagger }Y \text{Tr}[{YY}^{\dagger }\right]}{8}+\frac{9}{25} \text{F} g_1^4 q^2\right],
  \end{flalign*}
  \begin{flalign*}
 \beta_{\lambda_H}=&\beta^{SM}_{\lambda_H}+\frac{1}{16\pi^2}\left[4 \lambda _{H\sigma }^2+4\text{Tr}[{FF}^{\dagger }] \lambda _H-2\text{Tr}\left[{FF}^{\dagger }{FF}^{\dagger }\right]\right]+\frac{1}{(16\pi^2)^2}\left[\frac{18}{125} g_1^4 q^2 \left(25 \lambda _H-6 g_1^2\right.\right.\\
 &\left.-10 g_2^2\right)-40 \lambda _H \lambda _{H\sigma }^2-32 \lambda _{H\sigma }^3-24 y^2 \lambda _{H\sigma }^2+g_1^2 \left(\frac{3\text{Tr}[{FF}^{\dagger }] \lambda _H}{2}-\frac{3 g_2^2\text{Tr}[{FF}^{\dagger }]}{10}\right)\\
 &+\frac{15}{2} g_2^2\text{Tr}[{FF}^{\dagger }] \lambda _H-\frac{9g_1^4}{100}  \text{Tr}[{FF}^{\dagger }]-\frac{3 g_2^4\text{Tr}[{FF}^{\dagger }]}{4}-14\text{Tr}\left[{GG}^{\dagger }{FF}^{\dagger }\right] \lambda _H-48\text{Tr}[{FF}^{\dagger }] \lambda _H^2\\
 &-\text{Tr}\left[{FF}^{\dagger }{FF}^{\dagger }\right] \lambda _H-2\text{Tr}\left[{FF}^{\dagger }{GG}^{\dagger }{GG}^{\dagger }\right]-2\text{Tr}\left[{FF}^{\dagger }{FF}^{\dagger }{GG}^{\dagger }\right]+10\text{Tr}\left[{FF}^{\dagger }{FF}^{\dagger }{FF}^{\dagger }\right]\\
 &\left.-3 \text{Tr}[Y^{\dagger }{YF}^{\dagger }F] \lambda _H-4 \text{Tr}[Y^{\dagger }Y] \lambda _{H\sigma }^2+2 \text{Tr}[Y^{\dagger }F^{t }F^*{YF}^{\dagger }F]+2 \text{Tr}[Y^{\dagger }{YF}^{\dagger }{FF}^{\dagger }F]\right],
  \end{flalign*}
  \begin{flalign*}
 \beta_{\mu^2_H}=&\beta^{SM}_{\mu^2_H}+\frac{1}{16\pi^2}\left[4 \lambda _{H\sigma } \mu^2_\sigma+2\text{Tr}[{FF}^{\dagger }] \mu^2_H\right]+\frac{1}{(16\pi^2)^2}\left[\frac{9}{5} g_1^4 q^2 \mu ^2{}_H-\mu^2_\sigma \left(16 \lambda _{H\sigma }^2+24 y^2 \lambda _{H\sigma }\right)\right.\\
 &-4 \mu^2_H \lambda _{H\sigma }^2+\mu^2_H \left(\frac{3 g_1^2\text{Tr}[{FF}^{\dagger }]}{4}+\frac{15 g_2^2\text{Tr}[{FF}^{\dagger }]}{4}-24\text{Tr}[{FF}^{\dagger }] \lambda _H-7\text{Tr}\left[{GG}^{\dagger }{FF}^{\dagger }\right]\right.\\
 &\left.\left.-\frac{9\text{Tr}\left[{FF}^{\dagger }{FF}^{\dagger }\right]}{2}\right)-\frac{3}{2} \text{Tr}[Y^{\dagger }{YF}^{\dagger }F] \mu^2_H-4 \text{Tr}[Y^{\dagger }Y] \lambda _{H\sigma } \mu^2_{\sigma }\right],
 \end{flalign*}
 \begin{flalign*}
 \beta_{y}=&\frac{1}{16\pi^2}\left[4 y^3-8 g_3^2 y-\frac{18}{5} g_1^2 q^2 y+\frac{1}{2}\text{Tr}[Y^{\dagger }Y] y\right]+\frac{1}{(16\pi^2)^2}\left[y \left(4 \lambda _{H\sigma }^2+4 \lambda _{\sigma }^2-\frac{932}{9}  g_3^4\right)\right.\\
 &+y^3 \left(\frac{92 g_3^2}{3}-8 \lambda _{\sigma }\right)-\frac{29 y^5}{4}+\frac{3}{50} g_1^2 q^2 y \left[g_1^2 \left(102 q^2+211\right)-80 g_3^2+230 y^2\right]\\
 &\left.-y \left(\frac{3\text{Tr}\left[{YY}^{\dagger }{YY}^{\dagger }\right]}{4}+\frac{3 \text{Tr}[Y^{\dagger }{YF}^{\dagger }F]}{2}\right)-\frac{3 \text{Tr}[Y^{\dagger }Y] y^3}{4}\right],
  \end{flalign*}
 \begin{flalign*}
 \beta_{\lambda_\sigma}=&\frac{1}{16\pi^2}\left[8 \lambda _{H\sigma }^2+20 \lambda _{\sigma }^2-6 y^4+12 y^2 \lambda _{\sigma }+2 \text{Tr}[Y^{\dagger }Y] \lambda _{\sigma }-\text{Tr}[{YY}^{\dagger }{YY}^{\dagger }]\right]+\frac{1}{(16\pi^2)^2}\left[\lambda _{H\sigma }^2 \left(48 g_2^2\right.\right.\\
 &\left.+\frac{48 g_1^2}{5}-48 y_b^2-80 \lambda _{\sigma }-48 y_t^2\right)+y^4 \left(6 \lambda _{\sigma }-32 g_3^2\right)+y^2 \left(80 g_3^2 \lambda _{\sigma }-120 \lambda _{\sigma }^2\right)-64 \lambda _{H\sigma }^3-240 \lambda _{\sigma }^3\\
 &+24 y^6+\frac{36}{5} g_1^2 q^2 y^2 \left(5 \lambda _{\sigma }-2 y^2\right)-16\text{Tr}[{GG}^{\dagger }] \lambda _{H\sigma }^2-16\text{Tr}{FF}^{\dagger }] \lambda _{H\sigma }^2+4 \text{Tr}[F^{\dagger }{FY}^{\dagger }{YY}^{\dagger }Y]\\
 &\left.+\lambda _{\sigma } (\text{Tr}[{YY}^{\dagger }{YY}^{\dagger }]-6 \text{Tr}[Y^{\dagger }{YF}^{\dagger }F])+4 \text{Tr}[Y^{\dagger }{YY}^{\dagger }{YY}^{\dagger }Y]-20 \text{Tr}[Y^{\dagger }Y] \lambda _{\sigma }^2\right],
 \end{flalign*}
 \begin{flalign*}
 \beta_{\lambda_{H\sigma}}=&\frac{1}{16\pi^2}\left[\lambda _{H\sigma } \left(6 y_b^2-\frac{9 g_2^2}{2}-\frac{9 g_1^2}{10}+12 \lambda _H+8 \lambda _{\sigma }+6 y_t^2+6 y^2\right)
 +8 \lambda _{H\sigma }^2
 +\lambda _{H\sigma } (2\text{Tr}[{GG}^{\dagger }]\right.\\
 &\left.+2\text{Tr}[{FF}^{\dagger }])+\text{Tr}[Y^{\dagger }Y] \lambda _{H\sigma }-2 \text{Tr}[Y^{\dagger }{YF}^{\dagger }F]
 \right]+\frac{1}{(16\pi^2)^2}\left[\lambda _H \left[\lambda _{H\sigma } \left(\frac{72 g_1^2}{5}-72 y_b^2\right.\right.\right.\\
 &+72 g_2^2-72 y_t^2\bigg)-144 \lambda _{H\sigma }^2\bigg]+\lambda _{H\sigma }^2 \left(\frac{6 g_1^2}{5}-24 y_b^2+6 g_2^2-96 \lambda _{\sigma }-24 y_t^2-24 y^2\right)\\
 &+\lambda _{H\sigma } \bigg[g_3^2 \left(40 y_b^2+40 y_t^2+40 y^2\right)+g_1^2 \left(\frac{5 y_b^2}{4}+\frac{9 g_2^2}{8}+\frac{17 y_t^2}{4}\right)+g_2^2 \left(\frac{45 y_b^2}{4}+\frac{45 y_t^2}{4}\right)\\
 &-21 y_b^2 y_t^2-\frac{27 y_b^4}{2}+\frac{1671 g_1^4}{400}-\frac{145 g_2^4}{16}-40 \lambda _{\sigma }^2-\frac{27 y_t^4}{2}-9 y^4-48 y^2 \lambda _{\sigma }\bigg]-60 \lambda _H^2 \lambda _{H\sigma }\\
 &-44 \lambda _{H\sigma }^3-\frac{9}{25} g_1^2 q^2 \left[g_1^2 \left(18 y^2-5 \lambda_{H\sigma}\right)-50 y^2 \lambda_{H\sigma}\right]+\lambda _{H\sigma } \left[g_1^2 \left(\frac{15\text{Tr}[{GG}^{\dagger }]}{4}+\frac{3\text{Tr}[{FF}^{\dagger }]}{4}\right)\right.\\
 &\left.+g_2^2 \left(\frac{15\text{Tr}[{GG}^{\dagger }]}{4}+\frac{15\text{Tr}[{FF}^{\dagger }]}{4}\right)-\frac{9\text{Tr}\left[{GG}^{\dagger }{GG}^{\dagger }\right]}{2}-7\text{Tr}\left[{GG}^{\dagger }{FF}^{\dagger }\right]-\frac{9\text{Tr}\left[{FF}^{\dagger }{FF}^{\dagger }\right]}{2}\right.\\
 &-\lambda _H  (24\text{Tr}[{GG}^{\dagger }]+24\text{Tr}[{FF}^{\dagger }])+ \frac{7 \text{Tr}[Y^{\dagger }{YF}^{\dagger }F]}{2}-\frac{3 \left.\text{Tr}[{YY}^{\dagger }{YY}^{\dagger }\right]}{2}-8 \text{Tr}[Y^{\dagger }Y] \lambda _{\sigma }\bigg]\\
 &-\lambda^2_{H\sigma}\left[4 \text{Tr}[Y^{\dagger }Y] + 8\text{Tr}[{GG}^{\dagger }]+8\text{Tr}[{FF}^{\dagger }]\right]+5 \text{Tr}[F^{\dagger }{FY}^{\dagger }{YY}^{\dagger }Y]+4 \text{Tr}[Y^{\dagger }F^{t}F^*{YF}^{\dagger }F]\\
 &+7 \text{Tr}[Y^{\dagger }{YF}^{\dagger }{FF}^{\dagger }F]-\text{Tr}[Y^{\dagger }{YF}^{\dagger }{GG}^{\dagger }F]\bigg],
 \end{flalign*}
 \begin{flalign*}
 \beta_{\mu^2_\sigma}=&\frac{1}{16\pi^2}\left[8 \mu^2_H \lambda _{H\sigma }+\mu^2_\sigma \left(8 \lambda _{\sigma }+6 y^2+\text{Tr}[Y^{\dagger }Y]\right)\right]+\frac{1}{(16\pi^2)^2}\left[\mu^2_H \left[\lambda _{H\sigma } \left(\frac{48 g_1^2}{5}-48 y_b^2\right.\right.\right.\\
 &+48 g_2^2-48 y_t^2\bigg)-32 \lambda _{H\sigma }^2\bigg]+\mu^2_\sigma \left(40 g_3^2 y^2-8 \lambda _{H\sigma }^2-40 \lambda _{\sigma }^2-9 y^4-48 y^2 \lambda _{\sigma }\right)+18 g_1^2 q^2 y^2 \mu ^2{}_{\sigma }\\
 &\left.-\mu^2_H \lambda _{H\sigma } (16\text{Tr}[{GG}^{\dagger }]+16\text{Tr}[{FF}^{\dagger }])-\mu^2_{\sigma } \left(\frac{3 \left.\text{Tr}[{YY}^{\dagger }{YY}^{\dagger }\right]}{2}+3 \text{Tr}[Y^{\dagger }{YF}^{\dagger }F]\right.\right.\\
 &+8 \text{Tr}[Y^{\dagger }Y] \lambda _{\sigma }\bigg)\bigg],\\
 \end{flalign*}
 \begin{flalign*}
 \beta^{(2)}_{Y}=&\frac{1}{16\pi^2)}\left[F^tF^*Y+{YF}^{\dagger }F+{YY}^{\dagger }Y+3 {Y} y^2+\frac{Y \text{Tr}[Y^{\dagger }Y]}{2}\right]+\frac{1}{(16\pi^2)^2}\left[g_1^2 \left(\frac{51 F^tF^*Y}{40}\right.\right.\\
 &\left.+\frac{51 {YF}^{\dagger }F}{40}\right)-y_b^2 \left(\frac{9 F^tF^*Y}{2}+\frac{9 {YF}^{\dagger }F}{2}\right)+g_2^2 \left(\frac{51 F^tF^*Y}{8}+\frac{51 {YF}^{\dagger }F}{8}\right)+20 g_3^2 {Y} y^2 \\
 &-\lambda _{H\sigma } (8 F^tF^*Y+8 {YF}^{\dagger }F)+4 {Y} \lambda _{H\sigma }^2-\frac{F^tF^*F^tF^*Y}{4}+4 F^tF^*{YF}^{\dagger }F-\left(\frac{9 F^tF^*Y}{2}\right.\\
 &\left.+\frac{9 {YF}^{\dagger }F}{2}\right) y_t^2-\frac{3 F^tF^*Y \text{Tr}[{FF}^{\dagger }]}{2}-\frac{3 F^tF^*Y \text{Tr}[{GG}^{\dagger }]}{2}-\frac{F^tG^*G^tF^*Y}{4}-\frac{{YF}^{\dagger }FY^{\dagger }Y}{4}\\
 &-\frac{3 {YF}^{\dagger }F \text{Tr}[{FF}^{\dagger }]}{2}-\frac{3 {YF}^{\dagger }F \text{Tr}[{GG}^{\dagger }]}{2}-\frac{{YF}^{\dagger }{FF}^{\dagger }F}{4}-\frac{{YF}^{\dagger }{GG}^{\dagger }F}{4}-\frac{{YY}^{\dagger }F^tF^*Y}{4}\\
 &+\frac{7 {YY}^{\dagger }{YY}^{\dagger }Y}{4}-8 {YY}^{\dagger }Y \lambda _{\sigma }-\frac{9 {YY}^{\dagger }Y y^2}{2}-\frac{3 {YY}^{\dagger }Y \text{Tr}[Y^{\dagger }Y]}{4}+4 {Y} \lambda _{\sigma }^2-\frac{1}{2} 9 {Y} y^4\\
 &\left.-\frac{3 Y \text{Tr}[F^{\dagger }{FY}^{\dagger }Y]}{2}-\frac{3 Y\text{Tr}\left[{YY}^{\dagger }{YY}^{\dagger }\right]}{4}+9 g_1^2 q^2 \text{Y} y^2\right].
 \end{flalign*}
 
{
\section{\label{app:CPsources}Appendix: Level of degeneracy required for the RH neutrinos}

Stability in the $\sigma$ direction for approximately degenerate neutrinos requires (see equation \eqref{SSMY33})
\begin{align}
 M_1 < 0.8\lambda_\sigma^{1/4} f_A.
\end{align}
For the smallest $\lambda_\sigma$ in the window of equation \eqref{inflationconstraint}, and for the smallest $f_A$ in the interval \eqref{farangePQrestoration}, one has $M_1\lesssim2\times 10^7$ GeV, which lies below the vanilla leptogenesis bound of $5\times10^8$ GeV \cite{Buchmuller:2002rq}. The latter bound comes from examining the sources of
CP violation in the RH neutrino decays and assuming a hierarchical RH neutrino spectrum; in this case, fixing the light neutrino masses, the sources can be seen proportional to $M_1$, and requiring a sufficient asymmetry leads to the lower bound on $M_1$ \cite{Davidson:2002qv,Hamaguchi:2001gw}. However, the bound can be circumvented  with some degree of degeneracy between the RH neutrinos. The CP-asymmetry in the decays of the former is generated by one-loop self-energy and vertex corrections. The self-energy contributions can be resonantly enhanced in the limit of nearly degenerate RH neutrinos --with mass splittings of the order of the decay widths-- which is the mechanism underlying resonant leptogenesis  \cite{Pilaftsis:1997jf,Pilaftsis:2003gt}. This allows for TeV masses. However, even if the mass splittings are much larger than the decay width, one can still achieve some degree of enhancement and evade the vanilla leptogenesis bound. Far  from the resonance in the self-energy corrections, the sum of the flavoured CP asymmetries for the RH neutrino with flavour index $i$ is \cite{Covi:1996wh}
\begin{align}\label{eq:CP_source}
\varepsilon_{i} &\equiv \frac{\sum_k[\Gamma(N_i\rightarrow L_k H)-\Gamma(N_i\rightarrow \bar L_k H^*)]}{\sum_l\left[\Gamma(N_i\rightarrow L_l H)+\Gamma(N_i\rightarrow \bar L_l H^*)\right]}
\nonumber= \frac{1}{8\pi}\,\frac{1}{(F^\dagger F)_{ii}}\,\sum_{j\neq i}\mathrm{Im}\left[(F^\dagger F)^2_{ij}\right]\,g(x_{ij}),
\end{align}
with 
\begin{align}
x_{ij} = \frac{M_j^2}{M_i^2},\quad g(x) = \sqrt{x}\left[\frac{1}{1-x}+1-(1+x)\log\left(\frac{1+x}{x}\right)\right].
\end{align}
Equation \eqref{eq:Fboundary} implies
\begin{align}
F^2\sim \frac{m M}{v^2}.
\end{align}
Then  with 2 RH neutrinos we may estimate a typical order of magnitude
\begin{align}
 \varepsilon_{i} \sim \frac{m M}{8\pi v^2}g(x_{12}),
\end{align}
which shows the proportionality $\varepsilon_{i}\propto M$ that we alluded to before.
Getting the same CP-source while lowering the mass from $5\times 10^8$ GeV (as in vanilla models) to $10^7$ GeV (as needed in SMASH realizations with small $\lambda_\sigma,f_A$), requires a $50\times$ enhancement of $|g(x_{12})|$. Starting with a vanilla scenario with $M_2=3M_1\Rightarrow x_{12}=9$, such enhancement is achieved for $x=0.96$, which implies a degeneracy of 4\%.

}


\section{\label{tune}Appendix: Level of tuning in SMASH}

With SMASH, we have introduced a new scale beyond the SM: $v_\sigma=f_A\sim10^{11}$ GeV. This scale is the VEV of the new scalar, $\sigma$, and it also plays the role of the axion decay constant.  The large difference between $f_A$ and the electroweak scale can be seen as a tree-level tuning, because the Higgs mass arises from the difference of two much larger quantities. It is then natural to ask whether the hierarchy between $f_A$ and the electroweak scale is stable under quantum corrections. 

In order to illustrate this tree-level tuning more explicitly, let us write the mass parameters of the Higgs doublet and $\sigma$ in SMASH:
\begin{align}
\label{eq:ms}
 \begin{aligned}
  m^2_H=&\,-\lambda_H v^2-\lambda_{H\sigma}v^2_\sigma,\\
 m^2_\sigma=&\,-\lambda_\sigma v^2_\sigma-\lambda_{H\sigma} v^2.
 \end{aligned}
\end{align}
Matching SMASH with the SM by integrating out the heavy field $|\sigma|$ gives the
following constraint for the mass parameters:
\begin{align}
\label{eq:massmatching}
 m^2_H=\bar m^2_H+\frac{\lambda_{H\sigma}}{\lambda_\sigma}m^2_\sigma.
\end{align}
In this equation, $\bar m^2_H$ represents the tree-level mass parameter in the SM, which is of the order of the electroweak scale (and the Higgs mass $m_h^2$). Neglecting the 
electroweak scale in \eq{eq:ms} with respect to $f_A=v_\sigma$, we get
\begin{align}
m^2_\sigma\sim-\lambda_\sigma f^2_A\,.
\end{align} 
Thus, the identity \eqref{eq:massmatching} implies  at tree-level
\begin{align}
\label{eq:m2Htuning}
 -m^2_H-\lambda_{H\sigma} f^2_A\sim\frac{m^2_h}{2},
\end{align}
showing explicitly the aforementioned tuning. This tuning would not be such if $|\lambda_{H\sigma}|f_A$ and $|m_H|$ were both of the order
of the electroweak scale. Since $f_A\sim 10^{11}$GeV, this would require $|\lambda_{H\sigma}|\lesssim10^{-9}$ (for $m_H^2\sim m_h^2$). Such values of  $\lambda_{H\sigma}$ are incompatible with predictive inflation.\footnote{This means inflation free from the unitarity problem, see Section \ref{inflation}.} {For $\lambda_{H\sigma}\sim 10^{-6}$ as required by inflation (assuming that $\tilde\lambda_\sigma\sim\lambda_\sigma\sim10^{-9}$) and stability, one would conclude from \eqref{eq:m2Htuning} that $(m^2_H)^{1/2}$ needs to be tuned against $\lambda_{H\sigma}^{1/2} f_A$ to within one part in $10^6$.} In any case, the question we want to answer is whether the relation \eq{eq:massmatching} is radiatively stable.  To check if this is the case and, importantly, since SMASH is meant to be a complete model of particle physics which does not require any new physics scales other than $f_A$, we may approach the problem by looking at the behaviour of finite quantum corrections to the mass parameters 
$m^2_H$ and $m^2_\sigma$ in equation \eqref{eq:massmatching}.\footnote{The assumption of no new physics scales beyond $f_A$ allows us to go around the hierarchy problem up to $M_P$. At this scale, new physics may anyway be expected  for a variety of reasons.} Since couplings in different renormalisation schemes differ by finite redefinitions or renormalisation group transformations, keeping the finite corrections to \eqref{eq:massmatching} small (including the logarithmic corrections encoding scale-dependence) indicates that the required cancellation is similar in all possible schemes and for different choices of scale. This would indicate that the physical running parameters associated with the two scales (running with momentum as opposed to the arbitrary, unphysical RG scale) will retain their hierarchy at different energies, and the cancellation giving the appropriate Higgs mass will not be an accident happening at a single energy scale.

The most dangerous finite corrections of $m^2_H$ and $m^2_\sigma$ in this respect involve insertions of  $f_A$ at the vertices. These diagrams affect both $m^2_H$ and $m^2_\sigma$ as
\begin{align}
 \delta m^2_H\sim \delta m^2_\sigma \sim \frac{\lambda^2_{H\sigma}f^2_A}{16\pi^2}.
\end{align}
The effect of these corrections will not cancel in equation \eqref{eq:massmatching} unless $\lambda_{H\sigma}\sim \lambda_\sigma$, which is again not compatible with predictive inflation. It is interesting to note that $\lambda_{H\sigma}= \lambda_\sigma$ could be realised with an approximate symmetry mixing Higgs and $\sigma$ components,  broken  explicitly (and softly) by the gauge and Yukawa interactions. In the limit in which the symmetry is exact, the Higgs would in fact be a massless Goldstone boson. This opens the posibility of making the Higgs scale arising dynamically from the explicit breaking of the approximate symmetry by the gauge and Yukawa interactions.

In absence of such a dynamical mechanism, maintaining the level of tuning across all scales requires very small values of $|\lambda_{H\sigma}|$, independently of its sign. In this regime of parameter space, the threshold stabilisation mechanism would need one-loop threshold effects. The post-inflationary history could be different from that in the scenarios studied in Section \ref{sec:reheating}. This is because our reheating analysis relied on having $|\lambda_{H\sigma}/\lambda_\sigma|\gg1$, which hindered the production of particles different from the inflaton due to kinematic blocking. For very small $\lambda_{H\sigma}$, the coupling of the inflaton to the Higgs becomes weaker, but the kinematic blocking is lifted and resonant production of Higgses and gauge bosons could be efficient. This could have interesting consequences, as for example ameliorating the dark radiation problem for $\lambda_{H\sigma}>0$. 

Diagrams with loops of heavy particles, such as $N,Q,\tilde Q, \rho$ might also be problematic to maintain the relation \eq{eq:massmatching} across a large range of scales. However, the corrections that they induce can be kept under control if the masses of the heavy particles are sufficiently small, even while keeping $\lambda_\sigma\sim 10^{-10}$ and $|\lambda_{H\sigma}|\sim 10^{-6}$ (which are typical values compatible with predictive inflation). This requires RH neutrinos below $\sim10^6$ GeV (due to their effect in one-loop  corrections to $m^2_\sigma$ \cite{Vissani:1997ys}), and $Q,\tilde Q$ vector quarks below $\sim 10^5$ GeV (as required by two-loop corrections to $m^2_H$ involving hypercharge gauge bosons and $Q,\tilde Q$ quarks \cite{Farina:2013mla}).




\begin{thebibliography}{100}

\bibitem{Aad:2012tfa}
  G.~Aad {\it et al.} [ATLAS Collaboration],
  ``Observation of a new particle in the search for the Standard Model Higgs boson with the ATLAS detector at the LHC,''
  Phys.\ Lett.\ B {\bf 716} (2012) 1
  [arXiv:1207.7214 [hep-ex]].
  
\bibitem{Chatrchyan:2012xdj}
  S.~Chatrchyan {\it et al.} [CMS Collaboration],
  ``Observation of a new boson at a mass of 125 GeV with the CMS experiment at the LHC,''
  Phys.\ Lett.\ B {\bf 716} (2012) 30
  [arXiv:1207.7235 [hep-ex]].

\bibitem{Asaka:2005an}
  T.~Asaka, S.~Blanchet and M.~Shaposhnikov,
  ``The nuMSM, dark matter and neutrino masses,''
  Phys.\ Lett.\ B {\bf 631} (2005) 151
  [hep-ph/0503065].

\bibitem{Asaka:2005pn}
  T.~Asaka and M.~Shaposhnikov,
  ``The nuMSM, dark matter and baryon asymmetry of the universe,''
  Phys.\ Lett.\ B {\bf 620} (2005) 17
  [hep-ph/0505013].
  

\bibitem{Dodelson:1993je}
  S.~Dodelson and L.~M.~Widrow,
  ``Sterile-neutrinos as dark matter,''
  Phys.\ Rev.\ Lett.\  {\bf 72} (1994) 17
  [hep-ph/9303287].

\bibitem{Akhmedov:1998qx}
  E.~K.~Akhmedov, V.~A.~Rubakov and A.~Y.~Smirnov,
  ``Baryogenesis via neutrino oscillations,''
  Phys.\ Rev.\ Lett.\  {\bf 81} (1998) 1359
  [hep-ph/9803255].
  
\bibitem{Canetti:2012kh}
  L.~Canetti, M.~Drewes, T.~Frossard and M.~Shaposhnikov,
  ``Dark Matter, Baryogenesis and Neutrino Oscillations from Right Handed Neutrinos,''
  Phys.\ Rev.\ D {\bf 87} (2013) 093006
  [arXiv:1208.4607 [hep-ph]].

\bibitem{Schneider:2016uqi}
  A.~Schneider,
  ``Astrophysical constraints on resonantly produced sterile neutrino dark matter,''
  JCAP {\bf 1604} (2016) no.04,  059
  [arXiv:1601.07553 [astro-ph.CO]].


\bibitem{Perez:2016tcq}
  K.~Perez, K.~C.~Y.~Ng, J.~F.~Beacom, C.~Hersh, S.~Horiuchi and R.~Krivonos,
  ``(Almost) Closing the Sterile Neutrino Dark Matter Window with NuSTAR,''
  arXiv:1609.00667 [astro-ph.HE].


\bibitem{Bezrukov:2007ep}
  F.~L.~Bezrukov and M.~Shaposhnikov,
  ``The Standard Model Higgs boson as the inflaton,''
  Phys.\ Lett.\ B {\bf 659} (2008) 703
  [arXiv:0710.3755 [hep-th]].

\bibitem{Barbon:2009ya}
  J.~L.~F.~Barbon and J.~R.~Espinosa,
  ``On the Naturalness of Higgs Inflation,''
  Phys.\ Rev.\ D {\bf 79} (2009) 081302
  [arXiv:0903.0355 [hep-ph]].
 
\bibitem{Burgess:2009ea}
  C.~P.~Burgess, H.~M.~Lee and M.~Trott,
  ``Power-counting and the Validity of the Classical Approximation During Inflation,''
  JHEP {\bf 0909} (2009) 103
  [arXiv:0902.4465 [hep-ph]].




%

\bibitem{Degrassi:2012ry}
  G.~Degrassi, S.~Di Vita, J.~Elias-Miro, J.~R.~Espinosa, G.~F.~Giudice, G.~Isidori and A.~Strumia,
  ``Higgs mass and vacuum stability in the Standard Model at NNLO,''
  JHEP {\bf 1208} (2012) 098
  [arXiv:1205.6497 [hep-ph]].

\bibitem{Buttazzo:2013uya}
  D.~Buttazzo, G.~Degrassi, P.~P.~Giardino, G.~F.~Giudice, F.~Sala, A.~Salvio and A.~Strumia,
  ``Investigating the near-criticality of the Higgs boson,''
  JHEP {\bf 1312} (2013) 089
  [arXiv:1307.3536 [hep-ph]].

\bibitem{Bezrukov:2012sa}
  F.~Bezrukov, M.~Y.~Kalmykov, B.~A.~Kniehl and M.~Shaposhnikov,
  ``Higgs Boson Mass and New Physics,''
  JHEP {\bf 1210} (2012) 140
  [arXiv:1205.2893 [hep-ph]].

\bibitem{Alekhin:2012py}
  S.~Alekhin, A.~Djouadi and S.~Moch,
  ``The top quark and Higgs boson masses and the stability of the electroweak vacuum,''
  Phys.\ Lett.\ B {\bf 716} (2012) 214
  [arXiv:1207.0980 [hep-ph]].

\bibitem{Bednyakov:2015sca}
  A.~V.~Bednyakov, B.~A.~Kniehl, A.~F.~Pikelner and O.~L.~Veretin,
  ``Stability of the Electroweak Vacuum: Gauge Independence and Advanced Precision,''
  Phys.\ Rev.\ Lett.\  {\bf 115} (2015) 20,  201802
  [arXiv:1507.08833 [hep-ph]].


\bibitem{Lebedev:2012zw}
  O.~Lebedev,
  ``On Stability of the Electroweak Vacuum and the Higgs Portal,''
  Eur.\ Phys.\ J.\ C {\bf 72} (2012) 2058
  [arXiv:1203.0156 [hep-ph]].

\bibitem{EliasMiro:2012ay}
  J.~Elias-Miro, J.~R.~Espinosa, G.~F.~Giudice, H.~M.~Lee and A.~Strumia,
  ``Stabilization of the Electroweak Vacuum by a Scalar Threshold Effect,''
  JHEP {\bf 1206} (2012) 031
  [arXiv:1203.0237 [hep-ph]].
  
\bibitem{Gonderinger:2009jp}
  M.~Gonderinger, Y.~Li, H.~Patel and M.~J.~Ramsey-Musolf,
  ``Vacuum Stability, Perturbativity, and Scalar Singlet Dark Matter,''
  JHEP {\bf 1001} (2010) 053
  [arXiv:0910.3167 [hep-ph]].

\bibitem{Minkowski:1977sc}
  P.~Minkowski,
  ``$\mu \to e\gamma$ at a Rate of One Out of $10^{9}$ Muon Decays?,''
  Phys.\ Lett.\ B {\bf 67} (1977) 421.

\bibitem{GellMann:1980vs}
  M.~Gell-Mann, P.~Ramond and R.~Slansky,
  ``Complex Spinors and Unified Theories,''
  Conf.\ Proc.\ C {\bf 790927} (1979) 315
  [arXiv:1306.4669 [hep-th]].

\bibitem{Yanagida:1979as}
  T.~Yanagida,
  ``Horizontal Symmetry And Masses Of Neutrinos,''
  Conf.\ Proc.\ C {\bf 7902131} (1979) 95
   [Conf.\ Proc.\ C {\bf 7902131} (1979) 95].

\bibitem{Mohapatra:1979ia}
  R.~N.~Mohapatra and G.~Senjanovic,
  ``Neutrino Mass and Spontaneous Parity Violation,''
  Phys.\ Rev.\ Lett.\  {\bf 44} (1980) 912.

\bibitem{Chikashige:1980ui}
  Y.~Chikashige, R.~N.~Mohapatra and R.~D.~Peccei,
  ``Are There Real Goldstone Bosons Associated with Broken Lepton Number?,''
  Phys.\ Lett.\ B {\bf 98} (1981) 265.

\bibitem{Gelmini:1980re}
  G.~B.~Gelmini and M.~Roncadelli,
  ``Left-Handed Neutrino Mass Scale and Spontaneously Broken Lepton Number,''
  Phys.\ Lett.\ B {\bf 99} (1981) 411.

\bibitem{Schechter:1981cv}
  J.~Schechter and J.~W.~F.~Valle,
  ``Neutrino Decay and Spontaneous Violation of Lepton Number,''
  Phys.\ Rev.\ D {\bf 25} (1982) 774.

\bibitem{Fukugita:1986hr}
  M.~Fukugita and T.~Yanagida,
  ``Baryogenesis Without Grand Unification,''
  Phys.\ Lett.\ B {\bf 174} (1986) 45.
  
\bibitem{Dias:2014osa}
  A.~G.~Dias, A.~C.~B.~Machado, C.~C.~Nishi, A.~Ringwald and P.~Vaudrevange,
  ``The Quest for an Intermediate-Scale Accidental Axion and Further ALPs,''
  JHEP {\bf 1406} (2014) 037
  [arXiv:1403.5760 [hep-ph]].
  
  
\bibitem{Peccei:1977hh}
  R.~D.~Peccei and H.~R.~Quinn,
  ``CP Conservation in the Presence of Instantons,''
  Phys.\ Rev.\ Lett.\  {\bf 38} (1977) 1440.
  
  


\bibitem{Kim:1979if}
  J.~E.~Kim,
  ``Weak Interaction Singlet and Strong CP Invariance,''
  Phys.\ Rev.\ Lett.\  {\bf 43} (1979) 103.

\bibitem{Shifman:1979if}
  M.~A.~Shifman, A.~I.~Vainshtein and V.~I.~Zakharov,
  ``Can Confinement Ensure Natural CP Invariance of Strong Interactions?,''
  Nucl.\ Phys.\ B {\bf 166} (1980) 493.
  

\bibitem{Weinberg:1977ma}
  S.~Weinberg,
  ``A New Light Boson?,''
  Phys.\ Rev.\ Lett.\  {\bf 40} (1978) 223.

\bibitem{Wilczek:1977pj}
  F.~Wilczek,
  ``Problem of Strong p and t Invariance in the Presence of Instantons,''
  Phys.\ Rev.\ Lett.\  {\bf 40} (1978) 279.

\bibitem{diCortona:2015ldu}
  G.~Grilli di Cortona, E.~Hardy, J.~P.~Vega and G.~Villadoro,
  ``The QCD axion, precisely,''
  JHEP {\bf 1601} (2016) 034
  [arXiv:1511.02867 [hep-ph]].
  
\bibitem{Preskill:1982cy}
  J.~Preskill, M.~B.~Wise and F.~Wilczek,
  ``Cosmology of the Invisible Axion,''
  Phys.\ Lett.\ B {\bf 120} (1983) 127.
  

\bibitem{Abbott:1982af}
  L.~F.~Abbott and P.~Sikivie,
  ``A Cosmological Bound on the Invisible Axion,''
  Phys.\ Lett.\ B {\bf 120} (1983) 133.

\bibitem{Dine:1982ah}
  M.~Dine and W.~Fischler,
  ``The Not So Harmless Axion,''
  Phys.\ Lett.\ B {\bf 120} (1983) 137.


\bibitem{Borsanyi:2016ksw} 
  S.~Borsanyi {\it et al.},
  ``Calculation of the axion mass based on high-temperature lattice quantum chromodynamics,''
  Nature {\bf 539}, no. 7627, 69 (2016)
  [arXiv:1606.07494 [hep-lat]].


\bibitem{Kawasaki:2014sqa}
  M.~Kawasaki, K.~Saikawa and T.~Sekiguchi,
  ``Axion dark matter from topological defects,''
  Phys.\ Rev.\ D {\bf 91} (2015) 6,  065014
  [arXiv:1412.0789 [hep-ph]].

\bibitem{Fleury:2016xrz}
  L.~M.~Fleury and G.~D.~Moore,
  ``Axion String Dynamics I: 2+1D,''
  JCAP {\bf 1605} (2016) no.05,  005
  [arXiv:1602.04818 [hep-ph]].

\bibitem{Moore:2016itg}
  G.~D.~Moore,
  ``Intercommutation of U(1) global cosmic strings,''
  arXiv:1604.02356 [hep-ph].

\bibitem{Fleury:2015aca} 
  L.~Fleury and G.~D.~Moore,
  ``Axion dark matter: strings and their cores,''
  JCAP {\bf 1601}, 004 (2016)
  [arXiv:1509.00026 [hep-ph]].

\bibitem{Sikivie:1982qv}
  P.~Sikivie,
  ``Of Axions, Domain Walls and the Early Universe,''
  Phys.\ Rev.\ Lett.\  {\bf 48} (1982) 1156.

\bibitem{Nardi:1990ku}
  E.~Nardi and E.~Roulet,
  ``Are exotic stable quarks cosmologically allowed?,''
  Phys.\ Lett.\ B {\bf 245} (1990) 105.



\bibitem{Perl:2001xi}
  M.~L.~Perl, P.~C.~Kim, V.~Halyo, E.~R.~Lee, I.~T.~Lee, D.~Loomba and K.~S.~Lackner,
  ``The Search for stable, massive, elementary particles,''
  Int.\ J.\ Mod.\ Phys.\ A {\bf 16} (2001) 2137
  [hep-ex/0102033].


\bibitem{Perl:2004qc}
  M.~L.~Perl, E.~R.~Lee and D.~Loomba,
  ``A Brief review of the search for isolatable fractional charge elementary particles,''
  Mod.\ Phys.\ Lett.\ A {\bf 19} (2004) 2595.


\bibitem{Chuzhoy:2008zy}
  L.~Chuzhoy and E.~W.~Kolb,
  ``Reopening the window on charged dark matter,''
  JCAP {\bf 0907} (2009) 014
  [arXiv:0809.0436 [astro-ph]].


\bibitem{Perl:2009zz}
  M.~L.~Perl, E.~R.~Lee and D.~Loomba,
  ``Searches for fractionally charged particles,''
  Ann.\ Rev.\ Nucl.\ Part.\ Sci.\  {\bf 59} (2009) 47.



\bibitem{Kim:1981jw}
  J.~E.~Kim,
  ``Reason for SU(6) Grand Unification,''
  Phys.\ Lett.\ B {\bf 107} (1981) 69.

\bibitem{Mohapatra:1982tc}
  R.~N.~Mohapatra and G.~Senjanovic,
  ``The Superlight Axion and Neutrino Masses,''
  Z.\ Phys.\ C {\bf 17} (1983) 53.

  

\bibitem{Berezhiani:1983hm} 
  Z.~G.~Berezhiani,
  ``The Weak Mixing Angles in Gauge Models with Horizontal Symmetry: A New Approach to Quark and Lepton Masses,''
  Phys.\ Lett.\ B {\bf 129}, 99 (1983).

\bibitem{Berezhiani:1985in} 
  Z.~G.~Berezhiani,
  ``Horizontal Symmetry and Quark - Lepton Mass Spectrum: The SU(5) x SU(3)-h Model,''
  Phys.\ Lett.\ B {\bf 150}, 177 (1985).
  
  
\bibitem{Shafi:1984ek}
  Q.~Shafi and F.~W.~Stecker,
  ``Implications of a Class of Grand Unified Theories for Large Scale Structure in the Universe,''
  Phys.\ Rev.\ Lett.\  {\bf 53} (1984) 1292.

\bibitem{Langacker:1986rj}
  P.~Langacker, R.~D.~Peccei and T.~Yanagida,
  ``Invisible Axions and Light Neutrinos: Are They Connected?,''
  Mod.\ Phys.\ Lett.\ A {\bf 1} (1986) 541.

\bibitem{Shin:1987xc}
  M.~Shin,
  ``Light Neutrino Masses and Strong {CP} Problem,''
  Phys.\ Rev.\ Lett.\  {\bf 59} (1987) 2515
   [Phys.\ Rev.\ Lett.\  {\bf 60} (1988) 383].

\bibitem{He:1988dm}
  X.~G.~He and R.~R.~Volkas,
  ``Models Featuring Spontaneous {CP} Violation: An Invisible Axion and Light Neutrino Masses,''
  Phys.\ Lett.\ B {\bf 208} (1988) 261
   [Phys.\ Lett.\ B {\bf 218} (1989) 508].

\bibitem{Dias:2005dn}
  A.~G.~Dias and V.~Pleitez,
  ``The Invisible axion and neutrino masses,''
  Phys.\ Rev.\ D {\bf 73} (2006) 017701
  [hep-ph/0511104].

\bibitem{Celis:2014iua}
  A.~Celis, J.~Fuentes-Martin and H.~Serodio,
  ``An invisible axion model with controlled FCNCs at tree level,''
  Phys.\ Lett.\ B {\bf 741} (2015) 117
  [arXiv:1410.6217 [hep-ph]].

\bibitem{Celis:2014jua}
  A.~Celis, J.~Fuentes-Martín and H.~Serôdio,
  ``A class of invisible axion models with FCNCs at tree level,''
  JHEP {\bf 1412} (2014) 167
  [arXiv:1410.6218 [hep-ph]].

\bibitem{Bertolini:2014aia}
  S.~Bertolini, L.~Di Luzio, H.~Kolešová and M.~Malinský,
  ``Massive neutrinos and invisible axion minimally connected,''
  Phys.\ Rev.\ D {\bf 91} (2015) 5,  055014
  [arXiv:1412.7105 [hep-ph]].
  
\bibitem{Bertolini:2015boa}
  S.~Bertolini, L.~Di Luzio, H.~Kolešová, M.~Malinský and J.~C.~Vasquez,
  ``Neutrino-axion-dilaton interconnection,''
  Phys.\ Rev.\ D {\bf 93} (2016) no.1,  015009
  [arXiv:1510.03668 [hep-ph]].
  
\bibitem{Ng:2015eia}
  J.~N.~Ng and A.~de la Puente,
  ``Electroweak Vacuum Stability and the seesaw Mechanism Revisited,''
  Eur.\ Phys.\ J.\ C {\bf 76} (2016) no.3,  122
  [arXiv:1510.00742 [hep-ph]].

\bibitem{Salvio:2015cja}
  A.~Salvio,
  ``A Simple Motivated Completion of the Standard Model below the Planck Scale: Axions and Right-Handed Neutrinos,''
  Phys.\ Lett.\ B {\bf 743} (2015) 428
  [arXiv:1501.03781 [hep-ph]].


\bibitem{Carvajal:2015dxa}
  C.~D.~R.~Carvajal, A.~G.~Dias, C.~C.~Nishi and B.~L.~Sánchez-Vega,
  ``Axion Like Particles and the Inverse seesaw Mechanism,''
  JHEP {\bf 1505} (2015) 069
   [JHEP {\bf 1508} (2015) 103]
  [arXiv:1503.03502 [hep-ph]].

{
\bibitem{Barenboim:2015cqa}
  G.~Barenboim and W.~I.~Park,
  ``Peccei-Quinn field for inflation, baryogenesis, dark matter, and much more,''
  Phys.\ Lett.\ B {\bf 756} (2016) 317
  [arXiv:1508.00011 [hep-ph]].
}

\bibitem{Clarke:2015bea}
  J.~D.~Clarke and R.~R.~Volkas,
  ``Technically natural nonsupersymmetric model of neutrino masses, baryogenesis, the strong CP problem, and dark matter,''
  Phys.\ Rev.\ D {\bf 93} (2016) no.3,  035001
   [Phys.\ Rev.\ D {\bf 93} (2016) 035001]
  [arXiv:1509.07243 [hep-ph]].
  
\bibitem{Ahn:2015pia}
  Y.~H.~Ahn and E.~J.~Chun,
  ``Minimal Models for Axion and Neutrino,''
  Phys.\ Lett.\ B {\bf 752} (2016) 333
  [arXiv:1510.01015 [hep-ph]].
  

  
\bibitem{Boucenna:2014uma} 
  S.~M.~Boucenna, S.~Morisi, Q.~Shafi and J.~W.~F.~Valle,
  ``Inflation and majoron dark matter in the seesaw mechanism,''
  Phys.\ Rev.\ D {\bf 90}, no. 5, 055023 (2014)
  [arXiv:1404.3198 [hep-ph]].


\bibitem{Okada:2015zfa}
  N.~Okada and Q.~Shafi,
  ``Higgs Inflation, seesaw Physics and Fermion Dark Matter,''
  Phys.\ Lett.\ B {\bf 747} (2015) 223
  [arXiv:1501.05375 [hep-ph]].

\bibitem{Budhi:2015sha}
  R.~H.~S.~Budhi, S.~Kashiwase and D.~Suematsu,
  ``Inflation due to a nonminimal coupling of singlet scalars in the radiative seesaw model,''
  Phys.\ Rev.\ D {\bf 93} (2016) no.1,  013022
  [arXiv:1509.05841 [hep-ph]].


\bibitem{Lerner:2009xg}
  R.~N.~Lerner and J.~McDonald,
  ``Gauge singlet scalar as inflaton and thermal relic dark matter,''
  Phys.\ Rev.\ D {\bf 80} (2009) 123507
  [arXiv:0909.0520 [hep-ph]].

\bibitem{Lerner:2011ge}
  R.~N.~Lerner and J.~McDonald,
  ``Distinguishing Higgs inflation and its variants,''
  Phys.\ Rev.\ D {\bf 83} (2011) 123522
  [arXiv:1104.2468 [hep-ph]].

\bibitem{Khoze:2013uia}
  V.~V.~Khoze,
  JHEP {\bf 1311} (2013) 215
  doi:10.1007/JHEP11(2013)215
  [arXiv:1308.6338 [hep-ph]].
  
  
\bibitem{Kahlhoefer:2015jma}
  F.~Kahlhoefer and J.~McDonald,
  ``WIMP Dark Matter and Unitarity-Conserving Inflation via a Gauge Singlet Scalar,''
  JCAP {\bf 1511} (2015) 11,  015
  [arXiv:1507.03600 [astro-ph.CO]].

\bibitem{Aravind:2015xst}
  A.~Aravind, M.~Xiao and J.~H.~Yu,
  ``Higgs Portal to Inflation and Fermionic Dark Matter,''
  Phys.\ Rev.\ D {\bf 93} (2016) no.12,  123513
  [arXiv:1512.09126 [hep-ph]].


\bibitem{Haba:2014zda}
  N.~Haba and R.~Takahashi,
  ``Higgs inflation with singlet scalar dark matter and right-handed neutrino in light of BICEP2,''
  Phys.\ Rev.\ D {\bf 89} (2014) 11,  115009
   [Phys.\ Rev.\ D {\bf 90} (2014) 3,  039905]
  [arXiv:1404.4737 [hep-ph]].

\bibitem{Haba:2014zja}
  N.~Haba, H.~Ishida and R.~Takahashi,
  ``Higgs inflation and Higgs portal dark matter with right-handed neutrinos,''
  PTEP {\bf 2015} (2015) 5,  053B01
  [arXiv:1405.5738 [hep-ph]].

\bibitem{Fairbairn:2014zta}
  M.~Fairbairn, R.~Hogan and D.~J.~E.~Marsh,
  ``Unifying inflation and dark matter with the Peccei-Quinn field: observable axions and observable tensors,''
  Phys.\ Rev.\ D {\bf 91} (2015) 2,  023509
  [arXiv:1410.1752 [hep-ph]].
  
\bibitem{Khoze:2016zfi}
  V.~V.~Khoze and A.~D.~Plascencia,
  JHEP {\bf 1611} (2016) 025
  doi:10.1007/JHEP11(2016)025
  [arXiv:1605.06834 [hep-ph]].
  
  
\bibitem{Khoze:2014xha}
  V.~V.~Khoze, C.~McCabe and G.~Ro,
  JHEP {\bf 1408} (2014) 026
  doi:10.1007/JHEP08(2014)026
  [arXiv:1403.4953 [hep-ph]].

  
\bibitem{Davoudiasl:2004be}
  H.~Davoudiasl, R.~Kitano, T.~Li and H.~Murayama,
  ``The New minimal standard model,''
  Phys.\ Lett.\ B {\bf 609} (2005) 117
  [hep-ph/0405097].

\bibitem{Ballesteros:2016euj}
  G.~Ballesteros, J.~Redondo, A.~Ringwald and C.~Tamarit,
  ``Unifying inflation with the axion, dark matter, baryogenesis and the seesaw mechanism,''
  Phys.\ Rev.\ Lett.\  {\bf 118} (2017) no.7,  071802
  [arXiv:1608.05414 [hep-ph]].
 

\bibitem{Berezhiani:1992rk}
  Z.~G.~Berezhiani, A.~S.~Sakharov and M.~Y.~Khlopov,
  ``Primordial background of cosmological axions,''
  Sov.\ J.\ Nucl.\ Phys.\  {\bf 55} (1992) 1063
   [Yad.\ Fiz.\  {\bf 55} (1992) 1918].


\bibitem{Ringwald:2015dsf} 
  A.~Ringwald and K.~Saikawa,
  ``Axion dark matter in the post-inflationary Peccei-Quinn symmetry breaking scenario,''
  Phys.\ Rev.\ D {\bf 93}, no. 8, 085031 (2016)
  Addendum: [Phys.\ Rev.\ D {\bf 94}, no. 4, 049908 (2016)]
  [arXiv:1512.06436 [hep-ph]].
 
\bibitem{Kaplan:1985dv}
  D.~B.~Kaplan,
  ``Opening the Axion Window,''
  Nucl.\ Phys.\ B {\bf 260} (1985) 215.

\bibitem{Srednicki:1985xd}
  M.~Srednicki,
  ``Axion Couplings to Matter. 1. CP Conserving Parts,''
  Nucl.\ Phys.\ B {\bf 260} (1985) 689.

\bibitem{Crewther:1977ce}
  R.~J.~Crewther,
  ``Chirality Selection Rules and the U(1) Problem,''
  Phys.\ Lett.\ B {\bf 70} (1977) 349.

\bibitem{DiVecchia:1980ve}
  P.~Di Vecchia and G.~Veneziano,
  ``Chiral Dynamics in the Large n Limit,''
  Nucl.\ Phys.\ B {\bf 171} (1980) 253.

\bibitem{Leutwyler:1992yt}
  H.~Leutwyler and A.~V.~Smilga,
  ``Spectrum of Dirac operator and role of winding number in QCD,''
  Phys.\ Rev.\ D {\bf 46} (1992) 5607.

\bibitem{Raffelt:2006cw}
  G.~G.~Raffelt,
  ``Astrophysical axion bounds,''
  Lect.\ Notes Phys.\  {\bf 741} (2008) 51
  [hep-ph/0611350].



\bibitem{Fischer:2016cyd}
  T.~Fischer, S.~Chakraborty, M.~Giannotti, A.~Mirizzi, A.~Payez and A.~Ringwald,
  ``Probing axions with the neutrino signal from the next galactic supernova,''
  Phys.\ Rev.\ D {\bf 94} (2016) no.8,  085012
  [arXiv:1605.08780 [astro-ph.HE]].


\bibitem{Hannestad:2005ex}
  S.~Hannestad and G.~Raffelt,
  ``Constraining invisible neutrino decays with the cosmic microwave background,''
  Phys.\ Rev.\ D {\bf 72} (2005) 103514
  [hep-ph/0509278].

\bibitem{Kachelriess:2000qc}
  M.~Kachelriess, R.~Tomas and J.~W.~F.~Valle,
  ``Supernova bounds on Majoron emitting decays of light neutrinos,''
  Phys.\ Rev.\ D {\bf 62} (2000) 023004
  [hep-ph/0001039].

\bibitem{Tomas:2001dh}
  R.~Tomas, H.~Pas and J.~W.~F.~Valle,
  ``Generalized bounds on Majoron - neutrino couplings,''
  Phys.\ Rev.\ D {\bf 64} (2001) 095005
  [hep-ph/0103017].

\bibitem{Farzan:2002wx}
  Y.~Farzan,
  ``Bounds on the coupling of the Majoron to light neutrinos from supernova cooling,''
  Phys.\ Rev.\ D {\bf 67} (2003) 073015
  [hep-ph/0211375].





\bibitem{Kaiser:2010ps}
  D.~I.~Kaiser,
  ``Conformal Transformations with Multiple Scalar Fields,''
  Phys.\ Rev.\ D {\bf 81} (2010) 084044
  [arXiv:1003.1159 [gr-qc]].

\bibitem{Kaiser:2010yu}
  D.~I.~Kaiser and A.~T.~Todhunter,
  ``Primordial Perturbations from Multifield Inflation with non-minimal Couplings,''
  Phys.\ Rev.\ D {\bf 81} (2010) 124037
  [arXiv:1004.3805 [astro-ph.CO]].

\bibitem{Giudice:2010ka}
  G.~F.~Giudice and H.~M.~Lee,
  ``Unitarizing Higgs Inflation,''
  Phys.\ Lett.\ B {\bf 694} (2011) 294
  [arXiv:1010.1417 [hep-ph]].



\bibitem{Kaiser:2012ak}
  D.~I.~Kaiser, E.~A.~Mazenc and E.~I.~Sfakianakis,
  ``Primordial Bispectrum from Multifield Inflation with non-minimal Couplings,''
  Phys.\ Rev.\ D {\bf 87} (2013) 064004
  [arXiv:1210.7487 [astro-ph.CO]].

\bibitem{Greenwood:2012aj}
  R.~N.~Greenwood, D.~I.~Kaiser and E.~I.~Sfakianakis,
  ``Multifield Dynamics of Higgs Inflation,''
  Phys.\ Rev.\ D {\bf 87} (2013) 064021
  [arXiv:1210.8190 [hep-ph]].

\bibitem{Kaiser:2013sna}
  D.~I.~Kaiser and E.~I.~Sfakianakis,
  ``Multifield Inflation after Planck: The Case for non-minimal Couplings,''
  Phys.\ Rev.\ Lett.\  {\bf 112} (2014) 1,  011302
  [arXiv:1304.0363 [astro-ph.CO]].

\bibitem{Schutz:2013fua}
  K.~Schutz, E.~I.~Sfakianakis and D.~I.~Kaiser,
  ``Multifield Inflation after Planck: Isocurvature Modes from nonminimal Couplings,''
  Phys.\ Rev.\ D {\bf 89} (2014) 6,  064044
  [arXiv:1310.8285 [astro-ph.CO]].

  
\bibitem{Lebedev:2011aq}
  O.~Lebedev and H.~M.~Lee,
  ``Higgs Portal Inflation,''
  Eur.\ Phys.\ J.\ C {\bf 71} (2011) 1821
  [arXiv:1105.2284 [hep-ph]].
  
\bibitem{DeSimone:2008ei}
  A.~De Simone, M.~P.~Hertzberg and F.~Wilczek,
  Phys.\ Lett.\ B {\bf 678} (2009) 1
  doi:10.1016/j.physletb.2009.05.054
  [arXiv:0812.4946 [hep-ph]].


\bibitem{Burgess:2012dz}
  C.~P.~Burgess, M.~W.~Horbatsch and S.~P.~Patil,
  ``Inflating in a Trough: Single-Field Effective Theory from Multiple-Field Curved Valleys,''
  JHEP {\bf 1301} (2013) 133
  [arXiv:1209.5701 [hep-th]].

\bibitem{Ade:2015xua}
  P.~A.~R.~Ade {\it et al.} [Planck Collaboration],
  ``Planck 2015 results. XIII. Cosmological parameters,''
  Astron.\ Astrophys.\  {\bf 594} (2016) A13
  [arXiv:1502.01589 [astro-ph.CO]].


\bibitem{Ade:2013zuv}
  P.~A.~R.~Ade {\it et al.} [Planck Collaboration],
  ``Planck 2013 results. XVI. Cosmological parameters,''
  Astron.\ Astrophys.\  {\bf 571} (2014) A16
  [arXiv:1303.5076 [astro-ph.CO]].


\bibitem{Planck:2013jfk}
  P.~A.~R.~Ade {\it et al.} [Planck Collaboration],
  ``Planck 2013 results. XXII. Constraints on inflation,''
  Astron.\ Astrophys.\  {\bf 571} (2014) A22
  [arXiv:1303.5082 [astro-ph.CO]].

\bibitem{Ade:2015lrj}
  P.~A.~R.~Ade {\it et al.} [Planck Collaboration],
  ``Planck 2015 results. XX. Constraints on inflation,''
  Astron.\ Astrophys.\  {\bf 594} (2016) A20
  [arXiv:1502.02114 [astro-ph.CO]].


\bibitem{Array:2015xqh}
  P.~A.~R.~Ade {\it et al.} [BICEP2 and Keck Array Collaborations],
  ``Improved Constraints on Cosmology and Foregrounds from BICEP2 and Keck Array Cosmic Microwave Background Data with Inclusion of 95 GHz Band,''
  Phys.\ Rev.\ Lett.\  {\bf 116} (2016) 031302
  [arXiv:1510.09217 [astro-ph.CO]].


\bibitem{Martin:2013nzq}
  J.~Martin, C.~Ringeval, R.~Trotta and V.~Vennin,
  ``The Best Inflationary Models After Planck,''
  JCAP {\bf 1403} (2014) 039
  [arXiv:1312.3529 [astro-ph.CO]].

 
\bibitem{Ballesteros:2015noa}
  G.~Ballesteros and C.~Tamarit,
  ``Radiative Plateau Inflation,''
  JHEP {\bf 1602} (2016) 153
  [arXiv:1510.05669 [hep-ph]].
 

\bibitem{Kaiser:1994vs}
  D.~I.~Kaiser,
  Phys.\ Rev.\ D {\bf 52} (1995) 4295
  doi:10.1103/PhysRevD.52.4295
  [astro-ph/9408044].
 
 
 
\bibitem{Ballesteros:2014yva}
  G.~Ballesteros and J.~A.~Casas,
  ``Large tensor-to-scalar ratio and running of the scalar spectral index with Instep Inflation,''
  Phys.\ Rev.\ D {\bf 91} (2015) 043502
  [arXiv:1406.3342 [astro-ph.CO]].

  
  
\bibitem{Starobinsky:1980te}
  A.~A.~Starobinsky,
  ``A New Type of Isotropic Cosmological Models Without Singularity,''
  Phys.\ Lett.\  {\bf 91B} (1980) 99.

\bibitem{Gorbunov:2012ns}
  D.~Gorbunov and A.~Tokareva,
  JCAP {\bf 1312} (2013) 021
  doi:10.1088/1475-7516/2013/12/021
  [arXiv:1212.4466 [astro-ph.CO]].
  
\bibitem{Adshead:2010mc}
  P.~Adshead, R.~Easther, J.~Pritchard and A.~Loeb,
  ``Inflation and the Scale Dependent Spectral Index: Prospects and Strategies,''
  JCAP {\bf 1102} (2011) 021
  [arXiv:1007.3748 [astro-ph.CO]].

  
\bibitem{Shimabukuro:2014ava}
  H.~Shimabukuro, K.~Ichiki, S.~Inoue and S.~Yokoyama,
  ``Probing small-scale cosmological fluctuations with the 21 cm forest: Effects of neutrino mass, running spectral index, and warm dark matter,''
  Phys.\ Rev.\ D {\bf 90} (2014) no.8,  083003
  [arXiv:1403.1605 [astro-ph.CO]].
  
  
\bibitem{Barvinsky:2008ia}
  A.~O.~Barvinsky, A.~Y.~Kamenshchik and A.~A.~Starobinsky,
  ``Inflation scenario via the Standard Model Higgs boson and LHC,''
  JCAP {\bf 0811} (2008) 021
  [arXiv:0809.2104 [hep-ph]].

 
\bibitem{Barvinsky:2009ii}
  A.~O.~Barvinsky, A.~Y.~Kamenshchik, C.~Kiefer, A.~A.~Starobinsky and C.~F.~Steinwachs,
  ``Higgs boson, renormalization group, and naturalness in cosmology,''
  Eur.\ Phys.\ J.\ C {\bf 72} (2012) 2219
  [arXiv:0910.1041 [hep-ph]].

  
\bibitem{Liddle:2003as}
  A.~R.~Liddle and S.~M.~Leach,
  ``How long before the end of inflation were observable perturbations produced?,''
  Phys.\ Rev.\ D {\bf 68} (2003) 103503
  [astro-ph/0305263].
  
\bibitem{Bezrukov:2013fca}
  F.~Bezrukov and D.~Gorbunov,
  JHEP {\bf 1307} (2013) 140
  doi:10.1007/JHEP07(2013)140
  [arXiv:1303.4395 [hep-ph]].
 
\bibitem{Matsumura:2013aja}
  T.~Matsumura {\it et al.},
  ``Mission design of LiteBIRD,''
  J.\ Low.\ Temp.\ Phys.\  {\bf 176} (2014) 733
  [arXiv:1311.2847 [astro-ph.IM]].
  
\bibitem{Andre:2013afa}
  P.~Andre {\it et al.} [PRISM Collaboration],
  ``PRISM (Polarized Radiation Imaging and Spectroscopy Mission): A White Paper on the Ultimate Polarimetric Spectro-Imaging of the Microwave and Far-Infrared Sky,''
  arXiv:1306.2259 [astro-ph.CO].
  
  
\bibitem{Mao:2008ug}
  Y.~Mao, M.~Tegmark, M.~McQuinn, M.~Zaldarriaga and O.~Zahn,
  ``How accurately can 21 cm tomography constrain cosmology?,''
  Phys.\ Rev.\ D {\bf 78} (2008) 023529
  [arXiv:0802.1710 [astro-ph]].
  
\bibitem{Barbon:2015fla} 
  J.~L.~F.~Barbon, J.~A.~Casas, J.~Elias-Miro and J.~R.~Espinosa,
  JHEP {\bf 1509}, 027 (2015)
  doi:10.1007/JHEP09(2015)027
  [arXiv:1501.02231 [hep-ph]].


\bibitem{Bezrukov:2010jz}
  F.~Bezrukov, A.~Magnin, M.~Shaposhnikov and S.~Sibiryakov,
  ``Higgs inflation: consistency and generalisations,''
  JHEP {\bf 1101} (2011) 016
  [arXiv:1008.5157 [hep-ph]].

\bibitem{Aad:2015zhl}
  G.~Aad {\it et al.} [ATLAS and CMS Collaborations],
  ``Combined Measurement of the Higgs Boson Mass in $pp$ Collisions at $\sqrt{s}=7$ and 8 TeV with the ATLAS and CMS Experiments,''
  Phys.\ Rev.\ Lett.\  {\bf 114} (2015) 191803
  [arXiv:1503.07589 [hep-ex]].


\bibitem{ATLAS:2014wva}
  [ATLAS and CDF and CMS and D0 Collaborations],
  ``First combination of Tevatron and LHC measurements of the top-quark mass,''
  arXiv:1403.4427 [hep-ex].


\bibitem{Agashe:2014kda}
  K.~A.~Olive {\it et al.} [Particle Data Group Collaboration],
  ``Review of Particle Physics,''
  Chin.\ Phys.\ C {\bf 38} (2014) 090001.

\bibitem{Bezrukov:2014ipa}
  F.~Bezrukov, J.~Rubio and M.~Shaposhnikov,
  ``Living beyond the edge: Higgs inflation and vacuum metastability,''
  Phys.\ Rev.\ D {\bf 92} (2015) no.8,  083512
  [arXiv:1412.3811 [hep-ph]].




\bibitem{Espinosa:2007qp}
  J.~R.~Espinosa, G.~F.~Giudice and A.~Riotto,
  ``Cosmological implications of the Higgs mass measurement,''
  JCAP {\bf 0805} (2008) 002
  [arXiv:0710.2484 [hep-ph]].

\bibitem{Lebedev:2012sy}
  O.~Lebedev and A.~Westphal,
  ``Metastable Electroweak Vacuum: Implications for Inflation,''
  Phys.\ Lett.\ B {\bf 719} (2013) 415
  [arXiv:1210.6987 [hep-ph]].

\bibitem{Kobakhidze:2013tn}
  A.~Kobakhidze and A.~Spencer-Smith,
  ``Electroweak Vacuum (In)Stability in an Inflationary Universe,''
  Phys.\ Lett.\ B {\bf 722} (2013) 130
  [arXiv:1301.2846 [hep-ph]].

\bibitem{Fairbairn:2014zia}
  M.~Fairbairn and R.~Hogan,
  ``Electroweak Vacuum Stability in light of BICEP2,''
  Phys.\ Rev.\ Lett.\  {\bf 112} (2014) 201801
  [arXiv:1403.6786 [hep-ph]].

\bibitem{Enqvist:2014bua}
  K.~Enqvist, T.~Meriniemi and S.~Nurmi,
  ``Higgs Dynamics during Inflation,''
  JCAP {\bf 1407} (2014) 025
  [arXiv:1404.3699 [hep-ph]].

\bibitem{Hook:2014uia}
  A.~Hook, J.~Kearney, B.~Shakya and K.~M.~Zurek,
  ``Probable or Improbable Universe? Correlating Electroweak Vacuum Instability with the Scale of Inflation,''
  JHEP {\bf 1501} (2015) 061
  [arXiv:1404.5953 [hep-ph]].

\bibitem{Shkerin:2015exa}
  A.~Shkerin and S.~Sibiryakov,
  ``On stability of electroweak vacuum during inflation,''
  Phys.\ Lett.\ B {\bf 746} (2015) 257
  [arXiv:1503.02586 [hep-ph]].

\bibitem{Kearney:2015vba}
  J.~Kearney, H.~Yoo and K.~M.~Zurek,
  ``Is a Higgs Vacuum Instability Fatal for High-Scale Inflation?,''
  Phys.\ Rev.\ D {\bf 91} (2015) 12,  123537
  [arXiv:1503.05193 [hep-th]].

\bibitem{Herranen:2014cua}
  M.~Herranen, T.~Markkanen, S.~Nurmi and A.~Rajantie,
  ``Spacetime curvature and the Higgs stability during inflation,''
  Phys.\ Rev.\ Lett.\  {\bf 113} (2014) 21,  211102
  [arXiv:1407.3141 [hep-ph]].

\bibitem{Espinosa:2015qea}
  J.~R.~Espinosa, G.~F.~Giudice, E.~Morgante, A.~Riotto, L.~Senatore, A.~Strumia and N.~Tetradis,
  ``The cosmological Higgstory of the vacuum instability,''
  JHEP {\bf 1509} (2015) 174
  [arXiv:1505.04825 [hep-ph]].
  
\bibitem{East:2016anr}
  W.~E.~East, J.~Kearney, B.~Shakya, H.~Yoo and K.~M.~Zurek,
  ``Spacetime Dynamics of a Higgs Vacuum Instability During Inflation,''
  [arXiv:1607.00381 [hep-ph]].

  

  
\bibitem{Rose:2015lna}
  L.~Delle Rose, C.~Marzo and A.~Urbano,
  ``On the fate of the Standard Model at finite temperature,''
  arXiv:1507.06912 [hep-ph].


\bibitem{Buchmuller:2002rq}
  W.~Buchmuller, P.~Di Bari and M.~Plumacher,
  ``Cosmic microwave background, matter - antimatter asymmetry and neutrino masses,''
  Nucl.\ Phys.\ B {\bf 643} (2002) 367
   [Nucl.\ Phys.\ B {\bf 793} (2008) 362]
  [hep-ph/0205349].

\bibitem{Pilaftsis:1997jf}
  A.~Pilaftsis,
  ``CP violation and baryogenesis due to heavy Majorana neutrinos,''
  Phys.\ Rev.\ D {\bf 56} (1997) 5431
  [hep-ph/9707235].

  
  
\bibitem{Pilaftsis:2003gt}
  A.~Pilaftsis and T.~E.~J.~Underwood,
  ``Resonant leptogenesis,''
  Nucl.\ Phys.\ B {\bf 692} (2004) 303
  [hep-ph/0309342].

\bibitem{Ballesteros:2015iua}
  G.~Ballesteros and C.~Tamarit,
  ``Higgs portal valleys, stability and inflation,''
  JHEP {\bf 1509} (2015) 210
  [arXiv:1505.07476 [hep-ph]].

\bibitem{Casas:2001sr}
  J.~A.~Casas and A.~Ibarra,
  ``Oscillating neutrinos and muon $\to$ e gamma,''
  Nucl.\ Phys.\ B {\bf 618} (2001) 171
  [hep-ph/0103065].

  
\bibitem{Forero:2014bxa}
  D.~V.~Forero, M.~Tortola and J.~W.~F.~Valle,
  ``Neutrino oscillations refitted,''
  Phys.\ Rev.\ D {\bf 90} (2014) 9,  093006
  [arXiv:1405.7540 [hep-ph]].



\bibitem{CMS:2014hta}
CMS.
\newblock {Combination of the CMS top-quark mass measurements from Run 1 of the
  LHC}.
\newblock 2014.


  
\bibitem{Hempfling:1994ar}
  R.~Hempfling and B.~A.~Kniehl,
  ``On the relation between the fermion pole mass and MS Yukawa coupling in the standard model,''
  Phys.\ Rev.\ D {\bf 51} (1995) 1386
  [hep-ph/9408313].


\bibitem{Chetyrkin:1999qi}
  K.~G.~Chetyrkin and M.~Steinhauser,
  ``The Relation between the MS-bar and the on-shell quark mass at order alpha(s)**3,''
  Nucl.\ Phys.\ B {\bf 573} (2000) 617
  [hep-ph/9911434].

  
\bibitem{Melnikov:2000qh}
  K.~Melnikov and T.~v.~Ritbergen,
  ``The Three loop relation between the MS-bar and the pole quark masses,''
  Phys.\ Lett.\ B {\bf 482} (2000) 99
  [hep-ph/9912391].

\bibitem{Greene:1997fu}
  P.~B.~Greene, L.~Kofman, A.~D.~Linde and A.~A.~Starobinsky,
  ``Structure of resonance in preheating after inflation,''
  Phys.\ Rev.\ D {\bf 56} (1997) 6175
  [hep-ph/9705347].


  
\bibitem{Shtanov:1994ce}
  Y.~Shtanov, J.~H.~Traschen and R.~H.~Brandenberger,
  ``Universe reheating after inflation,''
  Phys.\ Rev.\ D {\bf 51} (1995) 5438
  [hep-ph/9407247].



\bibitem{GarciaBellido:2008ab}
  J.~Garcia-Bellido, D.~G.~Figueroa and J.~Rubio,
  ``Preheating in the Standard Model with the Higgs-Inflaton coupled to gravity,''
  Phys.\ Rev.\ D {\bf 79} (2009) 063531
  [arXiv:0812.4624 [hep-ph]].
  

  
  
  
\bibitem{Bezrukov:2008ut}
  F.~Bezrukov, D.~Gorbunov and M.~Shaposhnikov,
  ``On initial conditions for the Hot Big Bang,''
  JCAP {\bf 0906} (2009) 029
  [arXiv:0812.3622 [hep-ph]].


\bibitem{Tkachev:1998dc}
  I.~Tkachev, S.~Khlebnikov, L.~Kofman and A.~D.~Linde,
  ``Cosmic strings from preheating,''
  Phys.\ Lett.\ B {\bf 440} (1998) 262
  [hep-ph/9805209].

\bibitem{Micha:2002ey}
  R.~Micha and I.~I.~Tkachev,
  ``Relativistic turbulence: A Long way from preheating to equilibrium,''
  Phys.\ Rev.\ Lett.\  {\bf 90} (2003) 121301
  [hep-ph/0210202].

  
\bibitem{Micha:2004bv}
  R.~Micha and I.~I.~Tkachev,
  ``Turbulent thermalization,''
  Phys.\ Rev.\ D {\bf 70} (2004) 043538
  doi:10.1103/PhysRevD.70.043538
  [hep-ph/0403101].

\bibitem{Figueroa:2016dsc}
  D.~G.~Figueroa and C.~T.~Byrnes,
  arXiv:1604.03905 [hep-ph].

\bibitem{Davidson:2000er}
  S.~Davidson and S.~Sarkar,
  ``Thermalization after inflation,''
  JHEP {\bf 0011} (2000) 012
  [hep-ph/0009078].



\bibitem{Sikivie:2006ni}
  P.~Sikivie,
  ``Axion Cosmology,''
  Lect.\ Notes Phys.\  {\bf 741} (2008) 19
  [astro-ph/0610440].

\bibitem{Ringwald:2012hr}
  A.~Ringwald,
  ``Exploring the Role of Axions and Other WISPs in the Dark Universe,''
  Phys.\ Dark Univ.\  {\bf 1} (2012) 116
  [arXiv:1210.5081 [hep-ph]].

\bibitem{Kawasaki:2013ae}
  M.~Kawasaki and K.~Nakayama,
  ``Axions: Theory and Cosmological Role,''
  Ann.\ Rev.\ Nucl.\ Part.\ Sci.\  {\bf 63} (2013) 69
  [arXiv:1301.1123 [hep-ph]].
  
  
\bibitem{Kibble:1976sj}
  T.~W.~B.~Kibble,
  ``Topology of Cosmic Domains and Strings,''
  J.\ Phys.\ A {\bf 9} (1976) 1387.

  
  
\bibitem{Petreczky:2016vrs} 
  P.~Petreczky, H.~P.~Schadler and S.~Sharma,
  ``The topological susceptibility in finite temperature QCD and axion cosmology,''
  Phys.\ Lett.\ B {\bf 762}, 498 (2016)
  [arXiv:1606.03145 [hep-lat]].




\bibitem{Wantz:2009mi}
  O.~Wantz and E.~P.~S.~Shellard,
  ``The Topological susceptibility from grand canonical simulations in the interacting instanton liquid model: Chiral phase transition and axion mass,''
  Nucl.\ Phys.\ B {\bf 829} (2010) 110
  [arXiv:0908.0324 [hep-ph]].

\bibitem{Wantz:2009it}
  O.~Wantz and E.~P.~S.~Shellard,
  ``Axion Cosmology Revisited,''
  Phys.\ Rev.\ D {\bf 82} (2010) 123508
  [arXiv:0910.1066 [astro-ph.CO]].



\bibitem{Borsanyi:2015cka}
  S.~Borsanyi {\it et al.},
  ``Axion cosmology, lattice QCD and the dilute instanton gas,''
  Phys.\ Lett.\ B {\bf 752} (2016) 175
  [arXiv:1508.06917 [hep-lat]].

\bibitem{Buchoff:2013nra}
  M.~I.~Buchoff {\it et al.},
  ``QCD chiral transition, U(1)A symmetry and the dirac spectrum using domain wall fermions,''
  Phys.\ Rev.\ D {\bf 89} (2014) 5,  054514
  [arXiv:1309.4149 [hep-lat]].

\bibitem{Bonati:2015vqz}
  C.~Bonati, M.~D'Elia, M.~Mariti, G.~Martinelli, M.~Mesiti, F.~Negro, F.~Sanfilippo and G.~Villadoro,
  ``Axion phenomenology and $\theta$-dependence from $N_f = 2+1$ lattice QCD,''
  JHEP {\bf 1603} (2016) 155
  [arXiv:1512.06746 [hep-lat]].






\bibitem{Turner:1990uz}
  M.~S.~Turner and F.~Wilczek,
  ``Inflationary axion cosmology,''
  Phys.\ Rev.\ Lett.\  {\bf 66} (1991) 5.

\bibitem{Fox:2004kb}
  P.~Fox, A.~Pierce and S.~D.~Thomas,
  ``Probing a QCD string axion with precision cosmological measurements,''
  hep-th/0409059.



\bibitem{Beltran:2006sq}
  M.~Beltran, J.~Garcia-Bellido and J.~Lesgourgues,
  ``Isocurvature bounds on axions revisited,''
  Phys.\ Rev.\ D {\bf 75} (2007) 103507
  [hep-ph/0606107].

\bibitem{Hertzberg:2008wr}
  M.~P. Hertzberg, M.~Tegmark and F.~Wilczek,
  ``Axion Cosmology and the Energy Scale of Inflation,''
  Phys.\ Rev.\ D {\bf 78} (2008) 083507
  [arXiv:0807.1726 [astro-ph]].

\bibitem{Hamann:2009yf}
  J.~Hamann, S.~Hannestad, G.~G.~Raffelt and Y.~Y.~Y.~Wong,
  ``Isocurvature forecast in the anthropic axion window,''
  JCAP {\bf 0906} (2009) 022
  [arXiv:0904.0647 [hep-ph]].


\bibitem{Masso:2002np}
  E.~Masso, F.~Rota and G.~Zsembinszki,
  ``On axion thermalization in the early universe,''
  Phys.\ Rev.\ D {\bf 66} (2002) 023004
  [hep-ph/0203221].


\bibitem{Graf:2010tv}
  P.~Graf and F.~D.~Steffen,
  ``Thermal axion production in the primordial quark-gluon plasma,''
  Phys.\ Rev.\ D {\bf 83} (2011) 075011
  [arXiv:1008.4528 [hep-ph]].


\bibitem{Salvio:2013iaa}
  A.~Salvio, A.~Strumia and W.~Xue,
  ``Thermal axion production,''
  JCAP {\bf 1401} (2014) 011
  [arXiv:1310.6982 [hep-ph]].

\bibitem{Abazajian:2013oma}
  K.~N.~Abazajian {\it et al.} [Topical Conveners: K.N. Abazajian, J.E. Carlstrom, A.T. Lee Collaboration],
  ``Neutrino Physics from the Cosmic Microwave Background and Large Scale Structure,''
  Astropart.\ Phys.\  {\bf 63} (2015) 66
  [arXiv:1309.5383 [astro-ph.CO]].

\bibitem{Errard:2015cxa}
  J.~Errard, S.~M.~Feeney, H.~V.~Peiris and A.~H.~Jaffe,
  ``Robust forecasts on fundamental physics from the foreground-obscured, gravitationally-lensed CMB polarization,''
  JCAP {\bf 1603} (2016) no.03,  052
  [arXiv:1509.06770 [astro-ph.CO]].

\bibitem{Davidson:2002qv}
  S.~Davidson and A.~Ibarra,
  ``A Lower bound on the right-handed neutrino mass from leptogenesis,''
  Phys.\ Lett.\ B {\bf 535} (2002) 25
  [hep-ph/0202239].

\bibitem{Hamaguchi:2001gw}
  K.~Hamaguchi, H.~Murayama and T.~Yanagida,
  ``Leptogenesis from N dominated early universe,''
  Phys.\ Rev.\ D {\bf 65} (2002) 043512
  [hep-ph/0109030].




\bibitem{Giudice:2003jh}
  G.~F.~Giudice, A.~Notari, M.~Raidal, A.~Riotto and A.~Strumia,
  ``Towards a complete theory of thermal leptogenesis in the SM and MSSM,''
  Nucl.\ Phys.\ B {\bf 685} (2004) 89
  [hep-ph/0310123].

\bibitem{Buchmuller:2004nz}
  W.~Buchmuller, P.~Di Bari and M.~Plumacher,
  ``Leptogenesis for pedestrians,''
  Annals Phys.\  {\bf 315} (2005) 305
  [hep-ph/0401240].



\bibitem{Sahu:2005vu}
  N.~Sahu, P.~Bhattacharjee and U.~A.~Yajnik,
  ``Baryogenesis via leptogenesis in presence of cosmic strings,''
  Nucl.\ Phys.\ B {\bf 752} (2006) 280
  [hep-ph/0512350].


\bibitem{Jeannerot:2005ah}
  R.~Jeannerot and M.~Postma,
  ``Leptogenesis from reheating after inflation and cosmic string decay,''
  JCAP {\bf 0512} (2005) 006
  [hep-ph/0507162].


\bibitem{Sahu:2004ir}
  N.~Sahu, P.~Bhattacharjee and U.~A.~Yajnik,
  ``B - L cosmic strings and baryogenesis,''
  Phys.\ Rev.\ D {\bf 70} (2004) 083534
  [hep-ph/0406054].


\bibitem{CULTASK}
W. Chung, in Proceedings of the 12th AXION-WIMP
workshop (2016).


\bibitem{Majorovits:2016yvk} 
  B.~Majorovits {\it et al.} [MADMAX Working Group Collaboration],
  ``MADMAX: A new Dark Matter Axion Search using a Dielectric Haloscope,''
  arXiv:1611.04549 [astro-ph.IM].

  
\bibitem{TheMADMAXWorkingGroup:2016hpc} 
  The MADMAX Working Group {\it et al.},
  ``Dielectric Haloscopes: A New Way to Detect Axion Dark Matter,''
  arXiv:1611.05865 [physics.ins-det].

  




\bibitem{Rybka:2014cya}
  G.~Rybka, A.~Wagner, A.~Brill, K.~Ramos, R.~Percival and K.~Patel,
  ``Search for dark matter axions with the Orpheus experiment,''
  Phys.\ Rev.\ D {\bf 91} (2015) 1,  011701
  [arXiv:1403.3121 [physics.ins-det]].

\bibitem{Horns:2012jf}
  D.~Horns, J.~Jaeckel, A.~Lindner, A.~Lobanov, J.~Redondo and A.~Ringwald,
  ``Searching for WISPy Cold Dark Matter with a Dish Antenna,''
  JCAP {\bf 1304} (2013) 016
  [arXiv:1212.2970 [hep-ph]].




\bibitem{vanBibber:2013ssa}
  K.~van Bibber and G.~Carosi,
  ``Status of the ADMX and ADMX-HF experiments,''
  arXiv:1304.7803 [physics.ins-det].


\bibitem{ADMX}
G. Rybka, in Proceedings of the 12th AXION-WIMP
workshop (2016).  DESY-PROC-2016-03, 
ISBN 978-3-945931-06-6.


\bibitem{X3}
S. Lewis, in Proceedings of the 12th AXION-WIMP workshop
(2016). DESY-PROC-2016-03, 
ISBN 978-3-945931-06-6.



\bibitem{Armengaud:2014gea} 
  E.~Armengaud {\it et al.},
  ``Conceptual Design of the International Axion Observatory (IAXO),''
  JINST {\bf 9}, T05002 (2014)
  [arXiv:1401.3233 [physics.ins-det]].


\bibitem{Hogan:1988mp}
  C.~J.~Hogan and M.~J.~Rees,
  ``Axion Miniclusters,''
  Phys.\ Lett.\ B {\bf 205} (1988) 228.

\bibitem{Kolb:1993zz}
  E.~W.~Kolb and I.~I.~Tkachev,
  ``Axion miniclusters and Bose stars,''
  Phys.\ Rev.\ Lett.\  {\bf 71} (1993) 3051
  [hep-ph/9303313].

\bibitem{Kolb:1993hw}
  E.~W.~Kolb and I.~I.~Tkachev,
  ``Nonlinear axion dynamics and formation of cosmological pseudosolitons,''
  Phys.\ Rev.\ D {\bf 49} (1994) 5040
  [astro-ph/9311037].

\bibitem{Zurek:2006sy}
  K.~M.~Zurek, C.~J.~Hogan and T.~R.~Quinn,
  ``Astrophysical Effects of Scalar Dark Matter Miniclusters,''
  Phys.\ Rev.\ D {\bf 75} (2007) 043511
  [astro-ph/0607341].

\bibitem{Kolb:1995bu}
  E.~W.~Kolb and I.~I.~Tkachev,
  ``Femtolensing and picolensing by axion miniclusters,''
  Astrophys.\ J.\  {\bf 460} (1996) L25
  [astro-ph/9510043].

  

\bibitem{Iwazaki:2014wka} 
  A.~Iwazaki,
  ``Axion stars and fast radio bursts,''
  Phys.\ Rev.\ D {\bf 91}, no. 2, 023008 (2015)
  [arXiv:1410.4323 [hep-ph]].

\bibitem{Tkachev:2014dpa}
  I.~I.~Tkachev,
  ``Fast Radio Bursts and Axion Miniclusters,''
  JETP Lett.\  {\bf 101} (2015) no.1,  1
   [Pisma Zh.\ Eksp.\ Teor.\ Fiz.\  {\bf 101} (2015) no.1,  3]
  [arXiv:1411.3900 [astro-ph.HE]].
  

\bibitem{Iwazaki:2014wta}
  A.~Iwazaki,
  ``Fast Radio Bursts from Axion Stars,''
  arXiv:1412.7825 [hep-ph].

\bibitem{Pshirkov:2016bjr}
  M.~S.~Pshirkov,
  ``May axion clusters be sources of fast radio bursts?,''
  arXiv:1609.09658 [astro-ph.HE].

\bibitem{Tinyakov:2015cgg}
  P.~Tinyakov, I.~Tkachev and K.~Zioutas,
  ``Tidal streams from axion miniclusters and direct axion searches,''
  JCAP {\bf 1601} (2016) no.01,  035
  [arXiv:1512.02884 [astro-ph.CO]].








\bibitem{Dine:1981rt}
  M.~Dine, W.~Fischler and M.~Srednicki,
  ``A Simple Solution to the Strong CP Problem with a Harmless Axion,''
  Phys.\ Lett.\ B {\bf 104} (1981) 199.
  


\bibitem{Zhitnitsky:1980tq}
  A.~R.~Zhitnitsky,
  ``On Possible Suppression of the Axion Hadron Interactions," (In Russian),
  Sov.\ J.\ Nucl.\ Phys.\  { 31} (1980) 260
  [Yad.\ Fiz.\  {31} (1980) 497].

\bibitem{Giannotti:2015kwo}
  M.~Giannotti, I.~Irastorza, J.~Redondo and A.~Ringwald,
  ``Cool WISPs for stellar cooling excesses,''
  JCAP {\bf 1605} (2016) no.05,  057
  [arXiv:1512.08108 [astro-ph.HE]].

 

\bibitem{Barbieri:2016vwg}
  R.~Barbieri {\it et al.},
  ``Searching for galactic axions through magnetized media: the QUAX proposal,''
  arXiv:1606.02201 [hep-ph].

\bibitem{Arvanitaki:2014dfa} 
  A.~Arvanitaki and A.~A.~Geraci,
  ``Resonantly Detecting Axion-Mediated Forces with Nuclear Magnetic Resonance,''
  Phys.\ Rev.\ Lett.\  {\bf 113}, no. 16, 161801 (2014)
  [arXiv:1403.1290 [hep-ph]].

\bibitem{Seidel:2013sqa}
  K.~Seidel, F.~Simon, M.~Tesar and S.~Poss,
  ``Top quark mass measurements at and above threshold at CLIC,''
  Eur.\ Phys.\ J.\ C {\bf 73} (2013) no.8,  2530
  [arXiv:1303.3758 [hep-ex]].
  
\bibitem{Fan:2014vta} 
  J.~Fan, M.~Reece and L.~T.~Wang,
  ``Possible Futures of Electroweak Precision: ILC, FCC-ee, and CEPC,''
  JHEP {\bf 1509}, 196 (2015)
  [arXiv:1411.1054 [hep-ph]].

\bibitem{Deppisch:2015qwa} 
  F.~F.~Deppisch, P.~S.~Bhupal Dev and A.~Pilaftsis,
  ``Neutrinos and Collider Physics,''
  New J.\ Phys.\  {\bf 17}, no. 7, 075019 (2015)
  [arXiv:1502.06541 [hep-ph]].

\bibitem{Okada:2012gy} 
  Y.~Okada and L.~Panizzi,
  ``LHC signatures of vector-like quarks,''
  Adv.\ High Energy Phys.\  {\bf 2013}, 364936 (2013)
  [arXiv:1207.5607 [hep-ph]].

\bibitem{Branco:2011iw} 
  G.~C.~Branco, P.~M.~Ferreira, L.~Lavoura, M.~N.~Rebelo, M.~Sher and J.~P.~Silva,
  ``Theory and phenomenology of two-Higgs-doublet models,''
  Phys.\ Rept.\  {\bf 516}, 1 (2012)
  [arXiv:1106.0034 [hep-ph]].

\bibitem{Alekhin:2015byh}
  S.~Alekhin {\it et al.},
  ``A facility to Search for Hidden Particles at the CERN SPS: the SHiP physics case,''
  Rept.\ Prog.\ Phys.\  {\bf 79} (2016) no.12,  124201
  [arXiv:1504.04855 [hep-ph]].



  

\bibitem{Martin:2014bca}
  S.~P.~Martin,
  ``Taming the Goldstone contributions to the effective potential,''
  Phys.\ Rev.\ D {\bf 90} (2014) 1,  016013
  [arXiv:1406.2355 [hep-ph]].



\bibitem{Elias-Miro:2014pca}
  J.~Elias-Miro, J.~R.~Espinosa and T.~Konstandin,
  ``Taming Infrared Divergences in the Effective Potential,''
  JHEP {\bf 1408} (2014) 034
  [arXiv:1406.2652 [hep-ph]].




  
\bibitem{Quiros:1999jp}
  M.~Quiros,
  ``Finite temperature field theory and phase transitions,''
  hep-ph/9901312.



\bibitem{Kapusta:2006pm}
  J.~I.~Kapusta and C.~Gale,
  ``Finite-temperature field theory: Principles and applications,''


\bibitem{Martin:2003it}
  S.~P.~Martin,
  ``Two loop scalar self energies in a general renormalizable theory
at leading order in gauge couplings,''
  Phys.\ Rev.\ D {\bf 70} (2004) 016005
  [hep-ph/0312092].







  
\bibitem{Machacek:1983tz}
  M.~E.~Machacek and M.~T.~Vaughn,
  ``Two Loop Renormalization Group Equations in a General Quantum Field Theory. 1. Wave Function Renormalization,''
  Nucl.\ Phys.\ B {\bf 222} (1983) 83.

  
\bibitem{Machacek:1983fi}
  M.~E.~Machacek and M.~T.~Vaughn,
  ``Two Loop Renormalization Group Equations in a General Quantum Field Theory. 2. Yukawa Couplings,''
  Nucl.\ Phys.\ B {\bf 236} (1984) 221.

  
\bibitem{Machacek:1984zw}
  M.~E.~Machacek and M.~T.~Vaughn,
  ``Two Loop Renormalization Group Equations in a General Quantum Field Theory. 3. Scalar Quartic Couplings,''
  Nucl.\ Phys.\ B {\bf 249} (1985) 70.

  
\bibitem{Luo:2002ti}
  M.~x.~Luo, H.~w.~Wang and Y.~Xiao,
  ``Two loop renormalization group equations in general gauge field theories,''
  Phys.\ Rev.\ D {\bf 67} (2003) 065019
  [hep-ph/0211440].

  
\bibitem{Luo:2002ey}
  M.~x.~Luo and Y.~Xiao,
  ``Two loop renormalization group equations in the standard model,''
  Phys.\ Rev.\ Lett.\  {\bf 90} (2003) 011601
  [hep-ph/0207271].

\bibitem{Covi:1996wh}
  L.~Covi, E.~Roulet and F.~Vissani,
  ``CP violating decays in leptogenesis scenarios,''
  Phys.\ Lett.\ B {\bf 384} (1996) 169
  [hep-ph/9605319].
 


\bibitem{Vissani:1997ys} 
  F.~Vissani,
  ``Do experiments suggest a hierarchy problem?,''
  Phys.\ Rev.\ D {\bf 57}, 7027 (1998)
  [hep-ph/9709409].

\bibitem{Farina:2013mla}
  M.~Farina, D.~Pappadopulo and A.~Strumia,
  ``A modified naturalness principle and its experimental tests,''
 JHEP {\bf 1308} (2013) 022
  [arXiv:1303.7244 [hep-ph]].

  



  


\end{thebibliography}
\end{document}